\documentclass{article}

\usepackage[english]{babel}
\usepackage[T1]{fontenc}

\usepackage[a4paper,top=2.5cm,bottom=2.5cm,left=2.5cm,right=2.5cm,marginparwidth=1.75cm]{geometry}

\usepackage{calc,pstcol,multido,pst-node,amsgen,amsmath,amsfonts,amssymb,bm}

\usepackage{graphicx}
\usepackage[colorlinks=true, allcolors=blue]{hyperref}

\usepackage{setspace}
\usepackage{lscape}
\usepackage{authblk}
\usepackage{threeparttable}
\usepackage{booktabs}
\usepackage{array}

\newcommand{\ci}{ \mbox{$\perp\!\!\!\perp$} }
\newcommand{\notci}{ \mbox{$\not\perp\!\!\!\perp$} }
\newcommand{\Var}{ {\rm Var} }
\newcommand{\Cov}{ {\rm Cov} }

\newcommand{\alphahat}{ \hat{\alpha} }
\newcommand{\betahat}{ \hat{\beta} }
\newcommand{\tauhat}{ \hat{\tau} }
\newcommand{\Deltahat}{ \hat{\Delta} }

\newcommand{\mhat}{ \hat{m} }
\newcommand{\muhat}{ \hat{\mu} }
\newcommand{\piBtrue}{ \pi^B }
\newcommand{\mtrue}{ m }
\newcommand{\mbar}{ \bar{m}_0 }
\newcommand{\mbarhat}{ \hat{\bar{m}}_0 }
\newcommand{\mbartrue}{ \bar{m} }
\newcommand{\alphalim}{ \alpha_0 }
\newcommand{\betalim}{ \beta_0 }
\newcommand{\curlF}{ \mathcal{F}_X }
\newcommand{\curlY}{ \mathcal{F}_Y }

\newcommand{\DRone}{ \hat{\mu}_{\rm DR1} }
\newcommand{\DRtwo}{ \hat{\mu}_{\rm DR2} }
\newcommand{\DR}{ \hat{\mu}_{\rm DR} }

\newcommand{\KHone}{ \hat{\mu}_{\rm KH1} }
\newcommand{\KHtwo}{ \hat{\mu}_{\rm KH2} }
\newcommand{\KH}{ \hat{\mu}_{\rm KH} }
\newcommand{\KHtau}{ \hat{\mu}_{{\rm KH}\tau} }
\newcommand{\PLone}{ \hat{\mu}_{\rm PL1} }
\newcommand{\PLtwo}{ \hat{\mu}_{\rm PL2} }
\newcommand{\HT}{ \hat{\mu}_{\rm HT} }
\newcommand{\Haj}{ \hat{\mu}_{\rm Haj} }
\newcommand{\HTHaj}{ \hat{\mu}_{\rm H} }
\newcommand{\HTPL}{ \hat{\mu}_{\rm HT,PL1} }
\newcommand{\HTKH}{ \hat{\mu}_{\rm HT,KH1} }
\newcommand{\HajPL}{ \hat{\mu}_{\rm Haj,PL2} }
\newcommand{\HajKH}{ \hat{\mu}_{\rm Haj,KH2} }
\newcommand{\Ybar}{ \bar{Y} }
\newcommand{\IPWone}{ \hat{\mu}_{\rm IPW1} }
\newcommand{\IPWtwo}{ \hat{\mu}_{\rm IPW2} }

\newcommand{\IPWparam}{ \mu_{\rm IPW2} }

\newcommand{\bthree}{ b_2 }
\newcommand{\bfour}{ b_1 }
\newcommand{\bthreehat}{ \hat{b}_2 }
\newcommand{\bfourhat}{ \hat{b}_1 }
\newcommand{\bthreepart}{ c_2 }
\newcommand{\Hajparam}{ \mu_{\rm Haj} }
\newcommand{\Xp}{ X^* }
\newcommand{\Xpi}{ X^*_i }
\newcommand{\Xpitop}{ X^{*\top}_i }
\newcommand{\usepi}{ T }

\newcommand{\thetahat}{ \hat{\theta} }
\newcommand{\piBhat}{ \hat{\pi}^B }

\newcommand{\GYmuhat}{ \hat{\mu}_{\rm GY} }
\newcommand{\GYestfix}{ \hat{\mu}_{ c_{\gamma}, \lambda} }
\newcommand{\GYestfixopt}{ \hat{\mu}_{ c_{\gamma}^{\dagger}, \lambda^{\dagger}} }
\newcommand{\GYteststat}{ T_{GY} }
\newcommand{\muhatA}{ \hat{\mu}_A }
\newcommand{\muhatB}{ \hat{\mu}_B }
\newcommand{\Gammau}{ \Gamma_u }
\newcommand{\GYteststatstar}{ T^{*(b)}_{\rm GY} }
\newcommand{\GYteststatstarstar}{ T^{**(b_1,b_2)}_{\rm GY} }

\title{Doubly robust integration of nonprobability and probability survey data}

\author[1*]{Shaun~R~Seaman}
\author[1,2*]{Tommy~Nyberg}
\author[3,4,5]{Soazig~Clifton}
\author[3]{Elise~Paul}
\author[3]{Luke~Muschialli}
\author[3]{Claudia~Gascoyne}
\author[3]{Anika~Knuppel}
\author[3,5]{Katharine~Sadler}
\author[3]{Nigel~Field}
\author[3]{Catherine~H~Mercer}
\author[3]{Pam~Sonnenberg}
\author[3]{Andrew~Copas}
\author[1]{Anne~M~Presanis}

\affil[1]{MRC Biostatistics Unit, University of Cambridge, Cambridge, UK.}
\affil[2]{Pathogen Dynamics Unit, Department of Genetics, University of Cambridge, Cambridge, UK.}
\affil[3]{Institute for Global Health, University College London, London, UK.}
\affil[4]{School of Health and Wellbeing, University of Glasgow, Glasgow, UK.}
\affil[5]{Survey Research Centre, the National Centre for Social Research, London, UK.}
\affil[*]{Equal contribution.}
\date{\today}

\begin{document}
\maketitle

\begin{abstract}
  \noindent
Doubly robust (DR) estimators of the population mean of an outcome have been proposed for integrating outcome and covariate data from a nonprobability sample with covariate data from a probability survey.
These estimators combine inverse probability weighting with mass imputation.
We review these DR estimators and extend them to allow estimation of a domain (subpopulation) mean, possibly using data from individuals outside the domain to improve estimation when the domain is small.
For situations where an outcome is observed in both samples, we describe estimators that efficiently combine one of these DR estimators with a Horvitz-Thompson or Hajek estimator that uses only the probability survey data.
We investigate the asymptotic relative efficiency of these combined estimators compared to their two component estimators, and carry out a simulation study to assess relative efficiency in finite samples.
The relative efficiency depends on the ratio of the variances of the two component estimators and on how predictive the covariates are of the outcome.
We illustrate the use of all the estimators by analysing data from the Natsal-3 probability survey and four contemporaneous online nonprobability surveys.
Our extensions will enable greater uptake and practical use of these combination methods in the survey statistics field.
\end{abstract}

\section{Introduction}
Probability surveys remain the gold standard for the estimation of a population mean outcome or prevalence (note that prevalence is just the mean of a binary outcome).
Nonprobability surveys are limited by a lack of a defined sampling frame and may be associated with considerable selection bias. 
However, with the increasing challenges of conducting probability surveys and the increasing availability and ease of collecting nonprobability survey data with large sample sizes~\cite{cornesse2020review}, there has been increased interest in the feasibility of using nonprobability survey data in recent years.
Inverse probability weighting (IPW) and mass imputation (MI) are prominent methods for integrating outcome data from a nonprobability survey with covariate data from a probability survey to produce an estimate of the population mean outcome.
Both methods rely on the assumption that the probability survey provides covariate data that are representative of the covariate distribution in the population. 
IPW estimators involve specifying a model (called the selection model), for the conditional probability of being included in the nonprobability sample given covariates.
This model is fitted using covariate data from both surveys, and then the fitted probabilities are used as inverse weights for the nonprobability survey participants~\cite{chen2020doublyrobust}.
MI estimators involve fitting a model (called the outcome model) for the conditional expectation of the outcome given the covariates to the nonprobability sample data, and then using this model to impute the outcomes in the probability sample~\cite{beaumont2020nonprobsurveys,wu2022nonprobsurveys}.
Doubly robust (DR) estimators combine an IPW and an MI estimator, and so use both a selection model and an outcome model.
DR estimators are consistent when at least one, but not necessarily both, of these models is correctly specified~\cite{seaman2018doublyrobustgeneralreference}.

Several DR estimators have been proposed for estimating a population mean (or prevalence) when data on a set of covariates are available for both a nonprobability sample and a representative probability sample but data on the outcome of interest are available only for the nonprobability sample~\cite{kim2014doublyrobust,kim2019surveyanalysis,chen2020doublyrobust,yang2020doublyrobust}. 
However, methods that can be used when the outcome is also measured on the probability sample have received limited attention.
In this situation the population mean outcome can be estimated in two ways: using a Horvitz-Thompson or Hajek estimator based on just the probability sample outcome data; and using a DR estimator that integrates the covariate and outcome data from the nonprobability sample with the covariate data from the probability sample.
Because both these estimators use the same probability sample covariate data, they are not, in general, independent.
In this article, we consider the class of combined estimators that are weighted averages of the two estimators.
The efficient choice of weights for such combined estimators depends on the variances of the two component estimators and on the covariance between them.
Recently, Gao and Yang (2023)~\cite{gao2023pretest} proposed a method to combine the DR estimator that Chen et al.\ (2020)~\cite{chen2020doublyrobust} denote as $\DRone$ with a Hajek estimator that uses only the probability sample data when the samples are small compared to the population size.
Their combined estimator is consistent when at least one of the selection and outcome models is correctly specified, but they suggest their formula for the variance of this estimator relies on both models being correctly specified.

In this article we begin, in Sections~\ref{sect:framework} and~\ref{sect:ChenDR}, with a description of the problem and a review of Chen et al.'s (2020) DR estimators for estimating a population mean, clarifying the distinction between design-based and model-design-based inference and offering additional variance estimators.
We generalise previously proposed DR estimators to enable estimation of a domain (i.e.\ subpopulation) mean, an important task in applied survey statistics, and we further propose a Hajek-style DR estimator that has a double robust variance estimator and does not require the size of the domain to be known.
Then, in Section~\ref{sect:combining}, we propose estimators that combine a Horvitz-Thompson or Hajek estimator that uses only the probability sample data with a DR estimator that uses the outcome data from only the nonprobability sample.
The results in this section also apply to the combination of a Horvitz-Thompson or Hajek estimator with an IPW estimator, since IPW estimators are special cases of DR estimators with misspecified outcome models.
In Section~\ref{sect:efficiency.gain} we derive formulae for the asymptotic relative efficiency of the combined estimator compared to its two component estimators.
Section~\ref{sect:simstudy} describes a simulation study that investigates the relative efficiency of the combined estimator in finite samples and the coverage of confidence intervals based on our variance estimators.
Section~\ref{sect:natsal} illustrates the use of the methods to combine data from the third National Survey of Sexual Attitudes and Lifestyles (Natsal-3), which is a probability sample, with four contemporaneous nonprobability samples in order to estimate an outcome prevalence in the domain of men aged 18-44-years.
Finally, Section~\ref{sect:discussion} contains a discussion, which includes: allowing the probability of inclusion in the nonprobability sample to depend on the outcome; testing for consistency between DR and Horvitz-Thompson/Hajek estimators before combining them; handling unit response and/or post-stratification in the probability sample; and using data-adaptive estimators to fit the selection and outcome models.
We conclude with implications of our findings for the survey statistics field.

\section{Sampling framework} \label{sect:framework}

Suppose that a finite population of size $N$ is sampled from a superpopulation model~\cite{chen2020doublyrobust}.
Let $X_i$ denote variables (called `covariates') for individual $i$ ($i=1, \ldots, N$), and let $\curlF = (X_1, \ldots, X_N)$.
Let $Y_i$ denote an outcome of interest for individual $i$.
Assume that $Y_i$ is independent of $\curlF$ given $X_i$, with $Y_i = m(X_i) + \epsilon_i$ ($i=1, \ldots, n$), where $E(\epsilon_i \mid X_i) = 0$ and $\Var (\epsilon_i \mid X_i) = v(X_i)$ for some functions $m(x)$ and $v(x)$ of $x$.
So, $m(x) = E(Y_i \mid X_i = x)$.
Let $\curlY = (Y_1, \ldots, Y_N)$.

A probability sample, called Sample A, is drawn from the population, possibly using a complex multistage sampling scheme.
Let $R^A_i = 1$ if individual $i$ is in this sample, and $R^A_i = 0$ if not.
Let $\pi_i^A = P(R^A_i = 1 \mid \curlF, \curlY)$ denote the first-order sampling probability for individual $i$.
We observe the $X$ and $\pi_i^A$ values of individuals in Sample A.

A nonprobability sample, called Sample B, is also drawn from the population.
Let $R^B_i = 1$ if individual $i$ is in this sample, and $R^B_i = 0$ if not.
Assume that conditional on $\curlF$, $R^B_1, \ldots, R^B_N$ are independent of each other, of $(Y_1, \ldots, Y_N)$ and of $(R^A_1, \ldots, R^A_N)$.
Also assume that $P(R^B_i = 1 \mid \curlF) = P(R^B_i = 1 \mid X_i) = \pi^B (X_i)$ for some function $\piBtrue (.)$ that satisfies $\piBtrue (x) > 0$ for all $x$.
Note this implies $E(Y_i \mid X_i=x, R^B_i=1) = m(x)$.
We observe the $X$ and $Y$ values of individuals in Sample B.

Chen et al.\ (2020)~\cite{chen2020doublyrobust}, Yang et al.\ (2020)~\cite{yang2020doublyrobust} and Gao and Yang (2023)~\cite{gao2023pretest} considered the problem of estimating the mean of $Y$ in the entire population.
We shall consider the more general problem of estimating the mean of $Y$ in a subpopulation, called a `domain'.
Let $S_i = 1$ if individual $i$ belongs to the domain, and $S_i = 0$ otherwise ($i=1, \ldots, N$).
Note that to estimate the mean of $Y$ in the entire population, just define $S_i = 1$ for all $i$.
We assume that $S_i = S(X_i)$ is a function of $X_i$ and that $S_i$ is observed for all individuals in Samples A and B.
Let $N_s = \sum_{i=1}^N S_i$ denote the number of individuals in the domain.
Our aim is to estimate the mean of $Y$ in the domain, i.e.\
$
\Ybar = \frac{1}{N_s} \sum_{i=1}^N S_i Y_i.
$

Like Chen et al.\ (2020), we shall consider both design-based inference and model-design-based inference~\cite{molina2001complexsurveys}.
Design-based inference concerns the repeated sampling properties of estimators of $\Ybar$ when $\curlF$ and $\curlY$ are held fixed and $R^A_1, \ldots, R^A_N$, $R^B_1, \ldots, R^B_N$ are repeatedly sampled using the generating mechanism described above.
In model-design-based inference, only $\curlF$ is held fixed and all of $R^A_1, \ldots, R^A_N$, $R^B_1, \ldots, R^B_N$ and $Y_1, \ldots, Y_N$ are repeatedly sampled.
When considering model-design-based inference, we (like Chen et al.) shall make two additional assumptions: 1) $Y_1, \ldots, Y_N$ are conditionally independent given $\curlF$; and 2) $P(R^A_1 = r_1, \ldots, R^A_N = r_N \mid \curlF, \curlY) =  P(R^A_1 = r_1, \ldots, R^A_N = r_N \mid \curlF)$ for all possible values $(r_1, \ldots, r_N)$ of $(R^A_1, \ldots, R^A_N)$, i.e.\ the sampling indicators for Sample A are independent of the outcomes given the covariates.
Note that this second assumption implies $\pi_i^A = P(R^A_i = 1 \mid \curlF)$ does not change over repeated samples. 

Let $\pi^B(X; \alpha) = \mbox{expit} (\alpha^\top \Xp)$ be a logistic regression model for $\pi^B(X)$ with covariates $\Xp$ and parameters $\alpha$.
Here, $\Xp$ is a function of $X$.
Let $m(X; \beta)$ be a (semi)parametric model for $m(X)$ with parameters $\beta$.
The two models may be used separately to define an IPW or an MI estimator, respectively.
We shall refer to these two models as the 'nuisance models'.
Assume that we have estimators $\alphahat$ and $\betahat$ of $\alpha$ and $\beta$ such that $\alphahat = \alphalim + O_p (N^{-1/2})$ and $\betahat = \betalim + O_p (N^{-1/2})$ for some $\alphalim$ and $\betalim$, i.e.\ $\alphahat$ and $\betahat$ converge at the $N^{1/2}$ rate to $\alphalim$ and $\betalim$.
Let $\mbar = N_s^{-1} \sum_{i=1}^N S_i m(X_i; \betalim)$ and $\mbartrue = N_s^{-1} \sum_{i=1}^N S_i m(X_i)$.
If model $\pi^B (X; \alphalim)$ is correctly specified, $\pi^B (X; \alphalim) = \piBtrue (X)$.
Likewise, if model $m(X; \betalim)$ is correctly specified, $m(X; \betalim) = \mtrue(X)$ and $\mbar = \mbartrue$.
Note that we said earlier that the covariates $X$ are observed in both samples.
In fact, if a covariate is not used either in model $\pi^B(X; \alpha)$ nor in model $m(X; \beta)$, then we do not need this covariate to be observed.

We assume the regularity conditions C1--C6 of Chen et al.\ (2020)~\cite{chen2020doublyrobust}.
In particular, C1 states that we are considering an asymptotic framework where the sizes of the finite population and of Samples A and B increase at the same rate as $N \rightarrow \infty$.
We also assume that the conditional variances given $\curlF$ and $\curlY$ of the Horvitz-Thompson estimators~\cite{sarndal1992surveysamplingbook} of the domain means of $Y - \Gamma$ and $m(X; \beta) - \Delta_{mi}$, where $\Gamma$ equals $\Ybar$ or zero and $\Delta_{mi}$ is defined in Section~\ref{sect:designbased}, are $O_p(N^{-1})$.
Chauvet and Vallee (2020)~\cite{chauvet2020samplingdesigns} provide sufficient conditions for this to be true.

\section{DR estimators and their variances}

\label{sect:ChenDR}

Define $\hat{N}^A_s = \sum_{i=1}^N S_i R^A_i / \pi^A_i$ and $\hat{N}^B_s = \sum_{i=1}^N S_i R^B_i / \pi^B (X_i; \alphahat)$, which are both estimators of $N_s$.
Define the following two DR estimators of $\Ybar$ that use $Y$ values only from Sample B:
\begin{eqnarray*}
  \DRone (\alphahat, \betahat)
  & = &
  \frac{1}{N_s} \sum_{i=1}^N S_i \left[ \frac{ R^A_i }{ \pi^A_i } m(X_i; \betahat)
+
\frac{ R^B_i }{ \pi^B (X_i; \alphahat) } \{ Y_i - m(X_i; \betahat) \} \right]
\\
  \DRtwo (\alphahat, \betahat)
  & = &
\frac{1}{ \hat{N}^A_s }
\sum_{i=1}^N S_i \; \frac{ R^A_i }{ \pi^A_i } m(X_i; \betahat)
+
\frac{1}{ \hat{N}^*_s }
\sum_{i=1}^N S_i \; \frac{ R^B_i }{ \pi^B (X_i; \alphahat) } \{ Y_i - m(X_i; \betahat) \}
\end{eqnarray*}
where either $\hat{N}^*_s = \hat{N}^A_s$ or $\hat{N}^*_s = \hat{N}^B_s$.
If $S_i=1$ for all $i$, then $\DRone$ and $\DRtwo$ with $\hat{N}^*_s = \hat{N}^B_s$ reduce to the estimators described by Chen et al.\ (2020).

We shall use $\DRone$ and $\DRtwo$ as shorthand for $\DRone (\alphahat, \betahat)$ and $\DRtwo (\alphahat, \betahat)$.
Both $\DRone$ and $\DRtwo$ are doubly robust; that is, they are consistent (i.e.\ $\DRone - \Ybar = o_p(1)$ and $\DRtwo - \Ybar = o_p(1)$) when either nuisance model is correctly specified, even if the other nuisance model is misspecified.
Chen et al.\ (2020) proposed that maximum pseudo likelihood or calibration be used to calculate $\alphahat$, and maximum likelihood be used to calculate $\betahat$.
These methods use the $X$ values of individuals in Samples A and B to calculate $\alphahat$, and the $X$ and $Y$ values of individuals in Sample B to calculate $\betahat$.
Chen et al.\ also proposed, as an alternative for $\DRone$, to calculate $\alphahat$ and $\betahat$ using the method of Kim and Haziza (KH)~\cite{kim2014doublyrobust,chen2020doublyrobust}.
We shall use $\PLone$ to denote $\DRone$ when $\alphahat$ and $\betahat$ are calculated by maximum pseudo likelihood and maximum likelihood, respectively.
The corresponding estimator $\DRtwo$ with $\hat{N}^*_s = \hat{N}^B_s$ will be denoted $\PLtwo$.
We shall use $\KHone$ and $\KHtwo$ to denote $\DRone$ and $\DRtwo$ with $\hat{N}^*_s = \hat{N}^A_s$ when $\alphahat$ and $\betahat$ are calculated using the KH method.
Note that when $m(X; \beta)$ is a linear regression model and $\alphahat$ is calculated using the KH method, $\hat{N}^A = \hat{N}^B$ (see Appendix~\ref{sect:var.KH2}).
So, in that case $\KHtwo$ and $\PLtwo$ differ only in how $\alphahat$ and $\betahat$ are calculated.

Chen et al.\ provided a formula for the asymptotic variance of $\PLtwo$ when $S_i=1$ for all $i$ and model $\pi^B(X; \alpha)$ is correctly specified.
They also gave a formula for the asymptotic variance of $\KHone$ and an estimator of this variance.
An advantage of $\KHone$ is that its variance estimator is double robust for model-design-based inference, i.e.\ it is valid when either nuisance model is correctly specified.
However, a limitation is it requires the same set of covariates to be used in the two nuisance models.
Two advantages of $\PLtwo$ over $\PLone$ and $\KHone$ are that the latter two estimators require $N_s$ to be known and that $\PLtwo$ may be more efficient than $\PLone$ and $\KHone$, at least when both nuisance models are correctly specified, for the same reason that the Hajek estimator may be more efficient than the Horvitz-Thompson estimator.
In Chen et al.'s simulation study, $\PLtwo$ was indeed more efficient.
Chen et al.\ did not consider $\KHtwo$.

In this section, we generalise Chen et al.'s formulae to domain mean estimation and add the following results.
First, we introduce the $\KHtwo$ estimator and provide a formula for its asymptotic variance.
Like $\KHone$, $\KHtwo$ has a double robust variance estimator and, like $\PLtwo$, $\KHtwo$ can be used when $N_s$ is unknown.
As we shall see, $\KHtwo$ is approximately equal to $\PLtwo$ when both nuisance models are correctly specified and the population is large compared to Sample B.
Second, we provide a formula for the asymptotic variance of $\PLone$ when model $\pi^B (X; \alpha)$ is correctly specified and model $m(X; \beta)$ may be misspecified.
Third, we provide simpler formulae for the asymptotic variances of $\PLone$ and $\PLtwo$ when both models $\pi^B (X; \alpha)$ and $m(X; \beta)$ are correctly specified.
We also provide formulae for estimators of all these asymptotic variances.

We assume $\alphahat$ is calculated using data from sampled individuals with $\usepi_i=1$, where $\usepi_i = \usepi (X_i)$ is some function of $X_i$.
Obvious choices of $\usepi (X)$ are $\usepi(X)=1$ and $\usepi(X)=S$, which correspond to using, respectively, all sampled individuals or only the sampled individuals in the domain.
However, other choices are possible for $\PLone$ and $\PLtwo$.
For example, if $X$ includes age and the domain consists of individuals aged 25--29 years, one might choose to estimate $\alpha$ using all sampled individuals in a wider age range, say, 18--39 years, in order to have more information to estimate $\alpha$.
For $\PLone$ and $\PLtwo$, $\betahat$ may be calculated using all of Sample B or any subset of it defined by $X$.
For $\KHone$ and $\KHtwo$, there is no choice: $T$ must equal $S$ and $\betahat$ must be calculated using exactly the set of individuals with $S=1$ in Sample B.

\subsection{Design-based inference}

\label{sect:designbased}

In this section, we consider design-based inference and assume model $\pi^B (X; \alpha)$ is correctly specified.
Results reported in this section are proved in Appendix~\ref{sect:var.PL1}--\ref{sect:var.KH2}.
Let $\DR$ denote $\PLone$, $\KHone$, $\PLtwo$ or $\KHtwo$.
The repeated-sampling variance of $\DR$ is given by
\begin{eqnarray}
\Var ( \DR - \Ybar \mid \curlF, \curlY )
  & = &
\left( \frac{N}{N_s} \right)^2
\Var \left[ \frac{1}{N} \sum_{i=1}^N \frac{ R^A_i }{ \pi^A_i } \;
  \{ S_i m(X_i; \betalim) - \Delta_{mi} \} \mid \curlF, \curlY \right]
  \nonumber \\
  && +~
  \frac{1}{N_s^2} \sum_{i=1}^N
  \frac{ 1 - \pi^B (X_i) }{ \pi^B (X_i) } \;
  \left[ S_i \{ Y_i - m(X_i; \beta_0) \} - \Delta_{yi} \right]^2 \nonumber \\
  && +~o_p(N^{-1}).
  \label{eq:var.generalform}
\end{eqnarray}
If $\DR$ is $\PLone$,
\begin{eqnarray}
  \Delta_{mi}
  & = &
  - \usepi_i \; \pi^B (X_i; \alphalim) \; \bfour^\top \Xpi
  \label{eq:Delta.mi.DR1b}
  \\
  \Delta_{yi}
  & = &
  \usepi_i \; \pi^B (X_i; \alphalim) \; \bfour^\top \Xpi
  \label{eq:Delta.yi.DR1b}
  \\
  \bfour
  & = &
  \left[ \sum_{i=1}^N \usepi_i \; \pi^B (X_i; \alphalim) \{ 1 - \pi^B (X_i; \alphalim) \} \; \Xpi \Xpitop \right]^{-1} 
  \nonumber \\
&& \qquad \times
  \sum_{i=1}^N S_i \{ 1 - \pi^B (X_i; \alphalim) \} \{ Y_i - m(X_i; \betalim) \} \Xpi.
  \label{eq:bfour}
\end{eqnarray}
When model $m(X; \beta)$ is correctly specified, equations~(\ref{eq:Delta.mi.DR1b})--(\ref{eq:bfour}) simplify to $\Delta_{mi} = \Delta_{yi} = \bfour = 0$.
If $\DR$ is $\KHone$, then $\Delta_{mi} = \Delta_{yi} = \bfour = 0$, even when model $m(X; \beta)$ is misspecified.
In fact, $\PLone$ and $\KHone$ are asymptotically equivalent when both nuisance models are correctly specified.

If $\DR$ is $\PLtwo$,
\begin{eqnarray}
  \Delta_{mi}
  & = &
  S_i \mbar - \usepi_i \; \pi^B (X_i; \alphalim) \; \bthree^\top \Xpi
  \label{eq:Delta.mi.DR2b}
  \\
  \Delta_{yi}
  & = &
  S_i (\Ybar - \mbar) + \usepi_i \; \pi^B (X_i; \alphalim) \bthree^\top \Xpi
  \label{eq:Delta.yi.DR2b}
  \\
  \bthree
  & = &
\left[ \sum_{i=1}^N \usepi_i \; \pi^B (X_i; \alphalim) \{ 1 - \pi^B (X_i; \alphalim) \} \;
 \Xpi \Xpitop \right]^{-1}
\nonumber \\
&& \qquad \times
\sum_{i=1}^N S_i \{ 1 - \pi^B (X_i; \alphalim) \} \;
    \{ Y_i - m(X_i; \betalim) - \Ybar + \mbar \} \;
    \Xpi.
    \label{eq:bthree}
\end{eqnarray}
When model $m(X; \beta)$ is correctly specified, equations~(\ref{eq:Delta.mi.DR2b})--(\ref{eq:bthree}) simplify to $\Delta_{mi} = S_i \mbar$ and $\Delta_{yi} = S_i (\Ybar - \mbar)$ and $\bthree = 0$.
If $\DR$ is $\KHtwo$, then $\Delta_{mi} = S_i \mbartrue$ and $\Delta_{yi} = \bthree = 0$, even when model $m(X; \beta)$ is misspecified.
If model $m(X; \beta)$ is correctly specified, $\mbartrue = \mbar$, and so $\Delta_{mi}$ is the same for $\PLtwo$ and $\KHtwo$.
If, furthermore, the domain is large compared to the number of sampled individuals from the domain in Sample B, i.e.\ if $N_s$ is large compared to $\sum_{i=1}^N S_i R^B_i$, then $S_i (\Ybar - \mbar)$ is negligibly small, and so $\Delta_{yi}$ is almost the same for $\PLtwo$ and $\KHtwo$; in fact, $\PLtwo$ and $\KHtwo$ are then approximately asymptotically equivalent, i.e.\ $\sqrt{N} (\PLtwo - \Ybar) \approx \sqrt{N} (\KHtwo - \Ybar) + o_p(1)$.

We now consider estimation of the two terms in expression~(\ref{eq:var.generalform}) for the variance of $\DR$.
The first term is the variance of the Horvitz-Thompson estimator of a population mean (specifically, the mean of $S_i m(X_i; \betalim) - \Delta_{mi}$) using Sample A, multiplied by $(N / N_s)^2$.
When Sample A uses sampling without replacement, this variance equals~\cite{sarndal1992surveysamplingbook}
\begin{equation*}
\frac{1}{N^2} \sum_{i=1}^N \sum_{j=1}^N (\pi^A_{ij} - \pi^A_i \pi^A_j) \;
\frac{ S_i m(X_i; \betalim) - \Delta_{mi} }{ \pi^A_i } \times
\frac{ S_i m(X_j; \betalim) - \Delta_{mj} }{ \pi^A_j },
\end{equation*}
where $\pi^A_{ij} = P(R^A_i = R^B_i = 1 \mid \curlF)$.
This variance can be estimated using any appropriate method for estimating the variance of a Horvitz-Thompson estimator, replacing $\alphalim$, $\betalim$ and terms in $\Delta_{mi}$ by estimators.
Specifically, $\alphalim$, $\betalim$, $\Ybar$, $\mbartrue$, $\mbar$, $\bfour$ and $\bthree$ would be replaced by, respectively, $\alphahat$, $\betahat$, $\DR$, $\DR$,
\begin{eqnarray}
\mbarhat
& = &
\sum_{i=1}^N S_i \frac{ R^A_i }{ \pi^A_i } m(X_i; \betahat)
\left/ \sum_{i=1}^N S_i \frac{ R^A_i }{ \pi^A_i } \right.
\nonumber \\
\bfourhat
& = &
  \left[ \sum_{i=1}^N \usepi_i \; R^B_i \;
  \{ 1 - \pi^B (X_i; \alphahat) \} \; \Xpi \Xpitop \right]^{-1}
  \nonumber \\
  && \times
  \sum_{i=1}^N S_i \frac{ R^B_i }{ \pi^B (X_i; \alphahat) } \;
  \{ 1 - \pi^B (X_i; \alphahat) \} \{ Y_i - m(X_i; \betahat) \} \;
  \Xpi
  \label{eq:original.bfourhat} \\
  \bthreehat
  & = &
  \left[ \sum_{i=1}^N \usepi_i \; R^B_i \{ 1 - \pi^B (X_i; \alphahat) \} \;
    \Xpi \Xpitop \right]^{-1}
\nonumber \\
&& \times
\sum_{i=1}^N S_i \frac{ R^B_i }{ \pi^B (X_i; \alphahat) } \; \{ 1 - \pi^B (X_i; \alphahat) \} \;
     \left[ Y_i - m(X_i; \betahat) - \frac{1}{N_s} \sum_{i=1}^N S_i \frac{ R^B_i }{ \pi^B (X_i; \alphahat) } \{ Y_i - m (X_i; \betahat) \} \right]
     \Xpi,
     \nonumber
\end{eqnarray}
as needed.

The second term in expression~(\ref{eq:var.generalform}) for the variance of $\DR$ can be estimated by
\begin{eqnarray}
  &&
  \frac{1}{N_s^2} \sum_{i=1}^N R^B_i \; 
  \frac{ 1 - \pi^B (X_i; \alphahat) } { \{ \pi^B (X_i; \alphahat) \}^2 } \; [ S_i \{ Y_i - m(X_i; \betahat) \} - \Deltahat_{yi} ]^2,
  \label{eq:var.generalform.estimator.nonprob}
\end{eqnarray}
where $\Deltahat_{yi}$ equals $\Delta_{yi}$ but with $\alphalim$, $\betalim$, $\Ybar$, $\mbar$, $\bfour$ and $\bthree$ replaced by $\alphahat$, $\betahat$, $\DR$, $\mbarhat$, $\bthreehat$ and $\bfourhat$, as needed.
If $N_s$ is unknown, it can be replaced by $\hat{N}^A_s$ or $\hat{N}^B_s$.

Note that the two inverse probability weighting estimators $\IPWone = N_s^{-1} \sum_{i=1}^N S_i R^B_i Y_i / \pi^B (X_i; \alphahat)$
and
$\IPWtwo = (\hat{N}_s^B)^{-1} \sum_{i=1}^N S_i R^B_i Y_i / \pi^B (X_i; \alphahat)$
are special cases of, respectively, $\PLone$ and $\PLtwo$, setting $m(X; \beta)$ equal to zero.
Therefore, the formulae given above for the variances of $\PLone$ and $\PLtwo$ also give the variances $\Var (\IPWone - \Ybar \mid \curlF, \curlY)$ and $\Var (\IPWtwo - \Ybar \mid \curlF, \curlY)$.
Simply set $m(X; \beta) = 0$ in those formulae.
Note that Chen et al.\ (2020) gave these formulae for the special case where $S_i=1$ for all $i$.

\subsection{Model-design-based inference}

\label{sect:modeldesignbased}

Throughout the remainder of this article, we shall use model-design-based inference.
Chen et al.\ (2020) adopted model-design-based inference to derive a double-robust variance estimator for $\KHone$.
Yang et al.\ (2020) and Gao and Yang (2023) also used model-design-based inference.
Results stated in this section are proved in Appendices~\ref{sect:modeldesign.PL1}--\ref{sect:modeldesign.KH2}.

First, consider $\KHone$.
For this estimator, it follows from equation~(\ref{eq:var.generalform}) that when model $\pi^B (X; \alpha)$ is correctly specified,
\begin{eqnarray}
  \Var ( \KHone - \Ybar \mid \curlF )
  & = &
  \left( \frac{N}{N_s} \right)^2
  \Var \left\{ \frac{1}{N} \sum_{i=1}^N \frac{ R^A_i }{ \pi^A_i } \;
    S_i m(X_i; \betalim) \mid \curlF \right\}
  \nonumber \\
  && \qquad +
  \frac{1}{N_s^2} \sum_{i=1}^N S_i \;
  \frac{ 1 - \pi^B (X_i) }{ \pi^B (X_i) } \;
  E [ \{ Y_i - m(X_i; \beta_0) \}^2 \mid X_i ] \nonumber \\
  && \qquad + o_p(N^{-1}).
  \label{eq:var.DRKH.modeldesign.piBcorrect}
\end{eqnarray}
If, on the other hand, model $\pi^B (X; \alpha)$ may be misspecified but model $m(X; \beta)$ is correctly specified,
\begin{eqnarray}
  \Var ( \KHone - \Ybar \mid \curlF )
  & = &
  \left( \frac{N}{N_s} \right)^2
  \Var \left\{ \frac{1}{N} \sum_{i=1}^N \frac{ R^A_i }{ \pi^A_i } \;
    S_i m(X_i; \betalim) \mid \curlF \right\}
  \nonumber \\
  && +~
  \frac{1}{N_s^2} \sum_{i=1}^N S_i
    \frac{ \piBtrue (X_i) \{ 1 - \pi^B (X_i; \alphalim) \} }{ \{ \pi^B (X_i; \alphalim) \}^2 } \;
    E [ \{ Y_i - m (X_i; \betalim) \}^2 \mid X_i ]
  \nonumber \\
  && +~
  \frac{1}{N_s^2} \sum_{i=1}^N S_i \left\{
    \frac{ E(R^A_i \mid \curlF) }{ \pi^A_i } - \frac{ E(R^B_i \mid X_i) }{ \pi^B (X_i; \alphalim) }
    \right\}
    \Var \{ Y_i - m (X_i; \betalim) \mid X_i \}
  \nonumber \\
  && +~
  o_p(N^{-1}).
  \label{eq:var.DRKH.modeldesign.mcorrect}
\end{eqnarray}
It can seen that when model $\pi^B(X; \alpha)$ is correctly specified, equation~(\ref{eq:var.DRKH.modeldesign.mcorrect}) reduces to equation~(\ref{eq:var.DRKH.modeldesign.piBcorrect}).
Therefore, equation~(\ref{eq:var.DRKH.modeldesign.mcorrect}) holds whenever either nuisance model is correctly specified.
This observation led Chen et al.\ to propose (in the context where $S_i=1$ for all $i$) to estimate $\Var ( \KHone - \Ybar \mid \curlF )$ as an estimate of the variance of the Horvitz-Thompson estimator $N^{-1} \sum_{i=1}^N R^A_i S_i m(X_i; \betalim) / \pi^A_i$, multiplied by $(N/N_s)^2$, plus
\begin{equation}
  \frac{1}{N_s^2} \sum_{i=1}^N S_i R^B_i \; 
  \frac{ 1 - \pi^B (X_i; \alphahat) } { \{ \pi^B (X_i; \alphahat) \}^2 } \; \{ Y_i - m (X_i; \betahat) \}^2
  +
  \frac{1}{N_s^2} \sum_{i=1}^N S_i \left\{
    \frac{ R^A_i }{ \pi^A_i } - \frac{ R^B_i }{ \pi^B (X_i; \alphahat) }
    \right\}
  \hat{\sigma}^2 (X_i),
  \label{eq:var.bias.cor}
\end{equation}
where $\hat{\sigma}^2 (X_i)$ is an estimator of $\Var \{ Y_i - m (X_i; \betalim) \mid X_i \}$.
This estimator of $\Var ( \KHone - \Ybar \mid \curlF )$ is valid when either nuisance model is correctly specified.
Note that if model $m(X; \beta)$ is a generalised linear model with canonical link function, then the second term in expression~(\ref{eq:var.bias.cor}) equals zero, and so expression~(\ref{eq:var.bias.cor}) reduces to expression~(\ref{eq:var.generalform.estimator.nonprob}) with $\Delta_{yi}=0$.

For $\KHtwo$, if either nuisance model is correctly specified, equation~(\ref{eq:var.DRKH.modeldesign.mcorrect}) again applies but with the first term replaced by
\begin{equation}
\left( \frac{N}{N_s} \right)^2
\Var \left[
  \frac{1}{N} \sum_{i=1}^N 
  \frac{ R^A_i }{ \pi^A_i } S_i \left\{ m(X_i; \betalim) - \mbartrue \right\} \mid \curlF
  \right].
\label{eq:var.KH2.firstpart}
\end{equation}
Hence, $\Var ( \KHtwo - \Ybar \mid \curlF )$ can be estimated by an estimate of the variance of the Horvitz-Thompson estimator $N^{-1} \sum_{i=1}^N R^A_i S_i \{ m(X_i; \betalim) - \Ybar \} / \pi^A_i$, multiplied by $(N/N_s)^2$, plus expression~(\ref{eq:var.bias.cor}).

For $\DR$ equal to $\PLone$ or $\PLtwo$, if model $\pi^B (X; \alpha)$ is correctly specified,
\begin{eqnarray*}
  \Var ( \DR - \Ybar \mid \curlF )
  & = &
  \left( \frac{N}{N_s} \right)^2
  E \left( \Var \left[
    \frac{1}{N} \sum_{i=1}^N 
    \frac{ R^A_i }{ \pi^A_i } \{ S_i m(X_i; \betalim) - \Delta_{mi} \} \mid \curlF, \curlY
    \right]
  \mid \curlF \right)
   \\
  && +~
    \frac{1}{N_s^2} \sum_{i=1}^N
    \frac{ 1 - \pi^B (X_i) }{ \pi^B (X_i) } \;
    E \left( \left[ S_i \{ Y_i - m(X_i; \betalim) \} - \Delta_{yi} \right]^2 \mid \curlF \right) \\
    && +~o_p (N^{-1}).
  \end{eqnarray*}
If model $m(X; \beta)$ is also correctly specified, this simplifies to
\begin{eqnarray}
  \Var ( \DR - \Ybar \mid \curlF )
  & = &
  \left( \frac{N}{N_s} \right)^2
  \Var \left[
    \frac{1}{N} \sum_{i=1}^N 
    \frac{ R^A_i }{ \pi^A_i } \{ S_i m(X_i; \betalim) - \Delta_{mi} \} \mid \curlF
    \right]
  \nonumber \\
  && +~
    \frac{1}{N_s^2} \sum_{i=1}^N
    \frac{ 1 - \pi^B (X_i) }{ \pi^B (X_i) } \;
    E \left( \left[ S_i \{ Y_i - m(X_i; \betalim) \} - \Delta_{yi} \right]^2 \mid \curlF \right) \nonumber \\
  &&  +~o_p (N^{-1})
  \label{eq:var.DR.correctm}
  \end{eqnarray}
with $\Delta_{mi} = \Delta_{yi} = 0$ for $\PLone$, and $\Delta_{mi} = S_i \mbar$ and $\Delta_{yi} = S_i (\Ybar - \mbar)$ for $\PLtwo$.

The estimators of $\Var ( \PLone - \Ybar \mid \curlF )$ and $\Var ( \PLtwo - \Ybar \mid \curlF )$ are the same as those described for $\Var ( \PLone - \Ybar \mid \curlF, \curlY )$ and $\Var ( \PLtwo - \Ybar \mid \curlF, \curlY )$ in Section~\ref{sect:designbased}.

\section{Combining data on $Y$ from Samples A and B}

\label{sect:combining}

So far, we have assumed $Y$ is observed only on individuals in Sample B.
Now suppose $Y$ is observed also on Sample A.
Let $\HT$ and $\Haj$ denote, respectively, the Horvitz-Thompson and Hajek estimators of $\Ybar$ that use only the data on Sample A \cite{sarndal1992surveysamplingbook}, i.e.\
$\HT = N_s^{-1} \sum_{i=1}^N S_i R^A_i Y_i / \pi^A_i$
and
$\Haj =  (\hat{N}^A_s)^{-1} \sum_{i=1}^N S_i R^A_i Y_i / \pi^A_i$.

Let $\HTHaj$ denote either $\HT$ or $\Haj$.
Note that $\DR$ uses only the $Y$ values from Sample B and $\HTHaj$ uses only the $Y$ values from Sample A.
A more precise estimator may be obtained by combining these two estimators efficiently.
We shall consider the class of combined estimators that are a weighted average of $\HTHaj$ and $\DR$ (with a fixed choice of nuisance models $\pi^B (X; \alpha)$ and $m (X; \beta)$), i.e.\ the class $\{ (1-w) \; \HTHaj + w \; \DR: w \in {\rm I\!R} \}$.
The variance of such a combined estimator is
\begin{equation}
  \Var \{ (1-w) \; \HTHaj + w \; \DR - \Ybar \mid \curlF \}
  =
  (1-w)^2 \; V_H
  + 2(1-w)w \; C
  + w^2 \; V_D
  \label{eq:var.combest}
\end{equation}
where $V_H = \Var (\HTHaj - \Ybar \mid \curlF)$, $V_D = \Var (\DR - \Ybar \mid \curlF)$ and $C = \Cov (\HTHaj - \Ybar, \; \DR  - \Ybar\mid \curlF)$.

By differentiating this variance with respect to $w$, equating the result to zero, and solving for $w$, we see that the most efficient estimator in this class uses
\begin{equation}
  w =
  \frac{ V_H - C }
       { V_H + V_D - 2 C }.
  \label{eq:efficient.w}
\end{equation}
With this choice of $w$, equation~(\ref{eq:var.combest}) reduces to
\begin{equation}
  \Var \{ (1-w) \HTHaj + w \; \DR - \Ybar \mid \curlF \}
  =
  V_H
  - \frac{ (V_H - C)^2 }
  { V_H + V_D - 2 C }
  =
  V_D
  - \frac{ (V_D - C)^2 }
  { V_H + V_D - 2 C }.
\label{eq:var.combined}
\end{equation}

In the remainder of this section, we assume that if $\DR$ is $\PLone$ or $\PLtwo$ then model $\pi^B (X; \alpha)$ is correctly specified, and that if $\DR$ is $\KHone$ or $\KHtwo$ then either nuisance model is correctly specified.
To estimate expressions~(\ref{eq:efficient.w}) and~(\ref{eq:var.combined}), we require estimators of $V_D$, $V_H$ and $C$.
An estimator of $V_D$ was provided in Section~\ref{sect:modeldesignbased}.
As shown in Appendices~\ref{sect:modeldesign.PL1}--\ref{sect:modeldesign.KH2},
$V_H = E \{ \Var (\HTHaj - \Ybar \mid \curlF, \curlY) \mid \curlF \} + o_p (N^{-1})$ and $C = E\{ \Cov ( \DR - \Ybar, \; \HTHaj - \Ybar \mid \curlF, \curlY ) \mid \curlF \} + o_p (N^{-1})$.
If the population is large compared to Samples A and B, then $E \{ \Var (\HTHaj - \Ybar \mid \curlF, \curlY) \mid \curlF \} \approx \Var (\HTHaj - \Ybar \mid \curlF, \curlY)$ and $E\{ \Cov ( \DR - \Ybar, \; \HTHaj - \Ybar \mid \curlF, \curlY ) \mid \curlF \} \approx \Cov ( \DR - \Ybar, \; \HTHaj - \Ybar \mid \curlF, \curlY )$.
Hence, we shall approximate $V_H$ and $C$ by $\Var (\HTHaj - \Ybar \mid \curlF, \curlY)$ and $\Cov ( \DR - \Ybar, \; \HTHaj - \Ybar \mid \curlF, \curlY )$.

As shown in Appendices~\ref{sect:var.PL1}--\ref{sect:var.KH2},
\begin{eqnarray}
  &&
  \Cov ( \DR - \Ybar, \; \HTHaj - \Ybar \mid \curlF, \curlY )
  \nonumber \\
  && \hspace{.2cm} =
  \left( \frac{N}{N_s} \right)^2
  \Cov \left[
    \frac{1}{N} \sum_{i=1}^N
    \frac{ R^A_i }{ \pi^A_i } \{ S_i m (X_i; \betalim) - \Delta_{mi} \}, \;
    \frac{1}{N} \sum_{j=1}^N
    \frac{ R^A_j }{ \pi^A_j } S_i (Y_j - \Gamma)
    \mid \curlF, \curlY \right]
    \nonumber \\
    && \hspace{.7cm} + \;
    o_p (N^{-1}),
  \label{eq:cov.generalform}
\end{eqnarray}
where $\Gamma = 0$ if $\HTHaj$ denotes $\HT$ and $\Gamma = \Ybar$ if $\HTHaj$ denotes $\Haj$. 
The first term on the right-hand side of equation~(\ref{eq:cov.generalform}) is the covariance between the Horvitz-Thompson estimators of the population means of two random variables (specifically $S_i m (X_i; \betalim) - \Delta_{mi}$ and $S_j (Y_j - \Gamma)$) using Sample A, multiplied by $(N/N_s)^2$.
When Sample A is sampled without replacement, this covariance equals~\cite{sarndal1992surveysamplingbook}
\begin{equation*}
\frac{1}{N^2} \sum_{i=1}^N \sum_{j=1}^N (\pi^A_{ij} - \pi^A_i \pi^A_j) \;
\frac{ S_i m(X_i; \betalim) - \Delta_{mi} }{ \pi^A_i } \times
\frac{ S_j (Y_j - \Gamma) }{ \pi^A_j }.
\end{equation*}
It can be estimated using any appropriate method for estimating the covariance of two Horvitz-Thompson estimators, replacing $\betalim$ and terms in $\Delta_{mi}$ by estimators (and replacing $N_s$ by $\hat{N}_s^A$ or  $\hat{N}_s^B$ if $N_s$ is unknown), as described in Section \ref{sect:ChenDR}, and replacing $\Gamma$ by $\Haj$ if $\HTHaj$ denotes $\Haj$.

Finally, as noted in Section~\ref{sect:ChenDR}, the IPW estimators $\IPWone$ and $\IPWtwo$ are equivalent to, respectively, $\PLone$ and $\PLtwo$ with $m(X; \beta) = 0$.
Therefore, equations~(\ref{eq:var.combest})--(\ref{eq:cov.generalform}) can also be used to find the most efficient estimator in the class of estimators $\{ (1-w) \; \HTHaj + w \; \IPWone: w \in {\rm I\!R} \}$ or in the class $\{ (1-w) \; \HTHaj + w \; \IPWtwo: w \in {\rm I\!R} \}$, and its variance.

\section{Efficiency gain of combined estimator}

\label{sect:efficiency.gain}

For simplicity, in this section we focus on estimating the population mean, i.e.\ $S_i = 1$ for all $i$.
Results reported are proved in Appendix~\ref{sect:efficiency.proof}.
Define $\eta = V_H / V_D$ and $G = C / V_H = \rho / \sqrt{\eta}$ where $\rho = C  / \sqrt{V_D V_H}$ is the correlation.
Equation~(\ref{eq:var.combined}) can be rewritten as
\begin{equation}
  \Var \{ (1-w) \HTHaj + w \DR - \Ybar \mid \curlF \}
  = V_D (1 - Q_D)
  = V_H (1 - Q_H)
  \label{eq:var.eff.rewritten}
\end{equation}
where
\begin{eqnarray}
  Q_D & = & \frac{ 1 - 2 \eta G + \eta^2 G^2 }{ 1 - 2 \eta G + \eta } = \frac{ 1 - 2 \rho \sqrt{\eta} + \rho^2 \eta }{ 1 - 2 \rho \sqrt{\eta} + \eta }
  \label{eq:QD} \\
  Q_H & = & \frac{ 1 - 2 G + G^2 }{ 1 - 2 G + \eta^{-1} } = \frac{ 1 - 2 \rho \sqrt{\eta}^{-1} + \rho^2 \eta^{-1}  }{ 1 - 2 \rho \sqrt{\eta}^{-1} + \eta^{-1} }.
  \label{eq:QH}
\end{eqnarray}
So, $1 / (1 - Q_D)$ and $1 / (1 - Q_H)$ are the relative efficiencies (REs) of the combined estimator compared to, respectively, $\DR$ and $\HTHaj$.

We shall focus on two combinations of estimators --- i) $\HT$ and either $\PLone$ or $\KHone$, and ii) $\Haj$ and either $\PLtwo$ or $\KHtwo$ --- and assume that both nuisance models are correctly specified.

For combination i), it can be shown that $C = \Var \left\{ \frac{1}{N} \sum_{i=1}^N \frac{ R^A_i }{ \pi^A_i } m(X_i; \betalim) \mid \curlF \right\}$ $+ o_p(N^{-1})$ and $V_H = \Var \left\{ \frac{1}{N} \sum_{i=1}^N \frac{ R^A_i }{ \pi^A_i } m(X_i; \betalim) \mid \curlF \right\}$ $+ \Var \left[ \frac{1}{N} \sum_{i=1}^N \frac{ R^A_i }{ \pi^A_i } \{Y_i - m(X_i; \betalim) \} \mid \curlF \right]$.
Recall that $V_D$ was given by equations~(\ref{eq:var.DRKH.modeldesign.piBcorrect}) and~(\ref{eq:var.DR.correctm}).
Hence,
\begin{equation}
  G
  =
  \frac{ \Var \left\{ \frac{1}{N} \sum_{i=1}^N
  \frac{ R^A_i }{ \pi^A_i } m(X_i; \betalim) \mid \curlF \right\}
  }
  {\Var \left\{ \frac{1}{N} \sum_{i=1}^N
    \frac{ R^A_i }{ \pi^A_i } m(X_i; \betalim) \mid \curlF \right\}
    + \Var \left[ \frac{1}{N} \sum_{i=1}^N
      \frac{ R^A_i }{ \pi^A_i } \{Y_i - m(X_i; \betalim) \} \mid \curlF \right]
  }
  \label{eq:G.DR1}
\end{equation}
and
\begin{equation}
  \eta
  = 
  \frac{ \Var \left\{ \frac{1}{N} \sum_{i=1}^N
    \frac{ R^A_i }{ \pi^A_i } m(X_i; \betalim) \mid \curlF \right\}
    + \Var \left[ \frac{1}{N} \sum_{i=1}^N
      \frac{ R^A_i }{ \pi^A_i } \{Y_i - m(X_i; \betalim) \} \mid \curlF \right]
  }
  { \Var \left\{ \frac{1}{N} \sum_{i=1}^N
    \frac{ R^A_i }{ \pi^A_i } m(X_i; \betalim) \mid \curlF \right\}
    + \frac{1}{N^2} \sum_{i=1}^N
  \frac{ 1 - \pi^B (X_i; \alphalim) }{ \pi^B (X_i; \alphalim) } \;
  E \left[ \{ Y_i - m(X_i; \betalim) \}^2 \mid \curlF \right]
  }
  \label{eq:eta.DR1}
\end{equation}
asymptotically.

Similarly, for combination ii), if the population is large compared to Samples A and B (so that $\Ybar - \mbar$ is negligible), then
\begin{equation}
  G
  =
  \frac{ \Var \left[ \frac{1}{N} \sum_{i=1}^N
      \frac{ R^A_i }{ \pi^A_i } \{ m(X_i; \betalim) - \mbar \} \mid \curlF \right]
  }
  { \Var \left[ \frac{1}{N} \sum_{i=1}^N
      \frac{ R^A_i }{ \pi^A_i } \{ m(X_i; \betalim) - \mbar \} \mid \curlF \right]
    +
    \Var \left[ \frac{1}{N} \sum_{i=1}^N
      \frac{ R^A_i }{ \pi^A_i } \{ Y_i - m(X_i; \betalim) \} \mid \curlF \right]
  }
  \label{eq:G.DR2}
\end{equation}
and
\begin{equation}
  \eta
  =
  \frac  { \Var \left[ \frac{1}{N} \sum_{i=1}^N
      \frac{ R^A_i }{ \pi^A_i } \{ m(X_i; \betalim) - \mbar \} \mid \curlF \right]
    +
    \Var \left[ \frac{1}{N} \sum_{i=1}^N
      \frac{ R^A_i }{ \pi^A_i } \{ Y_i - m(X_i; \betalim) \} \mid \curlF \right]
  }
  { \Var \left[ \frac{1}{N} \sum_{i=1}^N
      \frac{ R^A_i }{ \pi^A_i } \{ m(X_i; \betalim) - \mbar \} \mid \curlF \right]
    +
    \frac{1}{ N^2} \sum_{i=1}^N
    \frac{ 1 - \pi^B (X_i; \alphalim) }{\pi^B (X_i; \alphalim)} \;
    E \left[ \{ Y_i - m(X_i; \betalim) \}^2 \mid \curlF \right]
  }
  \label{eq:eta.DR2}
\end{equation}
approximately.

It can be seen from equations~(\ref{eq:G.DR1})--(\ref{eq:eta.DR2}) that $G$ depends on the design of Sample A but not on the selection mechanism $\piBtrue$ for Sample B, whereas $\eta$ depends on both of these.
We also see that $0 \leq G \leq 1$, with $G \approx 0$ (for $\DRone$) or $G = 0$ (for $\DRtwo$) when $X$ is not at all predictive of $Y$ (i.e.\ $Y$ is independent of $X$), and $G=1$ when $X$ perfectly predicts $Y$ (i.e.\ $Y = m(X)$).
Also, $0 < \eta \leq 1 / G$.
Equation~(\ref{eq:eta.DR1}) can be rewritten as
\begin{equation*}
  \eta
  =
  \frac{ \Var \left\{ \frac{1}{N} \sum_{i=1}^N
    \frac{ R^A_i }{ \pi^A_i } m(X_i; \betalim) \mid \curlF \right\}
    + \Var \left[ \frac{1}{N} \sum_{i=1}^N
      \frac{ R^A_i }{ \pi^A_i } \{Y_i - m(X_i; \betalim) \} \mid \curlF \right]
  }
  { \Var \left\{ \frac{1}{N} \sum_{i=1}^N
    \frac{ R^A_i }{ \pi^A_i } m(X_i; \betalim) \mid \curlF \right\}
    + \Var \left[ \frac{1}{N} \sum_{i=1}^N
      \frac{ R^B_i }{ \pi^B (X_i; \alphalim) } \{Y_i - m(X_i; \betalim) \} \mid \curlF \right]
  }
\end{equation*}
and analogously for equation~(\ref{eq:eta.DR2}).
Thus, we see that $\eta$ depends on the ratio of the variances of the Horvitz-Thompson estimators of a population mean (specifically, the mean of $Y - m(X; \betalim)$) based on, respectively, Sample A and Sample B.
These variances depend on the designs of the two samples and on the mean and variability of their respective sampling weights, $1/\pi^A$ and $1/\piBtrue$.
Sample A may have used a complex multistage sampling approach, whereas Sample B uses Poisson sampling (i.e. $R^B_i$ and $R^B_j$ are independent given $\curlF$).
The mean of the sampling weights determines the expected sample size.
For a given mean sampling weight, the variance of the Horvitz-Thompson estimator increases with the variability of the sampling weights.
Thus, for example, if Samples A and B are the same size but the true selection probabilities $\piBtrue$ for Sample B are more variable than the sampling probabilities $\pi^A$ for Sample A, then $\eta$ may be less than one.

\begin{figure}
    \centering
\includegraphics[page=2,width=1\textwidth]{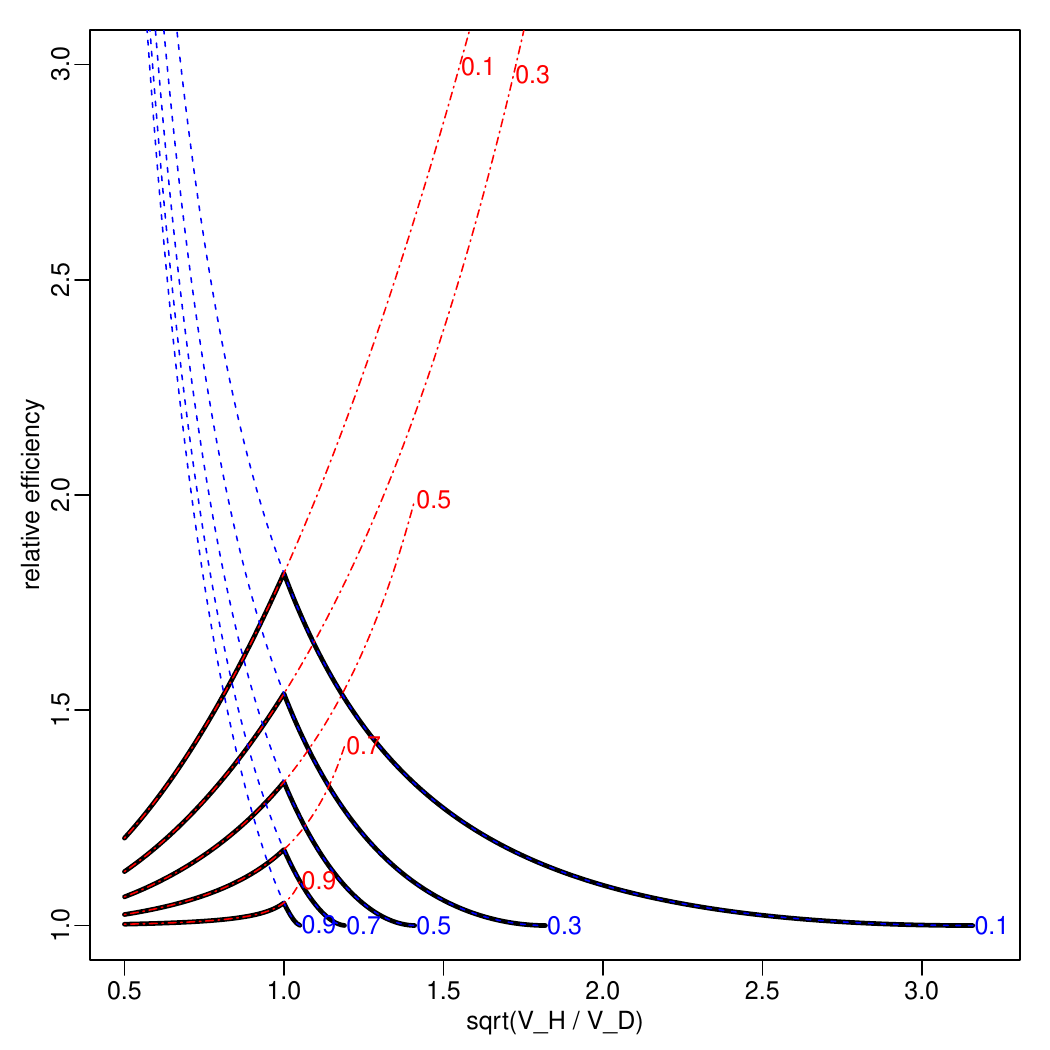}
\caption{Relative efficiency of the combined estimator compared to its two component estimators, $\HTHaj$ (red broken-dotted line) and $\DR$ (blue broken line), as a function of $\sqrt{\eta} = \sqrt{V_H / V_D}$ for each of five values (0.1, 0.3, 0.5, 0.7 and 0.9) of $G = C / V_H$.  Each black line is the minimum of the red and blue lines for that value of $G$.  It shows the relative efficiency of the combined estimator compared to the more efficient of the two component estimators.  Note that the y-axis is truncated at 3.0 to improve readibility of the graph.}
\label{fig:theoreticalRE}
\end{figure}

For any fixed value of $G$, it can be shown that $Q_D$ and $Q_H$ are, respectively, a decreasing function and an increasing function of $\eta$.
When $\eta = 1$, $Q_D = Q_H = (1-G) / 2$.
Hence, $\eta=1$ is the value of $\eta$ that maximises the minimum of $Q_D$ and $Q_H$, i.e.\ the efficiency gain of the combined estimator relative to the more efficient of $\DRone$ and $\HTHaj$.
This value $Q_D = Q_H = (1-G) / 2 = (1-\rho) / 2$ corresponds to a RE of $2 / (1+G) = 2 / (1+\rho)$, which, since $0 \leq G \leq 1$, must be between 1 and 2.
That is, the combined estimator cannot be more than twice as efficient as the more efficient of its component estimators.
Figure~\ref{fig:theoreticalRE} illustrates how the RE of the combined estimator compared to each of $\DR$, $\HTHaj$ and the more efficient of $\DR$ and $\HTHaj$ depends on $G$ and $\eta$.
Supplementary Figure~\ref{fig:theoreticalRE.Q} shows the same dependence but in terms of the proportion of variance reduction, i.e.\ $Q_D$ or $Q_H$ or their minimum.

\section{Simulation study}

\label{sect:simstudy}

\subsection{Methods}

We simulated a finite population with 1000 clusters.
Each cluster $j$ contains $H_{jq}$ households of $q$ individuals ($q=1, 2, 3$), where $H_{jq}$ is negatively binomially distributed with mean 100 and variance 400.
Hence, the expected number of individuals in a cluster is 600, and the expected population size is 600,000.
For each individual $i$ in cluster $j$, continuous variables $X_{1i}$ and $X_{2i}$ and binary variables $X_{3i}$ and $X_{4i}$ were generated independently, with expectations that depend on the total number of households in the cluster $j$, the size (1, 2 or 3) of the household to which individual $i$ belongs, and a cluster-level random effect.
See Appendix~\ref{sect:appendix.simstudy.dgm} for details.
This fixed population was used for all the simulated datasets.

We generated 1000 simulated datasets for each of 36 scenarios, corresponding to all combinations of three conditional distributions of $Y_i$ given $X_i$, two numbers of selected clusters for Sample A ($n_{\rm clust}$), two numbers of selected households per cluster for Sample A ($n_{\rm house}$), and three expected sizes of Sample B ($E(n^B)$).
For each simulated dataset, $Y_1, \ldots, Y_n$ were generated independently.
In Scenarios 1--12, $Y_i \mid X_i \sim \mbox{Normal} (5 + X_1 + X_2 + 2 X_3 + X_4, 1)$.
In Scenarios 13--24, $Y_i \mid X_i \sim \mbox{Normal} (X_1 + X_2 + 2 X_3 + X_4, 9)$.
In Scenarios 25--36, $Y_i$ was binary with $\mbox{logit } P(Y_i = 1 \mid X_i) = -4 + X_1 + X_2 + 2 X_3 + X_4$.
Sample A was drawn using a three-stage sampling design.
First, either $n_{\rm clust} = 200$ or $n_{\rm clust} = 50$ (depending on the scenario) clusters were selected using randomised systematic sampling, with the probability of selecting cluster $j$ being proportional to the total number of households in cluster $j$.
From each of these $n_{\rm clust}$ selected clusters, either $n_{\rm house} = 20$ or $n_{\rm house} = 5$  (depending on scenario) households were selected using simple random sampling without replacement.
Then one individual was sampled at random from each of these $n_{\rm house}$ sampled households.
Thus, Sample A contained $n_{\rm clust} \times n_{\rm house}$ individuals, i.e.\ 4000, 1000 or 250, depending on the scenario.

Each individual's probability of inclusion in Sample B was given by $\mbox{logit } \piBtrue (X) = \alpha_{\rm int} + 0.5 X_1 + X_2 + 0.5 X_3 + X_4$, with the $\alpha_{\rm int}$ chosen to make $E(n^B)$, the expected size of Sample B, either 4000, 1000 or 250 (depending on scenario).
Note that both Samples A and B are small compared to the population size.

Estimators $\HT$, $\Haj$, $\PLone$, $\KHone$, $\PLtwo$, $\KHtwo$, $\HTPL$, $\HTKH$, $\HajPL$ and $\HajKH$ of the population mean $\Ybar = N^{-1} \sum_{i=1}^N Y_i$ were calculated for each simulated dataset.
Here, $\HTPL$, $\HTKH$, $\HajPL$ and $\HajKH$ denote the efficient combined estimators using, respectively, $\HT$ and $\PLone$, $\HT$ and $\KHone$, $\Haj$ and $\PLtwo$, and $\Haj$ and $\KHtwo$.
Both models $\pi(X; \alpha)$ and $m(X; \beta)$ were correctly specified.
The maximum pseudo likelihood estimate of $\alpha$ was obtained using the nonprob function in the nonprobsvy R package~\cite{chrostowski2025nonprobsvyPreprint}.
The KH estimates of $\alpha$ and $\beta$ were obtained using the same function, with the maximum pseudo likelihood estimate of $\alpha$ and the maximum likelihood estimate of $\beta$ used as starting values for the iterative algorithm.

For each estimator of $\Ybar$, we estimated the standard error (SE) and calculated 95\% confidence intervals (CIs) using the variance estimators, given in Sections~\ref{sect:ChenDR} and~\ref{sect:combining}, that do not assume model $m (X; \alpha)$ is correctly specified.
For each scenario and each of the four combined estimators, we calculated its actual RE compared to each of its component estimators $\HTHaj$ and $\DR$.
We also used equations~(\ref{eq:var.eff.rewritten})--(\ref{eq:QH}) to calculate the corresponding predicted relative efficiencies, using the 1000 estimates $\HTHaj$ and $\DR$ to calculate $V_D$, $V_H$ and $C$.
In addition, we compared the mean estimated SE for each estimator with the corresponding empirical standard error of that estimator, and calculated the coverage of 95\% confidence intervals.

\subsection{Results}

\label{sect:results}

In scenarios where $Y$ is normally distributed, $\PLtwo$ can be much more efficient than $\PLone$ (Supplementary Figure~\ref{fig:releff.DR1.DR2}).
This is particularly true of scenarios where the error variance of $Y$ equals 1, Sample A is small and/or Sample B is large.
In scenarios where $Y$ is binary, the efficiency gain is much smaller.
The pattern of RE of $\PLtwo$ compared to $\PLone$ is reflected in the pattern of RE of $\Haj$ compared to $\HT$ (Supplementary Figure~\ref{fig:releff.HT.Haj}).

\begin{figure}
\begin{center}
 \includegraphics[width=1\textwidth,page=3]{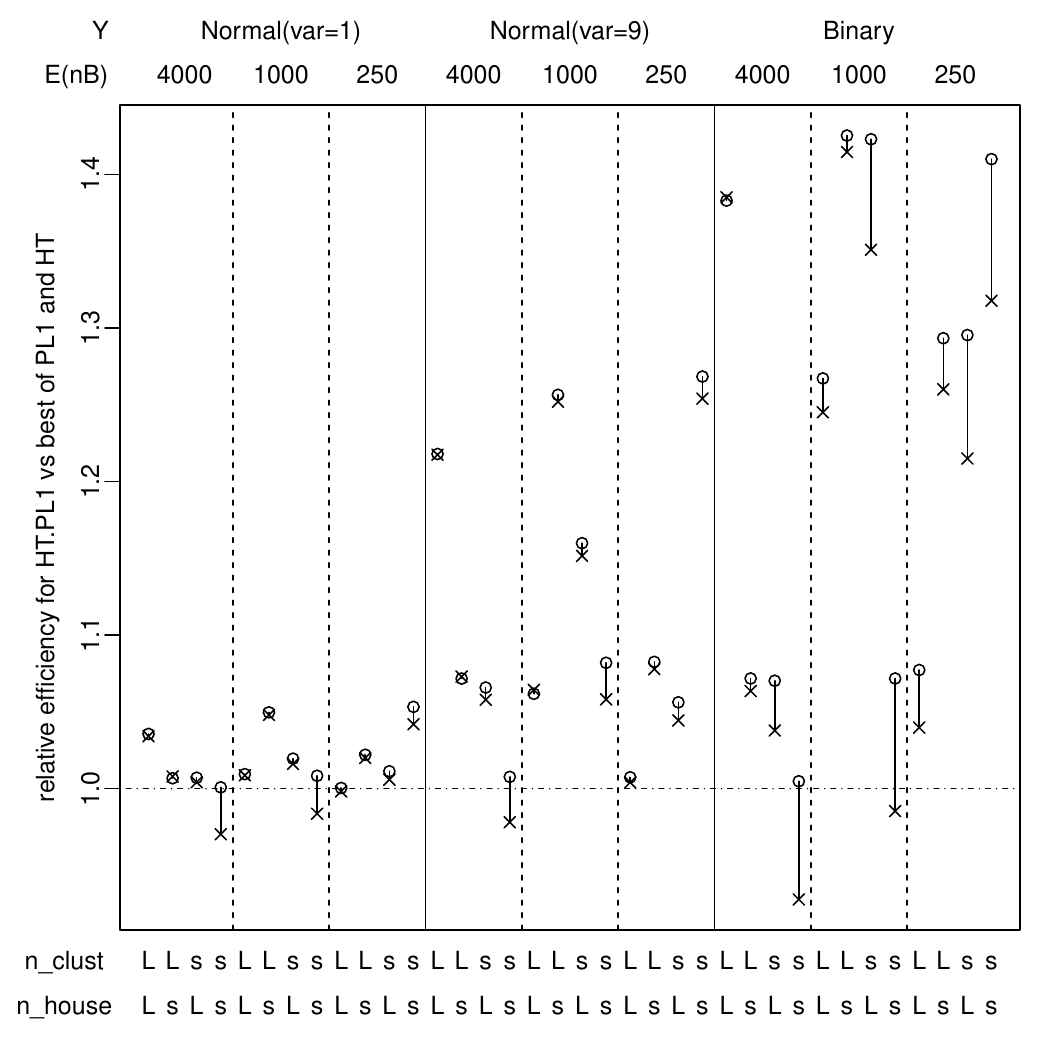}
\hspace*{-5mm}
\end{center}
\caption{Relative efficiency of $\HTPL$ compared to the most efficient of $\HT$ and $\PLone$ in 36 scenarios.  The dots indicate the theoretical asymptotic relative efficiency and the crosses indicate the empirical relative efficiency.  The 36 scenarios correspond to three conditional distributions of $Y$ given $X$, three expected size of Sample B, and two numbers of sampled clusters and of sampled households within sampled clusters in Sample A.  For the last two, `$L$' denotes large (200 clusters or 20 households) and `$s$' denotes small (50 clusters or 5 households).  The four combinations of 200 or 50 clusters and 20 or 5 households imply 4000, 1000, 1000 or 250 individuals in Sample A.} 
\label{fig:releff.DR1eff.best}
\end{figure}

Figure~\ref{fig:releff.DR1eff.best} shows the predicted and actual RE of $\HTPL$ compared to the more efficient of $\HT$ and $\PLone$.
In scenarios where $Y$ is continuous with error variance equal to 1, the predicted RE ranges from 1.00 to 1.05.
In scenarios where the error variance is 9, and so $X$ is less strongly predictive of $Y$, the predicted RE is greater, ranging from 1.00 to 1.27.
When $Y$ is binary, it ranges from 1.00 to 1.42.
For each of the three distributions of $Y$, the greatest REs are achieved in scenarios where Samples A and B are the same size.
Among the scenarios we studied, these are the ones where $V_D$ and $V_H$ are most similar to one another (see Supplementary Figure~\ref{fig:releff.DR1eff.DR1HT}).
So, this is consistent with the theoretical result that the RE is greatest when $V_D=V_H$ (Section \ref{sect:efficiency.gain}).
In scenarios where $Y$ is continuous, the predicted REs approximate well the actual REs, except in the six scenarios where Sample A is small. 
In three of those six scenarios, the actual RE is less than 1.00, probably reflecting inaccurate estimation of the optimal combining weight $w$.
The tendency of the predicted RE to overestimate the actual RE is more marked in scenarios with binary $Y$, especially when either Sample A or B is small.
However, only in two of these scenarios is the actual RE less than 1.00, with the lowest RE being 0.92.
Supplementary Figures~\ref{fig:releff.DR1eff.HT}--\ref{fig:releff.DR1eff.DR1HT} show the RE of $\HTPL$ compared to its individual components $\HT$ and $\PLone$.

\begin{figure}
\begin{center}
 \includegraphics[width=1\textwidth,page=9]{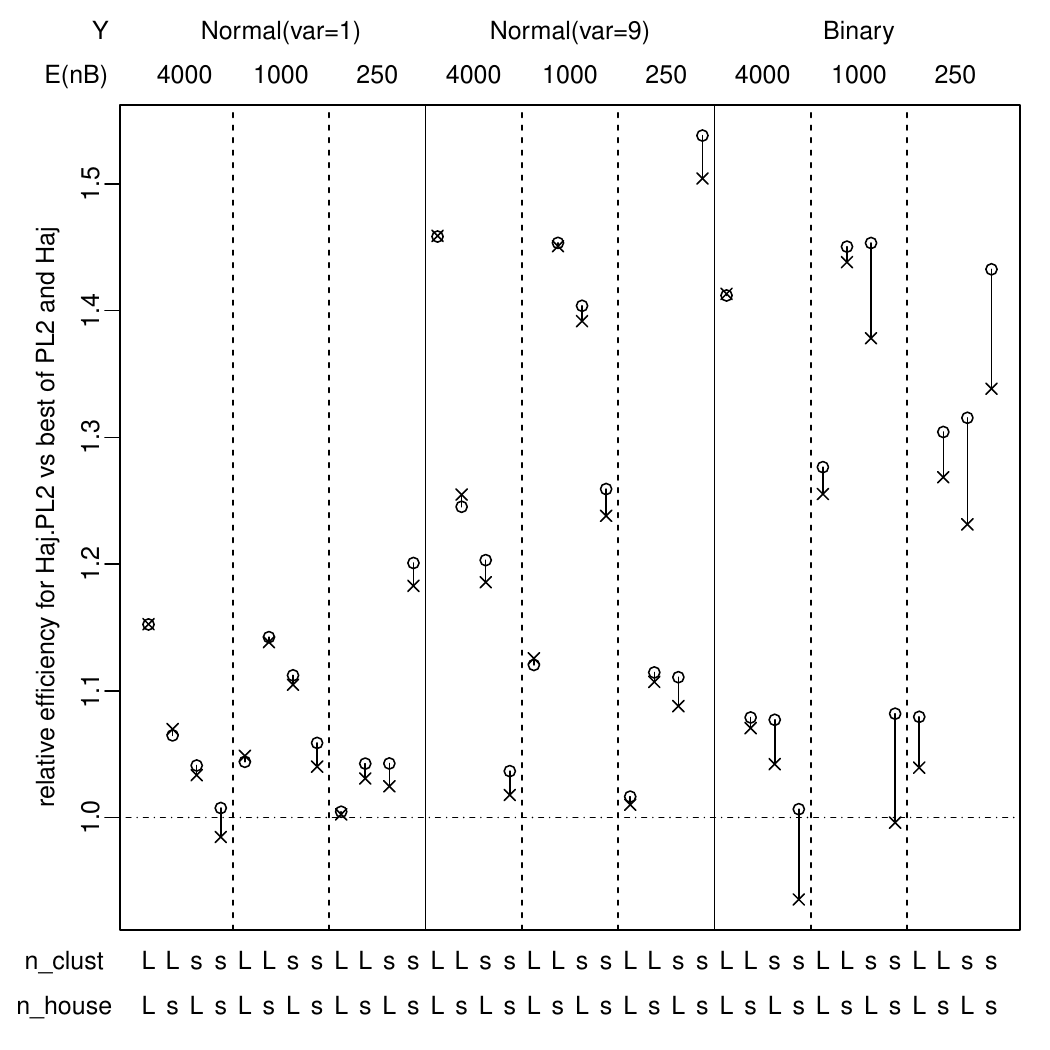}
\hspace*{-5mm}
\end{center}
\caption{Relative efficiency of $\HajPL$ compared to the most efficient of $\Haj$ and $\PLtwo$ in 36 scenarios.  The dots indicate the theoretical asymptotic relative efficiency and the crosses indicate the empirical relative efficiency.  The 36 scenarios correspond to three conditional distributions of $Y$ given $X$, three expected size of Sample B, and two numbers of sampled clusters and of sampled households within sampled clusters in Sample A.  For the last two, `$L$' denotes large (200 clusters or 20 households) and `$s$' denotes small (50 clusters or 5 households).  The four combinations of 200 or 50 clusters and 20 or 5 households imply 4000, 1000, 1000 or 250 individuals in Sample A.} 
\label{fig:releff.DR2eff.best}
\end{figure}

Figure~\ref{fig:releff.DR2eff.best} shows the predicted and actual REs of $\HajPL$ compared to the more efficient of $\Haj$ and $\PLtwo$.
When $Y$ is continuous, the predicted RE is greater than it was for $\HTPL$; it can be as large as 1.53.
The predicted REs are good approximations of the actual REs.
When $Y$ is binary, on the other hand, the predicted and actual REs for $\HajPL$ are very similar to those for $\HTPL$.
As was the case with $\HTPL$, the greatest REs for $\HajPL$ are seen in those of our scenarios where Samples A and B are the same size.
Supplementary Figures~\ref{fig:releff.DR2eff.Haj}--\ref{fig:releff.DR2eff.DR2Haj} show the RE of $\HajPL$ compared to its component estimators $\Haj$ and $\PLtwo$.

The RE of $\KHone$ compared to $\PLone$ and of $\KHtwo$ compared to $\PLtwo$ ranges from 0.99 to 1.03, except in one scenario when both samples are small, where the REs are 1.06 (Supplementary Figures~\ref{fig:releff.KH.DR1} and~\ref{fig:releff.KH2.DR2}).
The predicted and actual REs of $\HTKH$ compared to $\HT$ and $\KHone$  (Supplementary Figures~\ref{fig:releff.KHeff.best}--\ref{fig:releff.KHeff.KHHT}) are very similar to those of $\HTPL$ compared to $\HT$ and $\KHone$ (Figure~\ref{fig:releff.DR1eff.best} and Supplementary Figures~\ref{fig:releff.DR1eff.HT}--\ref{fig:releff.DR1eff.DR1HT}).
The same is true of $\HajKH$ and $\HajPL$ (compare Supplementary Figures~\ref{fig:releff.KH2eff.best}--\ref{fig:releff.KH2eff.KH2Haj} with Figure~\ref{fig:releff.DR2eff.best} and Supplementary Figures~\ref{fig:releff.DR2eff.Haj}--\ref{fig:releff.DR2eff.DR2Haj}).
Note that the estimators $\alphahat$ and $\betahat$ calculated using the KH method failed to converge for a small number (up to 2.5\%) of simulated datasets; see Appendix~\ref{sect:binaryKH.problem} for more details.
 
The ratio of the mean estimated SE to the empirical SE for the eight estimators of $\Ybar$ ranges from 0.94 to 1.05 (Supplementary Figures~\ref{fig:SEratio.HTHaj}--\ref{fig:SEratio.DReff}).
The coverage of 95\% confidence intervals for $\HT$, $\Haj$, $\PLone$, $\KHone$, $\PLtwo$ and $\KHtwo$ ranges from 93\% to 96.5\% (Supplementary Figures~\ref{fig:cover.HTHaj} and~\ref{fig:cover.DR}), except in four scenarios where Sample B is small, where it ranges from 91.5\% to 94\% for the DR estimators.
Coverage of confidence intervals for the combined estimators $\HTPL$, $\HTKH$, $\HajPL$ and $\HajKH$ ranges from 93\% to 96.5\% when $Y$ is continuous, and from 91.5\% to 95\% when $Y$ is binary (Supplementary Figure~\ref{fig:cover.DReff}).

\section{Application: Natsal-3}

\label{sect:natsal}

To illustrate the use of the methods in a real-world setting where both probability and nonprobability survey data on the same outcome $Y$ and covariates $X$ are available, we applied the methods to data from the third National Survey of Sexual Attitudes and Lifestyles (Natsal-3; $n=15162$, response rate 58\%), an address-based probability survey of the population aged 16--74 years in Britain carried out in 2010--2012 \cite{erens2014natsal3paf}. Nonprobability survey data were available from four online surveys targeting individuals aged 18--44 years.  These surveys were collected in 2012 and used a selection of questions from the Natsal-3 questionnaire ($n=8120$ participants in total) \cite{erens2014natsal3web,copas2020natsal3web}.

Natsal-3 and the online survey study was approved by the Oxfordshire Research Ethics Committee A (reference no.\ 09/H0604/27 and 12/SC/0070, respectively). Participants were informed about the study, use of the data, voluntary nature of participation, and right to withdraw, before proceeding with the survey. 

We used the STROBE reporting guideline\cite{elm2007strobe} and the STROBE reporting checklist\cite{elm2025checklist} when drafting and editing this manuscript, included in the Appendix.

\subsection{Methods}

As an example, we estimated the prevalence of the self-reported binary outcome ``having had sex with a same-gender partner at least once in the past five years'' in the domain (subpopulation) of men aged 18--44 years (probability sample $n=3628$, nonprobability sample $n=4074$). Because the nonprobability samples did not include individuals outside this age range, the DR estimators cannot be used to estimate the prevalence beyond this age range.  We applied the DR estimators $\PLtwo$ and $\KHtwo$, and then combined the resulting DR estimates with the corresponding Hajek estimate based on the probability sample alone, using the methods proposed in Section \ref{sect:combining}. The selection and outcome models, $\pi^B(X; \alpha)$ and $m(X; \beta)$, were fitted to the data on all participating men in the domain.
The covariates $X$ used in these models included 14 socio-demographic, sexual identity, health, attitudinal and behavioural factors (Table \ref{tab:N3covariates}).
Missing values in $X$ and $Y$ were multiply imputed using the R package mice.
In addition, we estimated the outcome prevalence in men separately in each of the (approximately) five-year age bands 18--24, 25--29, 30--34, 35--39 and 40--44 years.  For this, we fitted the nuisance models using the data on either (i) only the men in that age band, or (ii) all men aged 18--44 years.

\subsection{Results}

The Hajek estimate of the outcome prevalence based on the probability sample was 0.0323 (SE 0.0034, 95\% CI 0.0257--0.0389), which differed substantially from the prevalence in the (unweighted) nonprobability sample (0.1089, SE 0.0049, 95\% CI 0.0994--0.1185).
The nonprobability sample dataset came with calibration weights, designed to ensure that the means of basic demographic variables (age and region) in the weighted sample equal the corresponding population means.
We found that the Hajek estimate (0.1094, SE 0.0049, 95\% CI 0.0997--0.1191) of the outcome prevalence based on the nonprobability sample dataset and these calibration weights still differed substantially from the Hajek probability sample estimate.
This was also true of $\PLtwo$ when $X$ included only the basic demographic variables (0.1103, SE 0.0050, 95\% CI 0.1005--0.1201).
When $X$ included all 14 covariates, $\PLtwo$ (0.0363, SE 0.0033, 95\% CI 0.0299--0.0428) was consistent with the probability sample Hajek estimate (Figure \ref{fig:N3predictors}), as was $\KHtwo$ (0.0355, SE 0.0033, 95\% CI 0.0291--0.0420).
Inspection of the fitted nuisance models revealed that age and region were not significantly associated with either selection or outcome (conditional on the other covariates); most (11 of 14) covariates were significantly associated with selection, but only sexual identity and several of the attitudinal and behavioural covariates were significantly associated with the outcome (Table \ref{tab:N3waldtests}).
This may explain why adjustment for only the basic demographic variables was insufficient to make the estimates based on the outcome data from the nonprobability sample consistent with those from the probability sample.
The results presented in the remainder of this section were obtained using all 14 covariates.

\begin{figure}
    \centering
    \includegraphics[width=1\linewidth]{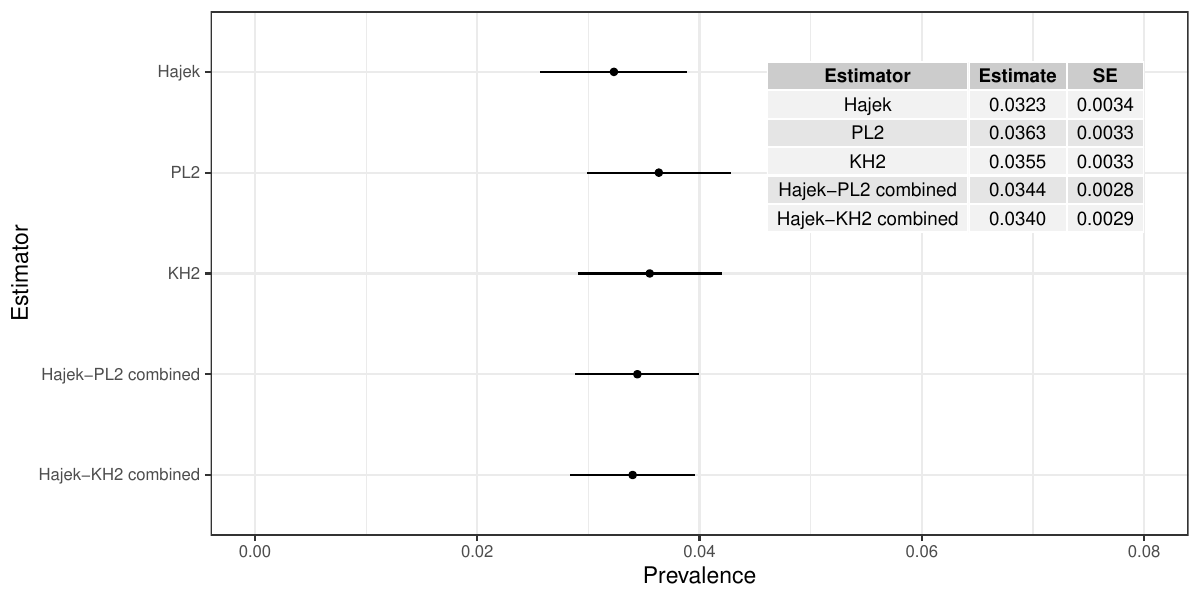}
    \caption{Application to data from the Natsal-3 study: estimated prevalences of the self-reported outcome ``having had sex with a same-gender partner at least once in the past five years'' for men aged 18-44 years in 2010-2012, by the estimators $\Haj$, $\PLtwo$, $\KHtwo$, $\HajPL$ and $\HajKH$. The error bars in the graph are 95\% confidence intervals. The table shows the point estimates and their standard errors.}
    \label{fig:N3main}
\end{figure}

The estimates from the combined estimators were $\HajPL=0.0344$ (SE 0.0028, 95\% CI 0.0288--0.0400) and $\HajKH=0.0340$ (SE 0.0029, 95\% CI 0.0284--0.0396) (Figure~\ref{fig:N3main}).
The estimated correlation between the Hajek and PL2 estimators is $\hat{\rho}=0.454$, and the estimated efficient weight given to $\PLtwo$ when combining the two estimators is $\hat{w}=0.522$.
The corresponding numbers for the combination of Hajek and KH2 estimates are $\hat{\rho}=0.483$ and $\hat{w}=0.523$. The standard errors of the $\Haj$, $\PLtwo$ and corresponding combined estimator $\HajPL$ are 0.0034, 0.0033 and 0.0028, respectively.
So, the apparent relative efficiency of the combined estimate compared to the PL2 estimate is $(0.0033 / 0.0028)^2 = 1.34$. This is consistent with the asymptotic relative efficiency predicted by the theory described in Section 5, which is also 1.34 (calculated using the estimated standard errors and correlation). 
Similarly, for the estimator that combines $\Haj$ and $\KHtwo$, the apparent and predicted asymptotic relative efficiencies compared to the $\KHtwo$ estimate are 1.32 and 1.32, respectively.

\begin{figure}
    \centering
    \includegraphics[width=1\linewidth]{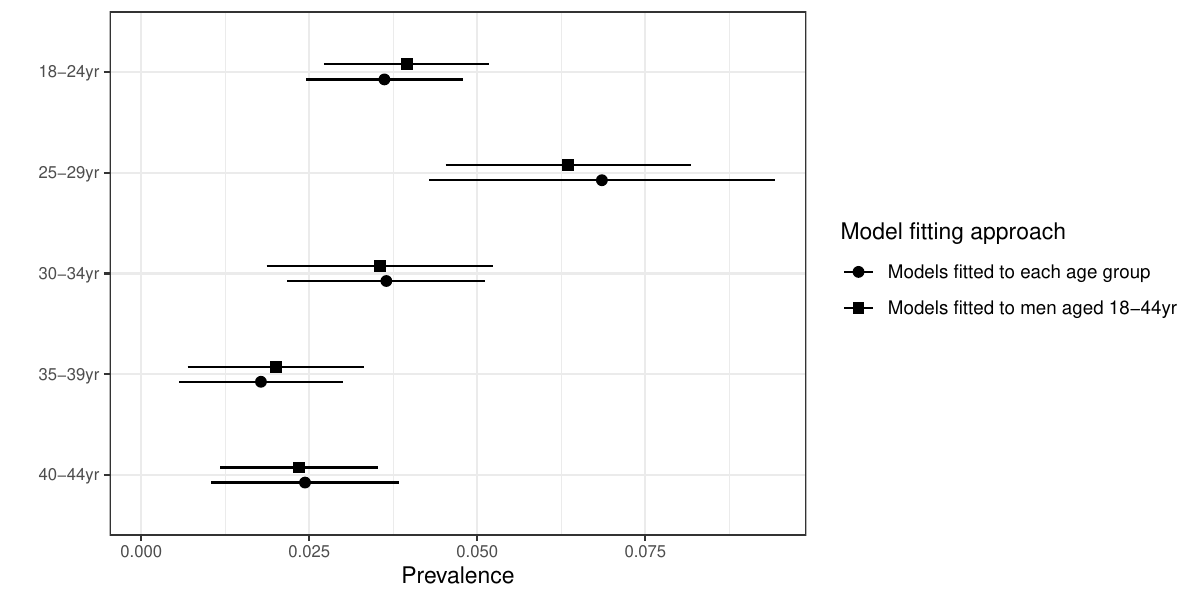}
    \caption{Application to data from the Natsal-3 study: estimated prevalences and 95\% confidence intervals of the outcome as in Figure \ref{fig:N3main}, in smaller age groups, by the approach used to fit the selection and outcome models of the $\PLtwo$ estimator.}
    \label{fig:N3agegroups}
\end{figure}

Figure \ref{fig:N3agegroups} shows $\PLtwo$ estimates for each of the age bands (the corresponding estimates combining $\PLtwo$ with $\Haj$ are given in Figure~\ref{fig:N3agegroups_combinedHajek}).
The $\PLtwo$ estimates obtained by fitting the nuisance models to only the data on that age band were similar to the corresponding estimates obtained by fitting those models to all the men aged 18--44 years. For two of the five age bands, the estimated standard errors were smaller when all men were used; in one case (25-29 years) the difference was substantial.  However, for two age bands the estimated standard errors were very similar, and for the remaining age band the estimated standard error was slightly larger when all men were used.

\section{Discussion}

\label{sect:discussion}

In this article, we have studied DR estimators that have been proposed for integrating nonprobability and probability survey data to estimate the population mean $\Ybar$ of an outcome $Y$ (which, if $Y$ is binary, is a population prevalence).
We have considered both the scenario where $Y$ is observed only on the nonprobability sample and the scenario where it is observed on both samples.
For both scenarios, we have generalised previously proposed DR estimators to enable mean estimation in a domain (i.e.\ subpopulation), allowing data on individuals outside the domain to be used when fitting the nuisance models $\pi^B (X; \alpha)$ and $m(X; \beta)$ (unless the KH method is used).
We have also proposed a Hajek-style DR estimator, $\KHtwo$, that has a double robust variance estimator and does not require the size of the domain to be known.

In the scenario where $Y$ is observed on both samples, $\Ybar$ can be estimated using a Horvitz-Thompson or Hajek estimator based on just the probability survey data or a potentially more efficient DR estimator that uses the $Y$ values from only the nonprobability survey.
We have addressed the question of how to combine efficiently these two estimators in order to make use of the data on $Y$ from both samples.
Gao and Yang~\cite{gao2023pretest} proposed methods for doing this.
However, they only provided formulae for when the DR estimator is $\DRone$, and they assumed that both nuisance models $\pi^B (X; \alpha)$ and $m(X; \beta)$ are correctly specified.
Our more general formulae allow other DR estimators to be used in place of $\DRone$ and allow model $m(X; \beta)$ to be misspecified.
As well as increasing robustness to model misspecification, this allows the use of an inverse probability weighting estimator in place of the DR estimator.
The efficient weighting of the two component estimators for the combined estimator depends on the covariance between these component estimators.
It has previously been claimed that this covariance equals zero \cite{rueda2023ipwcombined}, but we have shown that this is not true.

We have also investigated, both asymptotically and in a finite-sample simulation study, the size of the efficiency gain achieved by using the combined estimator rather than the Horvitz-Thompson/Hajek estimator or DR estimator alone.
We have shown that the relative efficiency of the combined estimator compared to the more efficient of these two component estimators is greatest when these component estimators have equal variance, but can never be greater than 200\%.
This is not surprising: if one of the component estimators is much more efficient than the other, it is implausible that making additional use of the less efficient estimator will add much efficiency.
The relative efficiency of the combined estimator also depends on how predictive the covariates $X$ are of the outcome $Y$.
The more predictive, the less efficiency can be gained by combining the component estimators.

Our analysis of Natsal-3 data corroborated these theoretical results.
Here, the apparent efficiencies were $34\%$ higher for the combined $\HajPL$ relative to the $\PLtwo$ estimator, and $32\%$ higher for the $\HajKH$ relative to the $\KHtwo$ estimator, and these apparent efficiency gains were consistent with the asymptotic efficiency gain predicted by the theory. 

Our Natsal-3 analysis has demonstrated the benefit of using covariate information available in both the probability and nonprobability samples. The Hajek estimate from the nonprobability sample does not use such information and was found to be not compatible with the Hajek estimate from the probability sample, demonstrating the non-representativeness of the nonprobability sample. The DR estimate uses this covariate information to account for this non-representativeness; it was found to be compatible with the estimate from the probability sample.
Earlier work, which used IPW rather than DR estimators, also showed the benefit of using covariate information for this outcome in Natsal-3 \cite{copas2020natsal3web}.

Our Natsal-3 analysis has also illustrated how data on individuals outside the domain can be used to improve precision of the $\PLtwo$ estimator of the domain mean, by using these data when estimating  the nuisance functions $\pi^B(X; \alpha)$ and $m(X; \beta)$.
This gain in precision was apparent for the 25--29 age band.
Notably, this is the age band with smallest number of participants in the nonprobability sample, and hence was the band for which improving precision was most needed.
However, the efficiency gain was less apparent, or not apparent, in the other age bands.
This may be because one of the properties of the DR estimator is that there is no {\it asymptotic} efficiency gain from fitting the nuisance models using data from outside the domain when both nuisance models are correctly specified.
The decision of whether to use data from outside the domain is not clear cut, because using such data does increase the risk of misspecifying the nuisance models (e.g.\ by not including required interaction terms).
Nevertheless, when the number of sampled individuals in the domain is small, it may be impractical to fit the nuisance models without using such data.

We now discuss possible directions for future research.

First, the methods we have described assume that after conditioning on covariates $X$ the probability of participation in the nonprobability sample does not depend on $Y$ (i.e.\ $R^B$ is independent of $Y$ given $X$).
In the scenario where $Y$ is observed only in the nonprobability sample, Liu et al.\ (2026)~\cite{LiuYuanLiWu2026} propose a DR method that allows for such dependence.
They identify the parameters in their logistic regression model for $P(R^B=1 \mid X, Y)$ by using an instrumental variable, i.e.\ an element of $X$ that is assumed to be conditionally independent of $R^B$ given $Y$ and the remaining elements of $X$.
Once more assuming that $R^B$ is independent of $Y$ given $X$, Chen et al.\ (2022)~\cite{ChenLiRaoWu2022} propose a DR method that uses maximum empirical likelihood. 
Their simulation study demonstrates improved CI coverage compared to $\DRtwo$ when the probability and nonprobability samples are both very small and single-stage sampling is used for the probability sample.
It would be of interest to extend these methods to the scenario where $Y$ is observed in both samples.

Second, in the scenario where $Y$ is observed on both samples, Gao and Yang (2023)~\cite{gao2023pretest} propose a method for testing for consistency between the Hajek estimate that uses only the probability sample data and the $\PLone$ estimate that uses only the outcome data from the nonprobability sample before deciding whether to combine these estimates.
The resulting estimator of $\Ybar$ is designed to minimise the asymptotic mean squared error under the assumption that nuisance models for $P(R^B = 1 \mid X)$ and $E(Y \mid X, R^B=1)$ are correctly specified.
In Appendix Section~\ref{sect:GaoYang}, we provide a detailed and accessible description of this method, a description which also covers the use of the Horvitz-Thompson, $\KHone$ and $\KHtwo$ estimators.
In further work, it would interesting to investigate the performance of this approach.

Third, Chen et al.\ (2020) and Gao and Yang (2023) assume there is no unit non-response in Sample A, i.e.\ $Y_i$ is always observed when $R^A_i = 1$, and that there is no post-stratification of weights $1 / \pi^A_i$.
In our article, we have followed this example.
However, in Appendix Section~\ref{sect:nonresponse.poststrat}, we relax these assumptions.
There we assume the Response Homogeneity Model of Sarndal et al.\ (1992)\cite{sarndal1992surveysamplingbook} and show how the DR and combined estimators can be used when there is unit non-response and/or post-stratification.
Further work is needed to assess the performance of these extended estimators.

Fourth, the methods we have described use parametric nuisance models.
In order to reduce the risk of misspecification of these models, Cobo et al.\ (2025)~\cite{Cobo2025} use data-adaptive estimators of $\pi^B$ and $m$, such as extreme gradient boosting, in a scenario where $Y$ is observed only in the nonprobability sample.
However, they note that ``there are hardly any results on the theoretical properties of variance estimators when more complex machine learning models and techniques are used, as in our case.''
Seaman (2025) \cite{Seaman2025dataadaptive} reviews methods that use data-adaptive estimators of the nuisance functions, and develops theory to justify the use of general data-adaptive nuisance function estimators with DR estimators of $\Ybar$ when $Y$ is observed only in the nonprobability sample.
When $Y$ is also observed in the probability sample, this data-adaptive approach can be used together with the combining method described in Section~\ref{sect:combining} of the current article.

Considering the current trend towards increased availability and ease of collecting nonprobability survey data~\cite{cornesse2020review}, methods that enable consistent and efficient estimates by combining nonprobability and probability survey data, such as those we have presented and expanded on, are of increasing importance. 
Our extensions to existing combination methods, to (a) enable combining either Horvitz-Thompson or Hajek estimators with the corresponding DR estimators, (b) demonstrate under which scenarios substantial efficiency gains can be achieved, and (c) allow subpopulation (domain) estimation, including when there are limited data on that domain, will enable greater uptake and practical use of these combination methods in the survey statistics field.

\section*{Acknowledgements}

We thank David Hussey at the National Centre for Social Research and Daniela De Angelis at the MRC Biostatistics Unit at University of Cambridge for helpful discussions about this work.

\section*{Funding}

This work is supported by the National Survey of Sexual Attitudes and Lifestyles (Natsal) grant from the Wellcome Trust (212931/Z/18/Z), with contributions from the Economic and Social Research Council (ESRC) and the National Institute for Health Research (NIHR).

\section*{Declaration of conflicting interest} 
No potential conflicts of interest.

\bibliographystyle{plain}
\bibliography{refsDRCombining.bib}

\begin{thebibliography}{10}

\bibitem{BeaumontEmond2022}
J-F Beaumont and N~Emond.
\newblock A bootstrap variance estimation method for multistage sampling and two-phase sampling when {P}oisson sampling is used at the second phase.
\newblock {\em Stat}, 5:339--357, 2022.

\bibitem{beaumont2020nonprobsurveys}
Jean-Francois Beaumont.
\newblock Are probability surveys bound to disappear for the production of official statistics?
\newblock {\em Survey Methodology}, 46(1):1--29, 2020.

\bibitem{chauvet2020samplingdesigns}
Guillaume Chauvet and Audrey-Anne Vall{\'e}e.
\newblock Inference for two-stage sampling designs.
\newblock {\em Journal of the Royal Statistical Society Series B: Statistical Methodology}, 82(3):797--815, 2020.

\bibitem{ChenLiRaoWu2022}
Y~Chen, P~Li, JNK Rao, and C~Wu.
\newblock Pseudo empirical likelihood inference for nonprobability survey samples.
\newblock {\em Canadian Journal of Statistics}, 50:1166--1185, 2022.

\bibitem{chen2020doublyrobust}
Yilin Chen, Pengfei Li, and Changbao Wu.
\newblock Doubly robust inference with nonprobability survey samples.
\newblock {\em Journal of the American Statistical Association}, 115(532):2011--2021, 2020.

\bibitem{chrostowski2025nonprobsvyPreprint}
{\L}ukasz Chrostowski, Piotr Chlebicki, and Maciej Beresewicz.
\newblock \texttt{nonprobsvy}--{A}n \texttt{R} package for modern methods for non-probability surveys.
\newblock {\em arXiv preprint}, arXiv:2504.04255, 2025.

\bibitem{Cobo2025}
B~Cobo, JL~Rueda-Sanchez, , R~Ferri-Garc\`{i}a, and M~{del Mar Rueda}.
\newblock A new technique for handling non-probability samples based on model-assisted kernel weighting.
\newblock {\em Mathematics and Computers in Simulation}, 227:272--281, 2025.

\bibitem{copas2020natsal3web}
Andrew Copas, Sarah Burkill, Fred Conrad, Mick~P Couper, and Bob Erens.
\newblock An evaluation of whether propensity score adjustment can remove the self-selection bias inherent to web panel surveys addressing sensitive health behaviours.
\newblock {\em BMC Medical Research Methodology}, 20:1--10, 2020.

\bibitem{cornesse2020review}
Carina Cornesse, Annelies~G Blom, David Dutwin, Jon~A Krosnick, Edith~D De~Leeuw, Stéphane Legleye, Josh Pasek, Darren Pennay, Benjamin Phillips, Joseph~W Sakshaug, Bella Struminskaya, and Alexander Wenz.
\newblock A review of conceptual approaches and empirical evidence on probability and nonprobability sample survey research.
\newblock {\em Journal of Survey Statistics and Methodology}, 8(1):4--36, February 2020.

\bibitem{elm2007strobe}
Erik~von Elm, Douglas~G. Altman, Matthias Egger, SJ~Pocock, PC~Gotzsche, JP~Vandenbroucke, and {for the STROBE Initiative}.
\newblock The {Strengthening the Reporting of Observational Studies in Epidemiology (STROBE)} statement: {G}uidelines for reporting observational studies.
\newblock {\em Annals of Internal Medicine}, 147(8):573--577, 2007.

\bibitem{elm2025checklist}
Erik~von Elm, Douglas~G. Altman, Matthias Egger, SJ~Pocock, PC~Gotzsche, JP~Vandenbroucke, and {for the STROBE Initiative}.
\newblock {The STROBE reporting checklist. In: Harwood J, Albury C, Beyer J de, Schlüssel M, Collins G, editors. The EQUATOR network reporting guideline platform}.
\newblock Technical report, {The UK EQUATOR Centre}, 2025.

\bibitem{erens2014natsal3web}
Bob Erens, Sarah Burkill, Mick~P Couper, Frederick Conrad, Soazig Clifton, Clare Tanton, Andrew Phelps, Jessica Datta, Catherine~H Mercer, Pam Sonnenberg, et~al.
\newblock Nonprobability web surveys to measure sexual behaviors and attitudes in the general population: a comparison with a probability sample interview survey.
\newblock {\em Journal of Medical Internet Research}, 16(12):e276, 2014.

\bibitem{erens2014natsal3paf}
Bob Erens, Andrew Phelps, Soazig Clifton, Catherine~H Mercer, Clare Tanton, David Hussey, Pam Sonnenberg, Wendy Macdowall, Nigel Field, Jessica Datta, et~al.
\newblock {Methodology of the third British national survey of sexual attitudes and lifestyles (Natsal-3)}.
\newblock {\em Sexually Transmitted Infections}, 90(2):84--89, 2014.

\bibitem{gao2023pretest}
Chenyin Gao and Shu Yang.
\newblock Pretest estimation in combining probability and non-probability samples.
\newblock {\em Electronic Journal of Statistics}, 17(1):1492--1546, 2023.

\bibitem{kim2014doublyrobust}
Jae~Kwang Kim and David Haziza.
\newblock Doubly robust inference with missing data in survey sampling.
\newblock {\em Statistica Sinica}, 24(1):375--394, 2014.

\bibitem{kim2019surveyanalysis}
Jae~Kwang Kim and Zhonglei Wang.
\newblock Sampling techniques for big data analysis.
\newblock {\em International Statistical Review}, 87:S177--S191, 2019.

\bibitem{LiuYuanLiWu2026}
Y~Liu, M~Yuan, P~Li, and C~Wu.
\newblock Statistical inference with nonignorable non-probability survey samples.
\newblock {\em Electronic Journal of Statistics}, 20:2009--2048, 2026.

\bibitem{molina2001complexsurveys}
EA~Molina, TMF Smith, and RA~Sugden.
\newblock Modelling overdispersion for complex survey data.
\newblock {\em International Statistical Review}, 69(3):373--384, 2001.

\bibitem{rueda2023ipwcombined}
Mar{\'\i}a del~Mar Rueda, Sara Pasadas-del Amo, Beatriz~Cobo Rodr{\'\i}guez, Luis Castro-Mart{\'\i}n, and Ram{\'o}n Ferri-Garc{\'\i}a.
\newblock {Enhancing estimation methods for integrating probability and nonprobability survey samples with machine-learning techniques. An application to a Survey on the impact of the COVID-19 pandemic in Spain}.
\newblock {\em Biometrical Journal}, 65(2):2200035, 2023.

\bibitem{sarndal1992surveysamplingbook}
Carl-Erik S{\"a}rndal, Bengt Swensson, and Jan Wretman.
\newblock {\em Model assisted survey sampling}.
\newblock Springer-Verlag, New York, NY, 1992.

\bibitem{seaman2018doublyrobustgeneralreference}
Shaun~R Seaman and Stijn Vansteelandt.
\newblock Introduction to double robust methods for incomplete data.
\newblock {\em Statistical Science}, 33(2):184, 2018.

\bibitem{Seaman2025dataadaptive}
SR~Seaman.
\newblock Debiased machine learning for combining probability and non-probability survey data.
\newblock {\em arXiv preprint}, arXiv:2508.08948v2, 2025.

\bibitem{ShaoTu2012}
J~Shao and D~Tu.
\newblock {\em The Jackknife and Bootstrap}, pages 247--248.
\newblock Springer, 2012.

\bibitem{wu2022nonprobsurveys}
Changbao Wu.
\newblock Statistical inference with non-probability survey samples.
\newblock {\em Survey Methodology}, 48(2):283--311, 2022.

\bibitem{yang2020doublyrobust}
Shu Yang, Jae~Kwang Kim, and Rui Song.
\newblock Doubly robust inference when combining probability and non-probability samples with high dimensional data.
\newblock {\em Journal of the Royal Statistical Society Series B: Statistical Methodology}, 82(2):445--465, 2020.

\end{thebibliography}

\newpage

\appendix
\renewcommand\thesection{A\arabic{section}}
\renewcommand\thetable{A\arabic{table}}
\renewcommand\thefigure{A\arabic{figure}}    
\renewcommand\theequation{A\arabic{equation}}    
\setcounter{table}{0}    
\setcounter{figure}{0}    
\setcounter{equation}{0}

\section*{Appendices}

Let $\HTHaj$ denote either $\HT$ or $\Haj$, and define $\Gamma = 0$ if $\HTHaj = \HT$ and $\Gamma = \Ybar$ if $\HTHaj = \Haj$.

As stated in the article, we assume that the conditional variances given $\curlF$ and $\curlY$ of the Horvitz-Thompson estimators~\cite{sarndal1992surveysamplingbook} of the domain means of $Y - \Gamma$ and $m(X; \beta) - \Delta_{mi}$, where $\Gamma$ equals $\Ybar$ or zero and $\Delta_{mi}$ is defined in Section~\ref{sect:designbased} of the main text, are $O_p(N^{-1})$, i.e.\
\begin{eqnarray}
  \Var \left\{ \frac{1}{N_s} \sum_{i=1}^N
    S_i \frac{ R^A_i }{ \pi^A_i } ( Y_i - \Gamma ) \mid \curlF, \curlY \right\}
  & = &
  O_p (N^{-1})
  \label{eq:HT.y.Op} \\
  \Var \left[ \frac{1}{N_s} \sum_{i=1}^N
    S_i \frac{ R^A_i }{ \pi^A_i } \{ m (X_i; \betalim) - \Delta_m \} \mid \curlF, \curlY \right]
  & = &
  O_p (N^{-1}).
 \label{eq:HT.m.Op}
\end{eqnarray}

\section{Design-based inference for $\PLone$}

\label{sect:var.PL1}

Suppose that model $\pi^B (X; \alpha)$ is correctly specified.
Using a Taylor series expansion of $\PLone (\alphahat, \betahat)$ about $(\alphalim, \betalim)$, we obtain
\begin{eqnarray}
  \PLone (\alphahat, \betahat)
  & = &
  \PLone (\alphalim, \betalim)
  + \left. \frac{ \partial \PLone (\alpha, \betalim)^\top }{ \partial \alpha } \right|_{\alpha=\alphalim}
  (\alphahat - \alphalim)
  \nonumber \\
  && +
  \left. \frac{ \partial \PLone (\alphalim, \beta)^\top }{ \partial \beta } \right|_{\beta=\betalim}
  (\betahat - \betalim)
  \; + o_p(N^{-1/2})
  \nonumber \\
  & = &
  \PLone (\alphalim, \betalim)
  + E \left\{ \left. \frac{ \partial \PLone (\alpha, \betalim)^\top }{ \partial \alpha } \right|_{\alpha=\alphalim} \mid \curlF, \curlY \right\}
  (\alphahat - \alphalim)
  \nonumber \\
  && +
  E \left\{ \left. \frac{ \partial \PLone (\alphalim, \beta)^\top }{ \partial \beta } \right|_{\beta=\betalim}
  \mid \curlF \right\}
  (\betahat - \betalim)
  \; + o_p(N^{-1/2})
  \nonumber
  \\
  & = &
  \PLone (\alphalim, \betalim)
  \nonumber \\
  && -
  E \left[ \frac{1}{N_s} \sum_{i=1}^N S_i R^B_i \frac{ 1 - \pi^B (X_i; \alphalim) }{ \pi^B (X_i; \alphalim) } \{ Y_i - m(X_i; \betalim) \} \Xpi \mid \curlF, \curlY \right]^\top
  (\alphahat - \alphalim)
  \nonumber \\
  && +
  E \left[ \frac{1}{N_s} \sum_{i=1}^N S_i \left\{ \frac{R^A_i}{\pi^A_i} - \frac{ R^B_i }{ \pi^B (X_i; \alphalim) } \right\}
    \left. \frac{ \partial m (X_i; \beta) }{ \partial \beta^\top } \right|_{\beta=\betalim} \mid \curlF \right]^\top
  (\betahat - \betalim)
  \nonumber \\
  && +
  o_p(N^{-1/2}).
  \label{eq:DR1expansion.starts}
\end{eqnarray}
Because model $\pi^B(X; \alpha)$ is correctly specified,
\begin{equation}
  E \left[ \frac{1}{N_s} \sum_{i=1}^N S_i \left\{ \frac{R^A_i}{\pi^A_i} - \frac{ R^B_i }{ \pi^B (X_i; \alphalim) } \right\}
    \left. \frac{ \partial m (X_i; \beta) }{ \partial \beta^\top } \right|_{\beta=\betalim}
    \mid \curlF \right]
  = 0.
  \label{eq:uses.C4}
\end{equation}
Hence, because $E(R^B \mid \curlF) = \pi^B (X; \alpha_0)$, equation~(\ref{eq:DR1expansion.starts}) implies
\begin{eqnarray}
  \PLone (\alphahat, \betahat)
  & = &
  \PLone (\alphalim, \betalim)
  -
  \left[ \frac{1}{N_s} \sum_{i=1}^N S_i \{ 1 - \pi^B (X_i; \alphalim) \} \{ Y_i - m(X_i; \betalim) \} \Xpi \right]^\top (\alphahat - \alphalim)
  \nonumber \\
  && +
  o_p(N^{-1/2}).
  \label{eq:DR1est.DR1true}
\end{eqnarray}
The maximum pseudo likelihood estimator of $\alpha$ solves the estimating equations
\begin{equation}
  U_{\alpha} (\alpha) \equiv \frac{1}{N} \sum_{i=1}^N
  \usepi_i \left\{
  R^B_i \Xpi - \frac{ R^A_i }{ \pi^A_i } \; \pi^B (X_i; \alpha) \; \Xpi
  \right\}
= 0.
\label{eq:pseudo.like.score}
\end{equation}
So,
\[
\left. \frac{ \partial U_{\alpha} (\alphalim) }{ \partial \alpha } \right|_{\alpha = \alphalim}
= - \frac{1}{N} \sum_{i=1}^N \usepi_i
  \frac{ R^A_i }{ \pi^A_i } \pi^B (X_i; \alpha_0) \; \{ 1 - \pi^B (X_i; \alpha_0) \} \; \Xpi \Xpitop
\]
and thus
\[
E \left\{ \left. \frac{ \partial U_{\alpha} (\alphalim) }{ \partial \alpha } \right|_{\alpha = \alphalim}
  \curlF, \curlY \right\}
= - \frac{1}{N} \sum_{i=1}^N \usepi_i \; \pi^B (X_i; \alpha_0) \; \{ 1 - \pi^B (X_i; \alpha_0) \} \; \Xpi \Xpitop.
\]
Hence, a Taylor series expansion of $U_{\alpha} (\alpha)$ around $\alphalim$ gives
\begin{eqnarray}
  \alphahat - \alphalim
  & = &
  - \left\{ \frac{1}{N} \sum_{i=1}^N \usepi_i \; \pi^B (X_i; \alphalim) \{ 1 - \pi^B (X_i; \alphalim) \} \Xpi \Xpitop \right\}^{-1}
  \nonumber \\
  && \times
  \frac{1}{N} \sum_{i=1}^N \usepi_i \left\{ R^B_i - \frac{ R^A_i }{ \pi^A_i } \pi^B (X_i; \alphalim) \right\} \Xpi
  + o_p(N^{-1/2}).
  \label{eq:alphahat.alphalim}
\end{eqnarray}

Define
\begin{equation}
\bfour =
  \left[ \sum_{i=1}^N \usepi_i \; \pi^B (X_i) \{ 1 - \pi^B (X_i) \} \; \Xpi \Xpitop \right]^{-1}
  \sum_{i=1}^N S_i \{ 1 - \pi^B (X_i) \} \{ Y_i - m(X_i; \betalim) \} \Xpi.
 \label{eq:b1.definition}
\end{equation}

Now, from equations~(\ref{eq:DR1est.DR1true}), (\ref{eq:alphahat.alphalim}) and~(\ref{eq:b1.definition}), we have
\begin{eqnarray}
  &&
  \PLone (\alphahat, \betahat) - \Ybar
  \nonumber \\
  && \hspace{.3cm} =
  \PLone (\alphalim, \betalim)
  - \bfour^\top \frac{1}{N_s} \sum_{i=1}^N \usepi_i \left\{ R^B_i - \frac{ R^A_i }{ \pi^A_i } \pi^B (X_i) \right\} \Xpi
  - \Ybar + o_p(N^{-1/2})
  \nonumber \\
  && \hspace{.3cm} =
    \frac{1}{N_s} \sum_{i=1}^N S_i
  \frac{ R^A_i }{ \pi^A_i } m(X_i; \betalim)
  + \frac{1}{N_s} \sum_{i=1}^N S_i
  \frac{ R^B_i }{ \pi^B (X_i) } \{ Y_i - m(X_i; \betalim) \}
  \nonumber \\
  && \hspace{.8cm} -
  \frac{1}{N_s} \sum_{i=1}^N \usepi_i \left\{ R^B_i - \frac{ R^A_i }{ \pi^A_i } \pi^B (X_i) \right\} \bfour^\top \Xpi
  - \Ybar + o_p(N^{-1/2})
  \nonumber \\
  && \hspace{.3cm} =
    \frac{1}{N_s} \sum_{i=1}^N
  \frac{ R^A_i }{ \pi^A_i } \left\{ S_i m(X_i; \betalim) + \usepi_i \; \pi^B (X_i) \; \bfour^\top \Xpi \right\}
  \nonumber \\
  && \hspace{.8cm} +
  \frac{1}{N_s} \sum_{i=1}^N
  \frac{ R^B_i }{ \pi^B (X_i) } \left[ S_i \{ Y_i - m(X_i; \betalim) \} - \usepi_i \; \pi^B (X_i) \; \bfour^\top \Xpi \right]
  - \Ybar + o_p(N^{-1/2})
  \nonumber \\
  && \hspace{.3cm} =
    \frac{1}{N_s} \sum_{i=1}^N
  \frac{ R^A_i }{ \pi^A_i } \left\{ S_i m(X_i; \betalim) + \usepi_i \; \pi^B (X_i) \; \bfour^\top \Xpi \right\}
  \nonumber \\
  && \hspace{.8cm} +
  \frac{1}{N_s} \sum_{i=1}^N
  \left\{ \frac{ R^B_i }{ \pi^B (X_i) } - 1 \right\}
  \left[ S_i \{ Y_i - m(X_i; \betalim) \} - \usepi_i \; \pi^B (X_i) \; \bfour^\top \Xpi \right]
  \nonumber \\
  && \hspace{.8cm} -
  \mbar - \bfour^\top 
  \frac{1}{N_s} \sum_{i=1}^N \usepi_i \; \pi^B (X_i) \; \Xpi
  + o_p(N^{-1/2}).
  \label{eq:DRone.Taylor.general}
\end{eqnarray}

It now follows that
\begin{eqnarray}
  &&
  \Var \{ \PLone (\alphahat, \betahat) - \Ybar \mid \curlF, \curlY \}
  \nonumber \\
  && \hspace{.3cm} =
  \Var \left(
      \frac{1}{N_s} \sum_{i=1}^N 
      \frac{ R^A_i }{ \pi^A_i } \left\{ S_i m(X_i; \betalim) + \usepi_i \; \pi^B (X_i) \; \bfour^\top \Xpi \right\}
      \right.
  \nonumber \\
  && \hspace{1.6cm} + \left.
  \frac{1}{N_s} \sum_{i=1}^N
  \left\{ \frac{ R^B_i }{ \pi^B (X_i) } - 1 \right\}
  \left[ S_i \{ Y_i - m(X_i; \betalim) \} - \usepi_i \; \pi^B (X_i) \; \bfour^\top \Xpi \right]
  \mid \curlF, \curlY \right)
  + o_p(N^{-1})
  \nonumber \\
  && \hspace{.3cm} =
  \Var \left[
      \frac{1}{N_s} \sum_{i=1}^N 
      \frac{ R^A_i }{ \pi^A_i } \left\{ S_i m(X_i; \betalim) + \usepi_i \; \pi^B (X_i) \; \bfour^\top \Xpi \right\}
    \mid \curlF, \curlY \right]
  \nonumber \\
  && \hspace{.8cm} +
  \Var \left(
    \frac{1}{N_s} \sum_{i=1}^N
    \left\{ \frac{ R^B_i }{ \pi^B (X_i) } - 1 \right\}
  \left[ S_i \{ Y_i - m(X_i; \betalim) \} - \usepi_i \; \pi^B (X_i) \; \bfour^\top \Xpi \right]
    \mid \curlF, \curlY \right)
  \nonumber \\
  && \hspace{.8cm} +
  o_p (N^{-1})
  \label{eq:use.zero.covariance} \\
  && \hspace{.3cm} =
  \Var \left[
      \frac{1}{N_s} \sum_{i=1}^N 
      \frac{ R^A_i }{ \pi^A_i } \left\{ S_i m(X_i; \betalim) + \usepi_i \; \pi^B (X_i) \; \bfour^\top \Xpi  \right\} \mid \curlF, \curlY
    \right]
  \nonumber \\
  && \hspace{.8cm} +
    \frac{1}{N_s^2} \sum_{i=1}^N
    \frac{ 1 - \pi^B (X_i) }{ \pi^B (X_i) } \;
    \left[ S_i \{ Y_i - m(X_i; \betalim) \} - \usepi_i \; \pi^B (X_i) \; \bfour^\top \Xpi \right]^2
    + o_p (N^{-1}).
    \label{eq:DR1.general}
\end{eqnarray}
Note that line~(\ref{eq:use.zero.covariance}) follows because the independence of $R^A_i$ and $R^B_j$ given $\curlF$ (and $\curlY$) implies
\begin{eqnarray*}
  &&
\Cov \left(
  \frac{1}{N_s} \sum_{i=1}^N
  \frac{ R^A_i }{ \pi^A_i } \{ S_i m(X_i; \betalim) + \usepi_i \; \pi^B (X_i) \; \bfour^\top \Xpi \},
  \right.
  \nonumber \\
  && \hspace{0.9cm} \left.
  \frac{1}{N_s} \sum_{i=1}^N
  \left\{ \frac{ R^B_i }{ \pi^B (X_i) } - 1 \right\}
  [S_i \{ Y_i - m(X_i; \betalim) \} - \usepi_i \; \pi^B (X_i) \; \bfour^\top \Xpi]
  \mid \curlF, \curlY \right)
=0.
\end{eqnarray*}

The second term in expression~(\ref{eq:DR1.general}) can be estimated using
\begin{equation}
  \frac{1}{N_s^2} \sum_{i=1}^N R^B_i \; 
  \frac{ 1 - \pi^B (X_i; \alphahat) } { \{ \pi^B (X_i; \alphahat) \}^2 } \; [ S_i \{ Y_i - m (X_i; \betahat) \} - \usepi_i \; \pi^B (X_i; \alphahat) \; \bfourhat^\top \Xpi ]^2
  \label{eq:second.term}
\end{equation}
where
\begin{eqnarray}
  \bfourhat
  & = &
  \left[ \sum_{i=1}^N \usepi_i \; R^B_i \;
  \{ 1 - \pi^B (X_i; \alphahat) \} \; \Xpi \Xpitop \right]^{-1}
  \nonumber \\
  && \hspace{0.3cm} \times
  \sum_{i=1}^N S_i \frac{ R^B_i }{ \pi^B (X_i; \alphahat) } \;
    \{ 1 - \pi^B (X_i; \alphahat) \} \{ Y_i - m(X_i; \betahat) \} \;
    \Xpi.
    \label{eq:b4.estimate}
\end{eqnarray}
Note that $\bfourhat$ is a consistent estimator of $\bfour$, because we are assuming model $\pi^B (X; \alpha)$ is correctly specified.

If model $m(X; \beta)$ is correctly specified, then $\bfour = o_p(1)$.
If we also assume the regularity conditions that $N_s^{-1} \sum_{i=1}^N \usepi_i R^B_i \Xpi = O_p (N^{-1/2})$ and $N_s^{-1} \sum_{i=1}^N \usepi_i R^A_i \pi^B (X_i) \Xpi / \pi^A_i = O_p (N^{-1/2})$, then equation~(\ref{eq:DRone.Taylor.general}) reduces to
\begin{eqnarray}
  &&
  \PLone (\alphahat, \betahat) - \Ybar
  \nonumber \\
  && \hspace{.3cm} =
  \frac{1}{N_s} \sum_{i=1}^N
  \frac{ R^A_i }{ \pi^A_i } S_i m(X_i; \betalim)
  +
  \bfour^\top \frac{1}{N_s} \sum_{i=1}^N
  \usepi_i \frac{ R^A_i }{ \pi^A_i } \pi^B (X_i) \Xpi
  \nonumber \\
  && \hspace{.8cm} +
  \frac{1}{N_s} \sum_{i=1}^N
  \left\{ \frac{ R^B_i }{ \pi^B (X_i) } - 1 \right\}
  S_i \{ Y_i - m(X_i; \betalim) \}
  -
  \bfour^\top \frac{1}{N_s} \sum_{i=1}^N
  \usepi_i \frac{ R^B_i }{ \pi^B (X_i) }
  \pi^B (X_i) \; \Xpi
  \nonumber \\
  && \hspace{.8cm} -
  \mbar
  + o_p(N^{-1/2})
  \nonumber \\
  && \hspace{.3cm} =
  \frac{1}{N_s} \sum_{i=1}^N
  \frac{ R^A_i }{ \pi^A_i } S_i m(X_i; \betalim)
  +
  o_p (1) \times O_p (N^{-1/2})
  \nonumber \\
  && \hspace{.8cm} +
  \frac{1}{N_s} \sum_{i=1}^N
  \left\{ \frac{ R^B_i }{ \pi^B (X_i) } - 1 \right\}
  S_i \{ Y_i - m(X_i; \betalim) \}
  -
  o_p (1) \times O_p (N^{-1/2})
  \nonumber \\
  && \hspace{.8cm} -
  \mbar
  + o_p(N^{-1/2})
  \nonumber \\
  && \hspace{.3cm} =
  \frac{1}{N_s} \sum_{i=1}^N
  \frac{ R^A_i }{ \pi^A_i } S_i m(X_i; \betalim)
  +
  \frac{1}{N_s} \sum_{i=1}^N
  \left\{ \frac{ R^B_i }{ \pi^B (X_i) } - 1 \right\}
  S_i \{ Y_i - m(X_i; \betalim) \}
  \nonumber \\
  && \hspace{.8cm} -
  \mbar
  + o_p(N^{-1/2})
  \label{eq:DR1.expansion.bothcorrect}
\end{eqnarray}
and hence equation~(\ref{eq:DR1.general}) reduces to
\begin{eqnarray}
  \Var \{ \PLone (\alphahat, \betahat) - \Ybar \mid \curlF, \curlY \}
  & = &
  \Var \left\{
      \frac{1}{N_s} \sum_{i=1}^N 
      S_i \frac{ R^A_i }{ \pi^A_i } m(X_i; \betalim) \mid \curlF, \curlY
    \right\}
  \nonumber \\
  && +
  \frac{1}{N_s^2} \sum_{i=1}^N
  S_i \frac{ 1 - \pi^B (X_i) }{ \pi^B (X_i) } \;
  \{ Y_i - m(X_i; \betalim) \}^2 \nonumber \\
  && + o_p (N^{-1}).
\label{eq:DR1.variance.bothcorrect}
\end{eqnarray}
The second term can be estimated as
\begin{equation*}
  \frac{1}{N_s^2} \sum_{i=1}^N S_i \; R^B_i \; 
  \frac{ 1 - \pi^B (X_i; \alphahat) } { \{ \pi^B (X_i; \alphahat) \}^2 } \; \{ Y_i - m (X_i; \betahat) \}^2.
\end{equation*}

Finally, consider the covariance between $\PLone$ and $\HTHaj$.
Using Taylor linearisation (see, for example, page 178 of~\cite{sarndal1992surveysamplingbook} or Appendix~\ref{sect:var.PL2}), we can write the Hajek estimator as
\begin{equation}
\Haj = \Ybar + \frac{1}{N_s} \sum_{i=1}^N S_i \frac{ R^A_i }{ \pi^A_i } (Y_i - \Ybar) + O_p(N^{-1}).
\label{eq:Taylor.linearisation}
\end{equation}
Hence, if Sample A uses sampling without replacement,
\begin{equation*}
\Var (\Haj \mid \curlF, \curlY)
=
\frac{1}{N_s^2} \sum_{i=1}^N \sum_{j=1}^N
S_i S_j (\pi^A_{ij} - \pi^A_i \pi^A_j) \; \frac{ (Y_i - \Ybar) }{ \pi^A_i } \frac{ (Y_j - \Ybar) }{ \pi^A_j }
+ o_p(N^{-1}).
\end{equation*}
This can be estimated using standard methods.
For example, if $\pi^A_{ij}>0$ for all $i$ and $j$, we can use
\begin{equation*}
\frac{1}{N_s^2} \sum_{i=1}^N \sum_{j=1}^N
\frac{ R^A_i R^A_j }{ \pi^A_{ij} } S_i S_j (\pi^A_{ij} - \pi^A_i \pi^A_j) \frac{ (Y_i - \Haj) }{ \pi^A_i } \frac{ (Y_j - \Haj) }{ \pi^A_j }.
\end{equation*}

Using equations~(\ref{eq:DRone.Taylor.general}) and (\ref{eq:Taylor.linearisation}), we have
\begin{eqnarray}
  &&
  \Cov ( \PLone - \Ybar, \; \HTHaj - \Ybar \mid \curlF, \curlY )
  \nonumber \\
  && =
  \Cov \left(
    \frac{1}{N_s} \sum_{i=1}^N
    \frac{ R^A_i }{ \pi^A_i } \left\{ S_i m(X_i; \betalim) + \usepi_i \; \pi^B (X_i) \; \bfour^\top \Xpi \right\}
    \right.
\nonumber \\
&& \hspace{1.3cm} +
\left.
\frac{1}{N_s} \sum_{i=1}^N
\left\{ \frac{ R^B_i }{ \pi^B (X_i) } - 1 \right\}
\left[ S_i \{ Y_i - m(X_i; \betalim) \} - \usepi_i \; \pi^B (X_i) \; \bfour^\top \Xpi \right]
    \right.
    + o_p(N^{-1/2}) \; , \;
\nonumber \\
  && \hspace{1.3cm} \left.
    \frac{1}{N_s} \sum_{i=1}^N
    S_i \frac{ R^A_i }{ \pi^A_i } (Y_i - \Gamma)
    + o_p(N^{-1/2})
    \mid \curlF, \curlY \right)
\nonumber \\
  && =
  \Cov \left[
    \frac{1}{N_s} \sum_{i=1}^N
    \frac{ R^A_i }{ \pi^A_i } \left\{ S_i m(X_i; \betalim) + \usepi_i \; \pi^B (X_i) \; \bfour^\top \Xpi \right\} , \; 
    \frac{1}{N_s} \sum_{i=1}^N
    S_i \frac{ R^A_i }{ \pi^A_i } (Y_i - \Gamma)
    \mid \curlF, \curlY \right]
\nonumber \\
  && \hspace{0.5cm} +
  \Cov \left[
    \frac{1}{N_s} \sum_{i=1}^N
    \left\{ \frac{ R^B_i }{ \pi^B (X_i) } - 1 \right\}
    \left[ S_i \{ Y_i - m(X_i; \betalim) \} - \usepi_i \; \pi^B (X_i) \; \bfour^\top \Xpi \right], \; \right. \nonumber \\
   && \hspace{1.3cm} \left.
    \frac{1}{N_s} \sum_{i=1}^N
    S_i \frac{ R^A_i }{ \pi^A_i } (Y_i - \Gamma)
    \mid \curlF, \curlY \right]
\nonumber \\
&& \hspace{0.5cm} +
  \Cov \left\{
    o_p(N^{-1/2}), \;
    \frac{1}{N_s} \sum_{i=1}^N
    S_i \frac{ R^A_i }{ \pi^A_i } (Y_i - \Gamma)
    \mid \curlF, \curlY \right\}
\nonumber \\
&& \hspace{.5cm} +
  \Cov \left[
    \frac{1}{N_s} \sum_{i=1}^N
    \frac{ R^A_i }{ \pi^A_i } \left\{ S_i m(X_i; \betalim) + \usepi_i \; \pi^B (X_i) \; \bfour^\top \Xpi \right\} , \;
    o_p(N^{-1/2})
    \mid \curlF, \curlY \right]
\nonumber \\
  && \hspace{.5cm} +
  \Cov \left[
    \frac{1}{N_s} \sum_{i=1}^N
    \left\{ \frac{ R^B_i }{ \pi^B (X_i) } - 1 \right\}
    \left[ S_i \{ Y_i - m(X_i; \betalim) \} - \usepi_i \; \pi^B (X_i) \; \bfour^\top \Xpi \right] , \;
    o_p(N^{-1/2})
    \mid \curlF, \curlY \right]
  \nonumber \\
  && \hspace{.5cm} +
  \Cov \left\{ o_p(N^{-1/2}), \;
    o_p(N^{-1/2})
    \mid \curlF, \curlY \right]
\nonumber \\
&& =
  \Cov \left[
    \frac{1}{N_s} \sum_{i=1}^N
    \frac{ R^A_i }{ \pi^A_i } \left\{ S_i m(X_i; \betalim) + \usepi_i \; \pi^B (X_i) \; \bfour^\top \Xpi \right\} , \;
    \frac{1}{N_s} \sum_{i=1}^N
    S_i \frac{ R^A_i }{ \pi^A_i } (Y_i - \Gamma)
    \mid \curlF, \curlY \right]
\nonumber \\
&& \hspace{0.5cm} +
0
\label{eq:why.zero} \\
&& \hspace{0.5cm} +
  \Cov \left\{
    o_p(N^{-1/2}), \;
    \Ybar - \Gamma + O_p(N^{-1/2})
    \mid \curlF, \curlY \right\}
  \label{eq:needsHT.y.Op} \\
  && \hspace{.5cm} +
  \Cov \left\{
  \mbar + \bfour^\top \frac{1}{N_s} \sum_{i=1}^N \usepi_i \; \pi^B (X_i) \Xpi + O_p(N^{-1/2}) , \;
    o_p(N^{-1/2})
    \mid \curlF, \curlY \right\}
  \label{eq:needsHT.m.Op} \\
  && \hspace{.5cm} +
  \Cov \left\{
    O_p(N^{-1/2}) , \;
    o_p(N^{-1/2})
    \mid \curlF, \curlY \right\}
\nonumber \\
  && \hspace{.5cm} +
  o_p(N^{-1})
\nonumber \\
&& =
  \Cov \left[
    \frac{1}{N_s} \sum_{i=1}^N
    \frac{ R^A_i }{ \pi^A_i } \left\{ S_i m(X_i; \betalim) + \usepi_i \; \pi^B (X_i) \; \bfour^\top \Xpi \right\} , \;
    \frac{1}{N_s} \sum_{i=1}^N
    S_i \frac{ R^A_i }{ \pi^A_i } (Y_i - \Gamma)
    \mid \curlF, \curlY \right] \nonumber \\
  && \hspace{.5cm} + o_p(N^{-1}).
  \label{eq:cov.HTHaj}
\end{eqnarray}
Note that line~(\ref{eq:why.zero}) follows because of the independence of $R^A_i$ and $R^B_j$ given $\curlF$ (and $\curlY$), and lines~(\ref{eq:needsHT.y.Op}) and~(\ref{eq:needsHT.m.Op}) follow from the assumption of equations~(\ref{eq:HT.y.Op}) and~(\ref{eq:HT.m.Op}), respectively.

If model $m(X; \beta)$ is correctly specified, then $\bfour = o_p(1)$ and equation~(\ref{eq:cov.HTHaj}) reduces to
\begin{eqnarray}
  &&
  \Cov ( \PLone - \Ybar, \; \HTHaj - \Ybar \mid \curlF, \curlY )
  \nonumber \\
  && \hspace{.4cm} =
  \Cov \left\{
    \frac{1}{N_s} \sum_{i=1}^N
    S_i \frac{ R^A_i }{ \pi^A_i } m(X_i; \betalim) , \;
    \frac{1}{N_s} \sum_{i=1}^N
    S_i \frac{ R^A_i }{ \pi^A_i } (Y_i - \Gamma)
    \mid \curlF, \curlY \right\}
  + o_p(N^{-1}).
  \label{eq:cov.HTHaj.correctm}
\end{eqnarray}

\section{Design-based inference for $\KHone$}

\label{sect:var.KH1}

Assume that at least one of the models $\pi^B (X; \alpha)$ and $m(X; \beta)$ is correctly specified.
Equation~(\ref{eq:DR1expansion.starts}) also applies to $\KHone$.
When the KH method is applied to sampled individuals with $S=1$, $\alphahat$ and $\betahat$ solve the estimating equations
\begin{eqnarray}
  \sum_{i=1}^N S_i R^B_i \Xpi \; \frac{ 1 - \pi^B (X_i; \alphahat) }{ \pi^B (X_i; \alphahat) } \{ Y_i - m(X_i; \betahat) \}
  & = & 0
  \label{eq:KH.estim.eq1}
  \\
  \sum_{i=1}^N S_i \left\{ \frac{R^A_i}{\pi^A_i} - \frac{ R^B_i }{ \pi^B (X_i; \alphahat) } \right\}
  \left. \frac{ \partial m (X_i; \beta) }{ \partial \beta^\top } \right|_{\beta=\betahat}
  & = & 0.
  \label{eq:KH.estim.eq2}
\end{eqnarray}

Equations~(\ref{eq:KH.estim.eq1}) and~(\ref{eq:KH.estim.eq2}) imply that equation~(\ref{eq:DR1expansion.starts}) with $\KHone$ in place of $\PLone$ reduces to $\KHone (\alphahat, \betahat) =  \KHone (\alphalim, \betalim) + o_p(N^{-1/2})$.
Thus,
\begin{eqnarray}
  \KHone (\alphahat, \betahat) - \Ybar
  & = &
    \frac{1}{N_s} \sum_{i=1}^N
  S_i \frac{ R^A_i }{ \pi^A_i } m(X_i; \betalim)
  +
  \frac{1}{N_s} \sum_{i=1}^N
  S_i \left\{ \frac{ R^B_i }{ \pi^B (X_i) } - 1 \right\}
  \{ Y_i - m(X_i; \betalim) \}
  \nonumber \\
  &&
  - \mbar + o_p(N^{-1/2}).
  \label{eq:KH1.expansion}
\end{eqnarray}
That is, equation~(\ref{eq:DR1.expansion.bothcorrect}) holds (with $\KHone$ in place of $\PLone$).
Therefore, if model $\pi^B (X; \alpha)$ is correctly specified, equation~(\ref{eq:DR1.variance.bothcorrect}) also applies, i.e.\
\begin{eqnarray*}
  &&
  \Var \{ \KHone (\alphahat, \betahat) - \Ybar \mid \curlF, \curlY \}
  \nonumber \\
  && \hspace{.3cm} =
  \Var \left\{
      \frac{1}{N_s} \sum_{i=1}^N 
      S_i \frac{ R^A_i }{ \pi^A_i } m(X_i; \betalim)
      \mid \curlF, \curlY
    \right\}
  +
  \frac{1}{N_s^2} \sum_{i=1}^N
    S_i \frac{ 1 - \pi^B (X_i) }{ \pi^B (X_i) } \;
    \{ Y_i - m(X_i; \betalim) \}^2
    + o_p (N^{-1}).
\end{eqnarray*}
Even if $\pi^B (X; \alpha)$ is misspecified (provided that one of $\pi^B (X; \alpha)$ and $m(X; \beta)$ is correctly specified),
\begin{eqnarray}
  &&
  \Cov ( \KHone - \Ybar, \; \HTHaj - \Ybar \mid \curlF, \curlY )
  \nonumber \\
  && \hspace{.4cm} =
  \Cov \left\{
    \frac{1}{N_s} \sum_{i=1}^N
    S_i \frac{ R^A_i }{ \pi^A_i } m(X_i; \betalim) , \;
    \frac{1}{N_s} \sum_{i=1}^N
    S_i \frac{ R^A_i }{ \pi^A_i } (Y_i - \Gamma)
    \mid \curlF, \curlY \right\}
  + o_p(N^{-1}).
  \label{eq:cov.KH}
\end{eqnarray}

\section{Design-based inference for $\PLtwo$}

\label{sect:var.PL2}

Suppose that model $\pi^B (X; \alpha)$ is correctly specified.
The ratio estimator $\IPWtwo$ that uses only $Y$ values from Sample B, i.e.\
\[
\IPWtwo =
\left. \sum_{i=1}^N S_i \; \frac{ R^B_i }{ \pi^B (X_i; \alphahat) } Y_i
\right/
\sum_{i=1}^N S_i \; \frac{ R^B_i }{ \pi^B (X_i; \alphahat) },
\]
solves the estimating equation
\[
\frac{1}{N} \sum_{i=1}^N
    S_i \; \frac{ R^B_i }{ \pi^B (X_i; \alpha) } (Y_i - \IPWparam) = 0.
\]
    The estimator $\Haj$ solves the estimating equation
\[
\frac{1}{N} \sum_{i=1}^N
    S_i \; \frac{ R^A_i }{ \pi^A_i } (Y_i - \Hajparam) = 0.
\]
The maximum pseudo likelihood estimator of $\alpha$ solves estimating equations~(\ref{eq:pseudo.like.score}).
We combine these three sets of estimating equations into a single vector of estimating equations.
Let $\theta = (\IPWparam, \alpha, \Hajparam)^\top$ and $\theta_0 = (\Ybar, \alpha_0, \Ybar)^\top$, and let $\thetahat$ denote the solution to the estimating equations $\Phi (\thetahat) = 0$, where
\begin{equation}
\Phi (\theta) =
\frac{1}{N} \sum_{i=1}^N
\left[ \begin{array}{l}
    S_i \; \frac{ R^B_i }{ \pi^B (X_i; \alpha) } (Y_i - \IPWparam) \\
    \usepi_i \{ R^B_i \Xpi - \frac{ R^A_i }{ \pi^A_i } \; \pi^B (X_i; \alpha) \; \Xpi \} \\
    S_i \; \frac{ R^A_i }{ \pi^A_i } (Y_i - \Hajparam)
  \end{array} \right].
\label{eq:IPW.estim.eqs.domain}
\end{equation}

Define
\[
\phi (\theta_0) = \left. \frac{ \partial \Phi (\theta) }{ \partial \theta^\top } \right|_{\theta=\theta_0}.
\]

Now,
\begin{eqnarray*}
  &&
  E \{ \phi (\theta_0) \mid \curlF, \curlY \}
  \nonumber \\
  && \hspace{0.5cm} =
  \left[ \begin{array}{lll}
      - \frac{N_s}{N} & - \frac{1}{N} \sum_{i=1}^N S_i \{ 1 - \pi^B (X_i; \alpha_0) \} (Y_i - \Ybar) \Xpitop & 0 \\
      0 & - \frac{1}{N} \sum_{i=1}^N \usepi_i \; \pi_i^B \{ 1 - \pi^B (X_i; \alpha_0) \} \Xpi \Xpitop & 0  \\
      0 & 0 & - \frac{N_s}{N}
    \end{array} \right].
\end{eqnarray*}
It is then easy to show that
\begin{eqnarray}
  &&
  [ E \{ \phi (\theta_0) \mid \curlF, \curlY \} ]^{-1}
  \nonumber \\
  && \hspace{0.5cm} =
  \left[ \begin{array}{lll}
      - \frac{N}{N_s} & \frac{N}{N_s} \bthreepart^\top & 0 \\
      0 & - \left[ \frac{1}{N} \sum_{i=1}^N \usepi_i \; \pi^B (X_i; \alpha_0) \{ 1 - \pi^B (X_1; \alpha_0) \} \Xpi \Xpitop \right]^{-1} & 0 \\
      0 & 0 & - \frac{N}{N_s} \\
    \end{array} \right]
  \label{eq:inv.expect.phi}
\end{eqnarray}
where
\begin{eqnarray*}
  \bthreepart
  & = &
  \left[ \sum_{i=1}^N \usepi_i \; \pi^B (X_i; \alpha_0) \{ 1 - \pi^B (X_i; \alpha_0) \} \; \Xpi \Xpitop \right]^{-1}
  \\
  && \times
  \sum_{i=1}^N S_i \{ 1 - \pi^B (X_i; \alpha_0) \} ( Y_i - \Ybar ) \Xpi.
\end{eqnarray*}
A Taylor series expansion of $\Phi(\theta)$ around $\theta_0$ gives
\begin{equation}
\thetahat - \theta_0 = - [ E \{ \phi (\theta_0) \mid \curlF, \curlY \} ]^{-1} \;
\Phi (\theta_0) + o_p (N^{-1/2}).
\label{eq:Taylor.Phi}
\end{equation}
It follows from equations~(\ref{eq:IPW.estim.eqs.domain}), (\ref{eq:inv.expect.phi}) and~(\ref{eq:Taylor.Phi}) that
\begin{eqnarray}
  \IPWtwo - \Ybar
  & = &
  \frac{1}{N_s} \sum_{i=1}^N
  S_i \; \frac{ R^B_i }{ \pi^B (X_i; \alpha_0) } (Y_i - \Ybar)
  - \bthreepart^\top \frac{1}{N_s} \sum_{i=1}^N \usepi_i \left\{ 
  R^B_i \Xpi - \frac{ R^A_i }{ \pi^A_i } \; \pi^B (X_i; \alpha_0) \; \Xpi \right\}
  \nonumber \\
  && + o_p(N^{-1/2})
  \nonumber \\
  & = &
  \frac{1}{N_s} \sum_{i=1}^N
  \; \frac{ R^B_i }{ \pi^B (X_i; \alpha_0) }
  \{ S_i (Y_i - \Ybar) - \usepi_i \; \pi^B (X_i; \alpha_0) \; \bthreepart^\top \Xpi \}
  + \frac{1}{N_s} \sum_{i=1}^N
   \usepi_i \frac{ R^A_i }{ \pi^A_i } \; \pi^B (X_i; \alpha_0) \; \bthreepart^\top \Xpi
   \nonumber \\
   && + o_p(N^{-1/2})
   \label{eq:IPW2s.minusYs}
\end{eqnarray}
and
\begin{equation}
  \Haj - \Ybar =
  \frac{1}{N_s} \sum_{i=1}^N S_i \; \frac{ R^A_i }{ \pi^A_i } (Y_i - \Ybar)
  + o_p(N^{-1/2}).
  \label{eq:Hajs.expansion}
\end{equation}

Following the Taylor expansion on page 5 of the appendix of Chen et al.\ (2020), we can write $\PLtwo$ as
\begin{eqnarray*}
  \PLtwo
  & = &
\left.
\sum_{i=1}^N S_i \; \frac{ R^A_i }{ \pi^A_i } \; m(X_i; \betalim)
\right/
\sum_{i=1}^N S_i \; \frac{ R^A_i }{ \pi^A_i }
\\
&& +
\left.
\sum_{i=1}^N S_i \; \frac{ R^B_i }{ \pi^B (X_i; \alphahat) } \{ Y_i - m(X_i; \betalim) \}
\right/
\sum_{i=1}^N S_i \; \frac{ R^B_i }{ \pi^B (X_i; \alphahat) }
+ o_p(N^{-1/2}).
\end{eqnarray*}

We recognise this as the ratio estimator $\Haj$ that uses only Sample A, but with $Y$ replaced by $m(X; \betalim)$, plus the IPW estimator $\IPWtwo$ that uses Sample B, but with $Y$ replaced by $Y - m(X; \betalim)$, plus an $o_p(N^{-1/2})$ term.
Hence, using equations~(\ref{eq:IPW2s.minusYs}) and~(\ref{eq:Hajs.expansion}), we have
\begin{eqnarray}
  \PLtwo - \Ybar
  & = &
  \frac{1}{N_s} \sum_{i=1}^N S_i \frac{ R^A_i }{ \pi^A_i } \{ m(X_i; \betalim) - \mbar \}
  \nonumber \\
&&
  + \frac{1}{N_s} \sum_{i=1}^N \frac{ R^B_i }{ \pi^B (X_i; \alpha_0) }
  [ S_i \{ Y_i - \Ybar - m(X_i; \betalim) + \mbar \} - \usepi_i \; \pi^B (X_i; \alpha_0) \; \bthree^\top \Xpi ]
  \nonumber \\
&&  
  + \frac{1}{N_s} \sum_{i=1}^N
   \usepi_i \frac{ R^A_i }{ \pi^A_i } \; \pi^B (X_i; \alpha_0) \; \bthree \Xpi
   + o_p(N^{-1/2})
   \nonumber \\
   & = &
   \frac{1}{N_s} \sum_{i=1}^N \frac{ R^A_i }{ \pi^A_i }
   \left[
     S_i \{ m(X_i; \betalim) - \mbar \}
     + \usepi_i \; \pi^B (X_i; \alpha_0) \; \bthree^\top \Xpi
     \right]
   \nonumber \\
   &&
   + \frac{1}{N_s} \sum_{i=1}^N
   \frac{ R^B_i }{ \pi^B (X_i; \alpha_0) }
   [ S_i \{ Y_i - \Ybar - m(X_i; \betalim) + \mbar \} - \usepi_i \; \pi^B (X_i; \alpha_0) \; \bthree^\top \Xpi ]
   + o_p(N^{-1/2})
   \nonumber \\
   &&
   \label{eq:DR2s.expansion}
\end{eqnarray}
where $\bthree$ is $\bthreepart$ with $Y$ replaced by $Y - m(X; \betalim)$, i.e.\
\begin{eqnarray}
  \bthree
  & = &
  \left[ \sum_{i=1}^N \usepi_i \; \pi^B (X_i; \alpha_0) \{ 1 - \pi^B (X_i; \alpha_0) \} \; \Xpi \Xpitop \right]^{-1}
  \nonumber \\
  && \times
  \sum_{i=1}^N S_i \{ 1 - \pi^B (X_i; \alpha_0) \} \{ Y_i - m(X_i; \betalim) - \Ybar + \mbar \} \Xpi.
\label{eq:bsformula}
\end{eqnarray}

The asymptotic expansion given by equations~(\ref{eq:DR2s.expansion}) and~(\ref{eq:Hajs.expansion}) enables us to derive formulae for the variances of $\PLtwo$ and $\Haj$ and their covariance, as well as the covariance between $\PLtwo$ and $\HT$.
We have
\begin{eqnarray}
  &&
  \Var (\PLtwo - \Ybar \mid \curlF, \curlY)
  \nonumber \\
  && \hspace{.3cm} =
  \Var \left( \frac{1}{N_s} \sum_{i=1}^N \frac{ R^A_i }{ \pi^A_i}
  [ S_i \{ m(X_i; \betalim) - \mbar \} + \usepi_i \; \pi^B (X_i) \; \bthree^\top \Xpi ]
  \mid \curlF, \curlY \right)
  \nonumber \\
  && \hspace{.8cm} +
  \frac{1}{ N_s^2} \sum_{i=1}^N
  \frac{ 1 - \pi^B (X_i; \alphalim) }{\pi^B (X_i; \alphalim)} \;
  \left[ S_i \{ Y_i - \Ybar - m(X_i; \betalim) + \mbar \}  - \usepi_i \; \pi^B (X_i; \alphalim) \; \bthree^\top \Xpi \right]^2
  \nonumber \\
  && \hspace{.8cm} +
  o_p(N^{-1})
  \label{eq:var.DR2.general}
\end{eqnarray}
and if Sample A is drawn without replacement,
\begin{equation*}
\Var (\Haj - \Ybar \mid \curlF, \curlY)
=
\frac{1}{ N_s^2 } \sum_{i=1}^N \sum_{j=1}^N
S_i S_j (\pi^A_{ij} - \pi^A_i \pi^A_j) \; \frac{ (Y_i - \Ybar) }{ \pi^A_i } \frac{ (Y_j - \Ybar) }{ \pi^A_j }
+ o_p(N^{-1}).
\end{equation*}
Just as in Appendix~\ref{sect:var.PL1}, it can be shown that
\begin{eqnarray}
  &&
  \Cov (\PLtwo - \Ybar, \; \HTHaj - \Ybar \mid \curlF, \curlY)
\nonumber \\
  && =
  \Cov \left(
  \frac{1}{N_s} \sum_{i=1}^N \frac{ R^A_i }{ \pi^A_i }
   \left[
     S_i \{ m(X_i; \betalim) - \mbar \}
     + \usepi_i \; \pi^B (X_i; \alpha_0) \; \bthree^\top \Xpi \right]
     , \;
     \frac{1}{N_s} \sum_{i=1}^N S_i \; \frac{ R^A_i }{ \pi^A_i } (Y_i - \Gamma) \mid \curlF, \curlY
     \right)
     \nonumber \\
     && \hspace{.5cm} +     
     o_p(N^{-1}).
     \label{eq:cov.DR2.general}
\end{eqnarray}

If model $m(X; \beta)$ is correctly specified, then $\bthree = o_p(1)$.
In that case, equations~(\ref{eq:var.DR2.general}) and~(\ref{eq:cov.DR2.general}) reduce to, respectively
\begin{eqnarray}
  \Var (\PLtwo - \Ybar \mid \curlF)
  & = &
  \Var \left[ \frac{1}{N_s} \sum_{i=1}^N S_i \; \frac{ R^A_i }{ \pi^A_i}
  \{ m(X_i; \betalim) - \mbar \}
  \mid \curlF, \curlY \right]
  \nonumber \\
  && +
  \frac{1}{N_s^2} \sum_{i=1}^N
  S_i \frac{ 1 - \pi^B (X_i; \alphalim) }{\pi^B (X_i; \alphalim)} \;
  \left\{ Y_i - \Ybar - m(X_i; \betalim) + \mbar \right\}^2
     + o_p(N^{-1}).
  \nonumber
\end{eqnarray}
and
\begin{eqnarray}
  &&
  \Cov (\PLtwo - \Ybar, \; \HTHaj - \Ybar \mid \curlF, \curlY)
  \nonumber \\
  && \hspace{.4cm} =
  \Cov \left[
  \frac{1}{N_s} \sum_{i=1}^N
  S_i \frac{ R^A_i }{ \pi^A_i } \{ m(X_i; \betalim) - \mbar \}
     , \;
     \frac{1}{N_s} \sum_{i=1}^N S_i \frac{ R^A_i }{ \pi^A_i } (Y_i - \Gamma) \mid \curlF, \curlY
     \right]
     + o_p(N^{-1}).
  \nonumber
\end{eqnarray}

\section{Design-based inference for $\KHtwo$}

\label{sect:var.KH2}

Assume that at least one of the models $\pi^B (X; \alpha)$ and $m(X; \beta)$ is correctly specified.
Define $\KHtau (\alpha, \beta, \tau) = \tau^{-1} \KHone (\alpha, \beta)$ and $\tauhat = N_s^{-1} \sum_{i=1}^N S_i R^A_i / \pi^A_i$ and $\mbartrue = N_s^{-1} \sum_{i=1}^N S_i m(X_i)$.
Note that if model $m(X; \beta)$ is correctly specified, then $\mbartrue = \mbar$.

Now,
\begin{eqnarray}
  \KHtwo - \Ybar
  & = &
  \KHtau (\alphahat, \betahat, \tauhat) - \Ybar
  \nonumber \\
  & = &
  \KHtau (\alphahat, \betahat, 1) + \frac{ \partial }{ \partial \tau }
  \left. \KHtau (\alphahat, \betahat, \tau) \right|_{\tau=1} (\tauhat - 1) - \Ybar + o_p (N^{1/2})
  \nonumber \\
  & = &
  \KHone + \frac{ \partial }{ \partial \tau }
  \left. \KHtau (\alphahat, \betahat, \tau) \right|_{\tau=1} (\tauhat - 1) - \Ybar + o_p (N^{1/2})
  \nonumber \\
  & = &
  \KHone - \KHone \times (\tauhat - 1) - \Ybar + o_p (N^{1/2})
  \nonumber \\
  & = &
  \KHone - \{ \mbartrue + o_p(1) \} (\tauhat - 1) - \Ybar + o_p (N^{1/2})
  \nonumber \\
  & = &
  \KHone - \{ \mbartrue + o_p(1) \} \frac{1}{N_s} \sum_{i=1}^N S_i \left( \frac{R^A_i}{\pi^A_i} - 1 \right) - \Ybar + o_p (N^{1/2})
  \nonumber \\
  & = &
  \frac{1}{N_s} \sum_{i=1}^N S_i \left[ \frac{R^A_i}{\pi^A_i} m (X_i; \betalim)
    + \frac{ R^B_i }{ \pi^B (X_i; \alphalim) } \{ Y_i - m (X_i; \betalim) \} \right]
  \nonumber \\
  &&  
  - \mbartrue \frac{1}{N_s} \sum_{i=1}^N S_i \left( \frac{R^A_i}{\pi^A_i} - 1 \right) - \Ybar + o_p (N^{1/2})
  \nonumber \\
  & = &
  \frac{1}{N_s} \sum_{i=1}^N S_i \left[ \frac{R^A_i}{\pi^A_i} \{ m (X_i; \betalim) - \mbartrue \}
    + \frac{ R^B_i }{ \pi^B (X_i; \alphalim) } \{ Y_i - m (X_i; \betalim) \} \right]
  \nonumber \\
  &&
  + \mbartrue - \Ybar + o_p (N^{1/2})
  \nonumber \\
  & = &
  \frac{1}{N_s} \sum_{i=1}^N S_i \left[ \frac{R^A_i}{\pi^A_i} \{ m (X_i; \betalim) - \mbartrue \}
    + \left\{ \frac{ R^B_i }{ \pi^B (X_i; \alphalim) } - 1 \right\} \{ Y_i - m (X_i; \betalim) \} \right]
  \nonumber \\
  &&
  + o_p (N^{1/2}).
  \label{eq:KH2.expansion}
\end{eqnarray}

Therefore, we obtain results analogous to those in Appendix~\ref{sect:var.KH1}.
Specifically, if model $\pi^B (X; \alpha)$ is correctly specified,
\begin{eqnarray*}
  \Var \{ \KHtwo (\alphahat, \betahat) - \Ybar \mid \curlF, \curlY \}
  & = &
  \Var \left\{
      \frac{1}{N_s} \sum_{i=1}^N 
      S_i \frac{ R^A_i }{ \pi^A_i } \{ m(X_i; \betalim) - \mbartrue \}
      \mid \curlF, \curlY
      \right\}
  \nonumber \\
  && +
  \frac{1}{N_s^2} \sum_{i=1}^N
    S_i \frac{ 1 - \pi^B (X_i) }{ \pi^B (X_i) } \;
    \{ Y_i - m(X_i; \betalim) \}^2
    + o_p (N^{-1}),
\end{eqnarray*}
and even if $\pi^B (X; \alpha)$ is misspecified (provided that one of $\pi^B (X; \alpha)$ and $m(X; \beta)$ is correctly specified),
\begin{eqnarray*}
  &&
  \Cov ( \KHtwo - \Ybar, \; \HTHaj - \Ybar \mid \curlF, \curlY )
  \nonumber \\
  && \hspace{.4cm} =
  \Cov \left\{
    \frac{1}{N_s} \sum_{i=1}^N
    S_i \frac{ R^A_i }{ \pi^A_i } \{ m(X_i; \betalim) - \mbartrue \} , \;
    \frac{1}{N_s} \sum_{i=1}^N
    S_i \frac{ R^A_i }{ \pi^A_i } (Y_i - \Gamma)
    \mid \curlF, \curlY \right\}
    + o_p(N^{-1}).
\end{eqnarray*}

We now compare the distribution of $\KHtwo$ with that of $\PLtwo$.
First, notice that when $m(X; \beta)$ is a linear regression model, $\partial m (X; \beta) / \partial \beta^\top = X$, and so (provided that $\beta$ includes an intercept term) the KH estimating equation~(\ref{eq:KH.estim.eq2}) implies $\hat{N}^A_s = \hat{N}^B_s$.
Hence, in this case, $\KHtwo$ equals $\PLtwo$ but with a different way of calculating $\alphahat$ and $\betahat$.
Second, irrespective of whether $m(X; \beta)$ is a linear regression model, when models $\pi^B (X; \alpha)$ and $m (X; \beta)$ are both correctly specified, equation~(\ref{eq:DR2s.expansion}) reduces to
\begin{eqnarray*}
  \PLtwo - \Ybar
   & = &
  \frac{1}{N_s} \sum_{i=1}^N S_i \left[
  \frac{ R^A_i }{ \pi^A_i }
  \{ m(X_i; \betalim) - \mbar \}
  +
   \frac{ R^B_i }{ \pi^B (X_i; \alpha_0) }
   \{ Y_i - m(X_i; \betalim) + \mbar - \Ybar \}
   \right]
   + o_p(N^{-1/2})
\end{eqnarray*}
and equation~(\ref{eq:KH2.expansion}) can be written as
\begin{eqnarray*}
  \KHtwo - \Ybar
  & = &
  \frac{1}{N_s} \sum_{i=1}^N S_i \left[ \frac{R^A_i}{\pi^A_i} \{ m (X_i; \betalim) - \mbar \}
    + \frac{ R^B_i }{ \pi^B (X_i; \alphalim) } \{ Y_i - m (X_i; \betalim) \} \right]
  + \mbar - \Ybar
  + o_p (N^{1/2}).
\end{eqnarray*}
If the domain is large compared to the number of sampled individuals from the domain in Sample B, i.e.\ if $N_s$ is large compared to $\sum_{i=1}^N S_i R^B_i$, then the $\mbar - \Ybar$ terms in these two expansions are practically negligible, and thus $\sqrt{N} (\PLtwo - \Ybar) \approx \sqrt{N} (\KHtwo - \Ybar) + o_p (N^{1/2})$.

\section{Model-design-based inference for $\PLone$}

\label{sect:modeldesign.PL1}

Suppose that model $\pi^B (X; \alpha)$ is correctly specified.
Recall that equation~(\ref{eq:DR1.general}) was
\begin{eqnarray*}
  &&
  \Var \{ \PLone (\alphahat, \betahat) - \Ybar \mid \curlF, \curlY \}
  \nonumber \\
  && \hspace{.3cm} =
  \Var \left[
      \frac{1}{N_s} \sum_{i=1}^N 
      \frac{ R^A_i }{ \pi^A_i } \left\{ S_i m(X_i; \betalim) + \usepi_i \; \pi^B (X_i) \; \bfour^\top \Xpi  \right\} \mid \curlF, \curlY
    \right]
  \nonumber \\
  && \hspace{.8cm} +
    \frac{1}{N_s^2} \sum_{i=1}^N
    \frac{ 1 - \pi^B (X_i) }{ \pi^B (X_i) } \;
    \left[ S_i \{ Y_i - m(X_i; \betalim) \} - \usepi_i \; \pi^B (X_i) \; \bfour^\top \Xpi \right]^2
    + o_p (N^{-1}).
\end{eqnarray*}

Now,
\begin{eqnarray}
  &&
  \Var \{ \PLone (\alphahat, \betahat) - \Ybar \mid \curlF \}
  \nonumber \\
  && =
E [ \Var \{ \PLone (\alphahat, \betahat) - \Ybar \mid \curlF, \curlY \} \mid \curlF ]
+
\Var [ E \{ \PLone (\alphahat, \betahat) - \Ybar \mid \curlF, \curlY \} \mid \curlF ]
  \nonumber \\
  && =
E [ \Var \{ \PLone (\alphahat, \betahat) - \Ybar \mid \curlF, \curlY \} \mid \curlF ]
+
\Var \{ o_p(N^{-1/2}) \mid \curlF \}
  \nonumber \\
  && =
E [ \Var \{ \PLone (\alphahat, \betahat) - \Ybar \mid \curlF, \curlY \} \mid \curlF ]
+ o_p(N^{-1})
  \nonumber \\
  && =
  E \left( \Var \left[
    \frac{1}{N_s} \sum_{i=1}^N 
    \frac{ R^A_i }{ \pi^A_i } \left\{ S_i m(X_i; \betalim) + \usepi_i \; \pi^B (X_i) \; \bfour^\top \Xpi  \right\} \mid \curlF, \curlY
    \right]
  \mid \curlF \right)
  \nonumber \\
  && \hspace{.8cm} +
    \frac{1}{N_s^2} \sum_{i=1}^N
    \frac{ 1 - \pi^B (X_i) }{ \pi^B (X_i) } \;
    E \left( \left[ S_i \{ Y_i - m(X_i; \betalim) \} - \usepi_i \; \pi^B (X_i) \; \bfour^\top \Xpi \right]^2 \mid \curlF \right)
    + o_p (N^{-1}).
  \nonumber \\    
    \label{eq:varDR1.onlycurlF}
\end{eqnarray}

This expression for $\Var \{ \PLone (\alphahat, \betahat) - \Ybar \mid \curlF \}$ is identical to the expression for $\Var \{ \PLone (\alphahat, \betahat) - \Ybar \mid \curlF, \curlY \}$ except for the two expectations with respect to $Y_1, \ldots, Y_N$ given $\curlF$.
Note that the first of these expectations is needed because $\bfour$ is a function of $Y_1, \ldots, Y_N$.
To estimate $\Var \{ \PLone (\alphahat, \betahat) - \Ybar \mid \curlF \}$, we can use exactly the same estimator that we used to estimate $\Var \{ \PLone (\alphahat, \betahat) - \Ybar \mid \curlF, \curlY \}$ in Appendix~\ref{sect:var.PL1}.
Note that if nuisance model $m(X; \beta)$ is correctly specified, then $\bfour = o_p(1)$ and equation~(\ref{eq:varDR1.onlycurlF}) reduces to
\begin{eqnarray}
  &&
  \Var \{ \PLone (\alphahat, \betahat) - \Ybar \mid \curlF \}
  \nonumber \\
  && =
\Var \left\{
    \frac{1}{N_s} \sum_{i=1}^N 
    \frac{ R^A_i }{ \pi^A_i } S_i m(X_i; \betalim) \mid \curlF \right\}
  \nonumber \\
  && \hspace{.8cm} +
    \frac{1}{N_s^2} \sum_{i=1}^N S_i \;
    \frac{ 1 - \pi^B (X_i) }{ \pi^B (X_i) } \;
    E [ \{ Y_i - m(X_i; \betalim) \}^2 \mid \curlF ]
    + o_p (N^{-1}).
    \label{eq:DR1.onlycurlF.bothcorrect}
\end{eqnarray}

Now we consider the covariance between $\PLone$ and $\HTHaj$.
Recall that equation~(\ref{eq:cov.HTHaj}) was
\begin{eqnarray*}
  &&
  \Cov ( \PLone - \Ybar, \; \HTHaj - \Ybar \mid \curlF, \curlY )
  \nonumber \\
&& =
  \Cov \left[
    \frac{1}{N_s} \sum_{i=1}^N
    \frac{ R^A_i }{ \pi^A_i } \left\{ S_i m(X_i; \betalim) + \usepi_i \; \pi^B (X_i) \; \bfour^\top \Xpi \right\} , \;
    \frac{1}{N_s} \sum_{i=1}^N
    S_i \frac{ R^A_i }{ \pi^A_i } (Y_i - \Gamma)
    \mid \curlF, \curlY \right]
  + o_p(N^{-1}).
\end{eqnarray*}
So, we have
\begin{eqnarray*}
  &&
  \Cov ( \PLone - \Ybar, \; \HTHaj - \Ybar \mid \curlF )
  \nonumber \\
  && =
  E\{ \Cov ( \PLone - \Ybar, \; \HTHaj - \Ybar \mid \curlF, \curlY ) \mid \curlF \} \nonumber \\
  && \qquad +
  \Cov \{ E ( \PLone - \Ybar \mid \curlF, \curlY ), \, E ( \HTHaj - \Ybar \mid \curlF, \curlY ) \mid \curlF \}
    \nonumber \\
  && =
  E\{ \Cov ( \PLone - \Ybar, \; \HTHaj - \Ybar \mid \curlF, \curlY ) \mid \curlF \}
  +
  \Cov \{ o_p(N^{-1/2}), \, o_p(N^{-1/2}) \mid \curlF \}
  \nonumber \\
  && =
  E\{ \Cov ( \PLone - \Ybar, \; \HTHaj - \Ybar \mid \curlF, \curlY ) \mid \curlF \}
  +
  o_p (N^{-1})
  \nonumber \\
  && =
  E \left( \Cov \left[
    \frac{1}{N_s} \sum_{i=1}^N
    \frac{ R^A_i }{ \pi^A_i } \left\{ S_i m(X_i; \betalim) + \usepi_i \; \pi^B (X_i) \; \bfour^\top \Xpi \right\} , \;
    \frac{1}{N_s} \sum_{i=1}^N
    S_i \frac{ R^A_i }{ \pi^A_i } (Y_i - \Gamma)
    \mid \curlF, \curlY \right]
  \mid \curlF \right)
  \nonumber \\
  && \hspace{.4cm} + 
  o_p(N^{-1}).
\end{eqnarray*}
This is identical to the expression for $\Cov ( \PLone - \Ybar, \; \HTHaj - \Ybar \mid \curlF, \curlY )$ except for the expectation with respect to $Y_1, \ldots, Y_N$ given $\curlF$.
Again, if model $m(X; \beta)$ is correctly specified, $\bfour = o_p(1)$ and the covariance formula simplifies to
\begin{eqnarray}
  &&
  \Cov ( \PLone - \Ybar, \; \HTHaj - \Ybar \mid \curlF )
  \nonumber \\
  && =
  E \left[ \Cov \left\{
    \frac{1}{N_s} \sum_{i=1}^N
    \frac{ R^A_i }{ \pi^A_i } S_i m(X_i; \betalim) , \;
    \frac{1}{N_s} \sum_{i=1}^N
    S_i \frac{ R^A_i }{ \pi^A_i } (Y_i - \Gamma)
    \mid \curlF, \curlY \right\}
    \mid \curlF \right] \nonumber \\
  && \qquad + o_p(N^{-1}).
  \label{eq:cov.dr1.hthajek}
\end{eqnarray}
To estimate $\Cov ( \PLone - \Ybar, \; \HTHaj - \Ybar \mid \curlF )$ we can use exactly the same estimator that we used to estimate estimate $\Cov ( \PLone - \Ybar, \; \HTHaj - \Ybar \mid \curlF, \curlY )$ in Appendix~\ref{sect:var.PL1}.

\section{Model-design-based inference for $\PLtwo$}

\label{sect:modeldesign.PL2}

Suppose that model $\pi^B (X; \alpha)$ is correctly specified.
We proceed as in Appendix~\ref{sect:modeldesign.PL1}.
Recall that equation~(\ref{eq:var.DR2.general}) was
\begin{eqnarray*}
  &&
  \Var (\PLtwo - \Ybar \mid \curlF, \curlY)
  \nonumber \\
  && \hspace{.3cm} =
  \Var \left( \frac{1}{N_s} \sum_{i=1}^N \frac{ R^A_i }{ \pi^A_i}
  [ S_i \{ m(X_i; \betalim) - \mbar \} + \usepi_i \; \pi^B (X_i) \; \bthree^\top \Xpi ]
  \mid \curlF, \curlY \right)
  \nonumber \\
  && \hspace{.8cm} +
  \frac{1}{ N_s^2} \sum_{i=1}^N
  \frac{ 1 - \pi^B (X_i; \alphalim) }{\pi^B (X_i; \alphalim)} \;
  \left[ S_i \{ Y_i - \Ybar - m(X_i; \betalim) + \mbar \}  - \usepi_i \; \pi^B (X_i; \alphalim) \; \bthree^\top \Xpi \right]^2
  \nonumber \\
  && \hspace{.8cm} +
  o_p(N^{-1}).
\end{eqnarray*}
As with $\PLone$, we therefore have
\begin{eqnarray*}
  &&
  \Var (\PLtwo - \Ybar \mid \curlF)
  \nonumber \\
  && \hspace{.3cm} =
  E \left\{ \Var \left( \frac{1}{N_s} \sum_{i=1}^N \frac{ R^A_i }{ \pi^A_i}
  [ S_i \{ m(X_i; \betalim) - \mbar \} + \usepi_i \; \pi^B (X_i) \; \bthree^\top \Xpi ]
  \mid \curlF, \curlY \right)
  \mid \curlF \right\}
  \nonumber \\
  && \hspace{.8cm} +
  \frac{1}{ N_s^2} \sum_{i=1}^N
  \frac{ 1 - \pi^B (X_i; \alphalim) }{\pi^B (X_i; \alphalim)} \;
  E \left( \left[ S_i \{ Y_i - \Ybar - m(X_i; \betalim) + \mbar \} - \usepi_i \; \pi^B (X_i; \alphalim) \; \bthree^\top \Xpi \right]^2 \mid \curlF \right)
  \nonumber \\
  && \hspace{.8cm} +
  o_p(N^{-1}).
\end{eqnarray*}

For the covariance, recall that equation~(\ref{eq:cov.DR2.general}) was
\begin{eqnarray*}
  &&
  \Cov (\PLtwo - \Ybar, \; \HTHaj - \Ybar \mid \curlF, \curlY)
\nonumber \\
  && =
  \Cov \left(
  \frac{1}{N_s} \sum_{i=1}^N \frac{ R^A_i }{ \pi^A_i }
   \left[
     S_i \{ m(X_i; \betalim) - \mbar \}
     + \usepi_i \; \pi^B (X_i; \alpha_0) \; \bthree^\top \Xpi \right]
     , \;
     \frac{1}{N_s} \sum_{i=1}^N S_i \; \frac{ R^A_i }{ \pi^A_i } (Y_i - \Gamma) \mid \curlF, \curlY
     \right)
     \nonumber \\
     && \hspace{.5cm} +     
     o_p(N^{-1}).
\end{eqnarray*}
Hence, analogously to the result in Appendix~\ref{sect:modeldesign.PL1}, we have
\begin{eqnarray*}
  &&
  \Cov (\PLtwo - \Ybar, \; \HTHaj - \Ybar \mid \curlF)
\nonumber \\
  && =
  E \left\{ \Cov \left(
  \frac{1}{N_s} \sum_{i=1}^N \frac{ R^A_i }{ \pi^A_i }
   \left[
     S_i \{ m(X_i; \betalim) - \mbar \}
     + \usepi_i \; \pi^B (X_i; \alpha_0) \; \bthree^\top \Xpi \right] \right. \right.
     , \; \nonumber \\
  && \qquad \left. \left. \frac{1}{N_s} \sum_{i=1}^N S_i \; \frac{ R^A_i }{ \pi^A_i } (Y_i - \Gamma) \mid \curlF, \curlY \right)
     \mid \curlF \right\}
     \nonumber \\
     && \hspace{.4cm} +     
     o_p(N^{-1}).
\end{eqnarray*}

To estimate $\Var \{ \PLtwo (\alphahat, \betahat) - \Ybar \mid \curlF \}$ and $\Cov (\PLtwo - \Ybar, \; \HTHaj - \Ybar \mid \curlF)$, we can use exactly the same estimators that we used to estimate $\Var \{ \PLtwo (\alphahat, \betahat) - \Ybar \mid \curlF, \curlY \}$ and $\Cov (\PLtwo - \Ybar, \; \HTHaj - \Ybar \mid \curlF, \curlY)$ in Appendix~\ref{sect:var.PL2}.

Note that if model $m(X; \beta)$ is correctly specified, these variance and covariance formulae reduce to
\begin{eqnarray}
  &&
  \Var (\PLtwo - \Ybar \mid \curlF)
  \nonumber \\
  && \hspace{.3cm} =
  \Var \left[ \frac{1}{N_s} \sum_{i=1}^N \frac{ R^A_i }{ \pi^A_i}
  S_i \{ m(X_i; \betalim) - \mbar \}
  \mid \curlF \right]
  \nonumber \\
  && \hspace{.8cm} +
  \frac{1}{ N_s^2} \sum_{i=1}^N S_i
  \frac{ 1 - \pi^B (X_i; \alphalim) }{\pi^B (X_i; \alphalim)} \;
  E \left[ \{ Y_i - \Ybar - m(X_i; \betalim) + \mbar \}^2 \mid \curlF \right]
  \nonumber \\
  && \hspace{.8cm} +
  o_p(N^{-1})
  \label{eq:DR2.onlycurlF.bothcorrect}
\end{eqnarray}
and
\begin{eqnarray}
  &&
  \Cov (\PLtwo - \Ybar, \; \HTHaj - \Ybar \mid \curlF)
\nonumber \\
  && =
  E \left( \Cov \left[
  \frac{1}{N_s} \sum_{i=1}^N \frac{ R^A_i }{ \pi^A_i }
     S_i \{ m(X_i; \betalim) - \mbar \}
     , \;
     \frac{1}{N_s} \sum_{i=1}^N S_i \; \frac{ R^A_i }{ \pi^A_i } (Y_i - \Gamma) \mid \curlF, \curlY \right]
     \mid \curlF \right)
     \nonumber \\
     && \hspace{.5cm} +     
     o_p(N^{-1}).
     \label{eq:DR2.Haj.cov.onlycurlF}
\end{eqnarray}

\section{Model-design-based inference for $\KHone$}

\label{sect:modeldesign.KH1}

Assume that at least of models $\pi^B (X; \alpha)$ and $m(X; \beta)$ is correctly specified.
As in Appendix~\ref{sect:var.KH1}, we have
\begin{eqnarray}
  \KHone (\alphahat, \betahat) - \Ybar
  & = &
    \frac{1}{N_s} \sum_{i=1}^N
  S_i \frac{ R^A_i }{ \pi^A_i } m(X_i; \betalim)
  +
  \frac{1}{N_s} \sum_{i=1}^N
  S_i \left\{ \frac{ R^B_i }{ \pi^B (X_i; \alphalim) } - 1 \right\}
  \{ Y_i - m(X_i; \betalim) \}
  \nonumber \\
  &&
  - \mbar + o_p(N^{-1/2}).
\end{eqnarray}
Therefore
\begin{eqnarray}
  &&
\Var \{ \KHone (\alphahat, \betahat) - \Ybar \mid \curlF \}
\nonumber \\
&& =
\Var \left[ \frac{1}{N_s} \sum_{i=1}^N
  S_i \frac{ R^A_i }{ \pi^A_i } m(X_i; \betalim)
  +
  \frac{1}{N_s} \sum_{i=1}^N
  S_i \left\{ \frac{ R^B_i }{ \pi^B (X_i; \alphalim) } - 1 \right\}
  \{ Y_i - m(X_i; \betalim) \}
  \mid \curlF \right]
  \nonumber \\
  && \hspace{0.4cm} +
  o_p(N^{-1}).
\nonumber \\
&& =
\Var \left\{ \frac{1}{N_s} \sum_{i=1}^N
  S_i \frac{ R^A_i }{ \pi^A_i } m(X_i; \betalim)
  \mid \curlF \right\}
  \nonumber \\
  && \hspace{0.4cm} +
  \Var \left[ \frac{1}{N_s} \sum_{i=1}^N
    S_i \left\{ \frac{ R^B_i }{ \pi^B (X_i; \alphalim) } - 1 \right\}
    \{ Y_i - m(X_i; \betalim) \}
    \mid \curlF \right]
  + o_p(N^{-1})
  \label{eq:because.cov.zero} \\
&& =
\Var \left\{ \frac{1}{N_s} \sum_{i=1}^N
  S_i \frac{ R^A_i }{ \pi^A_i } m(X_i; \betalim)
  \mid \curlF, \curlY \right\}
  \nonumber \\
  && \hspace{0.4cm} +
  \Var \left[ \frac{1}{N_s} \sum_{i=1}^N
    S_i \left\{ \frac{ R^B_i }{ \pi^B (X_i; \alphalim) } - 1 \right\}
    \{ Y_i - m(X_i; \betalim) \}
    \mid \curlF \right]
  + o_p(N^{-1}).
  \label{eq:var.DR1.KH}
\end{eqnarray}
Note that line~(\ref{eq:because.cov.zero}) follows because
\begin{eqnarray}
  &&
  \Cov \left[ \frac{1}{N_s} \sum_{i=1}^N S_i
    \left\{ \frac{ R^B_i }{ \pi^B(X_i; \alphalim) } - 1 \right\} \{ Y_i - m(X_i; \betalim) \}, \;
    \frac{1}{N_s} \sum_{j=1}^N
    S_j \frac{ R^A_j }{ \pi^A_j } m(X_j; \betalim)
    \mid \curlF \right]
  \nonumber \\
  && =
  E \left( \Cov \left[ \frac{1}{N_s} \sum_{i=1}^N S_i
    \left\{ \frac{ R^B_i }{ \pi^B(X_i; \alphalim) } - 1 \right\} \{ Y_i - m(X_i; \betalim) \},
    \right. \right.
  \nonumber \\
    && \hspace{1.8cm} \left. \left.
  \frac{1}{N_s} \sum_{j=1}^N S_j
  \frac{ R^A_j }{ \pi^A_j } m(X_j; \betalim)
    \mid \curlF, \curlY \right] \mid \curlF \right)
  \nonumber \\
&& \hspace{0.5cm} +
  \Cov \left( E \left[ \frac{1}{N_s} \sum_{i=1}^N S_i
    \left\{ \frac{ R^B_i }{ \pi^B(X_i; \alphalim) } - 1 \right\} \{ Y_i - m(X_i; \betalim) \} \mid \curlF, \curlY \right],
  \right.
  \nonumber \\
  && \hspace{1.9cm} \left.
  E \left\{ \frac{1}{N_s} \sum_{j=1}^N S_j
  \frac{ R^A_j }{ \pi^A_j } m(X_j; \betalim)
  \mid \curlF, \curlY \right\} \mid \curlF \right)
  \label{eq:expect.cov.plus.cov.expect}
  \nonumber \\
  && =
  E (0 \mid \curlF)
  +
  \Cov \left[ \frac{1}{N_s} \sum_{i=1}^N S_i
    \left\{ \frac{ \pi^B(X_i) }{ \pi^B(X_i; \alphalim) } - 1 \right\} \{ Y_i - m(X_i; \betalim) \} , \;
  \mbar
  \mid \curlF \right]
  \label{eq:because.RARB.indep} \\
  && =
  0.
  \nonumber
\end{eqnarray}
Note that line~(\ref{eq:because.RARB.indep}) follows from the conditional independence of $R^A$ and $R^B$ given $\curlF$ (and $\curlY$).

If model $\pi^B (X; \alpha)$ is correctly specified, the second term in expression~(\ref{eq:var.DR1.KH}) can be consistently estimated by
\begin{equation}
  \frac{1}{N_s^2} \sum_{i=1}^N S_i R^B_i \; 
  \frac{ 1 - \pi^B (X_i; \alphahat) } { \{ \pi^B (X_i; \alphahat) \}^2 } \; \{ Y_i - m (X_i; \betahat) \}^2.
  \label{eq:SampleB.var}
\end{equation}

If, on the other hand, model $\pi^B (X; \alpha)$ may be misspecified but model $m(X; \beta)$ is correctly specified, then the second term in expression~(\ref{eq:var.DR1.KH}) becomes
\begin{eqnarray}
  &&
  \Var \left[ \frac{1}{N_s} \sum_{i=1}^N S_i
    \left\{ \frac{R^B_i}{ \pi^B (X_i; \alphalim) } - 1 \right\} \{ Y_i - m (X_i; \betalim) \} \mid \curlF \right]
  \nonumber \\
  && \hspace{0.3cm} =
  \frac{1}{N_s^2} \sum_{i=1}^N S_i \left[
    \frac{ \piBtrue (X_i) \{ 1 - \piBtrue (X_i) \} }{ \{ \pi^B (X_i; \alphalim) \}^2 } + \left\{ \frac{ \piBtrue (X_i) }{ \pi^B (X_i; \alphalim) } - 1 \right\}^2
    \right]
  \Var \{ Y_i - m (X_i; \betalim) \mid X_i \}
  \nonumber \\
  && \hspace{0.3cm} =
  \frac{1}{N_s^2} \sum_{i=1}^N S_i \left[
    \frac{ \piBtrue (X_i) \{ 1 - \pi^B (X_i; \alphalim) \} }{ \{ \pi^B (X_i; \alphalim) \}^2 } - \frac{ \piBtrue (X_i) }{ \pi^B (X_i; \alphalim) } + 1
    \right]
  \Var \{ Y_i - m (X_i; \betalim) \mid X_i \}
  \nonumber \\
  && \hspace{0.3cm} =
  \frac{1}{N_s^2} \sum_{i=1}^N S_i
    \frac{ \piBtrue (X_i) \{ 1 - \pi^B (X_i; \alphalim) \} }{ \{ \pi^B (X_i; \alphalim) \}^2 } \;
    E [ \{ Y_i - m (X_i; \betalim) \}^2 \mid X_i ]
  \nonumber \\
  && \hspace{0.8cm} +
  \frac{1}{N_s^2} \sum_{i=1}^N S_i \left\{
    \frac{ E(R^A_i \mid \curlF) }{ \pi^A_i } - \frac{ E(R^B_i \mid X_i) }{ \pi^B (X_i; \alphalim) }
    \right\}
  \Var \{ Y_i - m (X_i; \betalim) \mid X_i \}.   
  \nonumber
\end{eqnarray}

This leads Chen et al.\ (2020) to propose that the following estimator of
\[
\Var \left[ \frac{1}{N_s} \sum_{i=1}^N S_i
  \left\{ \frac{R^B_i}{ \pi^B (X_i; \alphalim) } - 1 \right\} \{ Y_i - m (X_i; \betalim) \} \mid \curlF \right]
\]
should be used:
\begin{equation}
  \frac{1}{N_s^2} \sum_{i=1}^N S_i R^B_i \; 
  \frac{ 1 - \pi^B (X_i; \alphahat) } { \{ \pi^B (X_i; \alphahat) \}^2 } \; \{ Y_i - m (X_i; \betahat) \}^2
  +
  \frac{1}{N_s^2} \sum_{i=1}^N S_i \left\{
    \frac{ R^A_i }{ \pi^A_i } - \frac{ R^B_i }{ \pi^B (X_i; \alphahat) }
    \right\}
  \hat{\sigma}^2 (X_i)
  \label{eq:SampleB.var.bias.cor}
\end{equation}
where $\hat{\sigma}^2 (X_i)$ is an estimator of $\sigma^2 (X_i) = \Var \{ Y_i - m (X_i; \betalim) \mid X_i \}$.
For example, if $m(X; \beta)$ is a logistic regression model, then $\hat{\sigma}^2 (X_i) = m(X_i; \betahat) \{ 1 - m(X_i; \betahat) \}$.
Since the extra term in expression~(\ref{eq:SampleB.var.bias.cor}) relative to expression~(\ref{eq:SampleB.var}) has expectation zero (given $\curlF$) when model $\pi^B(X;\alpha)$ is correctly specified, this estimator is valid when the KH method is used and at least one of the two nuisance models is correctly specified.
Hence, we have a doubly robust variance estimator.

We note two things about expression~(\ref{eq:SampleB.var.bias.cor}).
First, if model $m(X; \beta)$ is a correctly specified generalised linear model with canonical link function, then
\[
\left. \frac{ \partial m(X; \beta) }{ \partial \beta } \right|_{\beta = \betalim} \propto X \; \Var (Y \mid X).
\]
In that case, equation~(\ref{eq:KH.estim.eq2}) can be rewritten as
\begin{equation}
\sum_{i=1}^N S_i \left\{ \frac{R^A_i}{\pi^A_i} - \frac{ R^B_i }{ \pi^B (X_i; \alphahat) } \right\}
X \; \sigma^2 (X_i)
= 0.
\label{eq:second.term.equals0}
\end{equation}
If the generalised linear model for $m(X; \beta)$ includes an intercept term, then one of the elements of vector $X$ equals one, and so equation~(\ref{eq:second.term.equals0}) immediately implies that the second term in expression~(\ref{eq:SampleB.var.bias.cor}) equals zero (and so expression~(\ref{eq:SampleB.var.bias.cor}) reduces to expression~(\ref{eq:SampleB.var})).
Second, if model $\pi^B (X; \alpha)$ is misspecified and model $m(X; \beta)$ is correctly specified but is not a generalised linear model with canonical link function, then the argument for validity of the double robust variance estimator depends on the model for $\sigma^2 (X_i) = \Var \{ Y_i - m (X_i; \betahat) \mid \curlF \}$ being correctly specified.

Finally, consider the conditional covariance between $\KHone$ and $\HT$ or $\Haj$ given $\curlF$.
Equation~(\ref{eq:cov.KH}) still gives the conditional covariance given $\curlF$ and $\curlY$.
Moreover,
\begin{eqnarray}
  &&
  \Cov ( \KHone - \Ybar, \; \HTHaj - \Ybar \mid \curlF )
  \nonumber \\
  && \hspace{.5cm} =
  E \left\{ \Cov ( \KHone - \Ybar, \; \HTHaj - \Ybar \mid \curlF, \curlY ) \mid \curlF \right\}
  \nonumber \\
  && \hspace{0.9cm} +
  \Cov \left \{ E ( \KHone - \Ybar \mid \curlF, \curlY ), \; E ( \HTHaj - \Ybar \mid \curlF, \curlY ) \mid \curlF \right\}
  \nonumber \\
  && \hspace{.5cm} =
  E \left\{ \Cov ( \KHone - \Ybar, \; \HTHaj - \Ybar \mid \curlF, \curlY ) \mid \curlF \right\}
  \nonumber \\
  && \hspace{0.9cm} +
 \Cov \left\{ o_p (N^{-1/2}), \; o_p (N^{-1/2}) \mid \curlF \right\}
  \nonumber \\
  && \hspace{.5cm} =
  E \left\{ \Cov ( \KHone - \Ybar, \; \HTHaj - \Ybar \mid \curlF, \curlY ) \mid \curlF \right\}
  +
  o_p (N^{-1}).
  \label{eq:covariance.decomposition}
\end{eqnarray}
If the population is large compared to Samples A and B, this covariance should be well approximated by the formula given by equation~(\ref{eq:cov.KH}).

\section{Model-design-based inference for $\KHtwo$}

\label{sect:modeldesign.KH2}

Assume that models $\pi^B (X; \alpha)$ or $m(X; \beta)$ is correctly specified.
Recall equations~(\ref{eq:KH1.expansion}) and~(\ref{eq:KH2.expansion}):
\begin{eqnarray*}
  \KHone - \Ybar
  & = &
    \frac{1}{N_s} \sum_{i=1}^N
  S_i \left[ \frac{ R^A_i }{ \pi^A_i } m(X_i; \betalim)
  +
  \left\{ \frac{ R^B_i }{ \pi^B (X_i) } - 1 \right\}
  \{ Y_i - m(X_i; \betalim) \}
  \right]
  - \mbar + o_p(N^{-1/2})
\end{eqnarray*}
and
\begin{eqnarray*}
  \KHtwo - \Ybar
  & = &
  \frac{1}{N_s} \sum_{i=1}^N S_i \left[ \frac{R^A_i}{\pi^A_i} \{ m (X_i; \betalim) - \mbartrue \}
    + \left\{ \frac{ R^B_i }{ \pi^B (X_i; \alphalim) } - 1 \right\} \{ Y_i - m (X_i; \betalim) \} \right]
  + o_p (N^{1/2}).
\end{eqnarray*}

We see that the only difference between these equations is the replacement of $m(X_i; \beta)$ by $m(X_i; \beta) - \mbartrue$ in the first term.

In particular, we have
\begin{eqnarray}
  &&
  \Cov (\KHtwo - \Ybar, \; \HTHaj - \Ybar \mid \curlF)
\nonumber \\
  && =
  E \left( \Cov \left[
  \frac{1}{N_s} \sum_{i=1}^N \frac{ R^A_i }{ \pi^A_i }
     S_i \{ m(X_i; \betalim) - \mbartrue \}
     , \;
     \frac{1}{N_s} \sum_{i=1}^N S_i \; \frac{ R^A_i }{ \pi^A_i } (Y_i - \Gamma) \mid \curlF, \curlY \right]
     \mid \curlF \right)
     \nonumber \\
     && \hspace{.5cm} +     
     o_p(N^{-1}).
\label{eq:KH2.Haj.cov.onlycurlF}
\end{eqnarray}

\section{Efficiency gain of combined estimator}
  
\label{sect:efficiency.proof}

In this section, we consider the setting where the task is to estimate the mean of $Y$ in the entire population, i.e.\ $S_i=1$ for all $i$.
Define $\rho = \mbox{Cor} (\DR - \Ybar, \HTHaj - \Ybar \mid \curlF) = C / \sqrt{V_H V_D}$.
In this section, we find it convenient to define $\psi = \sqrt{\eta} = \sqrt{V_H / V_D}$ and work with this instead of with $\eta$.
We can rewrite equation~(\ref{eq:var.combined}) as
\begin{eqnarray*}
  \Var \{ (1-w) \HTHaj + w \DR - \Ybar \mid \curlF \}
  & = &
  V_D - \frac{ (V_D - V_D \psi \rho )^2 }{ V_D \psi^2 + V_D - 2 V_D \psi \rho }
  \\
  & = &
  V_D \left( 1 -
  \frac{ (1 - \psi \rho)^2 }{ \psi^2 + 1 - 2 \psi \rho }
  \right)
  \\
  & = &
  V_D \left( 1 -
  \frac{ 1 - 2 \psi \rho + \psi^2 \rho^2 }{ 1 - 2 \psi \rho + \psi^2 }
  \right)
  \\
  & = &
  V_D (1 - Q_D)
\end{eqnarray*}
where
\begin{equation*}
Q_D = \frac{ 1 - 2 \psi \rho + \psi^2 \rho^2 }{ 1 - 2 \psi \rho + \psi^2 }.
\end{equation*}
Since, $G = C / V_H$, $Q_D$ can be rewritten in terms of $\psi$ and $G$ as
\begin{equation}
  Q_D = \frac{ 1 - 2 \psi^2 G + \psi^{4} G^2 }{ 1 - 2 \psi^2 G + \psi^2 }.
  \label{eq:QD.appendix}
\end{equation}
Similarly, $Q_H$ can be shown to be
\begin{eqnarray}
  Q_H
  & = &
  \frac{ 1 - 2 \psi^{-1} \rho + \psi^{-2} \rho^2 }{ 1 - 2 \psi^{-1} \rho + \psi^{-2} }.
\nonumber \\
& = &
\frac{ 1 - 2 G + G^2 }{ 1 - 2 G + \psi^{-2} }.
\label{eq:QH.appendix}
\end{eqnarray}

We shall consider the combination of $\DRone$ and $\HT$ and the combination of $\DRtwo$ and $\Haj$.
When $\DR$ is $\PLone$ or $\PLtwo$, we shall assume model $\pi^B (X; \alpha)$ is correctly specified.
When $\DR$ is $\KHone$ or $\KHtwo$, we shall assume either model $\pi^B (X; \alpha)$ or $m(X; \beta)$ is correctly specified.

As pointed out in Section~\ref{sect:efficiency.gain} of the main text, $\psi \leq G^{-1/2}$ for both the combination of $\DRone$ and $\HT$ and the combination of $\DRtwo$ and $\Haj$.
It is easy to see from equation~(\ref{eq:QH.appendix}) that for any fixed value of $G$, $Q_H$ is a increasing function of $\psi$.
we shall now prove that for any fixed value of $G$, $Q_D$ is a decreasing function of $\psi$ over the range of values $(0, G^{-1/2})$ for $\psi$.

From equation~(\ref{eq:QD.appendix}), we obtain
\begin{eqnarray*}
  \frac{ \partial Q_D }{ \partial \psi }
  & = &
  \frac{ (-4 \psi G + 4 \psi^3 G^2) (1 - 2 \psi^2 G +\psi^2)
    + (4 \psi G - 2 \psi) (1 - 2 \psi^2 G + \psi^4 G^2) }
  { (1 - 2 \psi^2 G + \psi^2)^2 }
  \\
  & = &
  \frac{ 2 \psi \{ -1 + 2 \psi^2 G^2 - \psi^4 G^2 (2G-1) \} }
  { (1 - 2 \psi^2 G + \psi^2)^2 }.
  \end{eqnarray*}
Therefore, to show that $Q_D$ is a decreasing function of $\psi$ over the range $\psi \in (0, G^{-1/2})$ we need to show that $-1 + 2 \psi^2 G^2 - \psi^4 G^2 (2G-1) < 0$ over that range.
We can regard the equation $-1 + 2 \psi^2 G^2 - \psi^4 G^2 (2G-1) = 0$ as a quadratic equation in $\psi^2$.
Its roots are
\begin{eqnarray*}
  \psi^2
  & = &
  \frac{ -2 G^2 \pm \sqrt{ 4 G^4 - 4 G^2 (2G-1) } }
       { -2 G^2 (2G-1) }
       \\
       & = &
       \frac{1}{G} \mbox{ or } \frac{1}{G (2G-1)}.
\end{eqnarray*}
If $\psi \in (0, G^{-1/2})$, then $\psi^2 \in (0, 1/G)$.
The root $\psi^2 = 1/G$ is not in that range.
When $0 \leq G < 0.5$, $\frac{1}{G (2G-1)}$ is negative, which is not in the range $\psi^2 \in (0, 1/G)$.
When $0.5 \leq G \leq 1$, $\frac{1}{G (2G-1)} > 1/G$, which again is not in the range $\psi^2 \in (0, 1/G)$.
Hence, there are no roots in the range $\psi^2 \in (0, 1/G)$.
Therefore, $Q_D$ must be either an increasing or a decreasing function of $\psi$ over that range.
Since, $Q_D = 1$ when $\psi = 0$ and $Q_D = 0$ when $\psi = G^{-1/2}$, it is evident that $Q_D$ must be a decreasing function of $\psi$.

\begin{figure}
    \centering
\includegraphics[page=1,width=1\textwidth]{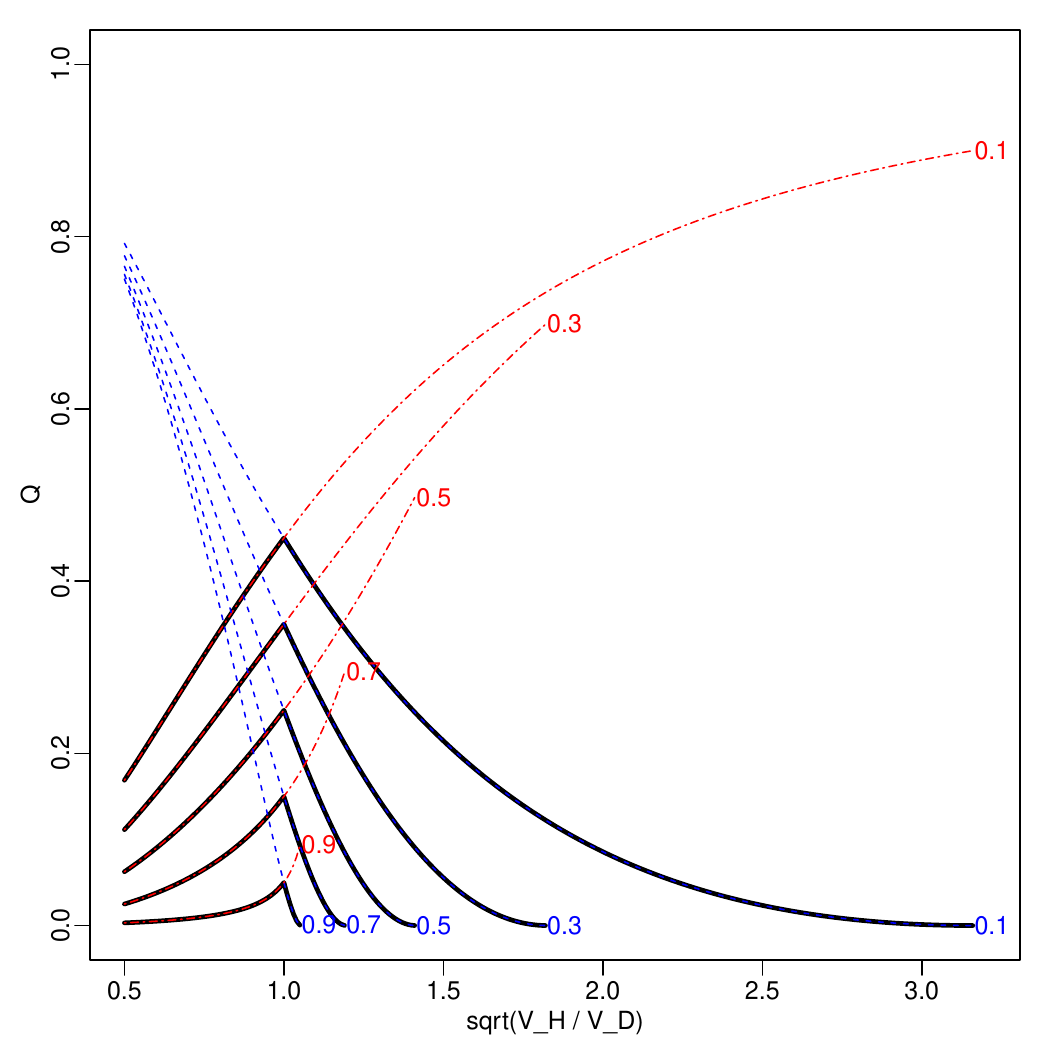}
\caption{Proportion of variance reduction for the combined estimator compared to its component estimator $\HTHaj$ ($Q_H$, red broken-dotted line) and its component $\DR$ estimator ($Q_D$, blue broken line), as a function of $\sqrt{V_H / V_D}$ for each of five values (0.1, 0.3, 0.5, 0.7 and 0.9) of $G = C / V_H$.  Each black line is the minimum of the red and blue lines for that value of $G$.  It shows the proportion of variance reduction for the combined estimator compared to the more efficient of the two component estimators.}
\label{fig:theoreticalRE.Q}
\end{figure}

Figure~\ref{fig:theoreticalRE.Q} shows the dependence of $Q_D$, $Q_H$ and their minimum on $\psi$ and $G$.

We now derive expressions for $C$ and $V_H$ for each of the two combinations of estimators.

\subsection{Combining Horvitz-Thompson and DR1 estimators}

\label{sect:comb.HT.DRone}

From equations~(\ref{eq:cov.KH}), (\ref{eq:cov.dr1.hthajek}) and~(\ref{eq:covariance.decomposition}), we have
\begin{eqnarray*}
  C
  & = &
  E \left[ \Cov \left\{
    \frac{1}{N} \sum_{i=1}^N
    \frac{ R^A_i }{ \pi^A_i } m (X_i; \betalim) , \;
    \frac{1}{N} \sum_{i=1}^N
    \frac{ R^A_i }{ \pi^A_i } Y_i
    \mid \curlF, \curlY \right\} \mid \curlF \right]
  + o_p(N^{-1})
  \nonumber \\
  & = &
  E \left[ \Cov \left\{
    \frac{1}{N} \sum_{i=1}^N
    \frac{ R^A_i }{ \pi^A_i } m (X_i; \betalim) , \;
    \frac{1}{N} \sum_{i=1}^N
    \frac{ R^A_i }{ \pi^A_i } m (X_i; \betalim)
    \mid \curlF, \curlY \right\} \mid \curlF \right]
  \nonumber \\
  && +
  E \left[ \Cov \left\{
    \frac{1}{N} \sum_{i=1}^N
    \frac{ R^A_i }{ \pi^A_i } m (X_i; \betalim) , \;
    \frac{1}{N} \sum_{i=1}^N
    \frac{ R^A_i }{ \pi^A_i } \{ Y_i - m (X_i; \betalim) \}
    \mid \curlF, \curlY \right\} \mid \curlF \right]
  + o_p(N^{-1})  
  \nonumber \\
  & = &
  E \left[ \Var \left\{
    \frac{1}{N} \sum_{i=1}^N
    \frac{ R^A_i }{ \pi^A_i } m (X_i; \betalim)
    \mid \curlF, \curlY \right\} \mid \curlF \right]
  \nonumber \\
  && +
  E \left[ \frac{1}{N^2} \sum_{i=1}^N \sum_{j=1}^N m (X_i; \betalim) \; \{ Y_j - m (X_j; \betalim) \} \;
    \Cov \left(
    \frac{ R^A_i }{ \pi^A_i }, \;
    \frac{ R^A_j }{ \pi^A_j }
    \mid \curlF, \curlY \right)
    \mid \curlF \right]
  + o_p(N^{-1})  
  \nonumber \\
  & = &
  \Var \left\{
    \frac{1}{N} \sum_{i=1}^N
    \frac{ R^A_i }{ \pi^A_i } m (X_i; \betalim)
    \mid \curlF \right\}
  \nonumber \\
  && +
  \frac{1}{N^2} \sum_{i=1}^N \sum_{j=1}^N m (X_i; \betalim) \;
  \Cov \left(
  \frac{ R^A_i }{ \pi^A_i }, \;
  \frac{ R^A_j }{ \pi^A_j }
  \mid \curlF \right) \;
 \{ E(Y_j \mid \curlF) - m (X_j; \betalim) \}
  + o_p(N^{-1})  
  \nonumber \\
  & = &
  \Var \left\{
    \frac{1}{N} \sum_{i=1}^N
    \frac{ R^A_i }{ \pi^A_i } m (X_i; \betalim)
    \mid \curlF \right\}
  \nonumber \\
  && +
  \frac{1}{N^2} \sum_{i=1}^N \sum_{j=1}^N m (X_i; \betalim) \;
  \Cov \left(
  \frac{ R^A_i }{ \pi^A_i }, \;
  \frac{ R^A_j }{ \pi^A_j }
  \mid \curlF \right) \times 0
  + o_p(N^{-1})
    \nonumber \\
  & = &
  \Var \left\{
  \frac{1}{N} \sum_{i=1}^N
  \frac{ R^A_i }{ \pi^A_i } m (X_i; \betalim)
  \mid \curlF \right\}
  + o_p(N^{-1}).
\end{eqnarray*}

Also,
\begin{eqnarray}
  V_H
  & = &
  \Var \left\{ \frac{1}{N} \sum_{i=1}^N
    \frac{ R^A_i }{ \pi^A_i } Y_i \mid \curlF \right\}
    \nonumber \\
    & = &
    \Var \left\{ \frac{1}{N} \sum_{i=1}^N
    \frac{ R^A_i }{ \pi^A_i } m(X_i; \betalim) \mid \curlF \right\}
    + \Var \left[ \frac{1}{N} \sum_{i=1}^N
      \frac{ R^A_i }{ \pi^A_i } \{Y_i - m(X_i; \betalim) \} \mid \curlF \right].
  \label{eq:VHT.expansion}
\end{eqnarray}
Note that equation~(\ref{eq:VHT.expansion}) follows because
\begin{eqnarray}
  &&
\Cov \left[ \frac{1}{N} \sum_{i=1}^N
\frac{ R^A_i }{ \pi^A_i } m(X_i; \betalim), \;
\frac{1}{N} \sum_{i=1}^N
\frac{ R^A_i }{ \pi^A_i } \{Y_i - m(X_i; \betalim) \} \mid \curlF \right]
\nonumber \\
&& \hspace{.4cm} =
\Cov \left( E \left\{ \frac{1}{N} \sum_{i=1}^N
\frac{ R^A_i }{ \pi^A_i } m(X_i; \betalim) \mid \curlF, R^A_1, \ldots, R^A_N \right\}, \right.
\nonumber \\
&& \hspace{1.6cm}
\left. E \left[ \frac{1}{N} \sum_{i=1}^N
  \frac{ R^A_i }{ \pi^A_i } \{Y_i - m(X_i; \betalim) \} \mid \curlF, R^A_1, \ldots, R^A_N \right]
\mid \curlF \right)
\nonumber \\
&& \hspace{.8cm} +
E \left(
\Cov \left[ \frac{1}{N} \sum_{i=1}^N
\frac{ R^A_i }{ \pi^A_i } m(X_i; \betalim), \;
\frac{1}{N} \sum_{i=1}^N
\frac{ R^A_i }{ \pi^A_i } \{Y_i - m(X_i; \betalim) \} \mid \curlF, R^A_1, \ldots, R^A_N \right]
\mid \curlF \right)
\nonumber \\
&& \hspace{.4cm} =
\Cov \left( E \left\{ \frac{1}{N} \sum_{i=1}^N
\frac{ R^A_i }{ \pi^A_i } m(X_i; \betalim) \mid \curlF, R^A_1, \ldots, R^A_N \right\}, \; 0 \mid \curlF \right)
+ E(0 \mid \curlF)
\nonumber \\
&& \hspace{.4cm} = 0.
\label{eq:cov.m.ym.equalzero}
\end{eqnarray}

\subsection{Combining Hajek and DR2 estimators}

Using equations~(\ref{eq:DR2.Haj.cov.onlycurlF}) and~(\ref{eq:KH2.Haj.cov.onlycurlF}), we have
\begin{eqnarray}
  C
  & = &
  E \left( \Cov \left[ \frac{1}{N} \sum_{i=1}^N
    \frac{ R^A_i }{ \pi^A_i } ( Y_i - \Ybar ), \;
    \frac{1}{N} \sum_{j=1}^N
    \frac{ R^A_j }{ \pi^A_j } \{ m(X_j; \betalim) - \mbar \}
    \mid \curlF, \curlY \right] \mid \curlF \right)
  + o_p(N^{-1})
  \nonumber \\
  & = &
  E \left( \Cov \left[
    \frac{1}{N} \sum_{i=1}^N
    \frac{ R^A_i }{ \pi^A_i } ( m(X_i; \betalim) - \mbar ), \;
    \frac{1}{N} \sum_{j=1}^N
    \frac{ R^A_j }{ \pi^A_j } \{ m (X_j; \betalim) - \mbar \}
    \mid \curlF, \curlY \right] \mid \curlF \right)
  \nonumber \\
  && +
  E \left( \Cov \left[
    \frac{1}{N} \sum_{i=1}^N
    \frac{ R^A_i }{ \pi^A_i } \{ Y_i - m(X_i; \betalim) - \Ybar + \mbar \}, \;
    \frac{1}{N} \sum_{j=1}^N
    \frac{ R^A_j }{ \pi^A_j } \{ m (X_j; \betalim) - \mbar \}
    \mid \curlF, \curlY \right] \mid \curlF \right)
  \nonumber \\
  && +  
  o_p(N^{-1})  
  \nonumber \\
  & = &
  E \left( \Var \left[
    \frac{1}{N} \sum_{i=1}^N
    \frac{ R^A_i }{ \pi^A_i } \{ m (X_i; \betalim) - \mbar \}
    \mid \curlF, \curlY \right] \mid \curlF \right)
  \nonumber \\
  && +
  E \left[ \frac{1}{N^2} \sum_{i=1}^N \sum_{j=1}^N
    \{ Y_i - m(X_i; \betalim) - \Ybar + \mbar \} \; \{ m (X_j; \betalim) - \mbar \} \; 
    \Cov \left(
    \frac{ R^A_i }{ \pi^A_i }, \;
    \frac{ R^A_j }{ \pi^A_j }
    \mid \curlF, \curlY \right)
    \mid \curlF \right]
  \nonumber \\
  && +  
  o_p(N^{-1})
  \nonumber \\
  & = &
  \Var \left\{
    \frac{1}{N} \sum_{i=1}^N
    \frac{ R^A_i }{ \pi^A_i } \{ m (X_i; \betalim) - \mbar \}
    \mid \curlF \right\}
  \nonumber \\
  && +
  \frac{1}{N^2} \sum_{i=1}^N \sum_{j=1}^N \{ m (X_j; \betalim) - \mbar \} \;
  \Cov \left(
  \frac{ R^A_i }{ \pi^A_i }, \;
  \frac{ R^A_j }{ \pi^A_j }
  \mid \curlF \right) \; \nonumber \\
  && \qquad \times
  \{ E(Y_i \mid \curlF) - m (X_i; \betalim) - E(\Ybar \mid \curlF) + \mbar \}
  \nonumber \\
  && +  
  o_p(N^{-1})
  \nonumber \\
  & = &
  \Var \left[
    \frac{1}{N} \sum_{i=1}^N
    \frac{ R^A_i }{ \pi^A_i } \{ m (X_i; \betalim) - \mbar \}
    \mid \curlF \right]
  \nonumber \\
  && +
  \frac{1}{N^2} \sum_{i=1}^N \sum_{j=1}^N \{ m (X_j; \betalim) - \mbar \} \;
  \Cov \left(
  \frac{ R^A_i }{ \pi^A_i }, \;
  \frac{ R^A_j }{ \pi^A_j }
  \mid \curlF \right) \times 0
  + o_p(N^{-1})
    \nonumber \\
  & = &
  \Var \left[
    \frac{1}{N} \sum_{i=1}^N
    \frac{ R^A_i }{ \pi^A_i } \{ m (X_i; \betalim) - \mbar \}
    \mid \curlF \right]
  + o_p(N^{-1}).
  \nonumber
\end{eqnarray}

Also,
\begin{eqnarray}
  V_H
  & = &
  \Var \left\{ \frac{1}{N} \sum_{i=1}^N
  \frac{ R^A_i }{ \pi^A_i } ( Y_i - \Ybar ) \mid \curlF \right\}
    + o_p(N^{-1})
    \label{eq:VHaj.preexpansion} \\
    & = &
    \Var \left[ \frac{1}{N} \sum_{i=1}^N
      \frac{ R^A_i }{ \pi^A_i } \{ m(X_i; \betalim) - \mbar \} \mid \curlF \right]
    +
    \Var \left[ \frac{1}{N} \sum_{i=1}^N
      \frac{ R^A_i }{ \pi^A_i } \{ Y_i - \Ybar - m(X_i; \betalim) + \mbar \} \mid \curlF \right]
    \nonumber \\
    && +
    o_p(N^{-1}).
  \label{eq:VHaj.expansion}
\end{eqnarray}
Note that line~(\ref{eq:VHaj.expansion}) follows from expanding $Y_i - \Ybar$ in the denominator of expression~(\ref{eq:VHaj.preexpansion}) as $\{ m(X_i; \betalim) - \mbar \} + \{ Y_i - \Ybar - m(X_i; \betalim) + \mbar \}$ and then using the fact that
\begin{equation}
  \Cov \left[
        \frac{1}{N} \sum_{j=1}^N
    \frac{ R^A_j }{ \pi^A_j } \{ m(X_j; \betalim) - \mbar \}, \;
    \frac{1}{N} \sum_{i=1}^N
    \frac{ R^A_i }{ \pi^A_i } \{ Y_i - \Ybar - m(X_i; \betalim) + \mbar \}
    \mid \curlF \right]
= 0.
\label{eq:zero.cov.Ymmm}
\end{equation}
The proof of equation~(\ref{eq:zero.cov.Ymmm}) is analogous to the proof of equation~(\ref{eq:cov.m.ym.equalzero}).

\section{Data-generating mechanism for simulation study}

\label{sect:appendix.simstudy.dgm}

Let $H_{jq} \sim \mbox{NegativeBinomial} (\mbox{mean}=100, \mbox{variance}=400)$ be the number of households of $q$ individuals in cluster $j$ ($q=1,2,3$ and $j=1, \ldots, 1000$).
Thus, the number of individuals in cluster $j$ is $N_j = \sum_{q=1}^3 q H_j$, which has expectation 600 and variance 2400, and the population size is $n = \sum_{j=1}^{1000} N_j$, which has expectation $1000 \times 600 = 600000$ and standard deviation $\sqrt{1000 \times 2400} = 1549$.

Let $H_j = \sum_{q=1}^3 H_{jq}$ be the total number of households in cluster $j$.
Let $$H_j^* = 1000 H_j \left/ \sum_{k=1}^{1000} H_k \right. + \mbox{Normal} (0, 0.1^2)$$ be the ratio of the number of households in cluster $j$ to the mean number of households per cluster plus a cluster-specific random effect.
Let $G_i$ denote the index of the cluster to which individual $i$ belongs, and let $B_i$ denote the number of individuals in the household to which individual $i$ belongs divided by the maximum household size.
So, $B=1/3$, 2/3 or 1.
Let $\mu_i = -1.3 + 0.5 H_{G_i}^* + 0.5 B_i$. 
Given $\mu_1, \ldots, \mu_n$, the $4n$ random variables $X_{11}, X_{21}, X_{31}, X_{41}, \ldots, X_{1n}, X_{2n}, X_{3n}, X_{4n}$ are independently distributed with
\begin{eqnarray*}
  X_{1i} \mid \mu_i & \sim & \mbox{Normal} (\mu_i, 0.4^2) \\
  X_{2i} \mid \mu_i & \sim & \mbox{Normal} (\mu_i, 0.2^2) \\
  X_{3i} \mid \mu_i & \sim & \mbox{Bernoulli} \big( \mbox{expit} (-1.35 + \mu_i) \big) \\
  X_{4i} \mid \mu_i & \sim & \mbox{Bernoulli} \big( \mbox{expit} (2 \mu_i) \big)
\end{eqnarray*}

Note that marginally, $P(X_{3i} = 1) = 0.2$ and $P(X_{3i} = 1) = 0.5$.

\section{Additional simulation study results}

\label{sect:appendix.simstudy.results}

\begin{figure}
\begin{center}
 \includegraphics[width=1\textwidth,page=13]{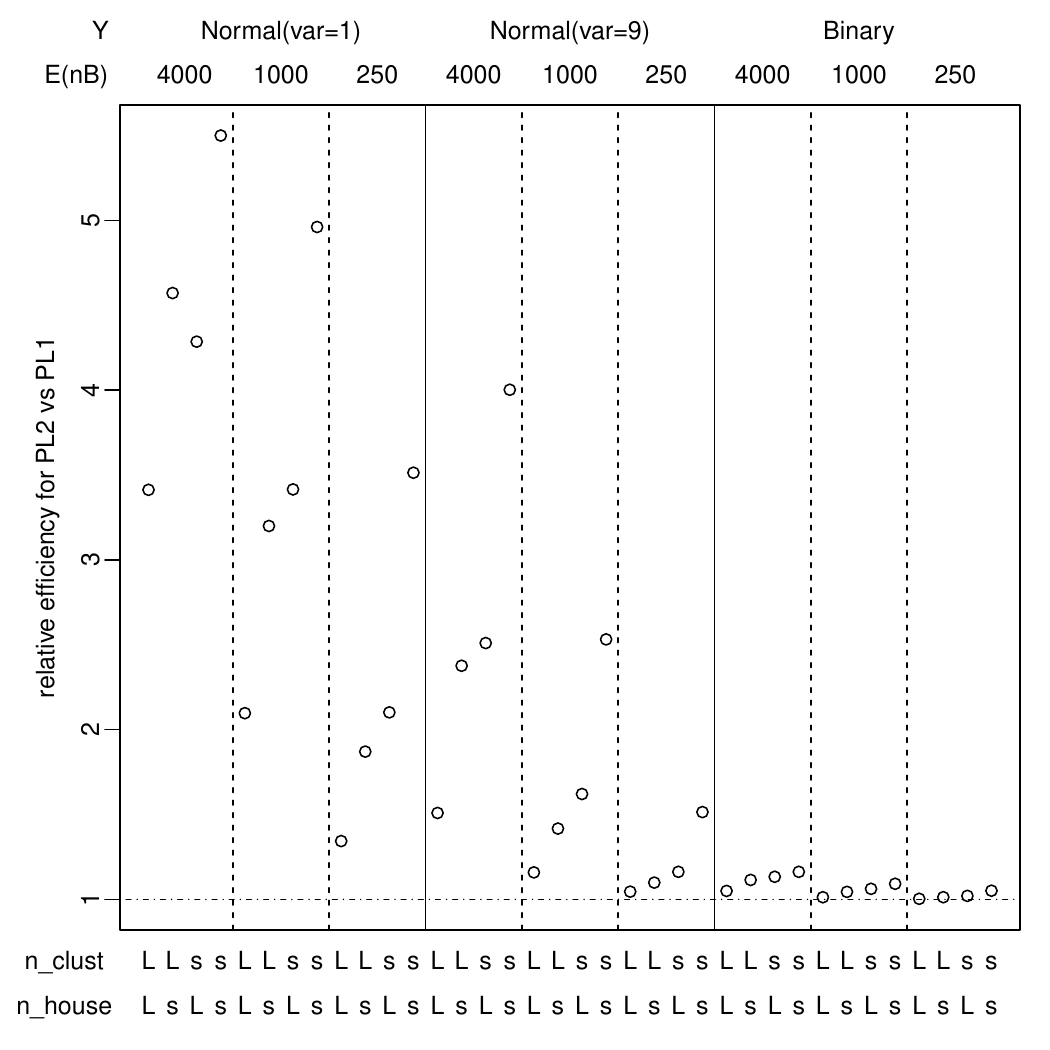}
\hspace*{-5mm}
\end{center}
\caption{Relative efficiency of $\PLtwo$ compared to $\PLone$ in 36 scenarios.  The 36 scenarios correspond to three conditional distributions of $Y$ given $X$, three expected size of Sample B, and two numbers of sampled clusters and of sampled households within sampled clusters in Sample A.  For the last two, `$L$' denotes large (200 clusters or 20 households) and `$s$' denotes small (50 clusters or 5 households).  The four combinations of 200 or 50 clusters and 20 or 5 households imply 4000, 1000, 1000 or 250 individuals in Sample A.}
\label{fig:releff.DR1.DR2}
\end{figure}

\begin{figure}
\begin{center}
 \includegraphics[width=1\textwidth,page=16]{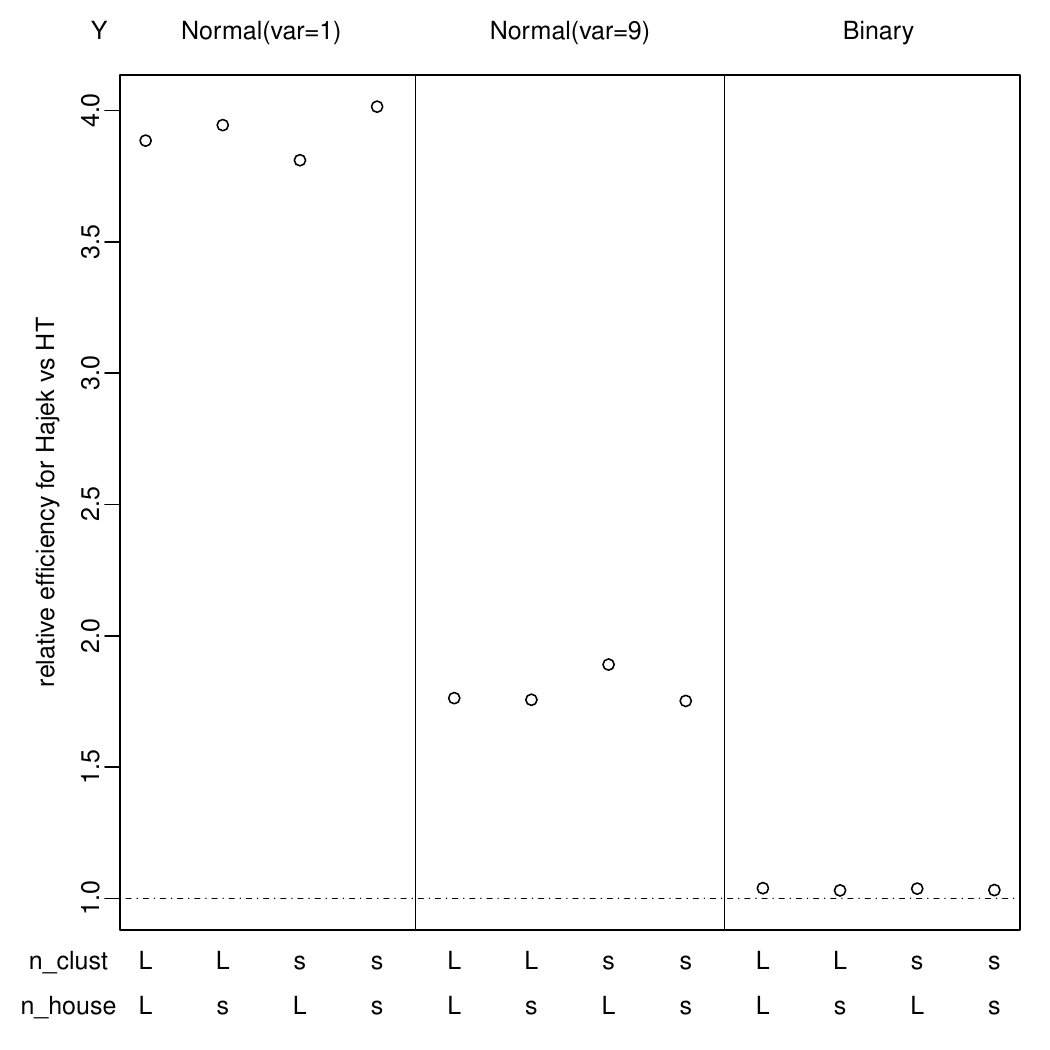}
\hspace*{-5mm}
\end{center}
\caption{Relative efficiency of $\Haj$ compared to $\HT$ in 12 scenarios.  The 12 scenarios correspond to three conditional distributions of $Y$ given $X$, and two numbers of sampled clusters and of sampled households within sampled clusters in Sample A.  For the last two, `$L$' denotes large (200 clusters or 20 households) and `$s$' denotes small (50 clusters or 5 households).  The four combinations of 200 or 50 clusters and 20 or 5 households imply 4000, 1000, 1000 or 250 individuals in Sample A.} 
\label{fig:releff.HT.Haj}
\end{figure}

\begin{figure}
\begin{center}
 \includegraphics[width=1\textwidth,page=2]{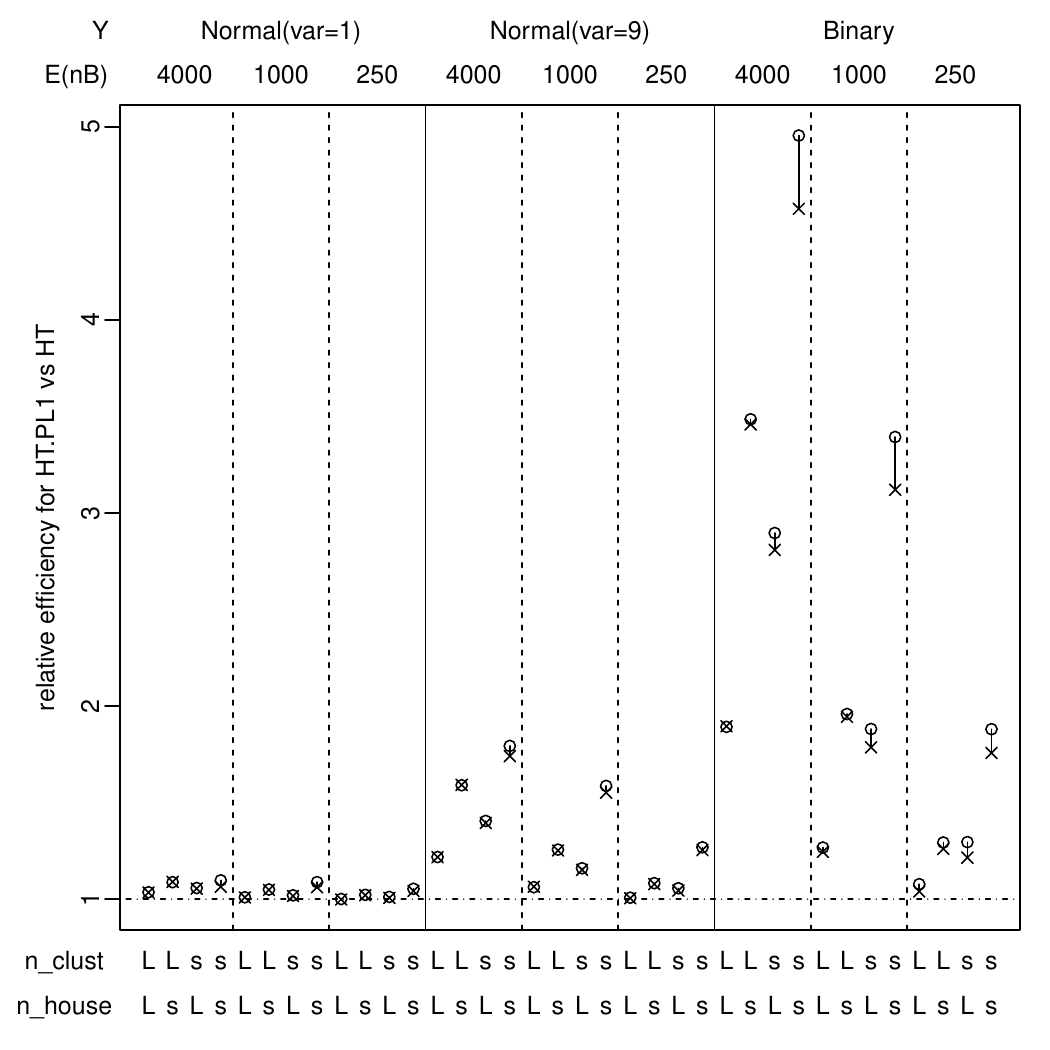}
\hspace*{-5mm}
\end{center}
\caption{Relative efficiency of $\HTPL$ compared to $\HT$ in 36 scenarios.  For each scenario.  The dots indicate the theoretical asymptotic relative efficiency and the crosses indicate the empirical relative efficiency.  The 36 scenarios correspond to three conditional distributions of $Y$ given $X$, three expected size of Sample B, and two numbers of sampled clusters and of sampled households within sampled clusters in Sample A.  For the last two, `$L$' denotes large (200 clusters or 20 households) and `$s$' denotes small (50 clusters or 5 households).} 
\label{fig:releff.DR1eff.HT}
\end{figure}

\begin{figure}
\begin{center}
 \includegraphics[width=1\textwidth,page=1]{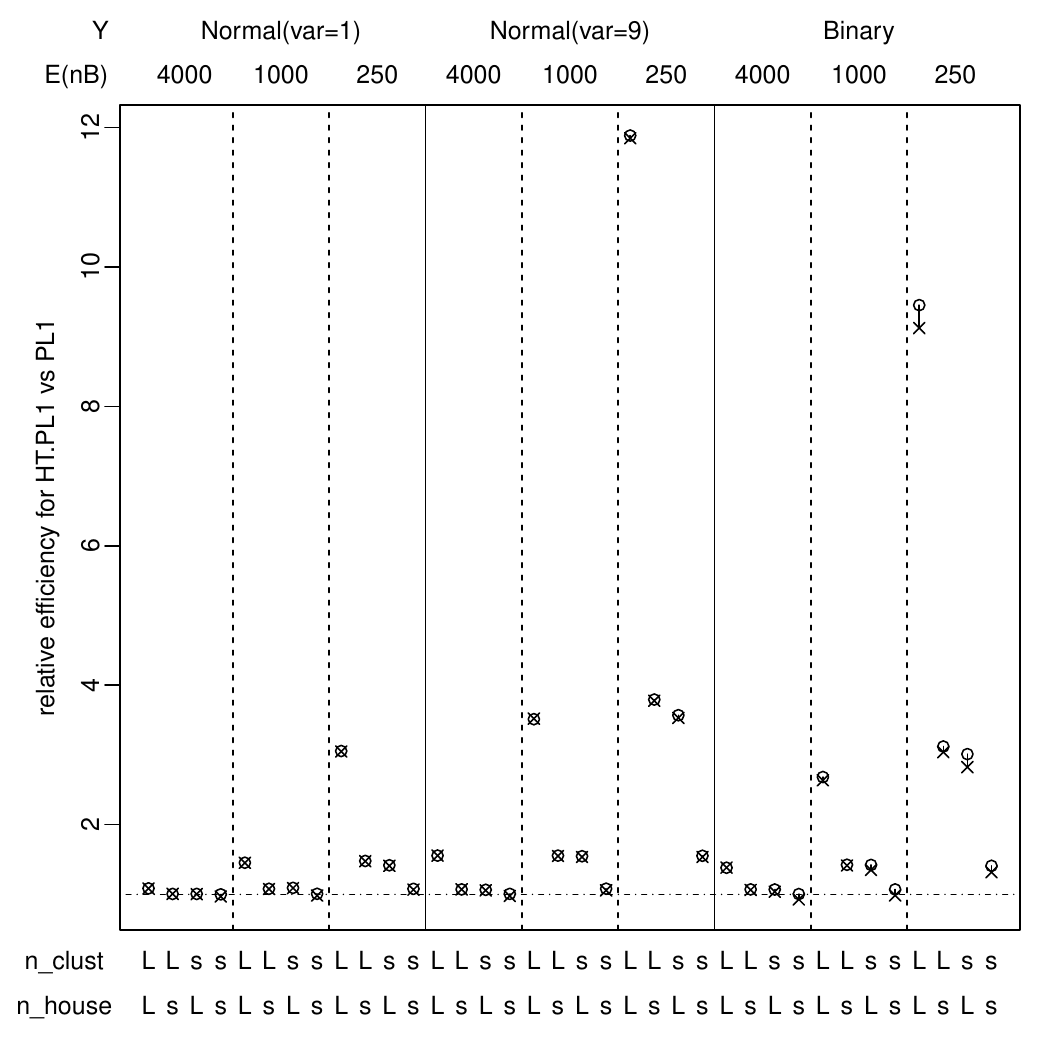}
\hspace*{-5mm}
\end{center}
\caption{Relative efficiency of $\HTPL$ compared to $\PLone$ in 36 scenarios.  The dots indicate the theoretical asymptotic relative efficiency and the crosses indicate the empirical relative efficiency. The 36 scenarios correspond to three conditional distributions of $Y$ given $X$, three expected size of Sample B, and two numbers of sampled clusters and of sampled households within sampled clusters in Sample A.  For the last two, `$L$' denotes large (200 clusters or 20 households) and `$s$' denotes small (50 clusters or 5 households).} 
\label{fig:releff.DR1eff.DR1}
\end{figure}

\begin{figure}
\begin{center}
 \includegraphics[width=1\textwidth,page=27]{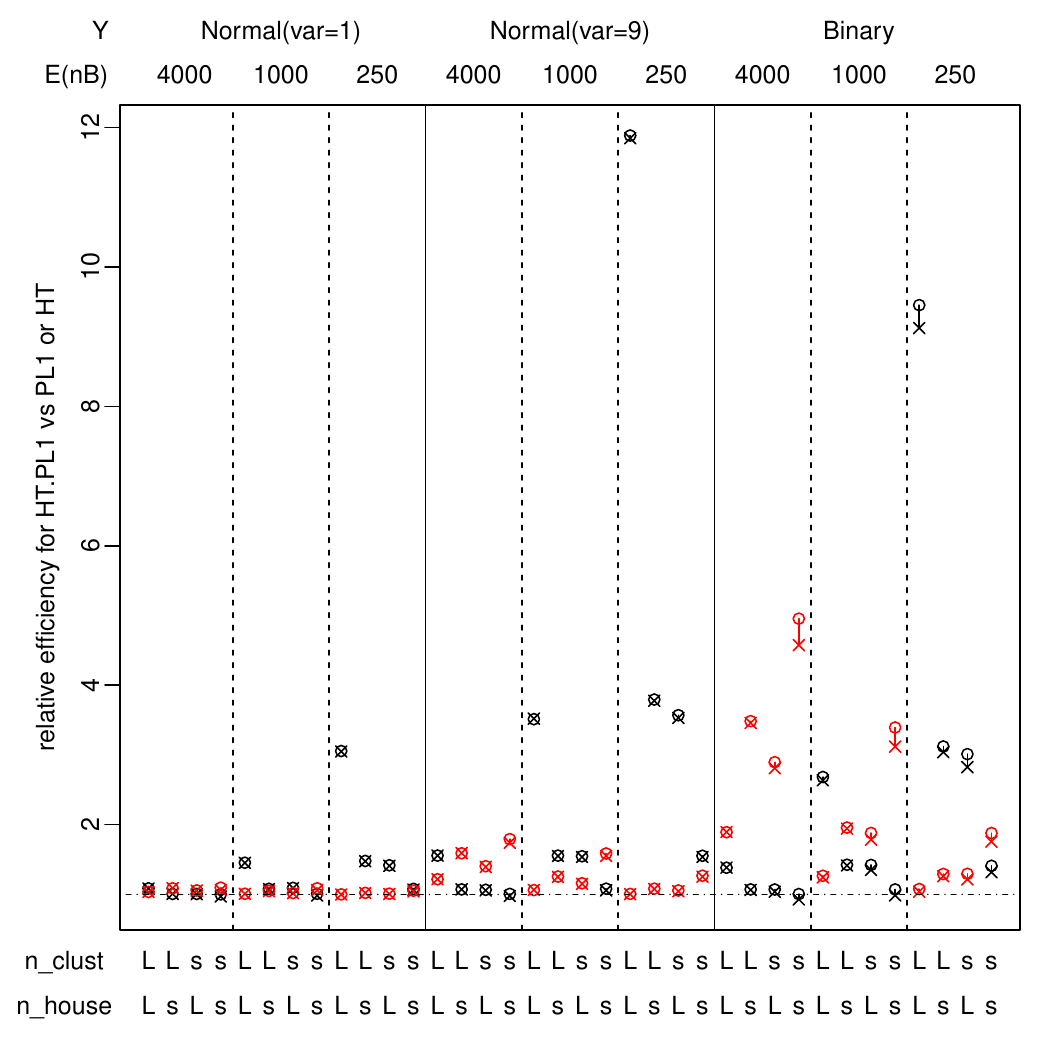}
\hspace*{-5mm}
\end{center}
\caption{Relative efficiency of $\HTPL$ compared to $\PLone$ (black) or $\HT$ (red) in 36 scenarios.  The dots indicate the theoretical asymptotic relative efficiency and the crosses indicate the empirical relative efficiency. The 36 scenarios correspond to three conditional distributions of $Y$ given $X$, three expected size of Sample B, and two numbers of sampled clusters and of sampled households within sampled clusters in Sample A.  For the last two, `$L$' denotes large (200 clusters or 20 households) and `$s$' denotes small (50 clusters or 5 households).}
\label{fig:releff.DR1eff.DR1HT}
\end{figure}

\begin{figure}
\begin{center}
 \includegraphics[width=1\textwidth,page=8]{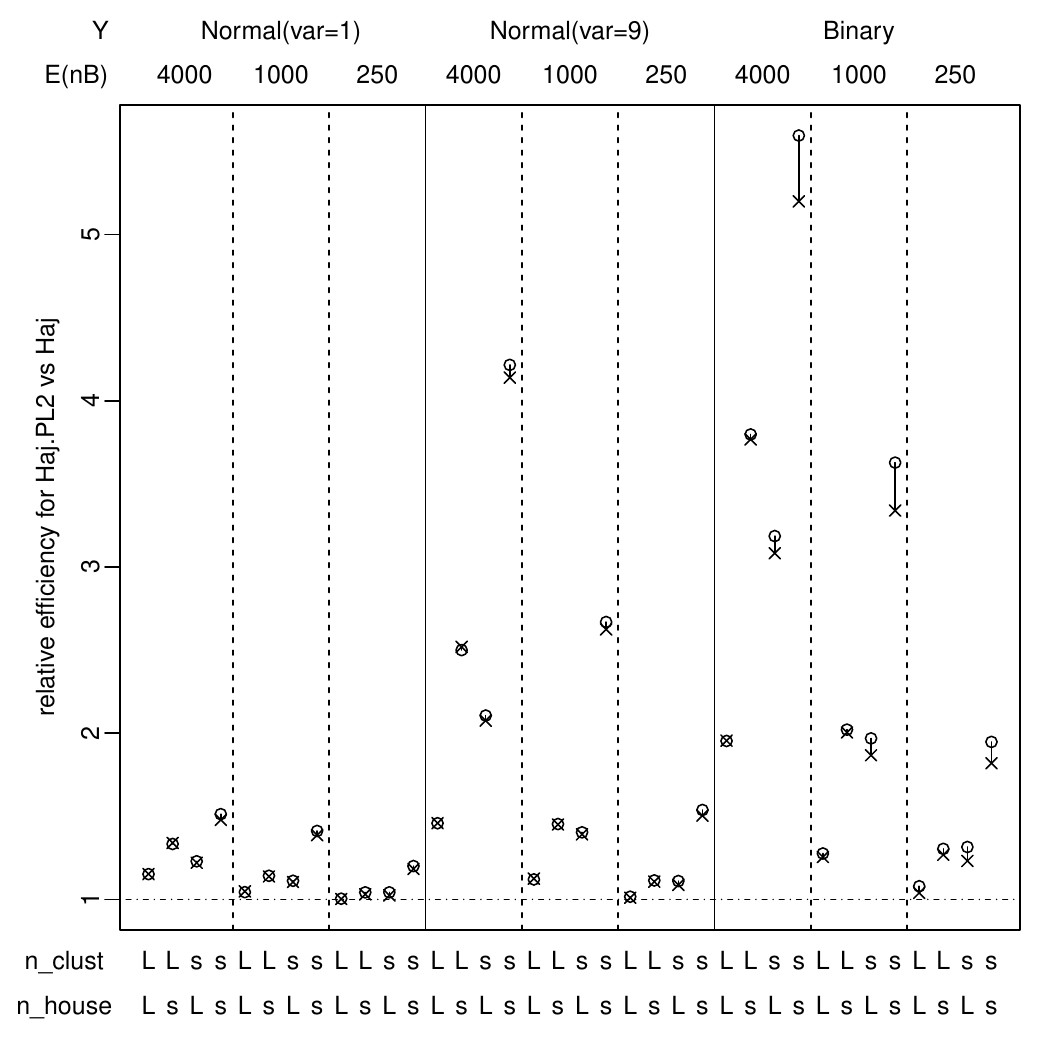}
\hspace*{-5mm}
\end{center}
\caption{Relative efficiency of $\HajPL$ compared to $\Haj$ in 36 scenarios.  The dots indicate the theoretical asymptotic relative efficiency and the crosses indicate the empirical relative efficiency.  The 36 scenarios correspond to three conditional distributions of $Y$ given $X$, three expected size of Sample B, and two numbers of sampled clusters and of sampled households within sampled clusters in Sample A.  For the last two, `$L$' denotes large (200 clusters or 20 households) and `$s$' denotes small (50 clusters or 5 households).} 
\label{fig:releff.DR2eff.Haj}
\end{figure}

\begin{figure}
\begin{center}
 \includegraphics[width=1\textwidth,page=7]{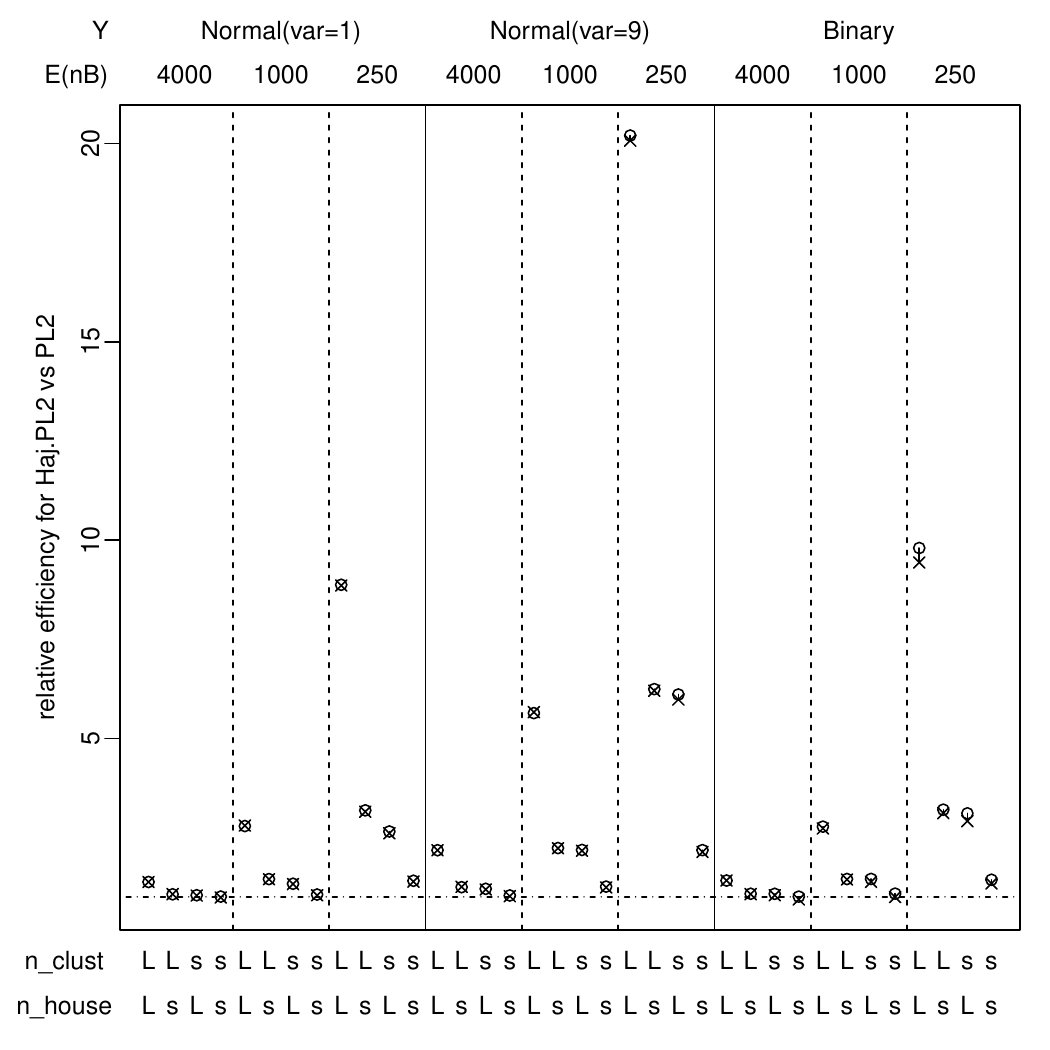}
\hspace*{-5mm}
\end{center}
\caption{Relative efficiency of $\HajPL$ compared to $\PLtwo$ in 36 scenarios.  The dots indicate the theoretical asymptotic relative efficiency and the crosses indicate the empirical relative efficiency.  The 36 scenarios correspond to three conditional distributions of $Y$ given $X$, three expected size of Sample B, and two numbers of sampled clusters and of sampled households within sampled clusters in Sample A.  For the last two, `$L$' denotes large (200 clusters or 20 households) and `$s$' denotes small (50 clusters or 5 households).} 
\label{fig:releff.DR2eff.DR2}
\end{figure}

\begin{figure}
\begin{center}
 \includegraphics[width=1\textwidth,page=30]{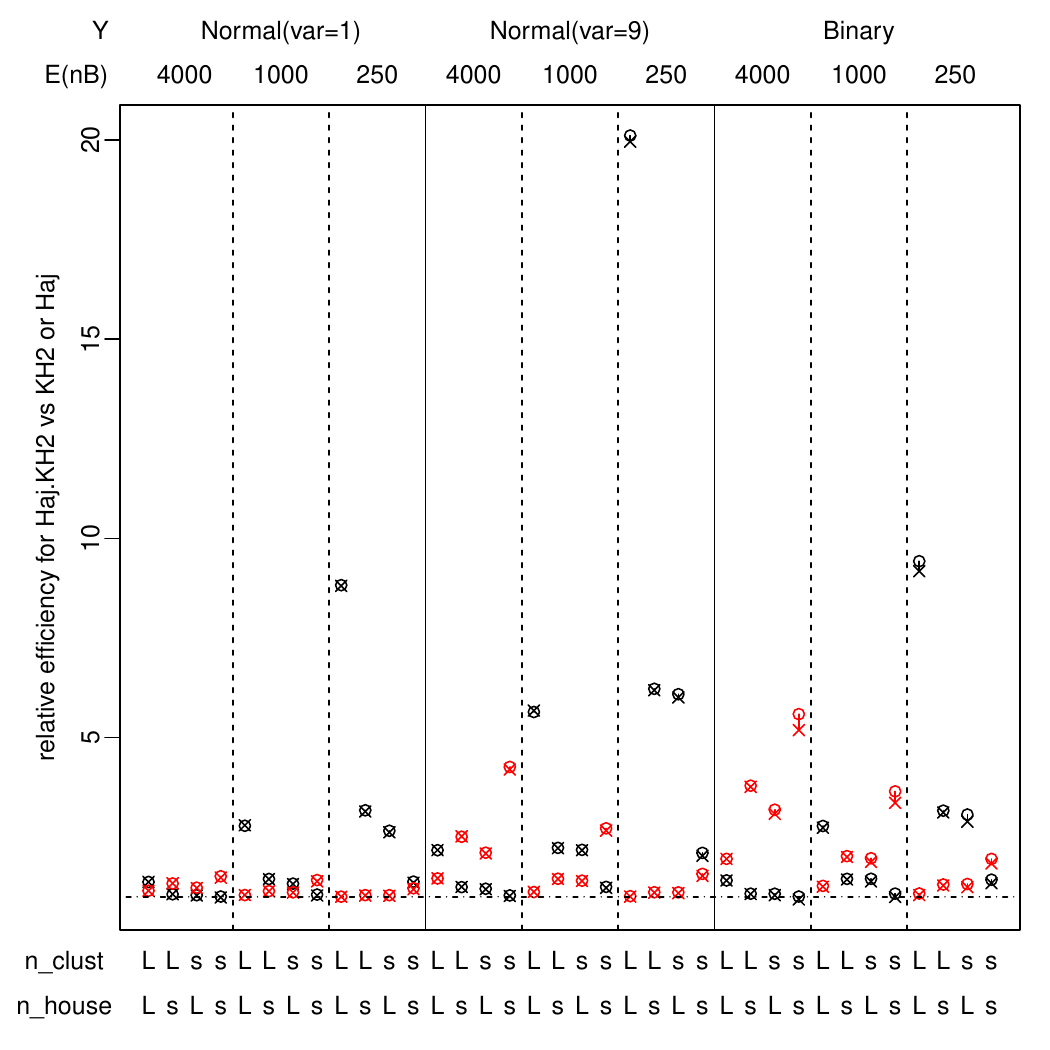}
\hspace*{-5mm}
\end{center}
\caption{Relative efficiency of $\HajPL$ compared to $\PLtwo$ (black) or $\Haj$ (red) in 36 scenarios.  The dots indicate the theoretical asymptotic relative efficiency and the crosses indicate the empirical relative efficiency.  The 36 scenarios correspond to three conditional distributions of $Y$ given $X$, three expected size of Sample B, and two numbers of sampled clusters and of sampled households within sampled clusters in Sample A.  For the last two, `$L$' denotes large (200 clusters or 20 households) and `$s$' denotes small (50 clusters or 5 households).} 
\label{fig:releff.DR2eff.DR2Haj}
\end{figure}

\begin{figure}
\begin{center}
 \includegraphics[width=1\textwidth,page=14]{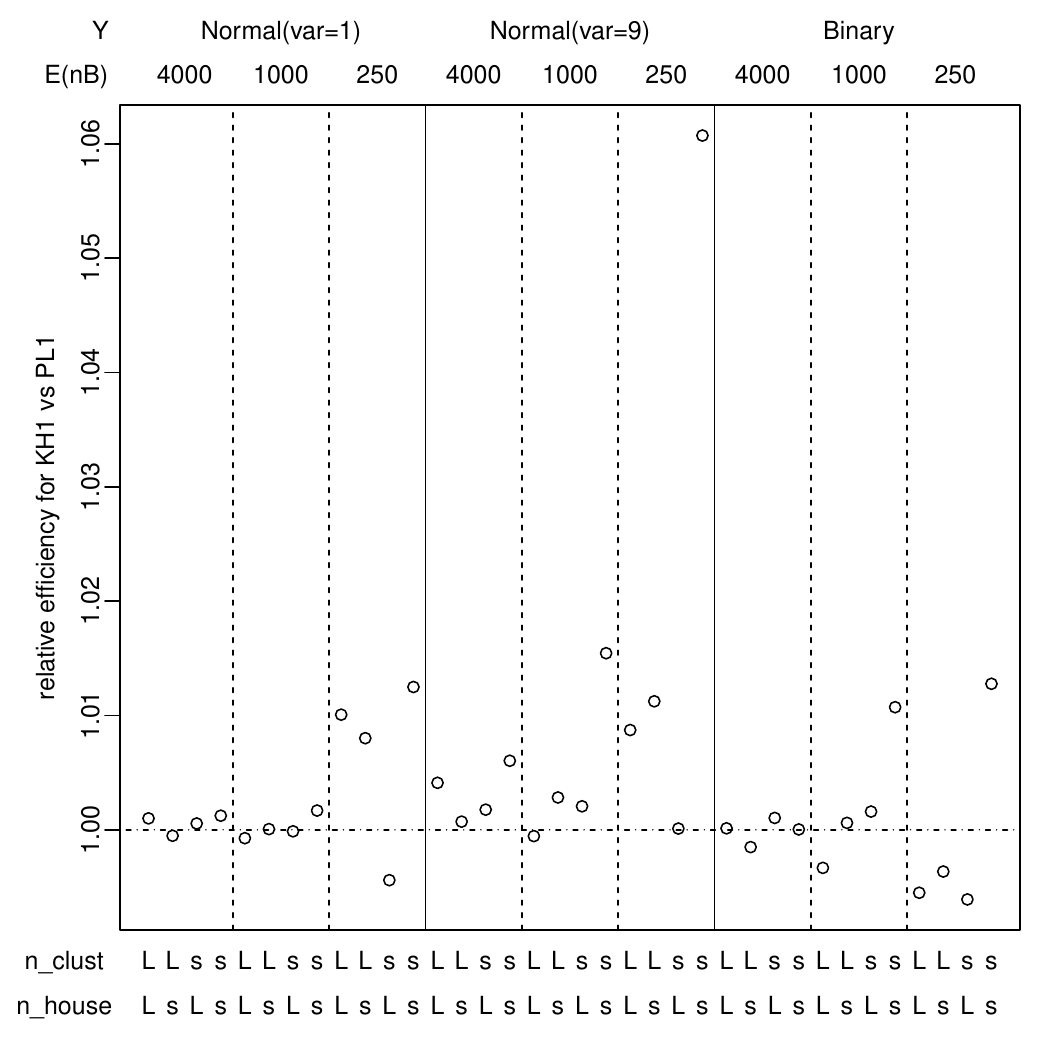}
\hspace*{-5mm}
\end{center}
\caption{Relative efficiency of $\KHone$ compared to $\PLone$ in 36 scenarios.  The 36 scenarios correspond to three conditional distributions of $Y$ given $X$, three expected size of Sample B, and two numbers of sampled clusters and of sampled households within sampled clusters in Sample A.  For the last two, `$L$' denotes large (200 clusters or 20 households) and `$s$' denotes small (50 clusters or 5 households).} 
\label{fig:releff.KH.DR1}
\end{figure}

\begin{figure}
\begin{center}
 \includegraphics[width=1\textwidth,page=15]{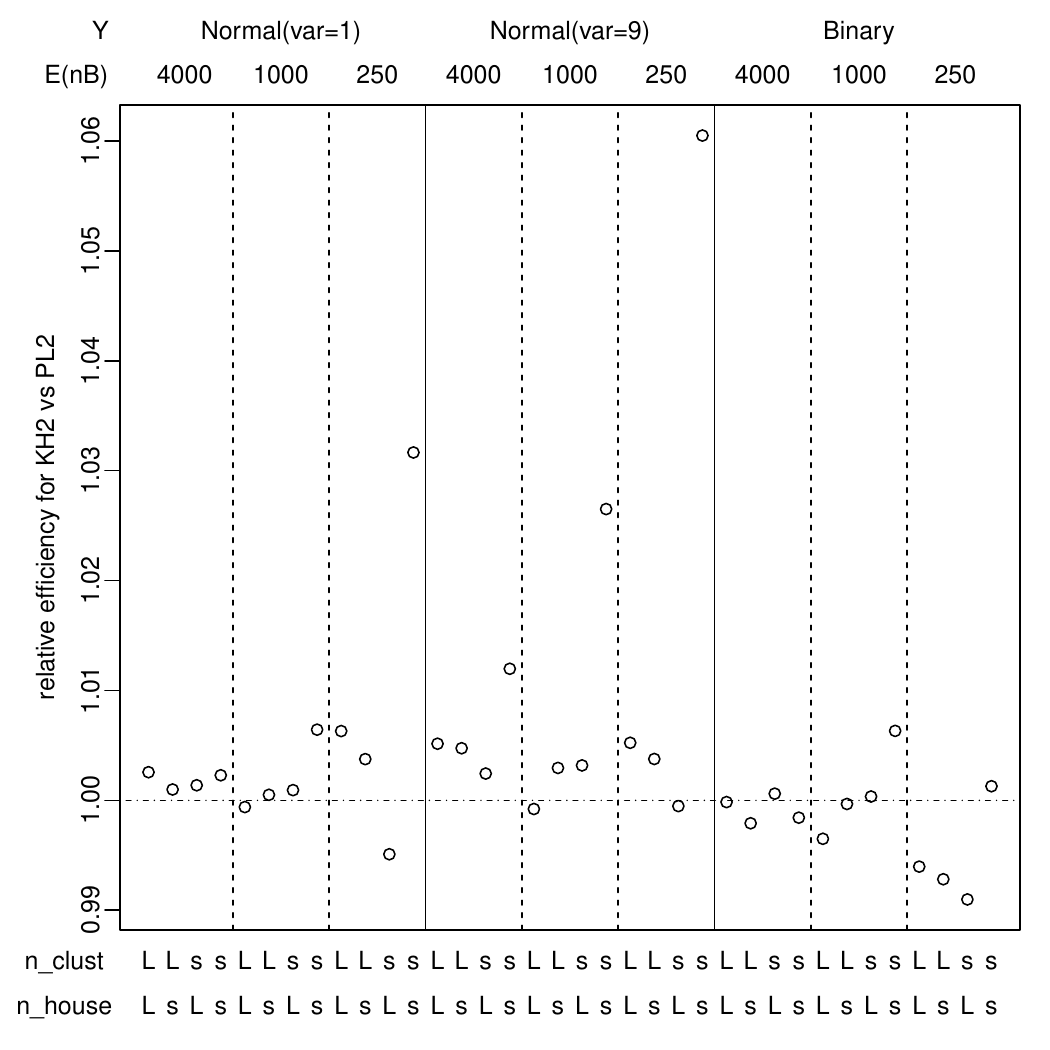}
\hspace*{-5mm}
\end{center}
\caption{Relative efficiency of $\KHtwo$ compared to $\PLtwo$ in 36 scenarios.  The 36 scenarios correspond to three conditional distributions of $Y$ given $X$, three expected size of Sample B, and two numbers of sampled clusters and of sampled households within sampled clusters in Sample A.  For the last two, `$L$' denotes large (200 clusters or 20 households) and `$s$' denotes small (50 clusters or 5 households).} 
\label{fig:releff.KH2.DR2}
\end{figure}

\begin{figure}
\begin{center}
 \includegraphics[width=1\textwidth,page=6]{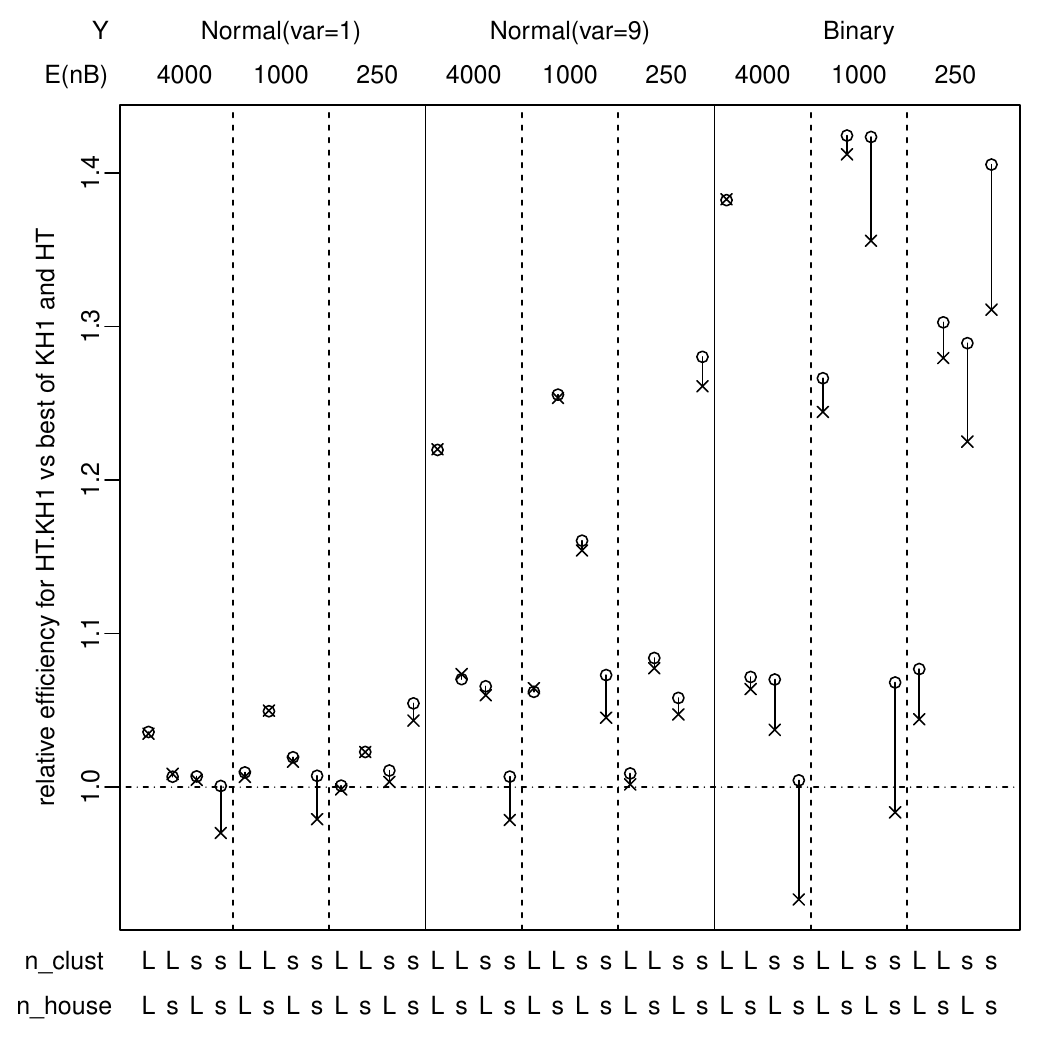}
\hspace*{-5mm}
\end{center}
\caption{Relative efficiency of $\HTKH$ compared to the most efficient of $\HT$ and $\KHone$ in 36 scenarios. The dots indicate the theoretical asymptotic relative efficiency and the crosses indicate the empirical relative efficiency.  The 36 scenarios correspond to three conditional distributions of $Y$ given $X$, three expected size of Sample B, and two numbers of sampled clusters and of sampled households within sampled clusters in Sample A.  For the last two, `$L$' denotes large (200 clusters or 20 households) and `$s$' denotes small (50 clusters or 5 households).} 
\label{fig:releff.KHeff.best}
\end{figure}

\begin{figure}
\begin{center}
 \includegraphics[width=1\textwidth,page=5]{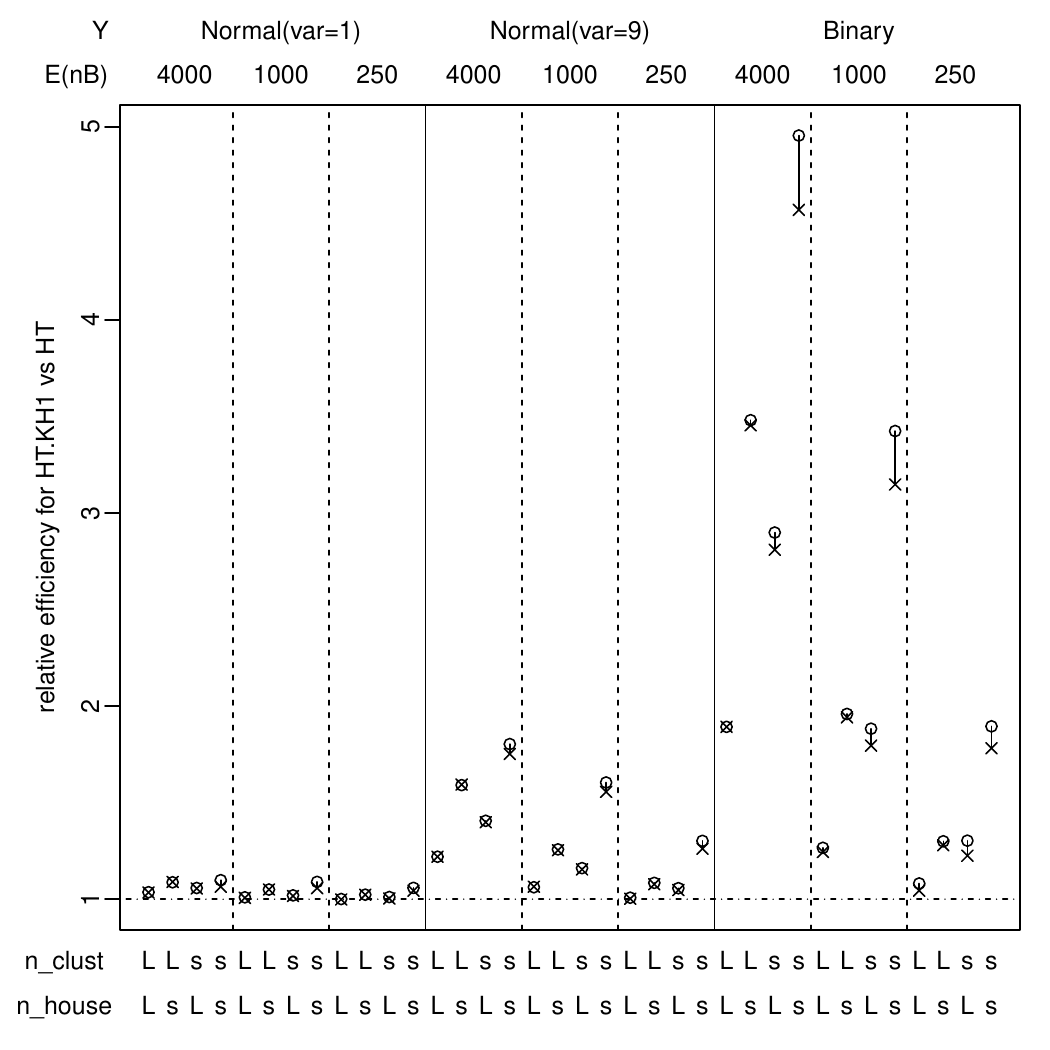}
\hspace*{-5mm}
\end{center}
\caption{Relative efficiency of $\HTKH$ compared to $\HT$ in 36 scenarios.  The dots indicate the theoretical asymptotic relative efficiency and the crosses indicate the empirical relative efficiency.  The 36 scenarios correspond to three conditional distributions of $Y$ given $X$, three expected size of Sample B, and two numbers of sampled clusters and of sampled households within sampled clusters in Sample A.  For the last two, `$L$' denotes large (200 clusters or 20 households) and `$s$' denotes small (50 clusters or 5 households).} 
\label{fig:releff.KHeff.HT}
\end{figure}

\begin{figure}
\begin{center}
 \includegraphics[width=1\textwidth,page=4]{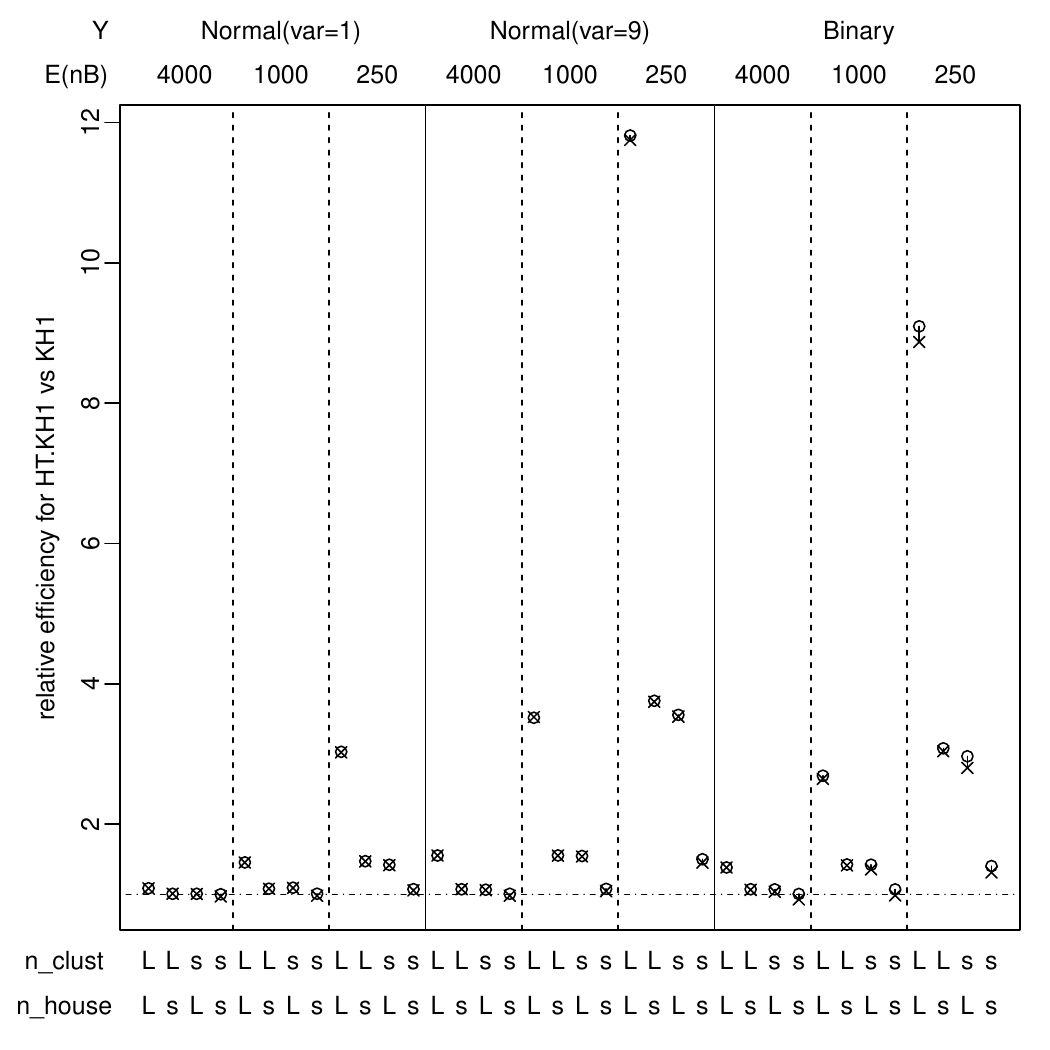}
\hspace*{-5mm}
\end{center}
\caption{Relative efficiency of $\HTKH$ compared to $\KHone$ in 36 scenarios.  The dots indicate the theoretical asymptotic relative efficiency and the crosses indicate the empirical relative efficiency.  The 36 scenarios correspond to three conditional distributions of $Y$ given $X$, three expected size of Sample B, and two numbers of sampled clusters and of sampled households within sampled clusters in Sample A.  For the last two, `$L$' denotes large (200 clusters or 20 households) and `$s$' denotes small (50 clusters or 5 households).} 
\label{fig:releff.KHeff.KH}
\end{figure}

\begin{figure}
\begin{center}
 \includegraphics[width=1\textwidth,page=28]{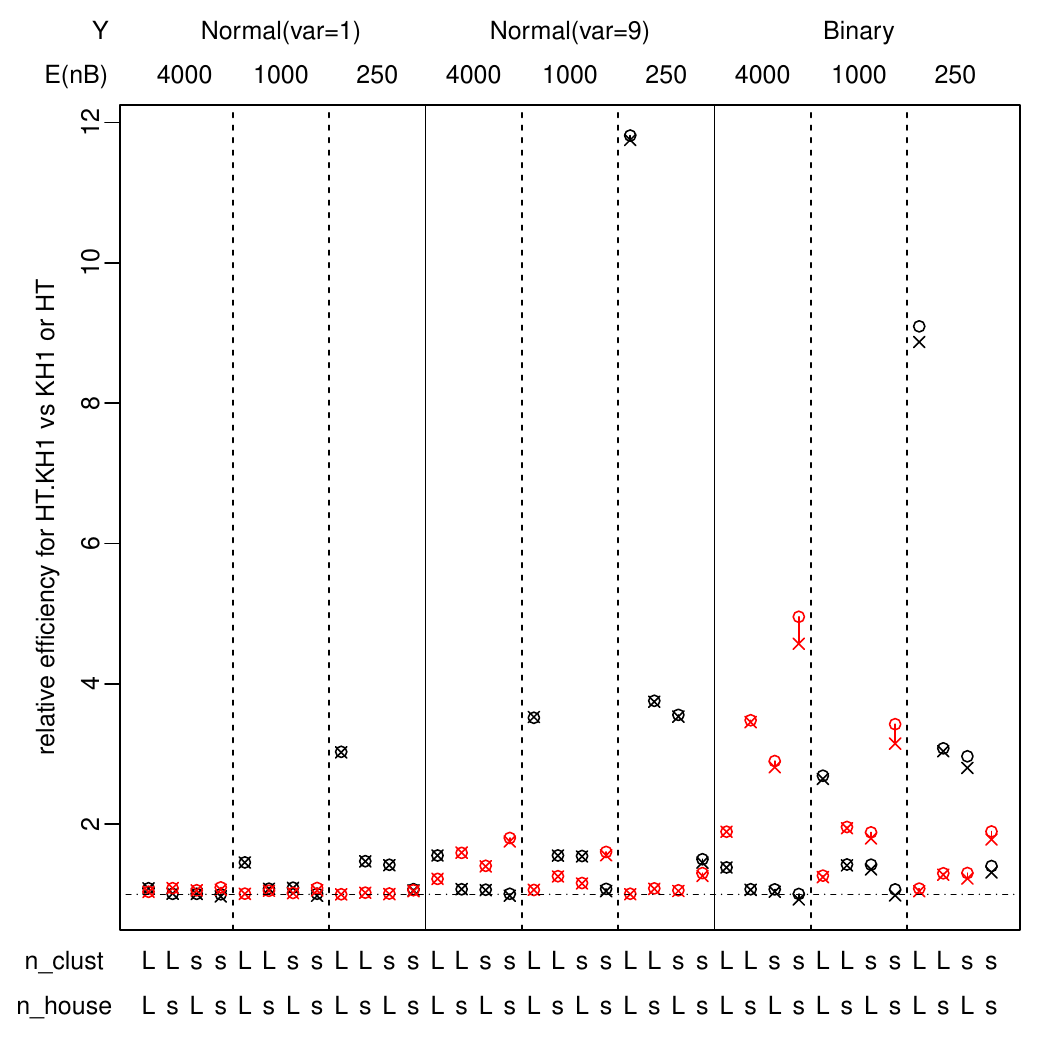}
\hspace*{-5mm}
\end{center}
\caption{Relative efficiency of $\HTKH$ compared to $\KHone$ (black) or $\HT$ (red) in 36 scenarios.  The dots indicate the theoretical asymptotic relative efficiency and the crosses indicate the empirical relative efficiency.  The 36 scenarios correspond to three conditional distributions of $Y$ given $X$, three expected size of Sample B, and two numbers of sampled clusters and of sampled households within sampled clusters in Sample A.  For the last two, `$L$' denotes large (200 clusters or 20 households) and `$s$' denotes small (50 clusters or 5 households).} 
\label{fig:releff.KHeff.KHHT}
\end{figure}

\begin{figure}
\begin{center}
 \includegraphics[width=1\textwidth,page=12]{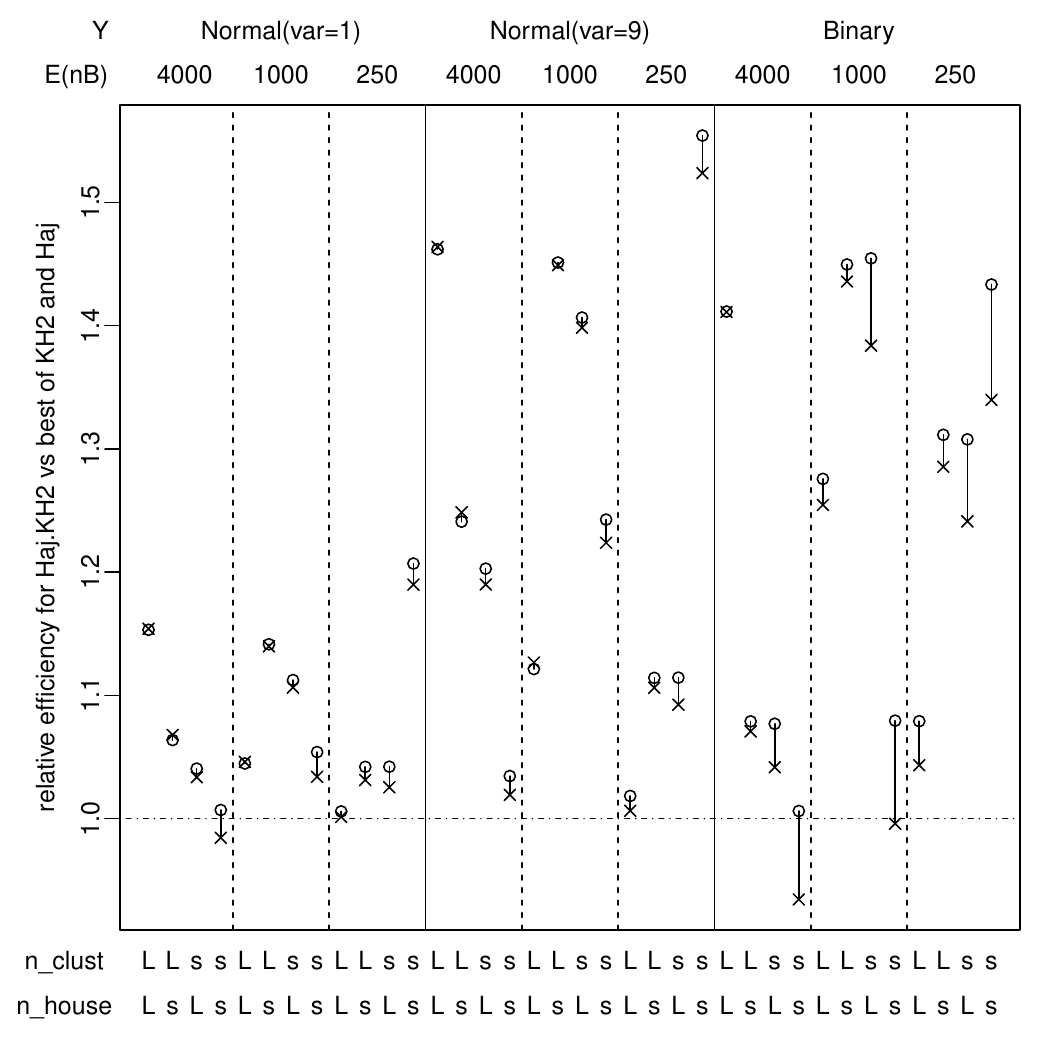}
\hspace*{-5mm}
\end{center}
\caption{Relative efficiency of $\HajKH$ compared to the most efficient of $\Haj$ and $\KHtwo$ in 36 scenarios. The dots indicate the theoretical asymptotic relative efficiency and the crosses indicate the empirical relative efficiency.  The 36 scenarios correspond to three conditional distributions of $Y$ given $X$, three expected size of Sample B, and two numbers of sampled clusters and of sampled households within sampled clusters in Sample A.  For the last two, `$L$' denotes large (200 clusters or 20 households) and `$s$' denotes small (50 clusters or 5 households).} 
\label{fig:releff.KH2eff.best}
\end{figure}

\begin{figure}
\begin{center}
 \includegraphics[width=1\textwidth,page=11]{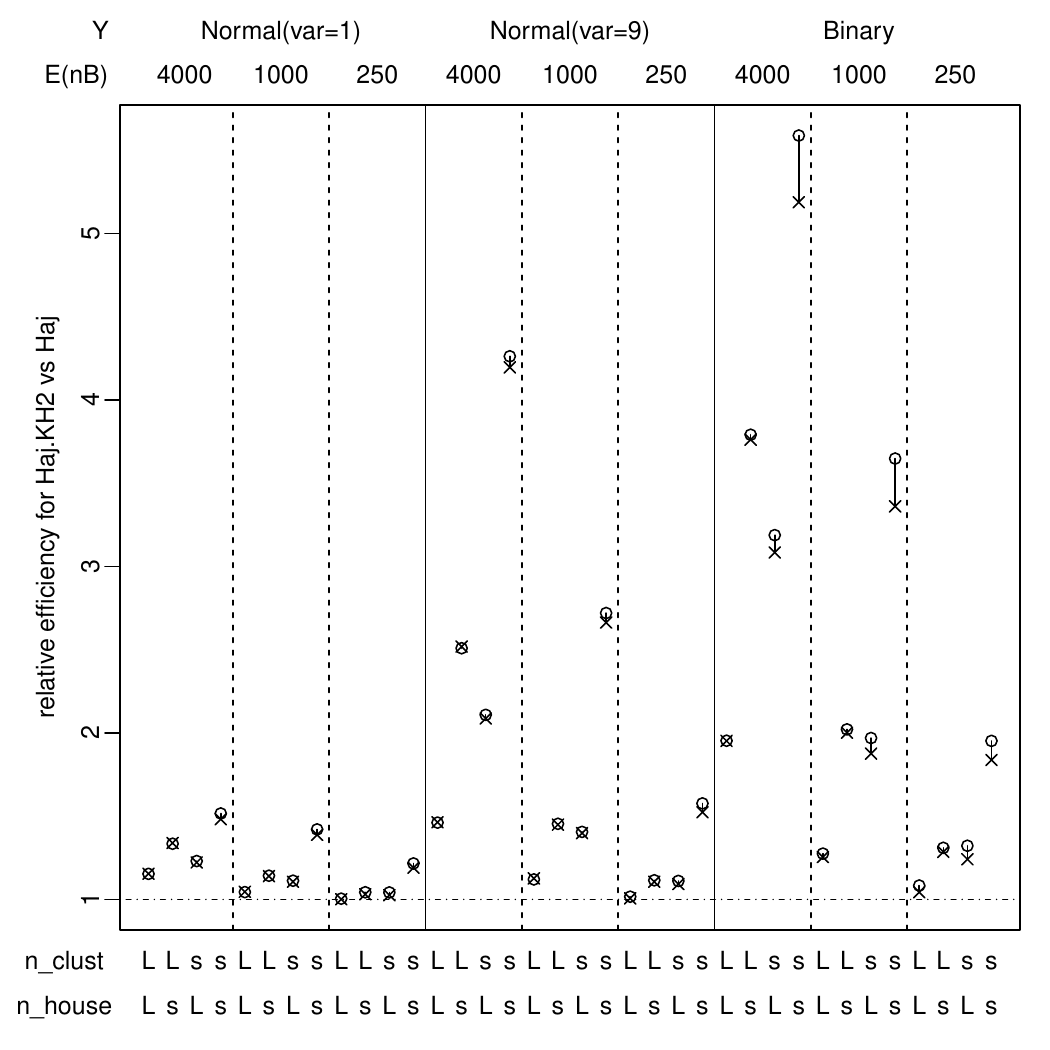}
\hspace*{-5mm}
\end{center}
\caption{Relative efficiency of $\HajKH$ compared to $\Haj$ in 36 scenarios.  The dots indicate the theoretical asymptotic relative efficiency and the crosses indicate the empirical relative efficiency.  The 36 scenarios correspond to three conditional distributions of $Y$ given $X$, three expected size of Sample B, and two numbers of sampled clusters and of sampled households within sampled clusters in Sample A.  For the last two, `$L$' denotes large (200 clusters or 20 households) and `$s$' denotes small (50 clusters or 5 households).} 
\label{fig:releff.KH2eff.HT}
\end{figure}

\begin{figure}
\begin{center}
 \includegraphics[width=1\textwidth,page=10]{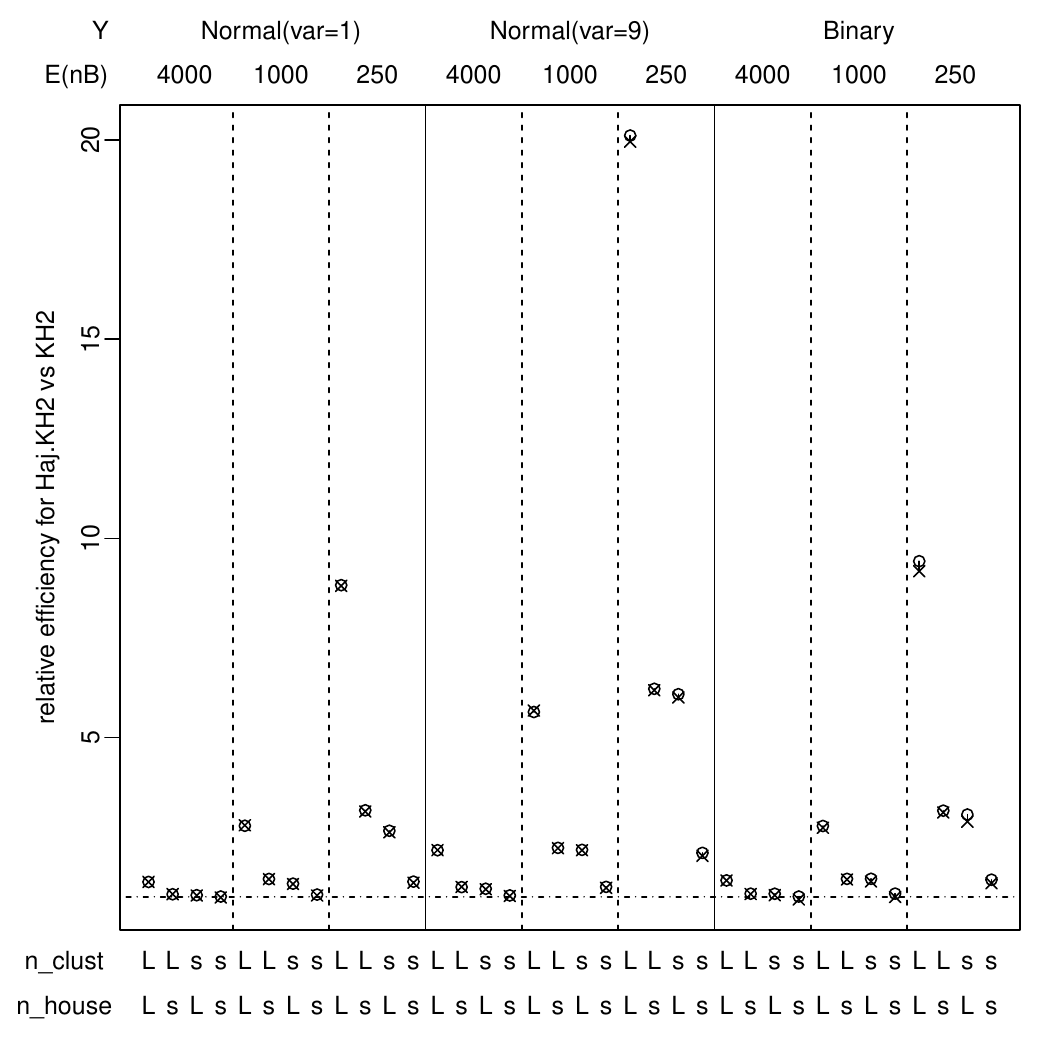}
\hspace*{-5mm}
\end{center}
\caption{Relative efficiency of $\HajKH$ compared to $\KHtwo$ in 36 scenarios.  The dots indicate the theoretical asymptotic relative efficiency and the crosses indicate the empirical relative efficiency.  The 36 scenarios correspond to three conditional distributions of $Y$ given $X$, three expected size of Sample B, and two numbers of sampled clusters and of sampled households within sampled clusters in Sample A.  For the last two, `$L$' denotes large (200 clusters or 20 households) and `$s$' denotes small (50 clusters or 5 households).} 
\label{fig:releff.KH2eff.KH}
\end{figure}

\begin{figure}
\begin{center}
 \includegraphics[width=1\textwidth,page=30]{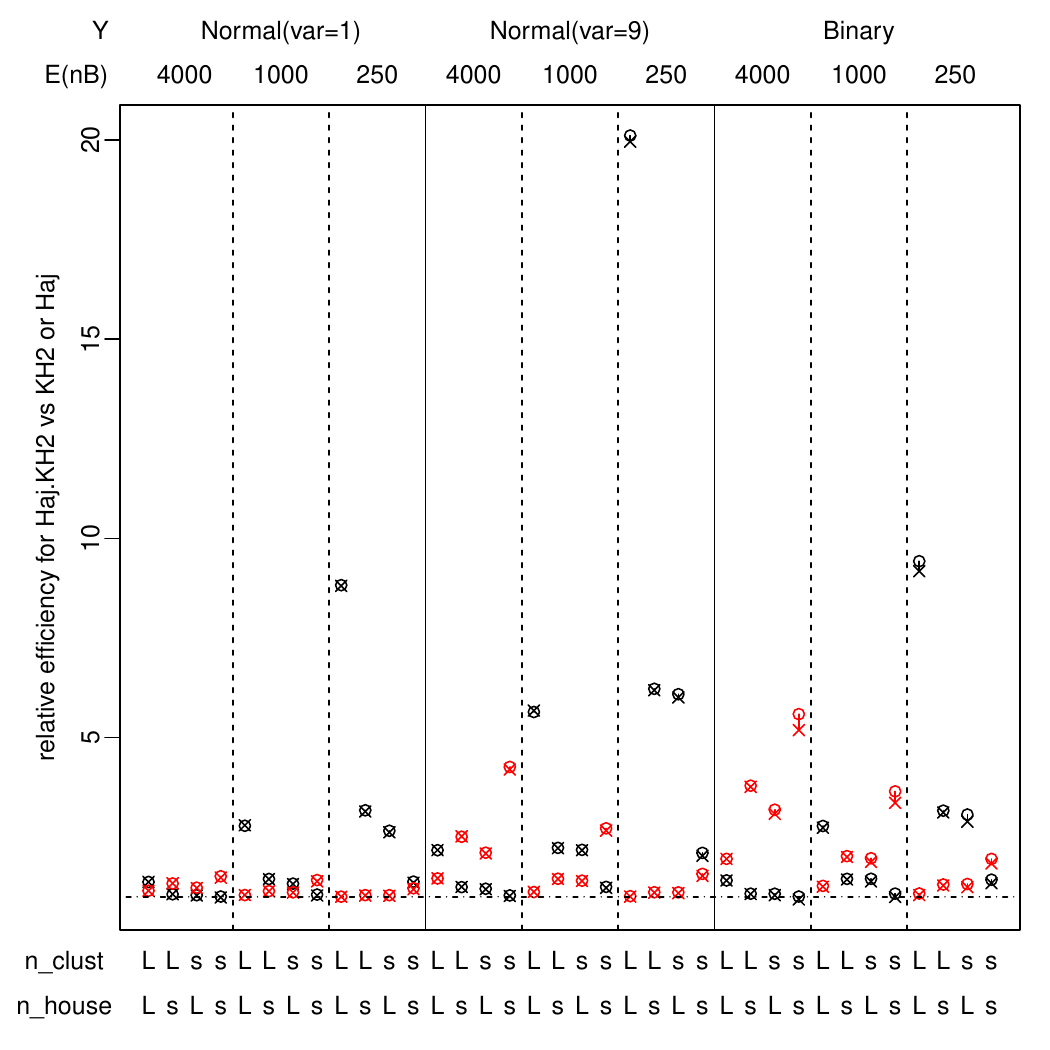}
\hspace*{-5mm}
\end{center}
\caption{Relative efficiency of $\HTKH$ compared to $\KHone$ (black) or $\HT$ (red) in 36 scenarios.  The dots indicate the theoretical asymptotic relative efficiency and the crosses indicate the empirical relative efficiency.  The 36 scenarios correspond to three conditional distributions of $Y$ given $X$, three expected size of Sample B, and two numbers of sampled clusters and of sampled households within sampled clusters in Sample A.  For the last two, `$L$' denotes large (200 clusters or 20 households) and `$s$' denotes small (50 clusters or 5 households).} 
\label{fig:releff.KH2eff.KH2Haj}
\end{figure}

\begin{figure}
\begin{center}
 \includegraphics[width=1\textwidth,page=24]{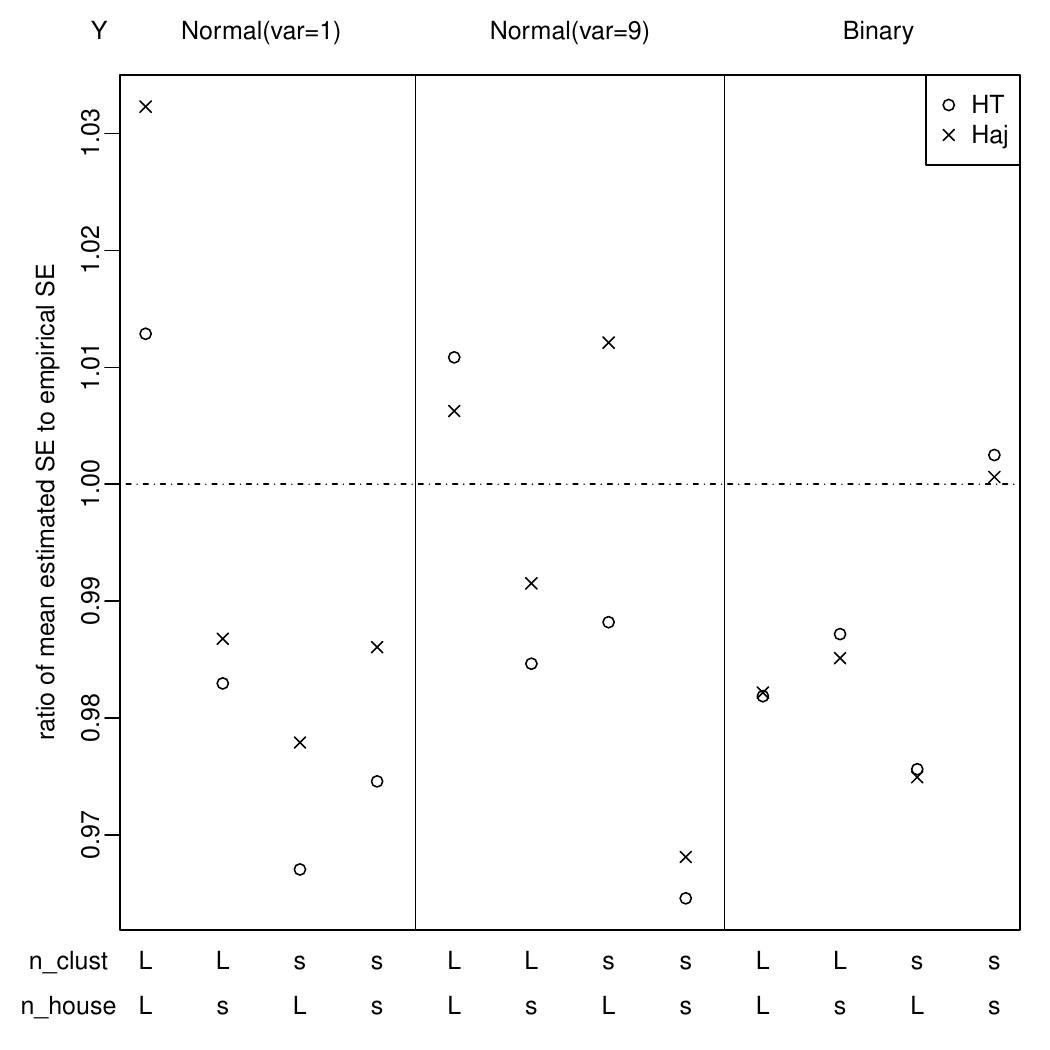}
\hspace*{-5mm}
\end{center}
\caption{Ratio of mean estimated SE to empirical SE for $\HT$ and $\Haj$ in 12 scenarios.  The 12 scenarios correspond to three conditional distributions of $Y$ given $X$, and two numbers of sampled clusters and of sampled households within sampled clusters in Sample A.  For the last two, `$L$' denotes large (200 clusters or 20 households) and `$s$' denotes small (50 clusters or 5 households).} 
\label{fig:SEratio.HTHaj}
\end{figure}

\begin{figure}
\begin{center}
 \includegraphics[width=1\textwidth,page=25]{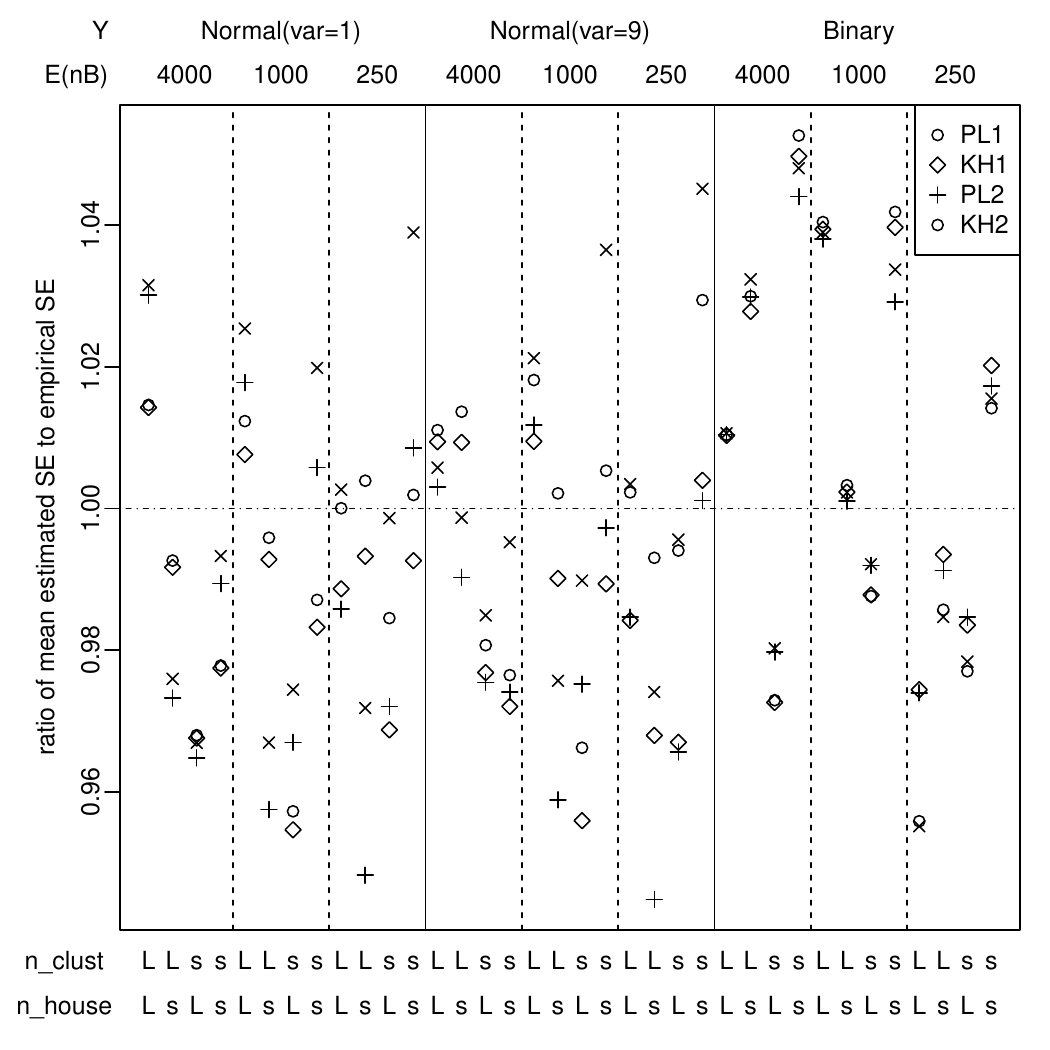}
\hspace*{-5mm}
\end{center}
\caption{Ratio of mean estimated SE to empirical SE for $\PLone$, $\KH$ and $\PLtwo$ in 36 scenarios.  The 36 scenarios correspond to three conditional distributions of $Y$ given $X$, three expected size of Sample B, and two numbers of sampled clusters and of sampled households within sampled clusters in Sample A.  For the last two, `$L$' denotes large (200 clusters or 20 households) and `$s$' denotes small (50 clusters or 5 households).} 
\label{fig:SEratio.DR}
\end{figure}

\begin{figure}
\begin{center}
 \includegraphics[width=1\textwidth,page=26]{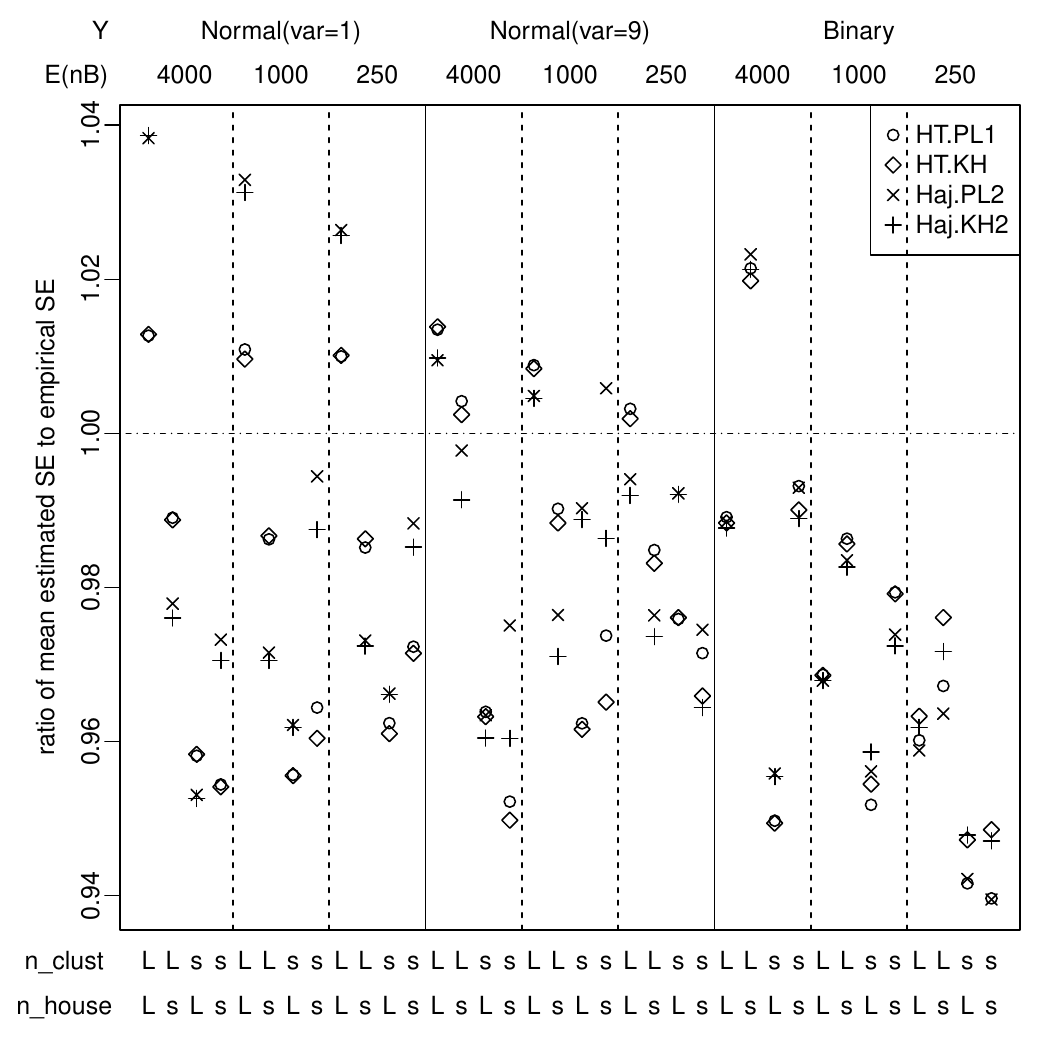}
\hspace*{-5mm}
\end{center}
\caption{Ratio of mean estimated SE to empirical SE for $\HTPL$, $\HTKH$ and $\HajPL$ in 36 scenarios.  The 36 scenarios correspond to three conditional distributions of $Y$ given $X$, three expected size of Sample B, and two numbers of sampled clusters and of sampled households within sampled clusters in Sample A.  For the last two, `$L$' denotes large (200 clusters or 20 households) and `$s$' denotes small (50 clusters or 5 households).} 
\label{fig:SEratio.DReff}
\end{figure}

\begin{figure}
\begin{center}
 \includegraphics[width=1\textwidth,page=21]{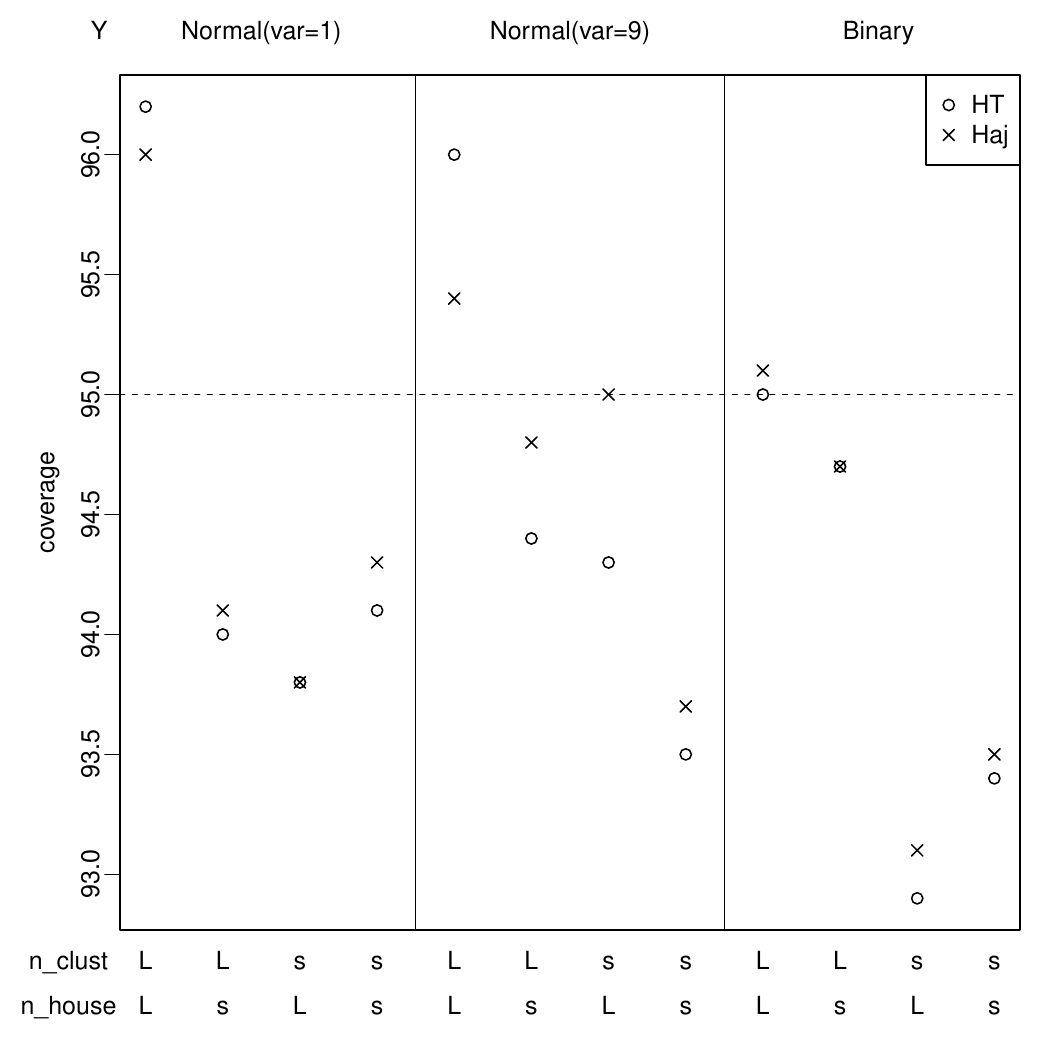}
\hspace*{-5mm}
\end{center}
\caption{Coverage of 95\% confidence intervals for $\HT$ and $\Haj$ in 12 scenarios.  The 12 scenarios correspond to three conditional distributions of $Y$ given $X$, and two numbers of sampled clusters and of sampled households within sampled clusters in Sample A.  For the last two, `$L$' denotes large (200 clusters or 20 households) and `$s$' denotes small (50 clusters or 5 households).} 
\label{fig:cover.HTHaj}
\end{figure}

\begin{figure}
\begin{center}
 \includegraphics[width=1\textwidth,page=22]{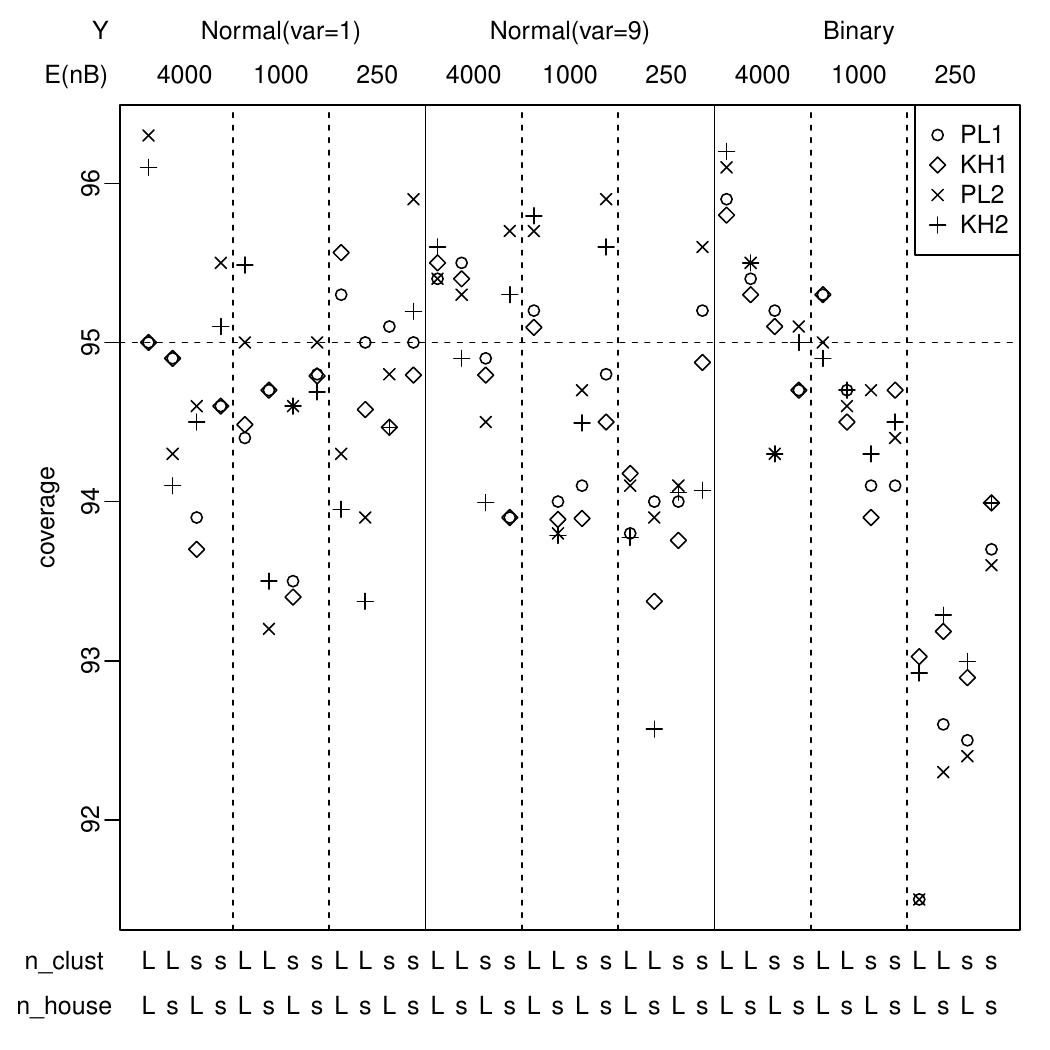}
\hspace*{-5mm}
\end{center}
\caption{Coverage of 95\% confidence intervals for $\PLone$, $\KH$ and $\PLtwo$ in 36 scenarios.  The 36 scenarios correspond to three conditional distributions of $Y$ given $X$, three expected size of Sample B, and two numbers of sampled clusters and of sampled households within sampled clusters in Sample A.  For the last two, `$L$' denotes large (200 clusters or 20 households) and `$s$' denotes small (50 clusters or 5 households).} 
\label{fig:cover.DR}
\end{figure}

\begin{figure}
\begin{center}
 \includegraphics[width=1\textwidth,page=23]{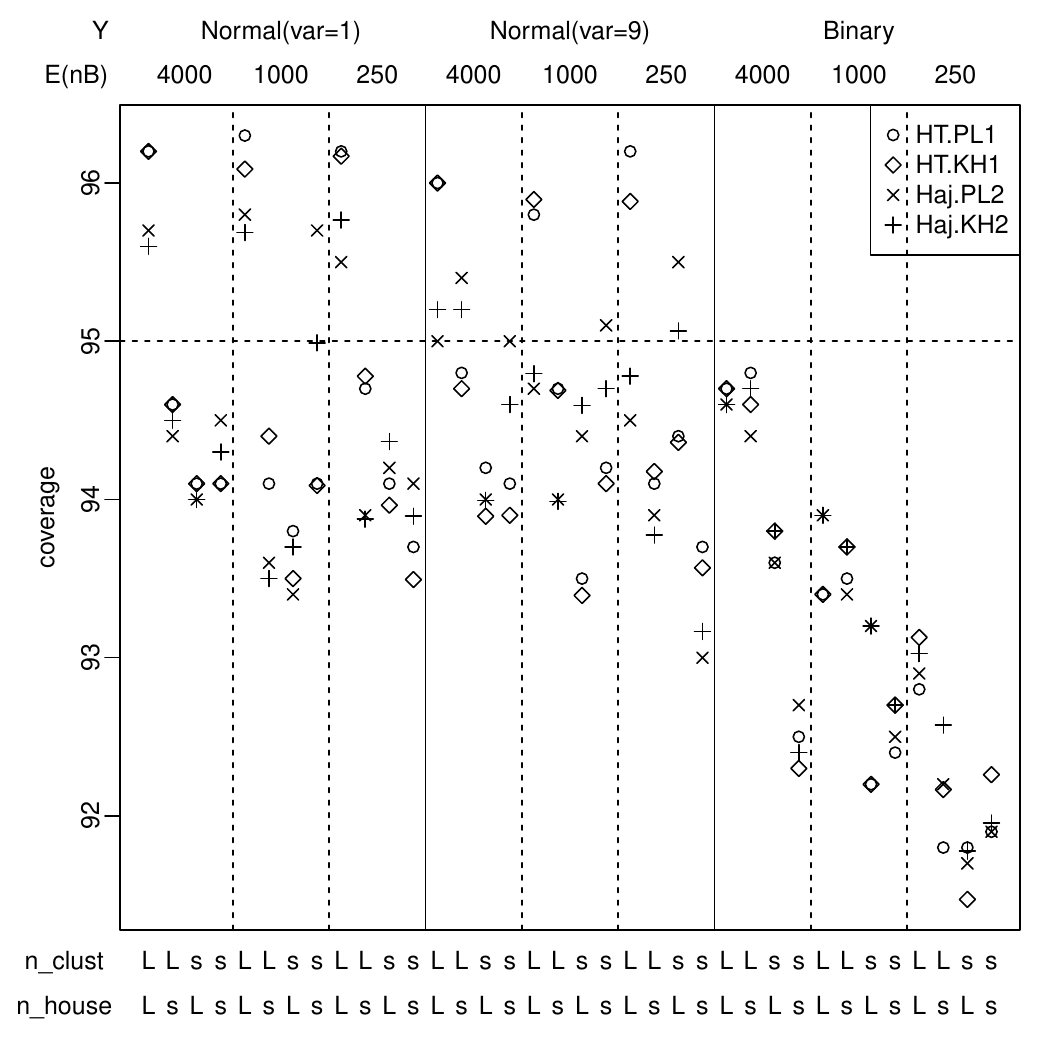}
\hspace*{-5mm}
\end{center}
\caption{Coverage of 95\% confidence intervals for $\HTPL$, $\HTKH$ and $\HajPL$ in 36 scenarios.  The 36 scenarios correspond to three conditional distributions of $Y$ given $X$, three expected size of Sample B, and two numbers of sampled clusters and of sampled households within sampled clusters in Sample A.  For the last two, `$L$' denotes large (200 clusters or 20 households) and `$s$' denotes small (50 clusters or 5 households).} 
\label{fig:cover.DReff}
\end{figure}

\begin{figure}
\begin{center}
 \includegraphics[width=1\textwidth,page=17]{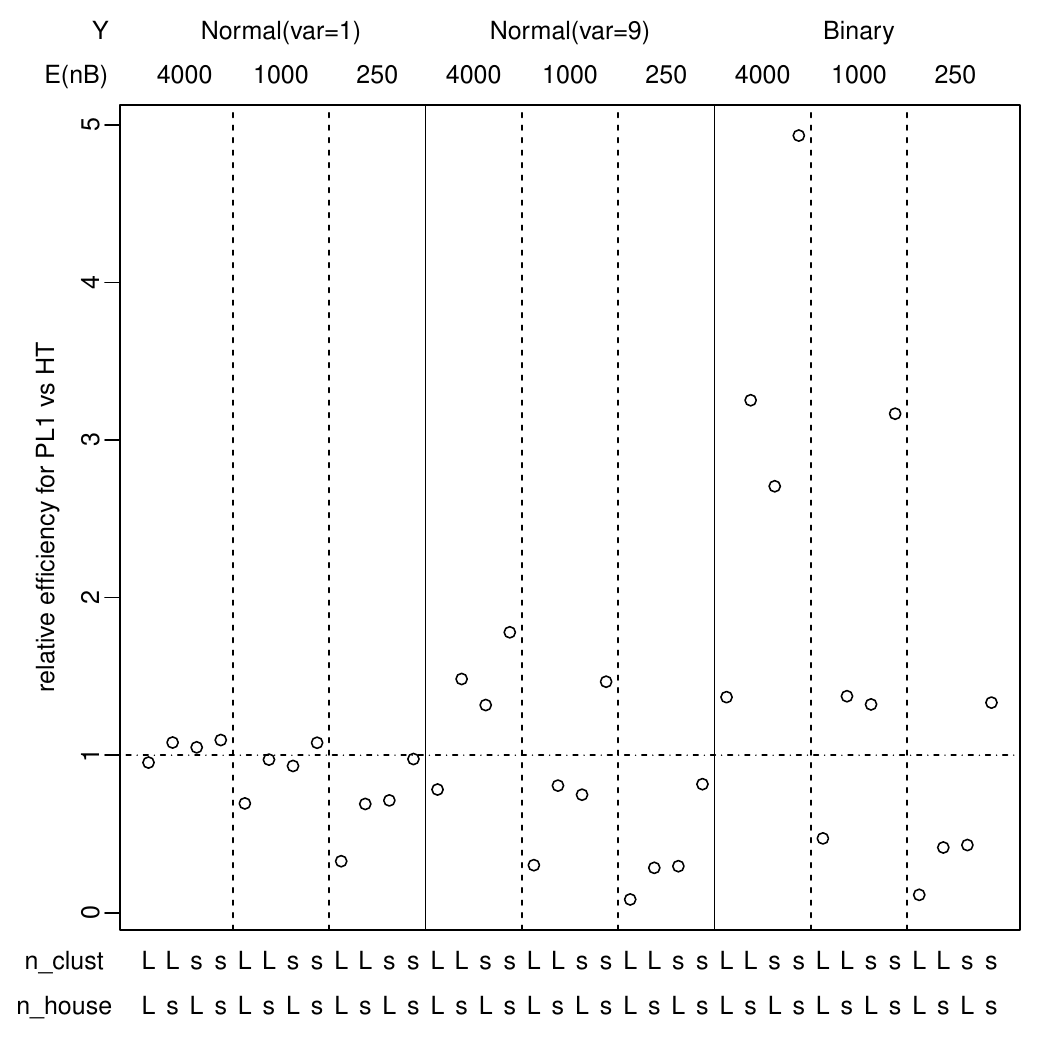}
\hspace*{-5mm}
\end{center}
\caption{Relative efficiency of $\PLone$ compared to $\HT$ in 36 scenarios.  The 36 scenarios correspond to three conditional distributions of $Y$ given $X$, three expected size of Sample B, and two numbers of sampled clusters and of sampled households within sampled clusters in Sample A.  For the last two, `$L$' denotes large (200 clusters or 20 households) and `$s$' denotes small (50 clusters or 5 households).} 
\label{fig:releff.DR1.HT}
\end{figure}

\begin{figure}
\begin{center}
 \includegraphics[width=1\textwidth,page=18]{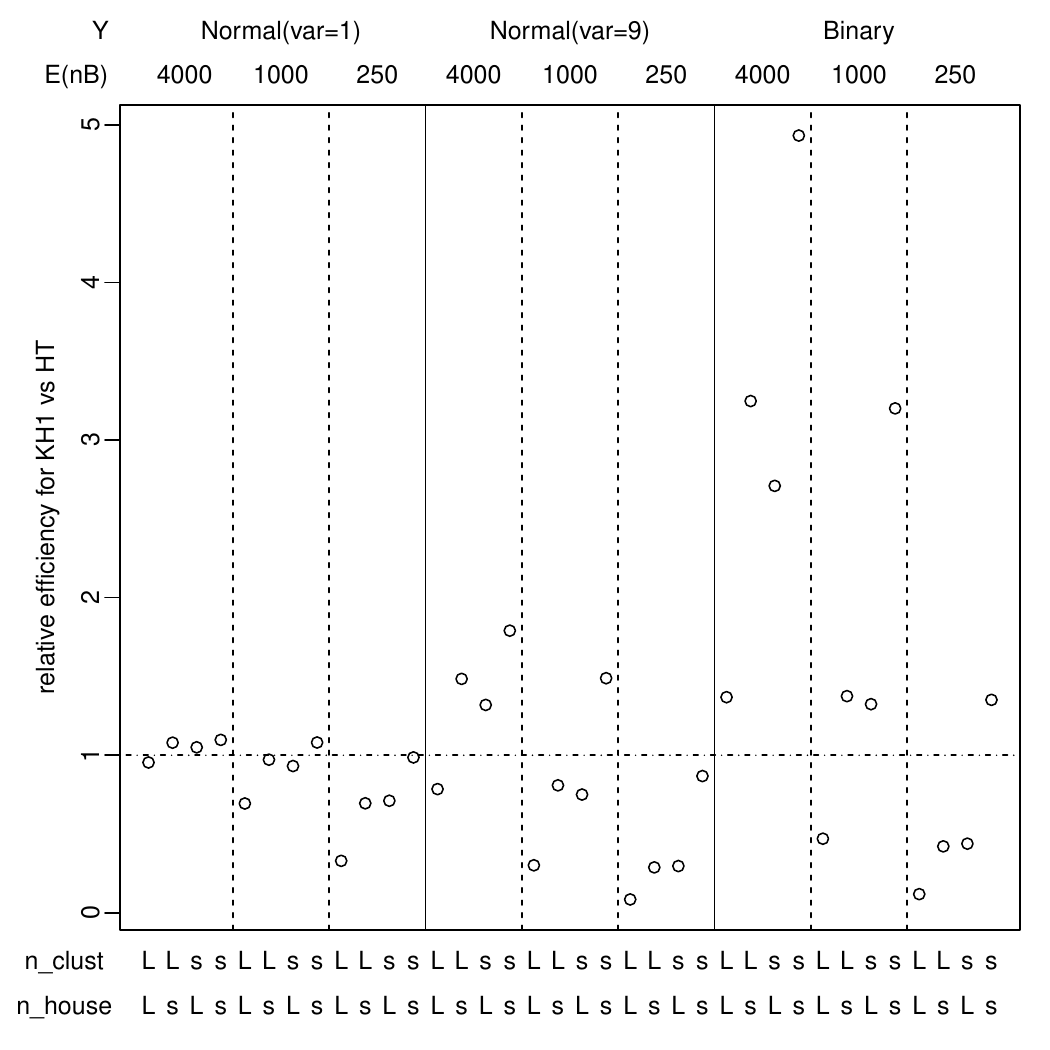}
\hspace*{-5mm}
\end{center}
\caption{Relative efficiency of $\KHone$ compared to $\HT$ in 36 scenarios.  The 36 scenarios correspond to three conditional distributions of $Y$ given $X$, three expected size of Sample B, and two numbers of sampled clusters and of sampled households within sampled clusters in Sample A.  For the last two, `$L$' denotes large (200 clusters or 20 households) and `$s$' denotes small (50 clusters or 5 households).} 
\label{fig:releff.KH.HT}
\end{figure}

\begin{figure}
\begin{center}
 \includegraphics[width=1\textwidth,page=19]{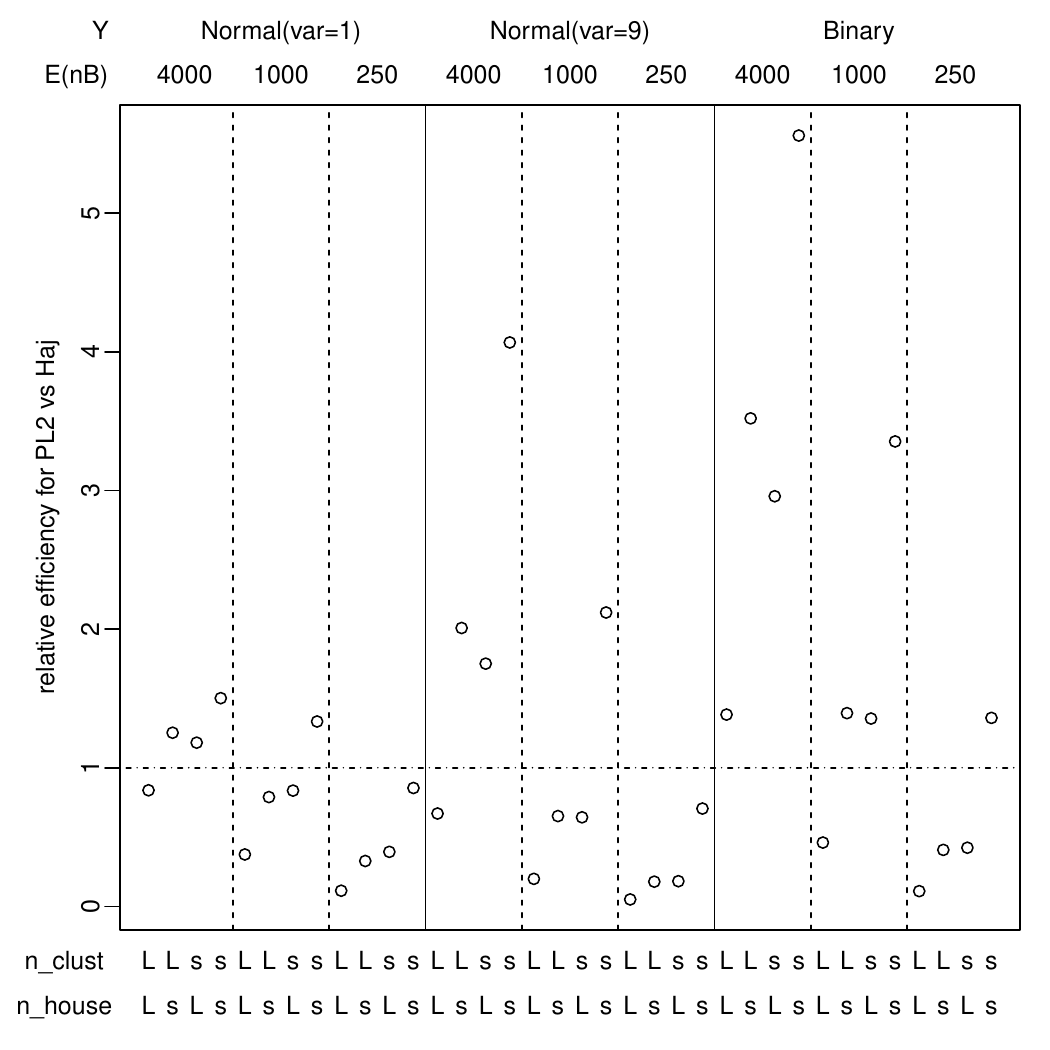}
\hspace*{-5mm}
\end{center}
\caption{Relative efficiency of $\PLtwo$ compared to $\Haj$ in 36 scenarios.  The 36 scenarios correspond to three conditional distributions of $Y$ given $X$, three expected size of Sample B, and two numbers of sampled clusters and of sampled households within sampled clusters in Sample A.  For the last two, `$L$' denotes large (200 clusters or 20 households) and `$s$' denotes small (50 clusters or 5 households).} 
\label{fig:releff.DR2.Haj}
\end{figure}

Figures~\ref{fig:releff.DR1.DR2}--\ref{fig:cover.DReff} are referred to in Section~\ref{sect:results} of the main text.
In addition, Figures~\ref{fig:releff.DR1.HT}, \ref{fig:releff.KH.HT} and \ref{fig:releff.DR2.Haj} show the relative efficiency of, respectively, $\PLone$ compared to $\HT$, $\KHone$ compared to $\HT$, and $\PLtwo$ compared to $\Haj$.


\section{Convergence of iterative algorithm for calculating Kim and Haziza estimates of $\alpha$ and $\beta$}

\label{sect:binaryKH.problem}

In the simulation study of Section~\ref{sect:simstudy} of the main text, we used the \texttt{nonprob} function of the \texttt{nonprobsvy} \texttt{R} package to solve the KH estimating equations~(\ref{eq:KH.estim.eq1}) and~(\ref{eq:KH.estim.eq2}) and thus obtain the KH estimates of $\alpha$ and $\beta$.
The \texttt{nonprob} function uses \texttt{R}'s \texttt{nleqslv} root-finding function, which employs an iterative algorithm.
The \texttt{nonprob} function uses a vector of zeroes as starting values for this iterative algorithm, which we found led to non-convergence in over 50\% of our simulated datasets in scenarios where $Y$ was binary.
When we instead used the pseudo-likelihood estimate of $\alpha$ and the maximum likelihood estimate of $\beta$ as starting values, this problem was greatly reduced: in scenarios where $Y$ is continuous and in scenarios where $Y$ is binary and $E(N^B) \geq 1000$, the algorithm failed to converge for only a tiny proportion ($\leq 0.8$\%) of simulated datasets, and in the four scenarios where $Y$ is binary and $E(N^B) = 250$, convergence failed for between 1.5\% and 2.5\% of datasets.

We now describe a cause of non-convergence when $Y$ is binary.

For simplicity, consider (as was done in Section~\ref{sect:simstudy} of the main text) the problem of estimating the mean of $Y$ in the whole population, i.e.\ $S_i = 1$ for all $i$.
Recall that when using the Kim and Haziza method, $\alphahat$ and $\betahat$ solve the estimating equations
\begin{eqnarray}
  \sum_{i=1}^N R^B_i X_i \; \frac{ 1 - \pi^B (X_i; \alphahat) }{ \pi^B (X_i; \alphahat) } \{ Y_i - m(X_i; \betahat) \}
  & = & 0
  \label{eq:KH.estim.all.eq1}
  \\
  \sum_{i=1}^N \left\{ \frac{ R^B_i }{ \pi^B (X_i; \alphahat) } - \frac{R^A_i}{\pi^A_i} \right\}
  \left. \frac{ \partial m (X_i; \beta) }{ \partial \beta^\top } \right|_{\beta=\betahat}
  & = & 0.
  \label{eq:KH.estim.all.eq2}
\end{eqnarray}
Let $\alpha_{\rm int}$ and $\beta_{\rm int}$ denote the intercept parameters in the models $\pi^B (X; \alpha)$ and $m (X; \beta)$ respectively.

When $m (X_i; \beta) = \beta^\top X_i$, as would typically be the case for a continuous outcome variable $Y$, the quantity $\left. \frac{ \partial m (X_i; \beta) }{ \partial \beta^\top } \right|_{\beta=\betahat}$ in equation~(\ref{eq:KH.estim.all.eq2}) equals $X_i$.
In this case, the estimating equation~(\ref{eq:KH.estim.all.eq2}) involves only $\alphahat$, and hence $\alphahat$ can be calculated first and then plugged into estimating equation~(\ref{eq:KH.estim.all.eq1}) to calculate $\betahat$.

When, instead, $m (X_i; \beta) = \exp (\beta^\top \Xpi) / \{ 1 + \exp (\beta^\top \Xpi) \}$, as is the case for a binary outcome $Y$ and a logistic regression model for $P(Y=1 \mid X)$, the quantity $\left. \frac{ \partial m (X_i; \beta) }{ \partial \beta^\top } \right|_{\beta=\betahat}$ equals $\Xpi \; m (X_i; \betahat) \{ 1 - m (X_i; \betahat) \}$.
In this case, both estimating equations~(\ref{eq:KH.estim.all.eq1}) and~(\ref{eq:KH.estim.all.eq2}) involve both $\alphahat$ and $\betahat$, and so must be solved simultaneously.
When both nuisance models are correctly specified, equations~(\ref{eq:KH.estim.all.eq1}) and~(\ref{eq:KH.estim.all.eq2}) are solved in expectation at the true values of $\alpha$ and $\beta$.
However, equation~(\ref{eq:KH.estim.all.eq1}) also holds when $\alphahat_{\rm int} = \infty$ (assuming that none of the other elements of $\alphahat$ equals $- \infty$), because then $\pi^B (X_i; \alphahat) = 1$.
This is irrespective of whether the nuisance models are correctly specified.
Similarly, equation~(\ref{eq:KH.estim.all.eq2}) holds when $\betahat_{\rm int} = \pm \infty$ (assuming that none of the other elements of $\betahat$ equals $\mp \infty$), because then $m (X_i; \betahat) \{ 1 - m (X_i; \betahat) \} = 0$.
Also, expression~(\ref{eq:KH.estim.all.eq1}) does not explode as $\betahat_{\rm int} \rightarrow \pm \infty$ (i.e.\ its derivative with respect to $\betahat_{\rm int}$ converges to zero as $\betahat_{\rm int} \rightarrow \pm \infty$), and expression~(\ref{eq:KH.estim.all.eq2}) does not explode as $\alphahat_{\rm int} \rightarrow \infty$ (i.e.\ its derivative with respect to $\alphahat_{\rm int}$ converges to zero as $\alphahat_{\rm int} \rightarrow \infty$).
All this suggests that, when a logistic regression model is used for $m(X; \beta)$, an iterative algorithm for finding a solution to estimating equations~(\ref{eq:KH.estim.all.eq1}) and~(\ref{eq:KH.estim.all.eq2}) may be at risk of finding a solution with $\alphahat_{\rm int} = \infty$ and either $\betahat_{\rm int} = \infty$ or $\betahat_{\rm int} = - \infty$.
Indeed, for most of the simulated datasets with binary $Y$ in which non-convergence occurred, this did appear to be happening.

Note that this issue does not affect the KH method when $Y$ is continuous and a linear regression model is used for $m(X; \beta)$, because $\betahat = \pm \infty$ is not a solution to equation~(\ref{eq:KH.estim.all.eq2}) in that case.

\section{Application to Natsal-3 data: supplementary information}

\begin{figure}
    \centering
    \includegraphics[width=1\linewidth]{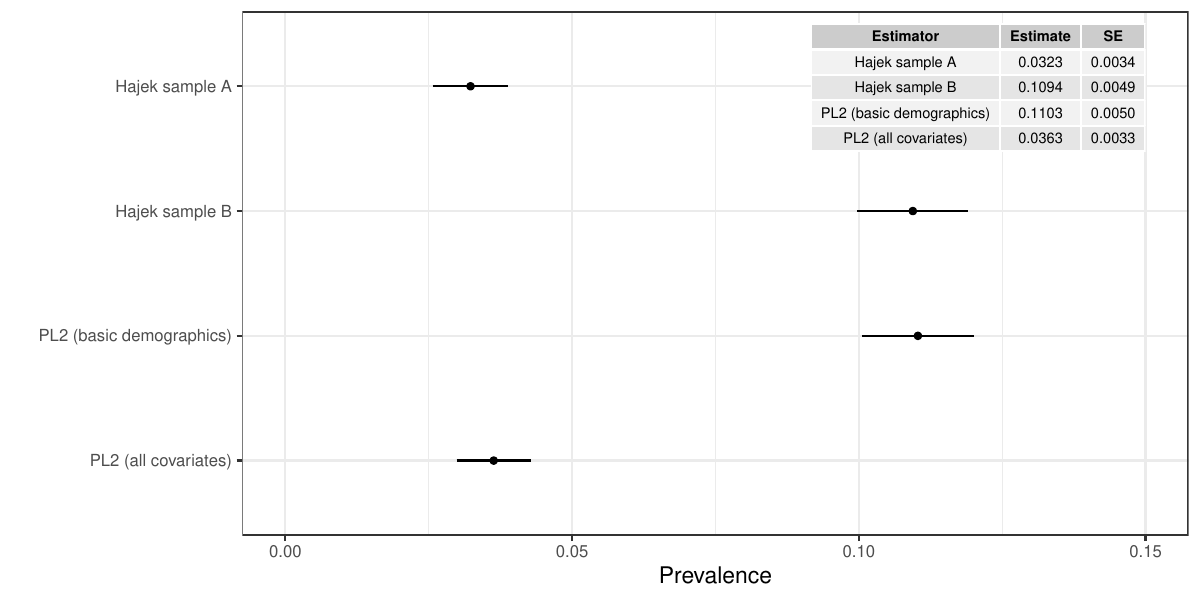}
    \caption{Estimated prevalences and 95\% confidence intervals based on Hajek estimates using sample A and sample B, and the PL2 estimator with two different sets of covariates used when fitting the selection and outcome models.}
    \label{fig:N3predictors}
\end{figure}

\begin{figure}
    \centering
    \includegraphics[width=1\linewidth]{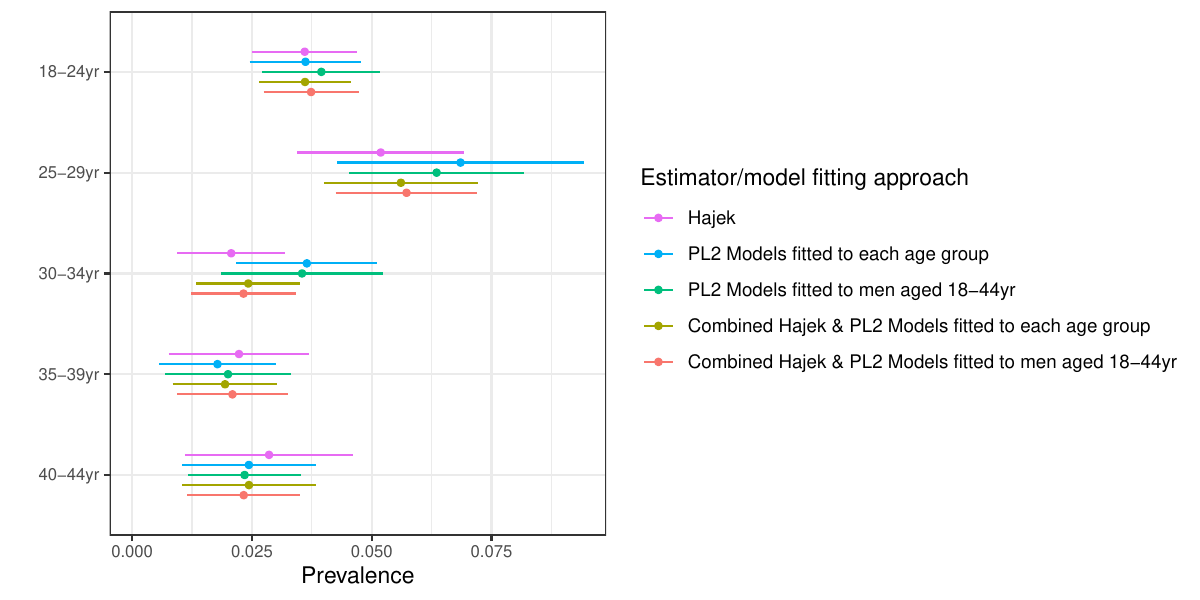}
    \caption{Estimated prevalences and 95\% confidence intervals of the outcome by smaller age groups as in Figure \ref{fig:N3agegroups} in the main text, but additionally showing the corresponding Hajek estimates and the combined estimates.}
    \label{fig:N3agegroups_combinedHajek}
\end{figure}

Tables~\ref{tab:N3covariates} and~\ref{tab:N3waldtests} and Figures~\ref{fig:N3predictors} and~\ref{fig:N3agegroups_combinedHajek} show additional information for the Natsal-3 application.

\begin{table}
\footnotesize
\begin{threeparttable}[t]
  \centering
    \caption{Covariates used for the doubly robust estimators in the applied analysis}
    \begin{tabular}{l>{\raggedright\arraybackslash}p{13cm}}
    \midrule    
    \textbf{Covariate group} & \textbf{Covariate} \\
    \midrule    
    \multicolumn{2}{l}{Basic demographics}       \\
    & Age (numerical) \\
    & Region of residence (7 regions; North \& Yorkshire,
                                             North West,
                                             Midlands,
                                             London,
                                             South East \& East,
                                             Wales South \& West,
                                             Scotland) \\ 
    \midrule
    \multicolumn{2}{l}{Other demographic covariates}       \\
    & Ethnic ancestry (white, non-white) \\
    & Cohabiting status (cohabiting, not cohabiting) \\
    & Educational attainment (degree level or above, other) \\
    & Employment status (in employment, other) \\
    \midrule
    \multicolumn{2}{l}{Sexual identity}       \\
    & Sexual identity (heterosexual, non-heterosexual) \\
    \midrule
    \multicolumn{2}{l}{Health}       \\
    & Self-perceived health (good/very good, other) \\
    \midrule
    \multicolumn{2}{l}{Attitudes}       \\
    & Opinion on same-sex relations (always/mostly acceptable for both male and female relationships, other) \\
    & Dissatisfied with sex life (yes, no) \\
    & Distressed or worried about sex life (yes, no) \\
    \midrule
    \multicolumn{2}{l}{Behaviours}       \\
    & Ever had sex  (yes, no) \\
    & Had first sex at age younger than 16yr (yes, no) \\
    & Attracted not only to opposite gender (yes, no) \\
    \midrule
    \end{tabular}
    \label{tab:N3covariates}
\end{threeparttable}
\end{table}

\begin{table}
\footnotesize
\begin{threeparttable}[t]
  \centering
    \caption{Wald tests of the coefficients of the nuisance models $\piBhat$ and $\mhat$ fitted with all 14 covariates, for the  $\PLtwo$ estimator.}
    \begin{tabular}{llc}
    \midrule
    \textbf{Model} & \textbf{Covariate} & \textbf{p} \\
    \midrule
    Selection model\tnote{a}, $\piBhat$
    & Age  & 0.156 \\
    & Region of residence  &  0.324 \\ 
    & Ethnic ancestry  &  $<$0.001 \\
    & Cohabiting status   &  0.052 \\
    & Educational attainment    &  $<$0.001 \\
    & Employment status    &  0.002 \\
    & Sexual identity  &   $<$0.001 \\
    & Self-perceived health   &  $<$0.001 \\
    & Opinion on same-sex relations   &   $<$0.001 \\
    & Dissatisfied with sex life    &   $<$0.001  \\
    & Distressed or worried about sex life     &   0.005 \\
    & Ever had sex      & 0.037  \\
    & Had first sex at age younger than 16yr   &   $<$0.001  \\
    & Attracted not only to opposite gender  &   $<$0.001  \\
    \midrule
    Outcome model\tnote{b}, $\mhat$ 
    & Age & 0.523 \\ 
    & Region of residence &  0.411  \\ 
    & Ethnic ancestry &  0.349 \\
    & Cohabiting status  &  0.064 \\
    & Educational attainment   &  0.293 \\
    & Employment status    &  0.209 \\
    & Sexual identity  &   $<$0.001 \\
    & Self-perceived health   &  0.274 \\
    & Opinion on same-sex relations   &  0.627 \\
    & Dissatisfied with sex life    &  0.008  \\
    & Distressed or worried about sex life     &  $<$0.001 \\
    & Ever had sex      &  $<$0.001   \\
    & Had first sex at age younger than 16yr   &  $<$0.001 \\
    & Attracted not only to opposite gender  &   $<$0.001  \\
    \midrule
    \end{tabular}
    \begin{tablenotes}
    \item[a] p values from generalised Wald tests, using a sandwich variance-covariance estimator due to the use of survey weights.
    \item[b] p values from Wald tests.
    \end{tablenotes}
    \label{tab:N3waldtests}
\end{threeparttable}
\end{table}

\section{Gao and Yang's test-and-pool procedure}

\label{sect:GaoYang}

\subsection{Description of the method}

\label{sect:GYdescription}

Gao and Yang (2023)\cite{gao2023pretest} (henceforth G\&Y) consider estimators $\muhat_A$ and $\muhat_B$, where $\muhat_A$ is calculated from the data on Sample A and $\muhat_A$ is calculated from the data on Samples A and B.
In particular, in their Appendix B they provide some description of how their method applies in the case where $\muhat_A = \Haj$, $\muhat_B = \PLone$ and Sample A is obtained using a single-stage without-replacement design.

In this section, we provide a more accessible account of G\&Y's method, apply it to more estimators (and without assuming that both nuisance models are correctly specified), and consider in more detail how the variances of $\muhatA$ and $\muhatB$ and covariance between them, all of which are required for the method, can be estimated.
We assume throughout that the aim is to estimate the population mean $\Ybar$, rather than a domain mean.

Let $\muhat_A$ denote $\HT$ or $\Haj$, and $\muhat_B$ denote $\DRone$, $\KHone$ or $\KHtwo$.
The formulae for $\DRone$, $\KHone$ and $\KHtwo$ involve the quantity $m(X; \beta)$, which is a model for $m(X)$.
Thus far in our article, $m(X)$ has been defined as $m(X) = E(Y \mid X)$.
Because we have been assuming $R^B \ci Y \mid X$, we could equivalently have defined $m(X)$ as $m(X) = E(Y \mid X, R^B=1)$.
In this section, because we are no longer assuming $R^B \ci Y \mid X$, $E(Y \mid X)$ and $E(Y \mid X, R^B=1)$ are, in general, different.
G\&Y define $m(X)$ as $m(X) = E(Y \mid X, R^B=1)$, and we use this definition $m(X)$ from now on in this section.

Let $f_B = \lim_{n \rightarrow \infty} n_B / (n_A + n_B)$, where $n_A = \sum_{i=1}^N R^A_i$ and $n_B = \sum_{i=1}^N R^B_i$ denote the sizes of Samples A and B, respectively.
Let $V_A$, $V_B$ and $\Gammau$ denote the asymptotic unconditional variances and covariance
\begin{eqnarray*}
  V_A & = &
  \lim_{n \rightarrow \infty} \Var \{ n^{1/2} (\muhatA - \Ybar ) \}
  \\
  V_B & = &
  \lim_{n \rightarrow \infty} \Var \{ n^{1/2} (\muhatB - \Ybar ) \}
  \\
  \Gammau & = &
  \lim_{n \rightarrow \infty} \Cov \{ n^{1/2} (\muhatA - \Ybar ), \; n^{1/2} (\muhatB - \Ybar ) \}.
\end{eqnarray*}

G\&Y assume that Samples A and B are very small compared to the finite population.
This enables them to consider the difference between the finite population mean $\Ybar$ of $Y$ and the superpopulation mean $\mu_{g0}$ of $Y$ to be negligible compared to the uncertainty in the estimators $\muhat_A$ and $\muhat_B$ of $\Ybar$.
It also enables them to consider the difference between the asymptotic unconditional variance $V_A$ and the corresponding asymptotic conditional variance $\lim_{n \rightarrow \infty} \Var \{ n^{1/2} (\muhatA - \Ybar) \mid \curlF \}$ to be negligible, and similarly for the differences between $V_B$ or $\Gammau$ and their corresponding asymptotic conditional variance or covariance.

Let $\eta = \lim_{n \rightarrow \infty} n_B^{1/2} (\muhatB - \muhatA)$ denote the asymptotic bias of $\muhat_B$.
Note that $\eta = 0$ when $R^B \ci Y \mid X$.
Let
\[
\GYteststat = \frac{1}{ V_A + V_B - 2 \Gammau } (\muhatB - \muhatA)^2
\]
be a test statistic for the null hypothesis that $\eta = 0$.
For any $c_{\gamma} > 0$ and $\lambda \geq 0$, define
\[
\GYestfix = \left\{ \begin{array}{ll}
  \frac{1}{1 + \lambda} \muhatA + \frac{\lambda}{1 + \lambda} \muhatB & \mbox{ if } \GYteststat < c_{\gamma} \\
    \muhatA & \mbox{ if } \GYteststat \geq c_{\gamma} \\
\end{array} \right.
\]
This is the estimator that uses a weighted average of $\muhatA$ and $\muhatB$  if the null hypothesis is not rejected (using test statistic $\GYteststat$ and critical value $c_{\gamma}$) and uses $\muhat_A$ if it is rejected.

The asymptotic bias of $\GYestfix$ is
\begin{equation*}
  \lim_{n \rightarrow \infty} n^{-1/2} (\GYestfix - \Ybar) =
  f_B^{-1/2} \eta \; 
  \frac{ \lambda }{ 1 + \lambda } \;
  F_3 (c_{\gamma}; \mu_2^2 / 2)
\end{equation*}
where $F_d (x ; q)$ denotes the cumulative distribution function evaluated at $x$ of a non-central chi-square distribution with $d$ degrees of freedom and non-centrality parameter $q$, i.e.\ the distribution of $\sum_{i=1}^d U_i^2$ where $U_1, \ldots, U_d$ are iid Normal$(2q, 1)$.\footnote{Note that the non-central chi-squared distribution with $d$ degrees of freedom and non-centrality parameter $q$ is sometimes defined as the distribution of $\sum_{i=1}^d U_i^2$ where $U_1, \ldots, U_d$ are iid Normal$(q, 1)$ (rather than Normal$(2q, 1)$).  We are using the notation of G\&Y.}

The asymptotic mean squared error (MSE) of $\GYestfix$ is
\begin{eqnarray}
  \lim_{n \rightarrow \infty} n^{-1} (\GYestfix - \Ybar)^2
  & = & 
  V_{\rm eff} (\mu_1^2 + 1)
  \nonumber \\
  && +
  \left\{
  F_3 (c_{\gamma}; \mu_2^2 / 2)
  +
  F_5 (c_{\gamma}; \mu_2^2 / 2) \mu_2^2
  \right\}
  (1+\lambda)^{-1} (V_{\rm A-eff}^{1/2} - \lambda V_{\rm B-eff}^{1/2})^2
  \nonumber \\
  && -
  2 \; F_3 (c_{\gamma}; \mu_2^2 / 2)
  V_{\rm eff}^{1/2} \mu_1 \mu_2
  (1+\lambda)^{-1} (V_{\rm A-eff}^{1/2} - \lambda V_{\rm B-eff}^{1/2})
  \nonumber \\
  && +
  V_{\rm A-eff}
  \left[
  \{ 1 - F_3 (c_{\gamma}; \mu_2^2 / 2) \}
  +
  \{ 1 - F_5 (c_{\gamma}; \mu_2^2 / 2) \} \mu_2^2
  \right]
  \nonumber \\
  && -
  2 \; \{ 1 - F_3 (c_{\gamma}; \mu_2^2 / 2) \}
  V_{\rm eff}^{1/2} V_{\rm A-eff}^{1/2} \mu_1 \mu_2
  \label{eq:asymp.MSE}
  \end{eqnarray}
where
\begin{eqnarray}
  \mu_1
  & = &
  f_B^{-1/2} \frac{ \Gammau - V_A }{ \sqrt{ (V_A + V_B - 2 \Gammau) (V_A V_B - \Gammau^2) } } \; \eta
  \nonumber \\
  \mu_2
  & = &
  f_B^{-1/2} \frac{1}{ \sqrt{ V_A + V_B - 2 \Gammau } } \; \eta
  \nonumber \\
  V_{\rm eff}^{1/2}
& = &
\frac{ \sqrt{V_A V_B - \Gammau^2} }{ \sqrt{V_A + V_B - 2 \Gammau} }
\nonumber \\
  V_{\rm A-eff}^{1/2}
  & = &
  \frac{ \Gammau - V_A }{ \sqrt{ V_A + V_B - 2 \Gammau } }
  \nonumber \\
  V_{\rm B-eff}^{1/2}
  & = &
  \frac{ \Gammau - V_B }{ \sqrt{ V_A + V_B - 2 \Gammau } }
  \nonumber
\end{eqnarray}

G\&Y's test-and-pool estimator $\GYmuhat$ of $\Ybar$ is obtained by finding the value $(c_{\gamma}^{\dagger}, \lambda^{\dagger})$ of $(c_{\gamma}, \lambda)$ that minimises this asymptotic MSE, and setting $\GYmuhat = \GYestfixopt$.

{\it Remark: If $\eta=0$, the formula for the asymptotic MSE given in equation~(\ref{eq:asymp.MSE}) reduces to
$
V_{\rm eff} + F_3 (c_{\gamma}; 0) (1+\lambda)^{-1} (V_{\rm A-eff}^{1/2} - \lambda V_{\rm B-eff}^{1/2})^2 + V_{\rm A-eff} \{ 1 - F_3 (c_{\gamma}; 0) \}.
$
The efficient combined estimator described in Section~\ref{sect:combining} of the main text does not use the pre-test and uses weights $(V_B - \Gammau) / (V_A + V_B - 2 \Gammau)$ and $(V_A - \Gammau) / (V_A + V_B - 2 \Gammau)$.
It thus corresponds to $\GYmuhat$ with $c_{\gamma} = \infty$ and $\lambda = (V_A - \Gammau) / (V_B - \Gammau)$.
It can be verified that with these values of $c_{\gamma}$ and $\lambda$, the formula for the asymptotic MSE further reduces to $V_{\rm eff}$, as expected.}

The value $(c_{\gamma}^{\dagger}, \lambda^{\dagger})$ can be found by using a numerical optimisation method.
The asymptotic MSE and the test statistic $\GYteststat$ depend on the unknown quantities $f_B$, $\eta$, $V_A$, $V_B$ and $\Gammau$.
So, these quantities are replaced by estimates.
In particular, $f_B$ is replaced by $n_B / (n_A + n_B)$, and $\eta$ by $n_B^{1/2} (\muhatB - \muhatA)$.
The variance $V_A$ can be estimated using ($n$ times) the estimator of $V_H$ given earlier.
However, because the estimators of the variance $V_D$ of $\DR$ and of the covariance $C$ between $\HTHaj$ and $\DR$ were derived under the assumption that $R^B \ci Y \mid X$, it is not obvious that ($n$ times) these variance/covariance estimators can be used for $V_B$ and $\Gammau$.
In Appendix~\ref{sect:using.closed}, we show that slightly modified versions of these estimators can be used under specified conditions.
Alternatively, a bootstrap procedure can be used to estimate $V_B$ and $\Gammau$ (and $V_A$).
This is what G\&Y did in the scenario they considered, where Sample A was sampled using a single-stage without-replacement design.
We discuss the bootstrap approach in Appendix~\ref{sect:using.boot}, both for G\&Y's scenario and for more general scenarios.

Next we describe the double bootstrap procedure proposed by G\&Y to obtain a $(1-\alpha)100$\% confidence interval (CI) for $\GYmuhat$.
Let $v_n > 0$ be some constant; we shall describe how to choose $v_n$ later.

Create $B_1$ (e.g.\ $B_1 = 1000$) bootstrap samples of the data.
Each of these bootstrap samples consists of data on Sample A and data on Sample B.
For the Sample B data, $n_B$ individuals can be resampled with replacement from the original Sample B.
For the Sample A data, an appropriate bootstrap sampling method will depend on the sampling design of Sample A.
In the simple scenario considered by G\&Y, where Sample A was small compared to the population and was sampled using a single-stage without-replacement design, it sufficed to sample $n_A$ individuals with replacement from Sample A.
Calculate the values of $\muhat_A$, $\muhatB$ and $\GYteststat$ from each bootstrap sample.
For the $b$th bootstrap sample, these will be denoted as $\muhat_A^{*(b)}$, $\muhatB^{*(b)}$ and $\GYteststat^{*(b)}$ ($b=1, \ldots, B_1$).
Then calculate
\begin{equation*}
  n^{-1/2} U_{v_n}^{*(b)} =
  \left\{ \begin{array}{ll}
  \muhatA^{*(b)} - \muhatA & \mbox{ if } \GYteststatstar \geq v_n \mbox{ and } \GYteststatstar > c_{\gamma} \\
  (1+\lambda)^{-1} (\muhatA^{*(b)} + \lambda \muhatB^{*(b)}) - \muhatA & \mbox{ if } \GYteststatstar \geq v_n \mbox{ and } \GYteststatstar \leq c_{\gamma} \\
  \infty  & \mbox{ if } \GYteststatstar < v_n
  \end{array} \right.
\end{equation*}
and
\begin{equation}
  n^{-1/2}L_{v_n}^{*(b)} =
  \left\{ \begin{array}{ll}
  n^{-1/2}U_{v_n}^{*(b)} & \mbox{ if } \GYteststatstar \geq v_n \\
  - \infty  & \mbox{ if } \GYteststatstar < v_n
  \end{array} \right.
\end{equation}
Here, $\muhat_A$ denotes the estimate of $\mu_A$ from the original dataset.\footnote{Note that we have included the $n^{-1/2}$ term when writing $n^{-1/2} U_{v_n}^{*(b)}$ and $n^{-1/2} L_{v_n}^{*(b)}$ in order to remain consistent with G\&Y's notation.}

Let $U_{v_n}^{{\rm Q}(1-\alpha/2)}$ denote the $(1 - \alpha/2)$th sample quantile of $U_{v_n}^{*(1)}, \ldots, U_{v_n}^{*(B_1)}$, and let $L_{v_n}^{{\rm Q}(\alpha/2)}$ denote the $(\alpha/2)$th quantile of $L_{v_n}^{*(1)}, \ldots, L_{v_n}^{*(B_1)}$.
Calculate the nominal $(1-\alpha)100$\% CI for $\Ybar$ as
\[
\big( \; \GYestfix - n^{-1/2} U_{v_n}^{{\rm Q}(1-\alpha/2)}, \;
\GYestfix - n^{-1/2} L_{v_n}^{{\rm Q}(\alpha/2)} \; \big).
\]

This procedure has only involved a single bootstrap round, and the {\it nominal} $(1-\alpha)100$\% CI that it yields depends on the choice of $v_n$.
The {\it actual} coverage of this nominal CI also depends on $v_n$.
So, this procedure requires the user to know the value of $v_n$ that ensures that the {\it actual} coverage of the CI is $(1-\alpha)100$\%.
We now describe how G\&Y find this value of $v_n$ by using two rounds of bootstrap.

First, create $B_1$ bootstrap samples (called `first-round' bootstrap samples) and calculate the values $\muhat_A^{*(b_1)}$, $\muhatB^{*(b_1)}$ and $\GYestfix^{*(b_1)}$ of $\muhat_A$, $\muhatB$ and $\GYestfix$ for each of these bootstrap samples.
Then each of these first-round bootstrap samples is treated in turn as the original sample and $B_2$ bootstrap samples (called `second-round' bootstrap samples) are created from this $b_1$th first-round bootstrap sample.
Let $\muhat_A^{**(b_1,b_2)}$, $\muhatB^{**(b_1,b_2)}$ and $\GYestfix^{**(b_1,b_2)}$ denote the values of $\muhat_A$, $\muhatB$ and $\GYestfix$ for the $b_2$th second-round bootstrap sample arising from the $b_1$th first-round bootstrap sample ($b_1 = 1, \ldots, B_1$; $b_2 = 1, \ldots, B_2$).
Let $n^{-1/2} U_{v_n}^{**(b_1,b_2)}$ denote the value of $n^{-1/2} U_{v_n}$ given by this second-round bootstrap sample.
It is defined as
\begin{equation*}
  n^{-1/2} U_{v_n}^{**(b_1,b_2)} =
  \left\{ \begin{array}{ll}
  \muhatA^{**(b_1,b_2)} - \muhatA^{*(b_1)} & \mbox{ if } \GYteststatstarstar \geq v_n \mbox{ and } \GYteststatstarstar > c_{\gamma} \\
  (1+\lambda)^{-1} (\muhatA^{**(b_1,b_2)} + \lambda \muhatB^{**(b_1,b_2)}) - \muhatA^{*(b_1)} & \mbox{ if } \GYteststatstarstar \geq v_n \mbox{ and } \GYteststatstarstar \leq c_{\gamma} \\
  \infty  & \mbox{ if } \GYteststatstarstar < v_n
  \end{array} \right.
\end{equation*}
Let $U_{v_n}^{*(b_1){\rm Q}(1-\alpha/2)}$ denote the $(1 - \alpha/2)$th sample quantile of $U_{v_n}^{**(b_1,1)}, \ldots, U_{v_n}^{**(b_1,B_2)}$.
Likewise, let
\begin{equation*}
  n^{-1/2} L_{v_n}^{**(b_1,b_2)} =
  \left\{ \begin{array}{ll}
  \muhatA^{**(b_1,b_2)} - \muhatA^{*(b_1)} & \mbox{ if } \GYteststatstarstar \geq v_n \mbox{ and } \GYteststatstarstar > c_{\gamma} \\
  (1+\lambda)^{-1} (\muhatA^{**(b_1,b_2)} + \lambda \muhatB^{**(b_1,b_2)}) - \muhatA^{*(b_1)} & \mbox{ if } \GYteststatstarstar \geq v_n \mbox{ and } \GYteststatstarstar \leq c_{\gamma} \\
  - \infty  & \mbox{ if } \GYteststatstarstar < v_n
  \end{array} \right.
\end{equation*}
and $L_{v_n}^{*(b_1){\rm Q}(\alpha/2)}$ denote the $(\alpha/2)$th sample quantile of $L_{v_n}^{**(b_1,1)}, \ldots, L_{v_n}^{**(b_1,B_2)}$.
For the $b_1$th first-round bootstrap sample, a nominal $(1-\alpha)100$\% CI is given by
\[
\big( \; \GYestfix^{*(b_1)} - n^{-1/2} U_{v_n}^{*(b_1){\rm Q}(1-\alpha/2)}, \;
\GYestfix^{*(b_1)} - n^{-1/2} L_{v_n}^{*(b_1){\rm Q}(\alpha/2)} \; \big).
\]

Now each of the $B_1$ first-round bootstrap samples has an estimate $\muhat_A^{*(b_1)}$ and a nominal $(1-\alpha)100$\% CI.
The proportion of these $B_1$ first-round bootstrap samples for which $\muhat_A^{*(b_1)}$ lies within the nominal $(1-\alpha)100$\% CI is calculated.
If this proportion is less than $1 - \alpha$, then the value of $v_n$ needs to be increased.
If the proportion is greater than $1 - \alpha$, then $v_n$ can be decreased.
G\&Y propose to use a grid search to find the value of $v_n$ for which the proportion equals $1 - \alpha$.

This double bootstrap approach requires a method for resampling Sample A (to get first-round bootstrap samples) and then for resampling another Sample A from each resulting resampled Sample A (to get second-round bootstrap samples).
Both of these rounds of resampling needs to be appropriate for the sampling design of Sample A.
In the scenario that G\&Y considered, where Sample A was small compared to the population and was sampled using single-stage sampling without replacement, both rounds of the double bootstrap could be performed using straightforward sampling with replacement (see their Algorithm B.1).

For readers who wish to understand why the method that we have described in this section is what G\&Y's general method reduces to when $\muhat_A$ is $\HT$ or $\Haj$, and $\muhat_B$ is $\DRone$, $\KHone$ or $\KHtwo$, Appendix~\ref{sect:showGYreduces} shows that this is the case.

\subsection{Estimating $V_A$, $V_B$ and $\Gammau$}

The procedure described in Appendix~\ref{sect:GYdescription} requires estimates of $V_A$, $V_B$ and $\Gammau$.
G\&Y describe a bootstrap procedure that can be used when Sample A is small compared to the population and is obtained by single-stage sampling without replacement.
In this subsection, we describe how estimates of $V_A$, $V_B$ and $\Gammau$ can be obtained using either closed formulae or bootstrap, and when using bootstrap we allow for a more general sampling design for Sample A.

Recall that $m(X)$ now means $m(X) = E(Y \mid X, R^B=1)$, and $m(X; \beta)$ is a model for this $m(X)$, but the definition of $\pi^B(X) = P(R^B=1 \mid X)$ remains unchanged.

\subsubsection{Using closed formulae}

\label{sect:using.closed}

Because the distribution of $\muhatA$ is not affected by whether or not $R^B \ci Y \mid X$, we could estimate $V_A$ using a standard estimator of the variance of a HT or Hajek estimator.
However, the estimators described in Sections~\ref{sect:ChenDR} and~\ref{sect:combining} of the main text for $V_D= \Var ( \DRone - \Ybar \mid \curlF )$ and $C = \Cov (\HTHaj - \Ybar, \; \DR  - \Ybar\mid \curlF)$ were derived under the assumption that $R^B \ci Y \mid X$, and so it is not obvious that they (multiplied by $n$) are valid estimators for $V_B$ and $\Gammau$ when $R^B \notci Y \mid X$.
In Appendix~\ref{sect:variance.validity} we show that they are still valid, provided that Samples A and B are small compared to the population, model $\pi^B (X; \alpha)$ is correctly specified, and two modifications are made to these estimators.
The first modification is that if $\muhatB = \PLone$, then $\bfour$ should be estimated by
\begin{eqnarray*}
  \bfourhat
  & = &
  \left[ \sum_{i=1}^N \usepi_i \; \frac{ R^A_i }{ \pi^A_i } \;
  \pi^B (X_i; \alphahat) \{ 1 - \pi^B (X_i; \alphahat) \} \; \Xpi \Xpitop \right]^{-1}
  \nonumber \\
  && \hspace{0.3cm} \times
  \sum_{i=1}^N S_i \frac{ R^A_i }{ \pi^A_i } \;
    \{ 1 - \pi^B (X_i; \alphahat) \} \{ Y_i - m(X_i; \betahat) \} \;
    \Xpi
\end{eqnarray*}
(rather than by equation~(\ref{eq:original.bfourhat})).
The second modification is that if $\muhatB = \KHone$ or $\muhatB = \KHtwo$, then expression~(\ref{eq:var.generalform.estimator.nonprob}) with $\hat{\Delta}_{yi} = 0$ (rather than expression~(\ref{eq:var.bias.cor})) should be used to estimate the second component of the variance.

If $\muhatB = \KHone$ or $\muhatB = \KHtwo$, this approach yields a closed-form estimator of $\Gammau$ that is double robust (i.e.\ it is valid when either of models $\pi^B (X; \alpha)$ and $m(X; \beta)$ is correctly specified) but a closed-form estimator of $V_B$ that is not double robust (it relies on model $\pi^B (X; \alpha)$ being correctly specified).

\subsubsection{Using bootstrap}

\label{sect:using.boot}

G\&Y propose to estimate $V_A$, $V_B$ and $\Gammau$ using a bootstrap procedure.
In their Appendix B.1, they consider the scenario where Sample A is sampled using a single-stage without-replacement scheme.
Since Sample A is assumed to be very small compared to the population, this enables them to approximate the sampling mechanism as sampling with replacement.
G\&Y use the with-replacement bootstrap procedure~\cite{ShaoTu2012}, which is suitable when Sample A is sampled with replacement or by simple random sampling without replacement.
A more general bootstrap approach is described by Beaumont and Emond (2022)~\cite{BeaumontEmond2022}.

An advantage of using bootstrap when $\muhatB = \KHone$ or $\muhatB = \KHtwo$ is that it provides a double-robust estimator of $V_B$ (i.e.\ an estimator that is valid when either of models $\pi^B (X; \alpha)$ and $m(X; \beta)$ is correctly specified).

For the purpose of illustrating Beaumont and Emond's bootstrap method, we consider the following two-stage sampling design for Sample A.
At the first stage, the clusters (primary sampling units) are sampled without replacement and with unequal probabilities.
To make the problem more interesting, suppose that the first-stage sampling probabilities are not small, so that treating this sampling as with replacement may be a poor approximation.
At the second-stage, a number of individuals are sampled from each sampled primary sampling unit, again using sampling without replacement and with unequal probabilities.
Suppose that the conditional second-stage sampling probabilities are small, and so we can approximate the second-stage sampling as sampling with replacement.
We shall assume that there is no stratification of the clusters.
However, we shall note the modifications needed when there is stratification.
Note that this scenario is just an example; Beaumont and Emond's method can be used more generally, including for multi-stage sampling designs that have more than two stages.

Let $J$ denote the number of clusters in the population.
Without loss of generality, index the clusters that are included in Sample A as $1, \ldots, n_1$ ($n_1 < J$).
Let $\pi_{1j}$ denote the probability that cluster $j$ is included in Sample A ($j=1, \ldots, n_1$).
Generate $(M_{11}^*, \ldots, M_{1n_1}^*) \sim \mbox{Multinomial} \big( n_1 - 1, (n_1^{-1}, \ldots, n_1^{-1}) \big)$ and let
\[
a_{1j}^{\rm raw} = 1 - \sqrt{ 1 - \pi_{1j} } + \sqrt{ 1 - \pi_{1j} } \; \frac{ n_1 }{ n_1 - 1 } \; M_{1j}^*
\hspace{1cm}
(j=1, \ldots, n_1).
\]
(Note that this is Beaumont and Emond's equation~(21) with $m_1 = n_1 - 1$.)
Now scale the $a_{1j}^{\rm raw}$'s so that they sum to $n_1$.
That is, set
\[
a_{1j} = a_{1j}^{\rm raw} \times n_1 \bigg/ \sum_{k=1}^{n_1} a_{1k}^{\rm raw}
\hspace{1cm}
(j=1, \ldots, n_1).
\]
If there are any clusters that have probability one of being included (i.e.\ $\pi_{1j} = 1$), then these are excluded from the scaling and their $a_{1j}$ values are set to one.

If the clusters are sampled within strata, then the calculation of $a_{1j}$ in the last paragraph is carried out separately for each stratum, $J$ denotes the number of clusters in the stratum, and $n_1$ denotes the number of these clusters that are included in Sample A.

Next, the following procedure is carried out separately in each sampled cluster $j$.
Without loss of generality, index the individuals in cluster $j$ who are included in Sample A as $1, \ldots, n_{2j}$.
Generate $(M_{2j1}^*, \ldots, M_{2jn_{2j}}^*) \sim \mbox{Multinomial} \big( n_{2j} - 1, (n_{2j}^{-1}, \ldots, n_{2j}^{-1}) \big)$ and let
\[
a_{2ji} = \frac{ n_{2j} }{ n_{2j} - 1 } \; M_{2ji}^*
\hspace{1cm}
(i=1, \ldots, n_{2j}).
\]
(Note that this is the procedure described in Beaumont and Emond's Section 3.2.)

Sample B was sampled using Poisson sampling.
If Sample B is large but is small compared to the population, we can approximate this Poisson sampling as sampling with replacement and again use the procedure described in Beaumont and Emond's Section 3.2.
Without loss of generality, index the individuals included in Sample B as $1, \ldots, n_B$.
Generate $(M_{B1}^*, \ldots, M_{Bn_B}^*) \sim \mbox{Multinomial} \big( n_B - 1, (n_B^{-1}, \ldots, n_B^{-1}) \big)$ and let
\[
a_{Bk} = \frac{ n_B }{ n_B - 1 } \; M_{Bk}^*
\hspace{1cm}
(k=1, \ldots, n_B).
\]
Alternatively, Beaumont and Emond's procedure for Poisson sampling could be used (see their Section 3.1).

Having generated the $a_{1j}$'s, $a_{2ji}$'s and $a_{Bk}$'s, calculate $\muhatA$ just as before but this time multiplying the weight $1 / \pi^A$ of individual $i$ in cluster $j$ in Sample A by $a_{1j} \times a_{2ji}$.
Also calculate $\muhatB$ just as before but this time multiplying the weight $1 / \pi^A$ of individual $i$ in cluster $j$ in Sample A by $a_{1j} \times a_{2ji}$ and weighting individual $k$ in Sample B by $a_{Bk}$.
Call these estimates $\muhatA^{*(1)}$ and $\muhatB^{*(1)}$.

Repeat this process $B_1$ times (e.g.\ $B_1 = 1000$) to get estimates $\muhatA^{*(b)}$ and $\muhatB^{*(b)}$ ($b=1, \ldots, B_1$).
Finally, estimate $V_A$, $V_B$ and $\Gammau$ as, respectively,
\begin{eqnarray*}
  &&
\frac{n}{B_1} \sum_{b=1}^{B_1} ( \muhatA^{*(b)} - \muhatA )^2
\\
  &&
\frac{n}{B_1} \sum_{b=1}^{B_1} ( \muhatB^{*(b)} - \muhatB )^2
\\
  &&
\frac{n}{B_1} \sum_{b=1}^{B_1} ( \muhatA^{*(b)} - \muhatA ) ( \muhatB^{*(b)} - \muhatB ).
\end{eqnarray*}

\subsection{Closed-form estimators of $V_B$ and $\Gammau$ when $R^B \notci Y \mid X$}

\label{sect:variance.validity}

We noted in Appendix~\ref{sect:using.closed} that the estimators of $V_D$ and $C$ given in Sections~\ref{sect:ChenDR} and~\ref{sect:combining} of the main text were derived under the assumption that $R^B \ci Y \mid X$, and so it is not obvious that they (multiplied by $n$) are valid estimators of $V_B$ and $\Gammau$ when $R^B \notci Y \mid X$.
In this subsection, we consider $\muhatB = \PLone$, $\muhatB = \KHone$ and $\muhatB = \KHtwo$ in turn and show that those estimators of $V_D$ and $C$ are valid estimators of $V_B$ and $\Gammau$ when $R^B \notci Y \mid X$, provided that Samples A and B are small compared to the population, model $\pi^B (X; \alpha)$ is correctly specified, and two modification are made.
First, when $\muhatB = \PLone$, $\bfour$ should be estimated by equation~(\ref{eq:modified.bfourhat}).
Second, when $\muhatB = \KHone$ or $\muhatB = \KHtwo$, equation~(\ref{eq:simplified.varB}) should be used to estimate the second component of the variance.

Note that $\PLone - \bar{m} \xrightarrow{p} 0$, $\KHone - \bar{m} \xrightarrow{p} 0$ and $\KHtwo - \bar{m} \xrightarrow{p} 0$, where $\bar{m} = N_s^{-1} \sum_{i=1}^N S_i m (X_i)$.

\subsubsection{When $\muhat_B = \PLone$}

\label{sect:considering.PLone}

Equation~(\ref{eq:DRone.Taylor.general}) shows that when model $\pi^B (X; \alpha)$ is correctly specified and $R^B \ci Y \mid X$,
\begin{eqnarray}
  &&
  \PLone (\alphahat, \betahat) - \Ybar
  \nonumber \\
  && \hspace{.3cm} =
    \frac{1}{N_s} \sum_{i=1}^N
  \frac{ R^A_i }{ \pi^A_i } \left\{ S_i m(X_i; \betalim) + \usepi_i \; \pi^B (X_i) \; \bfour^\top \Xpi \right\}
  \nonumber \\
  && \hspace{.8cm} +
  \frac{1}{N_s} \sum_{i=1}^N
  \left\{ \frac{ R^B_i }{ \pi^B (X_i) } - 1 \right\}
  \left[ S_i \{ Y_i - m(X_i; \betalim) \} - \usepi_i \; \pi^B (X_i) \; \bfour^\top \Xpi \right]
  \nonumber \\
  && \hspace{.8cm} -
  \mbar - \bfour^\top 
  \frac{1}{N_s} \sum_{i=1}^N \usepi_i \; \pi^B (X_i) \; \Xpi
  + o_p(N^{-1/2})
  \label{eq:PL1.expansion.GY}
\end{eqnarray}
where
\begin{eqnarray*}
  \bfour
  & = &
  \left[ \sum_{i=1}^N \usepi_i \; \pi^B (X_i; \alphalim) \{ 1 - \pi^B (X_i; \alphalim) \} \; \Xpi \Xpitop \right]^{-1}
  \\
  && \times
  \sum_{i=1}^N S_i \{ 1 - \pi^B (X_i; \alphalim) \} \{ Y_i - m(X_i; \betalim) \} \Xpi.
\end{eqnarray*}

This equation still holds when $R^B \notci Y \mid X$, with now $m(X)$ being $E(Y \mid X, R^B=1)$ and $m(X; \beta)$ being a model for $E(Y \mid X, R^B=1)$.

Furthermore, equation~(\ref{eq:use.zero.covariance}), i.e.\
\begin{eqnarray*}
  &&
  \Var \{ \PLone (\alphahat, \betahat) - \Ybar \mid \curlF, \curlY \}
  \nonumber \\
  && \hspace{.3cm} =
  \Var \left[
      \frac{1}{N_s} \sum_{i=1}^N 
      \frac{ R^A_i }{ \pi^A_i } \left\{ S_i m(X_i; \betalim) + \usepi_i \; \pi^B (X_i) \; \bfour^\top \Xpi \right\}
    \mid \curlF, \curlY \right]
  \nonumber \\
  && \hspace{.8cm} +
  \Var \left(
    \frac{1}{N_s} \sum_{i=1}^N
    \left\{ \frac{ R^B_i }{ \pi^B (X_i) } - 1 \right\}
  \left[ S_i \{ Y_i - m(X_i; \betalim) \} - \usepi_i \; \pi^B (X_i) \; \bfour^\top \Xpi \right]
  \mid \curlF, \curlY \right)
    \nonumber \\
  && \hspace{.8cm} +
    o_p (N^{-1})
\end{eqnarray*}
also still holds when $R^B \notci Y \mid X$.
However, the result that
\begin{eqnarray*}
  &&
  \Var \left(
    \frac{1}{N_s} \sum_{i=1}^N
    \left\{ \frac{ R^B_i }{ \pi^B (X_i) } - 1 \right\}
  \left[ S_i \{ Y_i - m(X_i; \betalim) \} - \usepi_i \; \pi^B (X_i) \; \bfour^\top \Xpi \right]
  \mid \curlF, \curlY \right)
  \\
  && =
    \frac{1}{N_s^2} \sum_{i=1}^N
    \frac{ 1 - \pi^B (X_i) }{ \pi^B (X_i) } \;
    \left[ S_i \{ Y_i - m(X_i; \betalim) \} - \usepi_i \; \pi^B (X_i) \; \bfour^\top \Xpi \right]^2
\end{eqnarray*}
no longer holds.
Instead,
\begin{eqnarray}
  &&
  \Var \left(
    \frac{1}{N_s} \sum_{i=1}^N
    \left\{ \frac{ R^B_i }{ \pi^B (X_i) } - 1 \right\}
  \left[ S_i \{ Y_i - m(X_i; \betalim) \} - \usepi_i \; \pi^B (X_i) \; \bfour^\top \Xpi \right]
  \mid \curlF, \curlY \right)
  \nonumber \\
  && =
    \frac{1}{N_s^2} \sum_{i=1}^N
    \frac{ P(R^B_i=1 \mid X_i, Y_i) \{ 1 - P(R^B_i=1 \mid X_i, Y_i) \} }{ \{ \pi^B (X_i) \}^2 }
  \nonumber \\
      && \hspace{1.8cm} \times
  \left[ S_i \{ Y_i - m(X_i; \betalim) \} - \usepi_i \; \pi^B (X_i) \; \bfour^\top \Xpi \right]^2
  \label{eq:var.PL1.B.GY}
\end{eqnarray}
If Sample B is small compared to the population, $1 - P(R^B=1 \mid X, Y) \approx 1$ and can be approximated by $1 - P(R^B=1 \mid X) = 1 - \pi^B(X)$.
Thus, when Sample B is small compared to the population and model $\pi^B (X; \alpha)$ is correctly specified, we can estimate expression~(\ref{eq:var.PL1.B.GY}) using equation~(\ref{eq:second.term}), i.e.
\begin{equation}
  \frac{1}{N_s^2} \sum_{i=1}^N R^B_i \; 
  \frac{ 1 - \pi^B (X_i; \alphahat) } { \{ \pi^B (X_i; \alphahat) \}^2 } \; [ S_i \{ Y_i - m (X_i; \betahat) \} - \usepi_i \; \pi^B (X_i; \alphahat) \; \bfourhat^\top \Xpi ]^2.
  \label{eq:var2.same.PL1}
\end{equation}
However, this expression involves $\bfourhat$, and the estimator $\bfourhat$ given in Section~\ref{sect:ChenDR} of the main text was
\begin{eqnarray*}
  &&
  \left[ \sum_{i=1}^N \usepi_i \; R^B_i \;
  \{ 1 - \pi^B (X_i; \alphahat) \} \; \Xpi \Xpitop \right]^{-1}
  \nonumber \\
  && \hspace{0.3cm} \times
  \sum_{i=1}^N S_i \; \frac{ R^B_i }{ \pi^B (X_i; \alphahat) } \;
    \{ 1 - \pi^B (X_i; \alphahat) \} \{ Y_i - m(X_i; \betahat) \} \;
    \Xpi.
\end{eqnarray*}
When $R^B \notci Y \mid X$, this needs to be modified to
\begin{eqnarray}
  \bfourhat
  & = &
  \left[ \sum_{i=1}^N \usepi_i \; \frac{ R^A_i }{ \pi^A_i } \;
  \pi^B (X_i; \alphahat) \{ 1 - \pi^B (X_i; \alphahat) \} \; \Xpi \Xpitop \right]^{-1}
  \nonumber \\
  && \hspace{0.3cm} \times
  \sum_{i=1}^N S_i \frac{ R^A_i }{ \pi^A_i } \;
    \{ 1 - \pi^B (X_i; \alphahat) \} \{ Y_i - m(X_i; \betahat) \} \;
    \Xpi.
    \label{eq:modified.bfourhat}
\end{eqnarray}

Recall that when $R^B \ci Y \mid X$, $\bfour = o_p(1)$ when both nuisance models are correctly specified.
Note that this is not the case when $R^B \notci Y \mid X$.

Looking now at the covariance, equation~(\ref{eq:cov.HTHaj}) shows that
\begin{eqnarray*}
  &&
  \Cov ( \PLone - \Ybar, \; \HTHaj - \Ybar \mid \curlF, \curlY )
  \nonumber \\
  && =
  \Cov \left[
    \frac{1}{N_s} \sum_{i=1}^N
    \frac{ R^A_i }{ \pi^A_i } \left\{ S_i m(X_i; \betalim) + \usepi_i \; \pi^B (X_i) \; \bfour^\top \Xpi \right\} , \;
    \frac{1}{N_s} \sum_{i=1}^N
    S_i \frac{ R^A_i }{ \pi^A_i } (Y_i - \Gamma)
    \mid \curlF, \curlY \right]
  \nonumber \\
  && \hspace{.5cm} +
  o_p(N^{-1}).
\end{eqnarray*}
This equation still holds when $R^B \notci Y \mid X$, and so the same estimator that was used in Section~\ref{sect:ChenDR} of the main text can be used.
As when estimating the variance, $\bfour$ should be estimated using the estimator $\bfourhat$ defined by equation~(\ref{eq:modified.bfourhat}).

We need to estimate $\Var \{ \PLone (\alphahat, \betahat) - \Ybar \mid \curlF \}$ and $\Cov ( \PLone - \Ybar, \; \HTHaj - \Ybar \mid \curlF )$.
As pointed out in Appendix~\ref{sect:modeldesign.PL1}, when $R^B \ci Y \mid X$,
\begin{eqnarray*}
  &&
  \Var \{ \PLone (\alphahat, \betahat) - \Ybar \mid \curlF \}
  \nonumber \\
  && =
E [ \Var \{ \PLone (\alphahat, \betahat) - \Ybar \mid \curlF, \curlY \} \mid \curlF ]
+
\Var [ E \{ \PLone (\alphahat, \betahat) - \Ybar \mid \curlF, \curlY \} \mid \curlF ]
  \nonumber \\
  && =
E [ \Var \{ \PLone (\alphahat, \betahat) - \Ybar \mid \curlF, \curlY \} \mid \curlF ]
+
\Var \{ o_p(N^{-1/2}) \mid \curlF \}
  \nonumber \\
  && =
E [ \Var \{ \PLone (\alphahat, \betahat) - \Ybar \mid \curlF, \curlY \} \mid \curlF ]
+
o_p(N^{-1}).
\end{eqnarray*}
However, when $R^B \notci Y \mid X$, $E \{ \PLone (\alphahat, \betahat) - \Ybar \mid \curlF, \curlY \}$ is not $o_p (N^{-1/2})$.
Instead, equation~(\ref{eq:PL1.expansion.GY}) gives us
\begin{eqnarray*}
  &&
  E \{ \PLone (\alphahat, \betahat) - \Ybar \mid \curlF, \curlY \}
  \\
  && \hspace{.4cm} =
    \frac{1}{N_s} \sum_{i=1}^N
  \left\{ S_i m(X_i; \betalim) + \usepi_i \; \pi^B (X_i) \; \bfour^\top \Xpi \right\}
  \nonumber \\
  && \hspace{.8cm} +
  \frac{1}{N_s} \sum_{i=1}^N
  \left\{ \frac{ E(R^B_i \mid X_i, Y_i) }{ \pi^B (X_i) } - 1 \right\}
  \left[ S_i \{ Y_i - m(X_i; \betalim) \} - \usepi_i \; \pi^B (X_i) \; \bfour^\top \Xpi \right]
  \nonumber \\
  && \hspace{.8cm} -
  \mbar - \bfour^\top 
  \frac{1}{N_s} \sum_{i=1}^N \usepi_i \; \pi^B (X_i) \; \Xpi
  - \Ybar
  + o_p(N^{-1/2})
  \\
  && \hspace{.4cm} =
    \frac{1}{N_s} \sum_{i=1}^N
  S_i m(X_i; \betalim)
  \nonumber \\
  && \hspace{.8cm} +
  \frac{1}{N_s} \sum_{i=1}^N
  \left\{ \frac{ E(R^B_i \mid X_i, Y_i) }{ \pi^B (X_i) } - 1 \right\}
  \left[ S_i \{ Y_i - m(X_i; \betalim) \} - \usepi_i \; \pi^B (X_i) \; \bfour^\top \Xpi \right]
  \nonumber \\
  && \hspace{.8cm} -
  \mbar - \Ybar
  + o_p(N^{-1/2})
\end{eqnarray*}
Hence
\begin{eqnarray}
  &&
\Var [ E \{ \PLone (\alphahat, \betahat) - \Ybar \mid \curlF, \curlY \} \mid \curlF ]
\nonumber \\
&& \hspace{.2cm} =
\Var \left[
  \frac{1}{N_s} \sum_{i=1}^N
  \left\{ \frac{ E(R^B_i \mid X_i, Y_i) }{ \pi^B (X_i) } - 1 \right\}
  \left[ S_i \{ Y_i - m(X_i; \betalim) \} - \usepi_i \; \pi^B (X_i) \; \bfour^\top \Xpi \right]
  - \Ybar
  \mid \curlF \right]
\nonumber \\
&&
\label{eq:ignore.GY}
\end{eqnarray}

If Samples A and B are small compared to the population, expression~(\ref{eq:ignore.GY}) will be small compared to $E [ \Var \{ \PLone (\alphahat, \betahat) - \Ybar \mid \curlF, \curlY \} \mid \curlF ]$, and hence
\begin{eqnarray*}
  \Var \{ \PLone (\alphahat, \betahat) - \Ybar \mid \curlF \}
  & \approx &
E [ \Var \{ \PLone (\alphahat, \betahat) - \Ybar \mid \curlF, \curlY \} \mid \curlF ]
  \nonumber \\
  & \approx &
\Var \{ \PLone (\alphahat, \betahat) - \Ybar \mid \curlF, \curlY \}.
\end{eqnarray*}

Similarly, we have
\begin{eqnarray*}
  &&
  \Cov \{ \PLone (\alphahat, \betahat) - \Ybar, \; \HTHaj - \Ybar \mid \curlF \}
  \nonumber \\
  && \hspace{.4cm} =
  E [ \Cov \{ \PLone (\alphahat, \betahat) - \Ybar, \; \HTHaj - \Ybar \mid \curlF, \curlY \} \mid \curlF ]
    \nonumber \\
  && \hspace{.9cm} +
\Cov [ E \{ \PLone (\alphahat, \betahat) - \Ybar \mid \curlF, \curlY \}, \; E ( \HTHaj - \Ybar \mid \curlF, \curlY )\mid \curlF].
\end{eqnarray*}

Again, when Samples A and B are small compared to the population, the second term would be expected to be small compared to the first term.
So, we have
\begin{eqnarray*}
  &&
  \Cov \{ \PLone (\alphahat, \betahat) - \Ybar, \; \HTHaj - \Ybar \mid \curlF \}
  \\
  && \hspace{.4cm} \approx
  E [ \Cov \{ \PLone (\alphahat, \betahat) - \Ybar, \; \HTHaj - \Ybar \mid \curlF, \curlY \} \mid \curlF ]
  \\
  && \hspace{.4cm} \approx
  \Cov \{ \PLone (\alphahat, \betahat) - \Ybar, \; \HTHaj - \Ybar \mid \curlF, \curlY \}.
\end{eqnarray*}

In summary, if model $\pi^B (X; \alpha)$ is correctly specified and Samples A and B are small compared to the population, we can use the same estimators of $V_B$ and $\Gammau$ that are valid when $R^B \ci Y \mid X$ but with $\bfourhat$ now estimated by equation~(\ref{eq:modified.bfourhat}).

\subsubsection{When $\muhat_B = \KHone$}

\label{sect:considering.KHone}

Equation~(\ref{eq:KH1.expansion}) shows that when i) either model $\pi^B (X; \alpha)$ or model $m(X; \beta)$ is correctly specified and ii) $R^B \ci Y \mid X$,
\begin{eqnarray*}
  \KHone (\alphahat, \betahat) - \Ybar
  & = &
    \frac{1}{N_s} \sum_{i=1}^N
  S_i \frac{ R^A_i }{ \pi^A_i } m(X_i; \betalim)
  +
  \frac{1}{N_s} \sum_{i=1}^N
  S_i \left\{ \frac{ R^B_i }{ \pi^B (X_i) } - 1 \right\}
  \{ Y_i - m(X_i; \betalim) \}
  \nonumber \\
  &&
  - \mbar + o_p(N^{-1/2}).
\end{eqnarray*}
This still holds true when $R^B \notci Y \mid X$, with now $m(X)$ being $E(Y \mid X, R^B=1)$ and $m(X; \beta)$ being a model for $E(Y \mid X, R^B=1)$.
It follows that
\begin{eqnarray}
  &&
  \Var \{ \KHone (\alphahat, \betahat) - \Ybar \mid \curlF, \curlY \}
  \nonumber \\
  && \hspace{.3cm} =
  \Var \left\{
      \frac{1}{N_s} \sum_{i=1}^N 
      S_i \frac{ R^A_i }{ \pi^A_i } m(X_i; \betalim)
      \mid \curlF, \curlY
      \right\}
  \nonumber \\
  && \hspace{0.7cm} +
    \frac{1}{N_s^2} \sum_{i=1}^N S_i
    \frac{ P(R^B_i=1 \mid X_i, Y_i) \{ 1 - P(R^B_i=1 \mid X_i, Y_i) \} }{ \{ \pi^B (X_i) \}^2 }
    \{ Y_i - m(X_i; \betalim) \}^2
    \nonumber
    \\
    && \hspace{0.7cm} +
    o_p (N^{-1}).
    \label{eq:var.KH1.B.GY}
\end{eqnarray}
As in Appendix~\ref{sect:considering.PLone}, if Sample B is small compared to the population and model $\pi^B (X; \alpha)$ is correctly specified, we can estimate the second term on the right-hand side of equation~(\ref{eq:var.KH1.B.GY}) using the following simplified version of expression~(\ref{eq:var2.same.PL1}):
\begin{equation}
  \frac{1}{N_s^2} \sum_{i=1}^N R^B_i \; 
  \frac{ 1 - \pi^B (X_i; \alphahat) } { \{ \pi^B (X_i; \alphahat) \}^2 } \; [ S_i \{ Y_i - m (X_i; \betahat) \} ]^2.
\label{eq:simplified.varB}
\end{equation}
Just as in Appendix~\ref{sect:considering.PLone}, when Samples A and B are small compared to the population and model $\pi^B (X; \alpha)$ is correctly specified, we can approximate $\Var \{ \KHone (\alphahat, \betahat) - \Ybar \mid \curlF \}$ by $\Var \{ \KHone (\alphahat, \betahat) - \Ybar \mid \curlF, \curlY \}$.

Equation~(\ref{eq:cov.KH}) shows that when i) either model $\pi^B (X; \alpha)$ or model $m(X; \beta)$ is correctly specified and ii) $R^B \ci Y \mid X$,
\begin{eqnarray*}
  &&
  \Cov ( \KHone - \Ybar, \; \HTHaj - \Ybar \mid \curlF, \curlY )
  \nonumber \\
  && \hspace{.4cm} =
  \Cov \left\{
    \frac{1}{N_s} \sum_{i=1}^N
    S_i \frac{ R^A_i }{ \pi^A_i } m(X_i; \betalim) , \;
    \frac{1}{N_s} \sum_{i=1}^N
    S_i \frac{ R^A_i }{ \pi^A_i } (Y_i - \Gamma)
    \mid \curlF, \curlY \right\}
  + o_p(N^{-1}).
\end{eqnarray*}
This equation still holds when $R^B \notci Y \mid X$.
Just as in Appendix~\ref{sect:considering.PLone}, when Samples A and B are small compared to the population and either model $\pi^B (X; \alpha)$ or $m(X; \beta)$ is correctly specified, we can approximate $\Cov ( \KHone - \Ybar, \; \HTHaj - \Ybar \mid \curlF )$ by $\Cov ( \KHone - \Ybar, \; \HTHaj - \Ybar \mid \curlF, \curlY )$.

As we saw earlier, when $R^B \ci Y \mid X$, a double robust estimator of $\Var \{ \KHone (\alphahat, \betahat) - \Ybar \mid \curlF, \curlY \}$ is available, i.e.\ an estimator that is valid when either model $\pi^B (X; \alpha)$ or model $m (X; \beta)$ is correctly specified.
Unfortunately, it is not obvious that such a variance estimator exists when $R^B \notci Y \mid X$.

\subsubsection{When $\muhat_B = \KHtwo$}

\label{sect:considering.KHtwo}

Appendix~\ref{sect:modeldesign.KH2} shows that when $R^B \ci Y \mid X$,
\begin{eqnarray*}
  \KHtwo - \Ybar
  & = &
  \frac{1}{N_s} \sum_{i=1}^N S_i \left[ \frac{R^A_i}{\pi^A_i} \{ m (X_i; \betalim) - \mbartrue \}
    + \left\{ \frac{ R^B_i }{ \pi^B (X_i; \alphalim) } - 1 \right\} \{ Y_i - m (X_i; \betalim) \} \right]
  \nonumber \\
  &&
  + o_p (N^{1/2}).
\end{eqnarray*}
This equation still holds when $R^B \notci Y \mid X$, with $m(X) = E(Y \mid X, R^B=1)$ and $m(X; \beta)$ being a model for $E(Y \mid X, R^B=1)$.

Therefore, just as in Appendix~\ref{sect:considering.KHone}, when model $\pi^B (X)$ is correctly specified and Samples A and B are small compared to the population, we can use the same estimators of $V_B$ and $\Gammau$ that we used for $\KHtwo$ when $R^B \ci Y \mid X$, except that we now use equation~(\ref{eq:simplified.varB}) for the second component of the variance $V_B$.
As we noted in Appendix~\ref{sect:considering.KHone}, the resulting estimator of $V_B$ is not double robust (the estimator of $\Gammau$ is double robust).

\subsection{Derivation of the description in Appendix~\ref{sect:GYdescription}}

\label{sect:showGYreduces}
  
In this section, we show that the method we described in Appendix~\ref{sect:GYdescription} is what G\&Y's general method reduces to when $\muhat_A$ is $\HT$ or $\Haj$, and $\muhat_B$ is $\DRone$, $\KHone$ or $\KHtwo$.
Appendix~\ref{sect:GYpointestimator} focuses on G\&Y's point estimator $\GYmuhat$, Appendix~\ref{sect:proof.Thm4.1} on their Theorem 4.1, and Appendix~\ref{eq:GYinference} on their method for calculating CIs.

\subsubsection{Gao and Yang's test-and-pool point estimator}

\label{sect:GYpointestimator}

Estimators $\HT$ and $\Haj$ are solutions to estimating equations of the form 
$$\sum_{i=1}^N \Phi_A (X_i, Y_i, R^A_i; \Ybar) = 0.$$
Estimators $\PLone$, $\KHone$ and $\KHtwo$ are solutions to estimating equations of the form $$\sum_{i=1}^N \Phi_B (X_i, Y_i, R^A_i, R^B_i; \Ybar, \alphahat, \betahat) = 0$$ and have the property that $E \{ \partial \Phi_B (\mu, \alphalim, \betalim) / \partial \mu |_{\mu=\Ybar} \} = -1$.
Note that Gao and Yang (G\&Y) consider the special case of $\muhatA = \Haj$ and $\muhatB = \PLone$ in Appendix B.2 (they also go some way to considering $\muhatB = \KHone$).

Like G\&Y, we assume the goal is to estimate the mean in the entire (finite) population, not in a domain.

The first step of G\&Y's method is to find the values of $c_{\gamma}$ and $\lambda$ that minimise the MSE given by
\begin{eqnarray}
  \mbox{MSE} & = &
  V_{\rm eff} (\mu_1^2 + 1)
  +
  \left\{
  F_3 (c_{\gamma}; \mu_2^2 / 2)
  +
  F_5 (c_{\gamma}; \mu_2^2 / 2) \mu_2^2
  \right\}
  (\omega_A V_{\rm A-eff}^{1/2} - \omega_B V_{\rm B-eff}^{1/2})^2
  \nonumber \\
  && -
  2 \; F_3 (c_{\gamma}; \mu_2^2 / 2)
  V_{\rm eff}^{1/2} \mu_1 \mu_2
  (\omega_A V_{\rm A-eff}^{1/2} - \omega_B V_{\rm B-eff}^{1/2})
  \nonumber \\
  && +
  V_{\rm A-eff}
  \left[
  \{ 1 - F_3 (c_{\gamma}; \mu_2^2 / 2) \}
  +
  \{ 1 - F_5 (c_{\gamma}; \mu_2^2 / 2) \} \mu_2^2
  \right]
  \nonumber \\
  && -
  2 \; \{ 1 - F_3 (c_{\gamma}; \mu_2^2 / 2) \}
  V_{\rm eff}^{1/2} V_{\rm A-eff}^{1/2} \mu_1 \mu_2
\end{eqnarray}
where
\begin{eqnarray*}
  \mu_1
  & = &
  \Sigma_S^{-1/2} \eta
  \\
  \mu_2
  & = &
  - \Sigma_T^{-1/2} \eta
  \\
  \Sigma_T^{1/2}
  & = &
  - f_B^{1/2} \sqrt{ V_A + V_B - 2 \Gammau }
  \\
  \Sigma_S^{1/2}
  & = &
  f_B^{1/2} \frac{ \sqrt{ (V_A + V_B - 2 \Gammau) (V_A V_B - \Gammau^2) } }
  { \Gammau - V_A }
  \\
  V_{\rm eff}^{1/2}
& = &
\frac{ \sqrt{V_A V_B - \Gammau^2} }{ \sqrt{V_A + V_B - 2 \Gammau} }
\\
  V_{\rm A-eff}^{1/2}
  & = &
  \frac{ \Gammau - V_A }{ \sqrt{ V_A + V_B - 2 \Gammau } }
  \\
  V_{\rm B-eff}^{1/2}
  & = &
  \frac{ \Gammau - V_B }{ \sqrt{ V_A + V_B - 2 \Gammau } }
  \\
  \omega_A
  & = &
  \frac{ 1 }{ 1 + \lambda }
  \\
  \omega_B
  & = &
  \frac{ \lambda }{ 1 + \lambda }.
 \end{eqnarray*}
Note that, by convention, $\sqrt{x}$ means the {\it non-negative} square root of $x$. 

Since $V_A$, $V_B$ and $\Gammau$ are unknown, they are replaced by estimators.
The unknown $f_B$ and $\eta$ are replaced by their estimators $n_B / (n_A + n_B)$ and $\sqrt{n_B} (\muhatB - \muhatA)$, respectively.\footnote{See penultimate line of G\&Y's page 1502 for this, remembering that $\Phi_{B,n}$ is defined in the middle of page 1500 as including a $n_B^{1/2}$ term.}

The second step of G\&Y's method is to calculate the test statistic
\[
\GYteststat = \hat{\Sigma}_T^{-1} (\muhatB - \muhatA)^2
\]
where $\hat{\Sigma}_T^{-1}$ is the estimator of $f_B \Sigma_T^{-1}$ obtained by plugging the estimators of $V_A$, $V_B$ and $\Gammau$ into the formula for $f_B \Sigma_T^{-1}$, i.e.\ $V_A + V_B - 2 \Gamma_u$. \footnote{Note that the definition of $\Sigma_T$ in G\&Y's equation (3.5) as the asymptotic variance of $\muhatB - \muhatB$ (which is $V_A + V_B - 2 \Gamma_u$) is different from the definition of $\Sigma_T$ in G\&Y's equation (A.15) (which is $f_B (V_A + V_B - 2 \Gamma_u)$).
Since we are using the definition given in their equation (A.15), we must multiply $\Sigma_T^{-1}$ by $f_B$ in order to get the $\Sigma_T$ suitable for their equation (3.5).}
If $\GYteststat \geq c_{\gamma}$, then $\GYmuhat$ is set equal to $\muhat_A$; otherwise $\GYmuhat$ is set equal to $(\muhat_A + \lambda \muhat_B) / (1 + \lambda)$.

\subsubsection{Summary of proof of Theorem 4.1 in special case where $\muhatA = \HT$ or $\Haj$, and $\muhat_B = \DRone$, $\KHone$ or $\KHtwo$}

\label{sect:proof.Thm4.1}

We now summarise the main steps in G\&Y's proof (given in their Appendix A.4) of their Theorem 4.1.
There they show that $\sqrt{n} (\muhat_A - \Ybar, \muhat_B - \Ybar)$ is asymptotically normally distributed with mean $(0, f_B^{-1/2} \eta)$, variances $V_A$ and $V_B$ and covariance $\Gammau$.
Our purpose in studying this proof is partly to establish what some of the notation means.

G\&Y consider the distribution of two random variables $(Z_1, Z_2)$ with this same bivariate normal distribution.
They transform $(Z_1, Z_2)$ to random variables $(U_1, U_2)$ and then to random variables $(W_1, W_2)$, where $W_1$ and $W_2$ are independent,
both with variance 1, and $E(W_j) = \mu_j$ $(j=1, 2)$, where
\begin{eqnarray*}
  \mu_1
  & = &
  \Sigma_S^{-1/2} \eta
  \\
  \mu_2
  & = &
  - \Sigma_T^{-1/2} \eta
  \\
  \Sigma_T
  & = &
  f_B (V_A + V_B - 2 \Gammau)
  \\
  \Sigma_S
  & = &
  f_B \Var \left( \frac{\Gammau - V_B}{\Gammau - V_A} Z_1 + Z_2 \right)
  \\
  & = &
  f_B \left\{ \left( \frac{\Gammau - V_B}{\Gammau - V_A} \right)^2 V_A + V_B + 2 \left( \frac{\Gammau - V_B}{\Gammau - V_A} \right) \Gammau \right\}
  \\
  & = &
  f_B \frac{ (V_A + V_B - 2 \Gammau) (V_A V_B - \Gammau^2) }
  { (\Gammau - V_A)^2 }.
\end{eqnarray*}

The actual relation between $(Z_1, Z_2)$ and $(W_1, W_2)$ is
\begin{eqnarray*}
  W_1
  & = &
  f_B^{1/2} \Sigma_S^{-1/2} \left( \frac{ \Gammau - V_B }{ \Gammau - V_A } Z_1 + Z_2 \right)
  \\
  W_2
  & = &
 - \frac{ Z_1 - Z_2 }{ (V_A + V_B - 2 \Gammau)^{1/2} }.
\end{eqnarray*}
In principle, each of $\Sigma_S^{1/2}$ could denote the positive or the negative square root of $\Sigma_S$, and similarly for $\Sigma_T^{1/2}$.
However, G\&Y implicitly define them as
\begin{eqnarray*}
  \Sigma_T^{1/2}
  & = &
  - f_B^{1/2} \sqrt{ V_A + V_B - 2 \Gammau) }
  \\
  \Sigma_S^{1/2}
  & = &
  f_B^{1/2} \frac{ \sqrt{ (V_A + V_B - 2 \Gammau) (V_A V_B - \Gammau^2) } }
  { \Gammau - V_A }.
\end{eqnarray*}
Note that this means $\Sigma_T^{1/2}$ is negative, and $\Sigma_S^{1/2}$ will be positive if $\Gammau > V_A$ and negative if $\Gammau < V_A$.

Let $\GYestfix$ denote G\&Y's efficient estimator that uses pre-testing with fixed critical value $c_{\gamma}$ for the test statistic $\GYteststat$ and pooling weights $\omega_A = 1 / (1 + \lambda)$ for $\muhat_A$ and $\omega_A = \lambda / (1 + \lambda)$ for $\muhat_B$ for a fixed value of $\lambda$.
Theorem 4.1 states that the asymptotic distribution of $n^{1/2} (\GYestfix - \Ybar)$ is the same as the distribution of $Q$, where $Q$ equals
\begin{eqnarray*}
  - V_{\rm eff}^{1/2} W_1 + ( \omega_A V_{A - {\rm eff}}^{1/2} - \omega_B V_{B - {\rm eff}}^{1/2} ) W_{ [ 0, c_{\gamma} ] }^t
  && \mbox{ with probability } \epsilon
  \\
  - V_{\rm eff}^{1/2} W_1 + V_{A - {\rm eff}}^{1/2} W_{ [ c_{\gamma}, \infty ] }^t
  && \mbox{ with probability } 1 - \epsilon.
\end{eqnarray*}
Here, $W_1 \sim N(\mu_1, 1)$, $W_{[a,b]}^t$ denotes a random variable that is independent of $W_1$ and whose distribution is the conditional distribution of $W_2$ given $a \leq W_2^2 \leq b$ when $W_2$ is independent of $W_1$ and has marginal distribution $W_2 \sim N (\mu_2, 1)$, and\footnote{Note that $V_{\rm eff}$ is defined in G\&Y's Lemma A.3 on page 1516.}
\begin{eqnarray*}
  V_{\rm eff}
  & = &
  \frac{ V_A V_B - \Gammau^2 }{ V_A + V_B - 2 \Gammau }
  \\
  V_{\rm A-eff}
  & = &
  V_A - V_{\rm eff} =
  \frac{ (\Gammau - V_A)^2 }{ V_A + V_B - 2 \Gammau }
  \\
  V_{\rm B-eff}
  & = &
  V_B - V_{\rm eff} =
  \frac{ (\Gammau - V_B)^2 }{ V_A + V_B - 2 \Gammau }
  \\
  \epsilon
  & = &
  P( W_2^2 < c_{\gamma} )
  =
  F_1 (c_{\gamma}; \mu_2^2 / 2)
  =
  F_1 (c_{\gamma}; \Sigma_T^{-1} \eta^2 / 2).
\end{eqnarray*}

Again, $V_{\rm eff}^{1/2}$, $V_{\rm A-eff}^{1/2}$ and $V_{\rm B-eff}^{1/2}$ could denote positive or negative square roots, but G\&Y implicitly define them as
\begin{eqnarray*}
V_{\rm eff}^{1/2}
& = &
\frac{ \sqrt{V_A V_B - \Gammau^2} }{ \sqrt{V_A + V_B - 2 \Gammau} }
\\
  V_{\rm A-eff}^{1/2}
  & = &
  \frac{ \Gammau - V_A }{ \sqrt{ V_A + V_B - 2 \Gammau } }
  \\
  V_{\rm B-eff}^{1/2}
  & = &
  \frac{ \Gammau - V_B }{ \sqrt{ V_A + V_B - 2 \Gammau } }.
\end{eqnarray*}
Note that $V_{\rm eff}^{1/2}$ is positive, but $V_{\rm A-eff}^{1/2}$ and $V_{\rm B-eff}^{1/2}$ can be positive or negative, depending on whether $\Gammau$ is greater or less than $V_A$ and $V_B$.

The bias and MSE of $\GYestfix$ as functions of $c_{\gamma}$, $\lambda$ and $\eta$ are derived in G\&Y's Appendix A.5.
They are
\begin{eqnarray*}
  \mbox{bias}
  & = &
  \epsilon \; \mbox{bias}_{T \leq c_{\gamma}} + (1 - \epsilon) \; \mbox{bias}_{T > c_{\gamma}}
  \\
  \mbox{MSE}
  & = &
  \epsilon \; \mbox{MSE}_{T \leq c_{\gamma}} + (1 - \epsilon) \; \mbox{MSE}_{T > c_{\gamma}}
\end{eqnarray*}
where\footnote{Note that the formula for $\mbox{bias}_{T > c_{\gamma}}$ follows from the final line on page 1520.}
\begin{eqnarray*}
  \mbox{bias}_{T \leq c_{\gamma}}
  & = &
  - V_{\rm eff}^{1/2} \mu_1 + (\omega_A V_{\rm A-eff}^{1/2} - \omega_B V_{\rm B-eff}^{1/2})
  \frac{ F_3 (c_{\gamma}; \mu_2^2 / 2) }
       { F_1 (c_{\gamma}; \mu_2^2 / 2) }
       \; \mu_2
  \\
  \mbox{bias}_{T > c_{\gamma}}
  & = &
  - V_{\rm eff}^{1/2} \mu_1 + V_{\rm A-eff}^{1/2} \;
  \frac{ 1 - F_3 (c_{\gamma}; \mu_2^2 / 2) }
       { 1 - F_1 (c_{\gamma}; \mu_2^2 / 2) }
       \; \mu_2
  \\
  \mbox{MSE}_{T \leq c_{\gamma}}
  & = &
  V_{\rm eff} (\mu_1^2 + 1) + 
  \left\{
  \frac{ F_3 (c_{\gamma}; \mu_2^2 / 2) }
       { F_1 (c_{\gamma}; \mu_2^2 / 2) }
  +
  \frac{ F_5 (c_{\gamma}; \mu_2^2 / 2) }
       { F_1 (c_{\gamma}; \mu_2^2 / 2) } \mu_2^2
  \right\}
  (\omega_A V_{\rm A-eff}^{1/2} - \omega_B V_{\rm B-eff}^{1/2})^2
  \\
  && -
  2 \; \frac{ F_3 (c_{\gamma}; \mu_2^2 / 2) }
       { F_1 (c_{\gamma}; \mu_2^2 / 2) }
       V_{\rm eff}^{1/2} \mu_1 \mu_2
  (\omega_A V_{\rm A-eff}^{1/2} - \omega_B V_{\rm B-eff}^{1/2})
  \\     
  \mbox{MSE}_{T > c_{\gamma}}
  & = &
  V_{\rm eff} (\mu_1^2 + 1)
  \\ && +
  V_{\rm A-eff}
  \left\{
  \frac{ 1 - F_3 (c_{\gamma}; \mu_2^2 / 2) }
       { 1 - F_1 (c_{\gamma}; \mu_2^2 / 2) }
  +
  \frac{ 1 - F_5 (c_{\gamma}; \mu_2^2 / 2) }
       { 1 - F_1 (c_{\gamma}; \mu_2^2 / 2) } \mu_2^2
  \right\}
  \\
  && -
  2 \; \frac{ 1 - F_3 (c_{\gamma}; \mu_2^2 / 2) }
       { 1 - F_1 (c_{\gamma}; \mu_2^2 / 2) }
       V_{\rm eff}^{1/2} V_{\rm A-eff}^{1/2} \mu_1 \mu_2.
\end{eqnarray*}

Hence,
\begin{eqnarray}
  \mbox{bias}
  & = &
  - V_{\rm eff}^{1/2} \mu_1
  + (\omega_A V_{\rm A-eff}^{1/2} - \omega_B V_{\rm B-eff}^{1/2}) F_3 (c_{\gamma}; \mu_2^2 / 2) \mu_2
  \nonumber \\
  && +
  V_{\rm A - eff}^{1/2} \{ 1 - F_3 (c_{\gamma}; \mu_2^2 / 2) \} \mu_2
  \label{eq:bias.formula} \\
  \mbox{MSE}
  & = &
  V_{\rm eff} (\mu_1^2 + 1)
  +
  \left\{
  F_3 (c_{\gamma}; \mu_2^2 / 2)
  +
  F_5 (c_{\gamma}; \mu_2^2 / 2) \mu_2^2
  \right\}
  (\omega_A V_{\rm A-eff}^{1/2} - \omega_B V_{\rm B-eff}^{1/2})^2
  \nonumber \\
  && -
  2 \; F_3 (c_{\gamma}; \mu_2^2 / 2)
  V_{\rm eff}^{1/2} \mu_1 \mu_2
  (\omega_A V_{\rm A-eff}^{1/2} - \omega_B V_{\rm B-eff}^{1/2})
  \nonumber \\
  && +
  V_{\rm A-eff}
  \left[
  \{ 1 - F_3 (c_{\gamma}; \mu_2^2 / 2) \}
  +
  \{ 1 - F_5 (c_{\gamma}; \mu_2^2 / 2) \} \mu_2^2
  \right]
  \nonumber \\
  && -
  2 \; \{ 1 - F_3 (c_{\gamma}; \mu_2^2 / 2) \}
  V_{\rm eff}^{1/2} V_{\rm A-eff}^{1/2} \mu_1 \mu_2.
  \label{eq:MSE.formula}
\end{eqnarray}

We now turn to G\&Y's Appendix B.2, where they use the formulae above but plug in the appropriate forms of $V_{\rm eff}$, $V_{{\rm A-eff}}$ and other terms into equation~(\ref{eq:bias.formula}) and show that this equation simplifies dramatically.

Notice that
\begin{eqnarray*}
  \mu_2
  & = &
  \frac{ f_B^{-1/2} \eta }{ \sqrt{ V_A + V_B - 2 \Gammau } }
  \\
  \Sigma_S^{-1/2}
  & = &
  f_B^{-1/2} \frac{ \Gammau - V_A }{ \sqrt{ (V_A + V_B - 2 \Gammau) (V_A V_B - \Gammau^2) } }
  \\
  V_{\rm A - eff}^{1/2}
  & = &
  \frac{ \Gammau - V_A }{ \sqrt{ V_A + V_B - 2 \Gammau } }
  \\
  \omega_A V_{\rm A-eff}^{1/2} - \omega_B V_{\rm B-eff}^{1/2}
  & = &
  \frac{1}{1+\lambda} \frac{ (\Gammau - V_A) - \lambda (\Gammau - V_B) }{ \sqrt{ V_A + V_B - 2 \Gammau } }
  \\
  V_{\rm eff}^{1/2} \mu_1
  & = &
  f_B^{-1/2} \eta \; \frac{ \Gammau - V_A }{ V_A + V_B - 2 \Gammau }
  \\
  V_{\rm A - eff}^{1/2} \mu_2
  & = &
  f_B^{-1/2} \eta \; \frac{ \Gammau - V_A }{ V_A + V_B - 2 \Gammau }  
  \\
  (\omega_A V_{\rm A-eff}^{1/2} - \omega_B V_{\rm B-eff}^{1/2}) \mu_2
  & = &
  f_B^{-1/2} \eta \;
  \frac{ (\Gammau - V_A) - \lambda (\Gammau - V_B) }{ (1 + \lambda) (V_A + V_B - 2 \Gammau) }.
\end{eqnarray*}

Plugging these expressions into equation~(\ref{eq:bias.formula}), we get
\begin{eqnarray*}
  \mbox{bias}
  & = &
  - f_B^{-1/2} \eta \; \frac{ \Gammau - V_A }{ V_A + V_B - 2 \Gammau }
  + f_B^{-1/2} \eta \;
  \frac{ (\Gammau - V_A) - \lambda (\Gammau - V_B) }{ (1 + \lambda) (V_A + V_B - 2 \Gammau) }
  F_3 (c_{\gamma}; \mu_2^2 / 2)
  \\
  && +
  f_B^{-1/2} \eta \; \frac{ \Gammau - V_A }{ V_A + V_B - 2 \Gammau } \{ 1 - F_3 (c_{\gamma}; \mu_2^2 / 2) \}
  \\
  & = &
  f_B^{-1/2} \eta \; \frac{ 1 }{ V_A + V_B - 2 \Gammau }
  \bigg[
    - (\Gammau - V_A)
    +
  \frac{ (\Gammau - V_A) - \lambda (\Gammau - V_B) }{ (1 + \lambda) }
  F_3 (c_{\gamma}; \mu_2^2 / 2)
  \\
  && +
  (\Gammau - V_A) \{ 1 - F_3 (c_{\gamma}; \mu_2^2 / 2) \}
\bigg]    
  \\
  & = &
  - f_B^{-1/2} \eta \; \frac{ 1 }{ V_A + V_B - 2 \Gammau } \;
  \frac{ \lambda }{ 1 + \lambda }
  F_3 (c_{\gamma}; \mu_2^2 / 2)
  \{ (\Gammau - V_B) + (\Gammau - V_A) \}
  \\
  & = &
  f_B^{-1/2} \eta \; 
  \frac{ \lambda }{ 1 + \lambda } \;
  F_3 (c_{\gamma}; \mu_2^2 / 2).
\end{eqnarray*}

This is the same formula for the bias as that which G\&Y give at the top of page 1538, except that their expression contains an extra multiplier that equals one.

The formula for the MSE that G\&Y give on the line immediately before their expression (B.13) is exactly what is obtained if we take expression~(\ref{eq:MSE.formula}) above and plug in the expressions given above for $V_{\rm eff}$, $V_{\rm eff}^{1/2}$, $V_{\rm A-eff}$, $V_{\rm A-eff}^{1/2}$ and $(\omega_A V_{\rm A-eff}^{1/2} - \omega_B V_{\rm B-eff}^{1/2})$.

\subsubsection{Inference}

\label{eq:GYinference}

Recall that $W_1$ and $W_2$ are independent, both with variance 1, and $E(W_j) = \mu_j$ $(j=1, 2)$, and that $\omega_A = 1 / (1 + \lambda)$ and $\omega_B = \lambda / (1 + \lambda)$.
Noting that $F_1 (c_{\gamma} ; \mu_2^2 / 2) = P(W_2^2 \leq \mu_2^2)$, G\&Y's Theorem 4.1 states that $\sqrt{n} (\GYmuhat - \Ybar)$ has the same distribution as the following random variable:
\[
- V_{\rm eff}^{1/2} W_1 + (\omega_A V_{\rm A-eff}^{1/2} - \omega_B V_{\rm B-eff}^{1/2}) W_2 I(W_2^2 \leq \mu_2^2) + V_{\rm A-eff}^{1/2} W_2 I(W_2^2 > \mu_2^2).
\]
This random variable can be rewritten as
\begin{eqnarray*}
  &&
  - V_{\rm eff}^{1/2} W_1 + (\omega_A V_{\rm A-eff}^{1/2} - \omega_B V_{\rm B-eff}^{1/2}) W_2
  - (\omega_A V_{\rm A-eff}^{1/2} - \omega_B V_{\rm B-eff}^{1/2}) W_2 I(W_2^2 > \mu_2^2)
  \\
  && \hspace{.8cm} +
   V_{\rm A-eff}^{1/2} W_2 I(W_2^2 > \mu_2^2)
  \\
  && \hspace{.4cm} =  
  - V_{\rm eff}^{1/2} W_1 + (\omega_A V_{\rm A-eff}^{1/2} - \omega_B V_{\rm B-eff}^{1/2}) W_2
  + (V_{\rm A-eff}^{1/2} -\omega_A V_{\rm A-eff}^{1/2} + \omega_B V_{\rm B-eff}^{1/2}) W_2 I(W_2^2 > \mu_2^2)
  \\
  && \hspace{.4cm} =  
  - V_{\rm eff}^{1/2} W_1 + (\omega_A V_{\rm A-eff}^{1/2} - \omega_B V_{\rm B-eff}^{1/2}) W_2
  + \omega_B (V_{\rm A-eff}^{1/2} + V_{\rm B-eff}^{1/2}) W_2 I(W_2^2 > \mu_2^2)
  \\
  && \hspace{.4cm} =  
  - V_{\rm eff}^{1/2} W_1 + (\omega_A V_{\rm A-eff}^{1/2} - \omega_B V_{\rm B-eff}^{1/2}) W_2
  + \omega_B (V_{\rm A-eff}^{1/2} + V_{\rm B-eff}^{1/2}) \mu_2 1_{\mu_2^2 > c_{\gamma}}
  \\
  && \hspace{.8cm} +
\omega_B (V_{\rm A-eff}^{1/2} + V_{\rm B-eff}^{1/2}) \{ W_2 I(W_2^2 > \mu_2^2) - \mu_2 1_{\mu_2^2 > c_{\gamma}} \}
  \\
  && \hspace{.4cm} =
  R_n + \omega_B (V_{\rm A-eff}^{1/2} + V_{\rm B-eff}^{1/2}) U_n
\end{eqnarray*}
where
\begin{eqnarray*}
  R_n & = &
  - V_{\rm eff}^{1/2} W_1 + (\omega_A V_{\rm A-eff}^{1/2} - \omega_B V_{\rm B-eff}^{1/2}) W_2 + \omega_B (V_{\rm A-eff}^{1/2} + V_{\rm B-eff}^{1/2}) \mu_2 1_{\mu_2^2 > c_{\gamma}}
  \\
  U_n & = &
  W_2 \; I(W_2^2 > c_{\gamma}) - \mu_2 1_{\mu_2^2 > c_{\gamma}}.
\end{eqnarray*}
This is what G\&Y's equation (4.1) says.

Recall from Appendix~\ref{sect:proof.Thm4.1} that
\begin{eqnarray}
  W_1
  & = &
  f_B^{1/2} \Sigma_S^{-1/2} \left( \frac{ \Gammau - V_B }{ \Gammau - V_A } Z_1 + Z_2 \right)
  \nonumber \\
  & = &
  \frac{ (\Gammau - V_B) Z_1 + (\Gammau - V_A) Z_2 }{ \sqrt{ (V_A + V_B - 2 \Gammau) (V_A V_B - \Gammau^2) } }
  \nonumber \\
  W_2
  & = &
  - \frac{ Z_1 - Z_2 }{ \sqrt{V_A + V_B - 2 \Gammau} }.
  \nonumber
\end{eqnarray}
Note that
\begin{eqnarray}
  V_{\rm eff}^{1/2} W_1
  & = &
  \frac{ (\Gammau - V_B) Z_1 + (\Gammau - V_A) Z_2 }{ V_A + V_B - 2 \Gammau }
  \nonumber \\
  V_{\rm A-eff}^{1/2} + V_{\rm B-eff}^{1/2}
  & = &
  - \sqrt{ V_A + V_B - 2 \Gammau }
  \nonumber \\
  \mu_2
  & = &
  \frac{ f_B^{-1/2} }{ \sqrt{V_A + V_B - 2 \Gammau} } \; \eta
  \nonumber \\
  (V_{\rm A-eff}^{1/2} + V_{\rm B-eff}^{1/2}) W_2
  & = &
  Z_1 - Z_2
  \nonumber \\
  (V_{\rm A-eff}^{1/2} + V_{\rm B-eff}^{1/2}) \mu_2
  & = &
  - f_B^{-1/2} \eta
  \nonumber
\end{eqnarray}

Thus, we can write $R_n$ and $U_n$ as
\begin{eqnarray*}
  R_n & = &
  \frac{ (V_B - \Gammau) Z_1 + (V_A - \Gammau) Z_2 }{ V_A + V_B - 2 \Gammau }
  + \frac{ \omega_B (\Gammau - V_B) - \omega_A (\Gammau - V_A) }{ V_A + V_B - 2 \Gammau } (Z_1 - Z_2)
\\
&& - 
\omega_B f_B^{-1/2} \eta \; 1_{\mu_2^2 > c_{\gamma}}
\\
& = &
\omega_A Z_1 + \omega_B
Z_2
  - \omega_B f_B^{-1/2} \eta \; 1_{\mu_2^2 > c_{\gamma}}
\\
  U_n & = &
  - \frac{ Z_1 - Z_2 }{ \sqrt{V_A + V_B - 2 \Gammau } } \; I \{ (Z_1 - Z_2)^2 > (V_A + V_B - 2 \Gammau ) c_{\gamma} \}
\\
&& -
  \frac{ f_B^{-1/2} }{ \sqrt{V_A + V_B - 2 \Gammau} } \; \eta \; 1_{\mu_2^2 > c_{\gamma}}
\end{eqnarray*}
and write $1_{\mu_2^2 > c_{\gamma}}$ as
\[
1_{\mu_2^2 > c_{\gamma}}
= 1_{ [ \eta^2 > f_B (V_A + V_B - 2 \Gammau) c_{\gamma} ] }.
\]

G\&Y define
\begin{eqnarray*}
  U_{v_n} & = &
  R_n + \omega_B (V_{\rm A-eff}^{1/2} + V_{\rm B-eff}^{1/2}) \; \{ W_2 \; I(W_2^2 > c_{\gamma}) - \mu_2 1_{\mu_2^2 > c_{\gamma}} \} \times 1_{\GYteststat \geq v_n}
  \nonumber \\
  && +
  \mbox{sup}_{\mu_2 \in \mathbb{R}} \left[ \omega_B (V_{\rm A-eff}^{1/2} + V_{\rm B-eff}^{1/2}) \; \{ W_2 \; I(W_2^2 > c_{\gamma}) - \mu_2 1_{\mu_2^2 > c_{\gamma}} \} \right] \times 1_{\GYteststat < v_n}.
\end{eqnarray*}
Recall that
\[
\GYteststat = \frac{1}{ V_A + V_B - 2 \Gammau } (\muhatB - \muhatA)^2.
\]
G\&Y define $L_{\rm v_n}$ as the same as $U_{\rm v_n}$ but with the supremum replaced by an infinum.

Note that we can write $U_{v_n}$ as
\begin{eqnarray*}
  U_{v_n}
  & = &
  R_n + \left\{ \omega_B (Z_1 - Z_2) \; I(W_2^2 > c_{\gamma})
  + \omega_B f_B^{-1/2} \eta \; 1_{\mu_2^2 > c_{\gamma}} \right\} \times 1_{\GYteststat \geq v_n}
  \nonumber \\
  && + \;
  \mbox{sup}_{\mu_2 \in \mathbb{R}} \left\{ \omega_B (Z_1 - Z_2) \; I(W_2^2 > c_{\gamma})
  + \omega_B f_B^{-1/2} \eta \; 1_{\mu_2^2 > c_{\gamma}} \right\} \times 1_{\GYteststat < v_n}
  \\
  & = &
  \omega_A Z_1 + \omega_B Z_2
  - \omega_B f_B^{-1/2} \eta \; 1_{\mu_2^2 > c_{\gamma}}
  \nonumber \\
  && +
  \left\{ \omega_B (Z_1 - Z_2) \; I(W_2^2 > c_{\gamma})
  + \omega_B f_B^{-1/2} \eta \; 1_{\mu_2^2 > c_{\gamma}} \right\} \times 1_{\GYteststat \geq v_n}
  \nonumber \\
  && + \;
  \mbox{sup}_{\mu_2 \in \mathbb{R}} \left\{ \omega_B (Z_1 - Z_2) \; I(W_2^2 > c_{\gamma})
  + \omega_B f_B^{-1/2} \eta \; 1_{\mu_2^2 > c_{\gamma}} \right\} \times 1_{\GYteststat < v_n}  \\
  & = &
  \omega_A Z_1 + \omega_B Z_2
  - \omega_B f_B^{-1/2} \eta \; 1_{ [ \eta^2 > f_B (V_A + V_B - 2 \Gammau) c_{\gamma} ] }
  \nonumber \\
  && +
  \bigg[ \omega_B (Z_1 - Z_2) \; I \{ (Z_1 - Z_2)^2 > (V_A + V_B - 2 \Gammau) c_{\gamma} \}
  \nonumber \\
  && \hspace{0.7cm} +
  \omega_B f_B^{-1/2} \eta \; 1_{ [ \eta^2 > f_B (V_A + V_B - 2 \Gammau) c_{\gamma} ] } \bigg] \times 1_{\GYteststat \geq v_n}
  \nonumber \\
  && + \;
  \mbox{sup}_{\eta \in \mathbb{R}} \bigg[ \omega_B (Z_1 - Z_2) \; I \{ (Z_1 - Z_2)^2 > (V_A + V_B - 2 \Gammau) c_{\gamma} \}
  \nonumber \\
  && \hspace{1.9cm} +
  \omega_B f_B^{-1/2} \eta \; 1_{ [ \eta^2 > f_B (V_A + V_B - 2 \Gammau) c_{\gamma} ] } \bigg] \times 1_{\GYteststat < v_n} 
  \\
  & = &
  \omega_A Z_1 + \omega_B Z_2
  - \omega_B f_B^{-1/2} \eta \; 1_{ [ \eta^2 > f_B (V_A + V_B - 2 \Gammau) c_{\gamma} ] }
  \nonumber \\
  && +
  \bigg[ \omega_B (Z_1 - Z_2) \; I \{ (Z_1 - Z_2)^2 > (V_A + V_B - 2 \Gammau) c_{\gamma} \}
  \nonumber \\
  && \hspace{0.7cm} +
  \omega_B f_B^{-1/2} \eta \; 1_{ [ \eta^2 > f_B (V_A + V_B - 2 \Gammau) c_{\gamma} ] } \bigg] \times 1_{\GYteststat \geq v_n}
  \nonumber \\
  && +
  \infty \times 1_{\GYteststat < v_n} 
 \\
  & = &
\left\{ \begin{array}{ll}
  Z_1 & \mbox{ if } \GYteststat \geq v_n \mbox{ and } (Z_1 - Z_2)^2 > (V_A + V_B - 2 \Gammau) c_{\gamma} \\
  \omega_A Z_1 + \omega_B Z_2 & \mbox{ if } \GYteststat \geq v_n \mbox{ and } (Z_1 - Z_2)^2 \leq (V_A + V_B - 2 \Gammau) c_{\gamma} \\
  \infty  & \mbox{ if } \GYteststat < v_n
  \end{array} \right.
\end{eqnarray*}
Similarly,
\begin{eqnarray*}
  L_{v_n}
  & = &
  \omega_A Z_1 + \omega_B Z_2
  - \omega_B f_B^{-1/2} \eta \; 1_{ [ \eta^2 > f_B (V_A + V_B - 2 \Gammau) c_{\gamma} ] }
  \nonumber \\
  && +
  \bigg[ \omega_B (Z_1 - Z_2) \; I \{ (Z_1 - Z_2)^2 > (V_A + V_B - 2 \Gammau) c_{\gamma} \}
  \nonumber \\
  && \hspace{0.7cm} +
  \omega_B f_B^{-1/2} \eta \; 1_{ [ \eta^2 > f_B (V_A + V_B - 2 \Gammau) c_{\gamma} ] } \bigg] \times 1_{\GYteststat \geq v_n}
  \nonumber \\
  && + \;
  \mbox{inf}_{\eta \in \mathbb{R}} \bigg[ \omega_B (Z_1 - Z_2) \; I \{ (Z_1 - Z_2)^2 > (V_A + V_B - 2 \Gammau) c_{\gamma} \}
  \nonumber \\
  && \hspace{1.9cm} +
  \omega_B f_B^{-1/2} \eta \; 1_{ [ \eta^2 > f_B (V_A + V_B - 2 \Gammau) c_{\gamma} ] } \bigg] \times 1_{\GYteststat < v_n}.
  \\
  & = &
  \omega_A Z_1 + \omega_B Z_2
  - \omega_B f_B^{-1/2} \eta \; 1_{ [ \eta^2 > f_B (V_A + V_B - 2 \Gammau) c_{\gamma} ] }
  \nonumber \\
  && +
  \bigg[ \omega_B (Z_1 - Z_2) \; I \{ (Z_1 - Z_2)^2 > (V_A + V_B - 2 \Gammau) c_{\gamma} \}
  \nonumber \\
  && \hspace{0.7cm} +
  \omega_B f_B^{-1/2} \eta \; 1_{ [ \eta^2 > f_B (V_A + V_B - 2 \Gammau) c_{\gamma} ] } \bigg] \times 1_{\GYteststat \geq v_n}
  \nonumber \\
  && -
 \infty \times 1_{\GYteststat < v_n}.
 \\
 & = &
\left\{ \begin{array}{ll}
  Z_1 & \mbox{ if } \GYteststat \geq v_n \mbox{ and } (Z_1 - Z_2)^2 > (V_A + V_B - 2 \Gammau) c_{\gamma} \\
  \omega_A Z_1 + \omega_B Z_2 & \mbox{ if } \GYteststat \geq v_n \mbox{ and } (Z_1 - Z_2)^2 \leq (V_A + V_B - 2 \Gammau) c_{\gamma} \\
  - \infty  & \mbox{ if } \GYteststat < v_n
  \end{array} \right.
\end{eqnarray*}

To construct a $(1 - \alpha) 100$\% CI using G\&Y's equation (4.4), it is necessary to find the $(1 - \alpha/2)$th quantile of $n^{-1/2} U_{v_n}$ and the $(\alpha / 2)$th quantile of $n^{-1/2} L_{v_n}$.
In order to do that, G\&Y use non-parametric bootstrap.
Recall that $n^{-1/2} Z_1$ and $n^{-1/2} Z_2$ represent $\muhat_A - \Ybar$ and $\muhat_A - \Ybar$, respectively (as explained in Appendix~\ref{sect:proof.Thm4.1}).
So, having obtained a bootstrap sample and calculated values $\muhat_A^{*(b)}$, $\muhatB^{*(b)}$ and $\GYteststat^{*(b)}$ of $\muhat_A$, $\muhatB$ and $\GYteststat$, the corresponding values $n^{-1/2} U_{v_n}^{*(b)}$ and $n^{-1/2} L_{v_n}^{*(b)}$ of $U_{v_n}$ and $L_{v_n}$ are
\begin{eqnarray*}
  &&
  n^{-1/2} U_{v_n}^{*(b)}
  \\
  && \hspace{.4cm} =
  \left\{ \begin{array}{ll}
  \muhatA^{*(b)} - \muhatA & \mbox{ if } \GYteststatstar \geq v_n \mbox{ and } (\muhatB^{*(b)} - \muhatA^{*(b)})^2 > (V_A + V_B - 2 \Gammau) c_{\gamma} \\
  \omega_A \muhatA^{*(b)} + \omega_B \muhatB^{*(b)} - \muhatA & \mbox{ if } \GYteststatstar \geq v_n \mbox{ and } (\muhatB^{*(b)} - \muhatA^{*(b)})^2 \leq (V_A + V_B - 2 \Gammau) c_{\gamma} \\
  \infty  & \mbox{ if } \GYteststatstar < v_n
  \end{array} \right.
  \\
  && \hspace{.4cm} =
  \left\{ \begin{array}{ll}
  \muhatA^{*(b)} - \muhatA & \mbox{ if } \GYteststatstar \geq v_n \mbox{ and } \GYteststatstar > c_{\gamma} \\
  \omega_A \muhatA^{*(b)} + \omega_B \muhatB^{*(b)} - \muhatA & \mbox{ if } \GYteststatstar \geq v_n \mbox{ and } \GYteststatstar \leq c_{\gamma} \\
  \infty  & \mbox{ if } \GYteststatstar < v_n
  \end{array} \right.
\end{eqnarray*}
and
\begin{equation*}
  n^{-1/2} L_{v_n}^{*(b)} =
  \left\{ \begin{array}{ll}
  n^{-1/2} U_{v_n}^{*(b)} & \mbox{ if } \GYteststatstar \geq v_n \\
  - \infty  & \mbox{ if } \GYteststatstar < v_n
  \end{array} \right.
\end{equation*}
where $\muhat_A$ is the estimate of $\mu_A$ from the original dataset.

\section{Unit non-response and post-stratification}

\label{sect:nonresponse.poststrat}

In this section, we extend the estimators described in the main article to allow for adjustment for unit non-response in Sample A.
This is done by viewing the selection of Sample A as the first phase of a two-phase design and the response of the selected individuals as the second phase.
We also consider post-stratification of Sample A in the scenario where response is complete, which is a special case of this extension.
For simplicity, we assume the aim is to estimate the mean of $Y$ in the entire population, i.e.\ $S_i=1$ for all $i$.

\subsection{Response homogeneity model}

Let $R^{A1}_i$ be an indicator that individual $i$ is selected for inclusion in Sample A.
Let $\pi^{A1}_i = P(R^{A1}_i = 1 \mid \curlF, \curlY)$ denote the probability that individual $i$ is selected for inclusion, and let $\pi^{A1}_{ij} = P(R^{A1}_i = R^{A1}_j = 1 \mid \curlF, \curlY)$ denote the probability that the pair of individuals $i$ and $j$ is selected for inclusion.
Let $R^A_i$ be an indicator that individual $i$ participates in Sample A.
Individuals with $R^{A1} = R^A = 1$ are `responders'; those with $R^{A1} = 1$ and $R^A = 0$ are `non-responders'; and $R^{A1} = 0$ implies that $R^A=0$.
We observe $X_i$ and $Y_i$ if and only if $R^A_i = 1$.

Suppose that the population is divided into $L$ groups.
Let $H_i$ be a categorical variable denoting the group to which individual $i$ belongs ($1 \leq H_i \leq L$), and assume that $H_i$ is a deterministic function of $X_i$.
Let $\pi^{A2}_h$ denote the probability of response in group $h$ ($h=1, \ldots, L$), which is assumed to be the same for all individuals in group $h$.
Assume $R^A_1, \ldots, R^A_N$ are conditionally independent given $(R^{A1}_1, \ldots, R^{A1}_N, \curlF,\curlY)$ with $P(R^A_i=1 \mid R^{A1}_1, \ldots, , R^{A1}_{i-1}, R^{A1}_i = 1, R^{A1}_{i+1}, \ldots, R^{A1}_N, \curlF, \curlY) = \pi^{A2}_{H_i}$.
This response model is called by Sarndal et al.\ (1992)\cite{sarndal1992surveysamplingbook} the Response Homogeneity (RHG) Model.

Let $N_h = \sum_{i=1}^N I(H_i=h)$, $n_h = \sum_{i=1}^N R^{A1}_i I(H_i=h)$ and $m_h = \sum_{i=1}^N R^A_i I(H_i=h)$ denote the numbers of individuals in group $h$ in the population, selected Sample A, and responding Sample A.
Like Sarndal et al., we assume that the probability that $m_h \leq 1$ is negligibly small.

In Appendix~\ref{sect:nonresponse} we study the estimator of $\Ybar$ that uses the ratio $m_h / n_h$ as an estimate of the response probability $\pi^{A2}_h$ and accounts for non-response by multiplying the weight $1 / \pi^{A1}_i$ of each responder $i$ by $1 / \pi^{A2}_{H_i}$.
This estimator of $\Ybar$ requires $n_h$ to be known (but does not need $N_h$ to be known).

In Appendix~\ref{sect:calibration} we instead consider the post-stratification estimator of $\Ybar$.
This is the estimator that uses weights that have been calibrated so that the weighted proportion of the responders in group $h$ equals the proportion of the population in group $h$.
It does this by multiplying the selection weight $1 / \pi^{A1}_i$ of each responder $i$ by a factor that is the same for all the responders in the same group.
The post-stratification estimator requires $N_h$ to be known (but does not need $n_h$ to be known).
Even when there is complete response (i.e.\ $\pi^{A2}_h = 1$ for all $h$), and hence no adjustment for non-response is necessary, calibrating the weights in this way can reduce the variance of the estimator of $\Ybar$.
Indeed, the Hajek estimator can be seen as the special case of this post-stratification estimator with $L=1$ (in this special case, $N$ does not need to be known).

In both Sections~\ref{sect:nonresponse} and~\ref{sect:calibration}, we begin by considering estimators that use only data on Sample A, drawing heavily on results described in Sections 9.7 and 15.6.2--15.6.4 of Sarndal et al.\ (1992) for two-phase studies and non-response.
We then go on to consider the double robust estimators $\PLone$, $\PLtwo$, $\KHone$ and $\KHtwo$ and the combined estimators $\HTPL$, $\HTKH$, $\HajPL$ and $\HajKH$, all of which make use of data from both Samples A and B.

\subsection{Adjusting for non-response using $n_h$}

\label{sect:nonresponse}

In this subsection, we assume $n_h$ ($h=1, \ldots, L$) is known and we define
\begin{eqnarray}
  \pi^A_i & = & \pi^{A1}_i \; \frac{ m_{H_i} }{ n_{H_i} }
  \label{eq:piAi.nonresponse}
    \\
  \pi^A_{ij} & = & \pi^{A1}_{ij} \; \frac{ m_{H_i} ( m_{H_i} - 1 ) }{ n_{H_i} ( n_{H_i} - 1 ) } \hspace{.8cm} \mbox{if $i \neq j$ and $H_i = H_j$}
  \label{eq:piAij.nonresponse1}
     \\
  \pi^A_{ij} & = & \pi^{A1}_{ij} \; \frac{ m_{H_i} m_{H_j} }{ n_{H_i} n_{H_j} } 
   \hspace{1.8cm} \mbox{if $i \neq j$ and $H_i \neq H_j$}
  \label{eq:piAij.nonresponse2}
  \\
  \pi^A_{ii} & = & \pi^{A1}_i \; \frac{ m_{H_i} }{ n_{H_i} }.
  \label{eq:piAii.nonresponse}
\end{eqnarray}

Define the following estimator of $\Ybar$:
\[
\hat{\mu}_{{\rm nr},Y} =
\frac{1}{N} \sum_{i=1}^N \frac{ R^A_i }{ \pi^A_i } Y_i.
\]

The variance of $\hat{\mu}_{{\rm nr},Y}$ is given by Sarndal et al.'s equation~(15.6.9), which is
\begin{eqnarray}
&&
  \Var ( \hat{\mu}_{{\rm nr},Y} \mid \curlF, \curlY )
  \nonumber \\
&& \hspace{.4cm} =
\frac{1}{N^2} \sum_{i=1}^N \sum_{j=1}^N (\pi^{A1}_{ij} - \pi^{A1}_i \pi^{A1}_j) \;
\frac{ Y_i }{ \pi^{A1}_i } \;
\frac{ Y_j }{ \pi^{A1}_j }
  \nonumber \\
  && \hspace{.8cm} +
  \frac{ 1 }{ N^2 }
  E \left\{ E \left( \sum_{h=1}^L n_h^2 \; \frac{ 1 - m_h / n_h }{ m_h} \; S_{A1,h}^2 \; \mid R^{A1}_1, \ldots, R^{A1}_N, \curlF, \curlY \right) \mid \curlF, \curlY \right\}
  \nonumber \\
  &&
  \label{eq:nonresponse.var}
\end{eqnarray}
where $S_{A1,h}^2$ denotes the variance of $Y_i / \pi^{A1}_i$ in the set of individuals with $R^{A1}_i=1$ in group $h$.
The variance~(\ref{eq:nonresponse.var}) can be estimated using Sarndal et al.'s expression~(15.6.10), i.e.\
\begin{equation}
  \frac{1}{N^2} \sum_{i=1}^N \sum_{j=1}^N R^A_i R^A_j \;
  \frac{ \pi^{A1}_{ij} - \pi^{A1}_i \pi^{A1}_j }{ \pi^A_{ij} } \;
\frac{ Y_i }{ \pi^{A1}_i }
\frac{ Y_j }{ \pi^{A1}_j }
+ \frac{1}{N^2} \sum_{h=1}^L n_h^2 \; \frac{ 1 - m_h / n_h }{ m_h} \; S_{A,h}^2
\label{eq:nonresponse.varest}
\end{equation}
where $S_{A,h}^2$ denotes the variance of $Y_i / \pi^{A1}_i$ in the set of individuals with $R^A_i=1$ in group $h$:
\[
S_{A,h}^2 =
\frac{ \sum_{i=1}^N I(H_i = h) R^A_i \left(
  \frac{ Y_i }{ \pi^{A1}_i } - \hat{\bar{Y}}_{A,h} \right)^2
}
{
  \sum_{i=1}^N I(H_i = h) R^A_i
}
\]
with
\begin{equation}
  \hat{\bar{Y}}_{A,h} =
  \frac{ \sum_{i=1}^N I(H_i=h) \frac{ R^A_i }{ \pi^{A1}_i } Y_i }
       { \sum_{i=1}^N I(H_i=h) \frac{ R^A_i }{ \pi^{A1}_i } }
  = \frac{ \sum_{i=1}^N I(H_i=h) \frac{ R^A_i }{ \pi^A_i } Y_i }
  { \sum_{i=1}^N I(H_i=h) \frac{ R^A_i }{ \pi^A_i } }.
  \label{eq:Yhatbar.nonresponse}
\end{equation}

Note that when response is complete (i.e.\ $\pi^{A2}_h = 1$ for $h=1, \ldots, L$), expressions~(\ref{eq:nonresponse.var}) and~(\ref{eq:nonresponse.varest}) reduce to the usual variance of a Horvitz-Thompson estimator and an estimator of this variance, since in that case $m_h = n_h$ for all $h$.

Now consider the double robust estimators $\PLone$, $\PLtwo$, $\KHone$ and $\KHtwo$ and the combined estimators $\HTPL$, $\HTKH$, $\HajPL$ and $\HajKH$.
In Appendix~\ref{sect:proof.expect.rA.piA} we show that $E(R^A_i / \pi^A_i \mid \curlF, \curlY) = 1$.
Having established this, it is straightforward to verify that the proofs in Appendix Sections~\ref{sect:var.PL1}--\ref{sect:modeldesign.KH2} still apply.
However, the variances of the form
\[
\Var \left( \frac{1}{N} \sum_{i=1}^N \frac{ R^A_i }{ \pi^A_i } Z_i \mid \curlF, \curlY \right)
\]
for some variable $Z$ (e.g.\ $Z=m(X; \beta_0)$) that appear in Appendices~\ref{sect:var.PL1}--\ref{sect:modeldesign.KH2} are no longer variances of simple Horvitz-Thompson estimators of the population mean of $Z$.
Instead, they are given by equation~(\ref{eq:nonresponse.varest}) with $Y$ replaced by $Z$.
Similarly, covariances of the form
\[
\Cov \left( \frac{1}{N} \sum_{i=1}^N \frac{ R^A_i }{ \pi^A_i } Z_i, \;
\frac{1}{N} \sum_{i=1}^N \frac{ R^A_i }{ \pi^A_i } V_i  \mid \curlF, \curlY \right)
\]
for some variables $Z$ and $V$ (e.g.\ $Z=m(X; \beta_0)$ and $V=Y - \Ybar$) are given by
\begin{eqnarray*}
&&
\frac{1}{N^2} \sum_{i=1}^N \sum_{j=1}^N (\pi^{A1}_{ij} - \pi^{A1}_i \pi^{A1}_j) \;
\frac{ Z_i }{ \pi^{A1}_i } \;
\frac{ V_j }{ \pi^{A1}_j }
  \nonumber \\
  && \hspace{.4cm} +
  \frac{ 1 }{ N^2 }
  E \left\{ E \left( \sum_{h=1}^L n_h^2 \; \frac{ 1 - m_h / n_h }{ m_h} \; C_{A1,h} \; \mid R^{A1}_1, \ldots, R^{A1}_N, \curlF, \curlY \right) \mid \curlF, \curlY \right\}
  \nonumber
\end{eqnarray*}
where $C_{A1,h}$ denotes the covariance between $Z_i / \pi^{A1}_i$ and $V_i / \pi^{A1}_i$ in the set of individuals with $R^{A1}_i=1$ in group $h$.
This can be estimated by
\begin{equation}
  \frac{1}{N^2} \sum_{i=1}^N \sum_{j=1}^N R^A_i R^A_j \;
  \frac{ \pi^{A1}_{ij} - \pi^{A1}_i \pi^{A1}_j }{ \pi^A_{ij} } \;
\frac{ Z_i }{ \pi^{A1}_i }
\frac{ V_j }{ \pi^{A1}_j }
+ \frac{1}{N^2} \sum_{h=1}^L n_h^2 \; \frac{ 1 - m_h / n_h }{ m_h} \; C_{A,h}
\label{eq:nonresponse.covest}
\end{equation}
where $C_{A,h}$ denotes the covariance between $Z_i / \pi^{A1}_i$ and $V_i / \pi^{A1}_i$ in the set of individuals with $R^A_i=1$ in group $h$:
\[
C_{A,h} =
\frac{ \sum_{i=1}^N I(H_i = h) R^A_i \left(
  \frac{ Z_i }{ \pi^{A1}_i } - \hat{\bar{Z}}_{A,h} \right)   \left(
  \frac{ V_i }{ \pi^{A1}_i } - \hat{\bar{V}}_{A,h} \right)
}
{
  \sum_{i=1}^N I(H_i = h) R^A_i
}.
\]

In conclusion, the double robust estimators $\PLone$, $\PLtwo$, $\KHone$ and $\KHtwo$ and the combined estimators $\HTPL$, $\HTKH$, $\HajPL$ and $\HajKH$ can be used when there is non-response, provided that: i) the RHG model holds; ii) $\pi^A_i$ and $\pi^A_{ij}$ are defined by equations~(\ref{eq:piAi.nonresponse})--(\ref{eq:piAii.nonresponse}); and iii) expressions~(\ref{eq:nonresponse.varest}) and~(\ref{eq:nonresponse.covest}) are used for estimating variances and covariances of the population mean estimators that use only Sample A (with $Y$, $Z$ and $V$ replaced by the variable(s) whose variance or covariance is being estimated).
For example, the term 
\[
\left( \frac{N}{N_s} \right)^2
\Var \left[ \frac{1}{N} \sum_{i=1}^N \frac{ R^A_i }{ \pi^A_i } \;
  \{ S_i m(X_i; \betalim) - \Delta_{mi} \} \mid \curlF, \curlY \right]
\]
in equation~(\ref{eq:var.generalform}) of Section~\ref{sect:designbased}
and the term
\[
  \left( \frac{N}{N_s} \right)^2
  \Cov \left[
    \frac{1}{N} \sum_{i=1}^N
    \frac{ R^A_i }{ \pi^A_i } \{ S_i m (X_i; \betalim) - \Delta_{mi} \}, \;
    \frac{1}{N} \sum_{j=1}^N
    \frac{ R^A_j }{ \pi^A_j } S_i (Y_j - \Gamma)
    \mid \curlF, \curlY \right]
\]
in equation~(\ref{eq:cov.generalform}) of Section~\ref{sect:combining} would be estimated using expressions~(\ref{eq:nonresponse.varest}) and~(\ref{eq:nonresponse.covest}) with $Y_i$, $Z_i$ and $V_i$ replaced by $m (X_i; \betalim) - \Delta_{mi}$, $m (X_i; \betalim) - \Delta_{mi}$ and $Y_j - \Gamma$ respectively.

\subsection{Post-stratification}

\label{sect:calibration}

In this section, we assume $N_h$ ($h=1,\ldots,H$) are known, define the calibrated weight
\begin{equation}
  \pi^A_i = \pi^{A1}_i \times \frac{ 1 }{ N_h } \sum_{j=1}^N I(H_j=h) \frac{ R^A_j }{ \pi^{A1}_j }
  \label{eq:piAi.calibration}
\end{equation}
and continue to define $\pi^A_{ij}$ by equations~(\ref{eq:piAij.nonresponse1})--(\ref{eq:piAii.nonresponse}).
Note that it follows from equation~(\ref{eq:piAi.calibration}) that
\[
\sum_{i=1}^N I(H_i=h) \frac{ R^A_i }{ \pi^A_i }
= \sum_{i=1}^N I(H_i=h) \frac{ R^A_i }{ \pi^{A1}_i } \times \left\{ \frac{ 1 }{ N_h } \sum_{j=1}^N I(H_j=h) \frac{ R^A_j }{ \pi^{A1}_j } \right\}^{-1}
= N_h.
\]
So, the calibrated weights $1 / \pi^A_i$ ensure that the weighted proportion of the responders in group $h$ equals the proportion of the population in group $h$.

Define the following estimator of $\Ybar$:
\[
\hat{\mu}_{{\rm reg},Y} = \sum_{i=1}^N \frac{ R^A_i }{ \pi^A_i } Y_i
\]

and let
\begin{eqnarray}
  \bar{Y}_h
  & = &
  N_h^{-1} \sum_{i=1}^N I(H_i=h) Y_i
  \nonumber \\
  \hat{\bar{Y}}_{A1,h}
  & = &
  \frac{ \sum_{i=1}^N I(H_i=h) \frac{ R^{A1}_i }{ \pi^{A1}_i } Y_i }
    { \sum_{i=1}^N I(H_i=h) \frac{ R^{A1}_i }{ \pi^{A1}_i } }
  \nonumber \\
  \hat{\bar{Y}}_{A,h}
  & = &
  \frac{ \sum_{i=1}^N I(H_i=h) \frac{ R^A_i }{ \pi^{A1}_i } Y_i }
       { \sum_{i=1}^N I(H_i=h) \frac{ R^A_i }{ \pi^{A1}_i } }
  =
  \frac{ \sum_{i=1}^N I(H_i=h) \frac{ R^A_i }{ \pi^A_i } Y_i }
       { \sum_{i=1}^N I(H_i=h) \frac{ R^A_i }{ \pi^A_i } }.
  \label{eq:Yhatbar.calibration}
  \end{eqnarray}
Note that equation~(\ref{eq:Yhatbar.calibration}) is the same as equation~(\ref{eq:Yhatbar.nonresponse}); it is reproduced here just for convenience.
Note also that $\hat{\mu}_{{\rm reg},Y}$ can be written equivalently as (see Appendix~\ref{sect:proof.calibration} for proof)
\[
\hat{\mu}_{{\rm reg},Y} = \sum_{h=1}^L N_k \hat{\bar{Y}}_{A,h}.
\]

In Appendix~\ref{sect:proof.calibration} we show that $\hat{\mu}_{{\rm reg},Y}$ is a special case of the general regression estimator given by equation~(15.6.19b) of Sarndal et al.
Hence, it follows from their Result 15.6.2 (or Result 9.7.1 and Case 1 on page 364) that $\hat{\mu}_{{\rm reg},Y}$ is approximately unbiased and that its approximate variance is
\begin{eqnarray}
  &&
  \Var ( \hat{\mu}_{{\rm reg},Y} \mid \curlF, \curlY )
  \approx
  \nonumber \\
  && \hspace{0.2cm}  
  \frac{1}{N^2} \sum_{i=1}^N \sum_{j=1}^N (\pi^{A1}_{ij} - \pi^{A1}_i \pi^{A1}_j) \;
  \frac{ Y_i - \bar{Y}_{H_i} }{ \pi^{A1}_i } \;
  \frac{ Y_j - \bar{Y}_{H_j} }{ \pi^{A1}_j }
  \nonumber \\
  && \hspace{0.2cm} +
  \frac{1}{N^2} E \left[ E \left\{ \sum_{h=1}^L n_h^2 \; \frac{ 1 - m_h / n_h }{ m_h} \; S_{E1,h}^2 \; \mid R^{A1}_1, \ldots, R^{A1}_N, \curlF, \curlY \right\} \mid \curlF, \curlY \right]
  \nonumber \\
  &&
  \label{eq:calibration.var}
\end{eqnarray}
where $S_{E1,h}^2$ denotes the variance of $(Y_i - \hat{\bar{Y}}_{A1,H_i}) / \pi^{A1}_i$ in the set of individuals with $R^{A1}_i=1$ in group $h$.
An estimator of the variance~(\ref{eq:calibration.var}) is
\begin{eqnarray}
  &&
  \frac{1}{N^2} \sum_{i=1}^N \sum_{j=1}^N R^A_i R^A_j \;
  \frac{ \pi^{A1}_{ij} - \pi^{A1}_i \pi^{A1}_j }{ \pi^A_{ij} } \;
  \frac{ Y_i - \hat{\bar{Y}}_{A,H_i} }{ \pi^{A1}_i } \;
  \frac{ Y_j - \hat{\bar{Y}}_{A,H_j} }{ \pi^{A1}_j }
  \nonumber \\
  && +
  \frac{1}{N^2} \sum_{h=1}^L n_h^2 \; \frac{ 1 - m_h / n_h }{ m_h} \; S_{E,h}^2
  \label{eq:calibration.varest}
\end{eqnarray}
where $S_{E,h}$ denotes the variance of $(Y_i - \hat{\bar{Y}}_{A,H_i}) / \pi^{A1}_i$ in the set of individuals with $R^A_i=1$ in group $h$, i.e.\
\[
S_{E,h}^2 =
\left\{ \sum_{i=1}^N I(H_i = h) R^A_i \right\}^{-1}
  \sum_{i=1}^N I(H_i = h) R^A_i
    \left( \frac{ Y_i - \hat{\bar{Y}}_{A,H_i} }{ \pi^{A1}_i } \right)^2.
\]

In Appendix~\ref{sect:nonresponse}, where $\pi^A_i$ was defined by equation~(\ref{eq:piAi.nonresponse}), $E(R^A_i / \pi^A_i \mid \curlF, \curlY) = 1$.
However, with $\pi^A_i$ now defined by equation~(\ref{eq:piAi.calibration}), this is no longer the case in general.
This means that the proofs in Appendix~\ref{sect:var.PL1}, \ref{sect:var.PL2}, \ref{sect:modeldesign.PL1} and~\ref{sect:modeldesign.PL2} for $\PLone$, $\PLtwo$, $\HTPL$ and $\HajPL$ are now longer valid.
However, the proofs in Appendix~\ref{sect:var.KH1}, \ref{sect:var.KH2}, \ref{sect:modeldesign.KH1} and~\ref{sect:modeldesign.KH2} for $\KHone$, $\KHtwo$, $\HTKH$ and $\HajKH$ do not require $E(R^A_i / \pi^A_i \mid \curlF, \curlY) = 1$ and it is easy to verify that these proofs do remain valid.

Therefore, the estimators $\KHone$, $\KHtwo$, $\HTKH$ and $\HajKH$ can be used with post-stratification, provided that: i) the RHG model holds; ii) $\pi^A_i$ and $\pi^A_{ij}$ are defined by equations~(\ref{eq:piAi.calibration}) and(\ref{eq:piAij.nonresponse1})--(\ref{eq:piAii.nonresponse}); and iii) expression~(\ref{eq:calibration.varest}) and the corresponding expression for a covariance is used for estimating variances and covariances (with $Y$ replaced by the variable(s) whose variance or covariance is being estimated).
In particular, to obtain an estimate of the covariance between $\hat{\mu}_{{\rm cal},Z}$ and $\hat{\mu}_{{\rm cal},V}$ for two outcomes $Z$ and $V$, simply replace the term $\frac{ Y_i - \hat{\bar{Y}}_{A,H_i} }{ \pi^{A1}_i } \; \frac{ Y_j - \hat{\bar{Y}}_{A,H_j} }{ \pi^{A1}_j }$ in expression~(\ref{eq:calibration.varest}) by $\frac{ Z_i - \hat{\bar{Z}}_{A,H_i} }{ \pi^{A1}_i } \; \frac{ V_j - \hat{\bar{V}}_{A,H_j} }{ \pi^{A1}_j }$ and the variance $S_{E,h}^2$ by the corresponding covariance between $(Z_i - \hat{\bar{Z}}_{A,H_i}) / \pi^{A1}_i$ and $(V_i - \hat{\bar{V}}_{A,H_i}) / \pi^{A1}_i$ in the set of individual with $R^A_i=1$ in group $h$.

As noted earlier, post-stratification can be used even when there is complete response.
In this case, $m_h = n_h$, $R^A_i = R^{A1}_i$, $\pi^A_i = \pi^{A1}_i$, $\pi^A_{ij} = \pi^{A1}_{ij}$ and equations~(\ref{eq:calibration.var}) and~(\ref{eq:calibration.varest}) simplify to, respectively,
\begin{equation*}
  \Var ( \hat{\mu}_{{\rm reg},Y} \mid \curlF, \curlY )
  \approx
  \frac{1}{N^2} \sum_{i=1}^N \sum_{j=1}^N (\pi^A_{ij} - \pi^A_i \pi^A_j) \;
  \frac{ Y_i - \bar{Y}_{H_i} }{ \pi^A_i } \;
  \frac{ Y_j - \bar{Y}_{H_j} }{ \pi^A_j }
\nonumber
\end{equation*}
and
\begin{equation*}
  \frac{1}{N^2} \sum_{i=1}^N \sum_{j=1}^N R^A_i R^A_j \;
  \frac{ \pi^A_{ij} - \pi^A_i \pi^A_j }{ \pi^A_{ij} } \;
  \frac{ Y_i - \hat{\bar{Y}}_{A,H_i} }{ \pi^A_i } \;
  \frac{ Y_j - \hat{\bar{Y}}_{A,H_j} }{ \pi^A_j }.
\end{equation*}

\subsection{Proof that $E(R^A_i / \pi^A_i \mid \curlF, \curlY) = 1$ when $\pi^A_i = \pi^{A1}_i m_{H_i} / n_{H_i}$}

\label{sect:proof.expect.rA.piA}

Let $h$ be the group to which individual $i$ belongs.
Then
\begin{eqnarray*}
  &&
  E \left( \frac{ R^A_i }{ \pi^A_i } \mid \curlF, \curlY \right)
  \nonumber \\
  && \hspace{.4cm} =
P(R^{A1}_i = 1 \mid \curlF, \curlY) \times E \left( \frac{ R^A_i }{ \pi^A_i } \mid \curlF, \curlY, R^{A1}_i=1 \right)
  \nonumber \\
  && \hspace{.4cm} =
  \pi^{A1}_i \times E \left( \frac{ R^A_i }{ \pi^A_i } \mid \curlF, \curlY, R^{A1}_i=1 \right)
  \nonumber \\
  && \hspace{.4cm} =
  \pi^{A1}_i \times E \left\{ E \left( \frac{ R^A_i }{ \pi^A_i } \mid \curlF, \curlY, R^{A1}_i=1, m_h \right) \mid \curlF, \curlY, R^{A1}_i=1 \right\}
  \nonumber \\
  && \hspace{.4cm} =
  \pi^{A1}_i \times E \left\{ E \left( \frac{ R^A_i }{ \pi^{A1}_i } \frac{ n_h }{ m_h} \mid \curlF, \curlY, R^{A1}_i=1, m_h \right) \mid \curlF, \curlY, R^{A1}_i=1 \right\}
  \nonumber \\
  && \hspace{.4cm} =
  E \left\{ \frac{ n_h }{ m_h} E ( R^A_i \mid \curlF, \curlY, R^{A1}_i=1, m_h ) \mid \curlF, \curlY, R^{A1}_i=1 \right\}
  \nonumber \\
  && \hspace{.4cm} =
  E \left( \frac{ n_h }{ m_h} \frac{ m_h }{ h_h} \mid \curlF, \curlY, R^{A1}_i=1 \right)
  \nonumber \\
  && \hspace{.4cm} =
  1.  \nonumber
\end{eqnarray*}

\subsection{Proof that $\hat{\mu}_{{\rm reg},Y}$ is special case of general regression estimator}

\label{sect:proof.calibration}

In this subsection, we borrow the notation of Sarndal et al.\ (1992) pages 583--584.

Let $x_{1k} = \left( I(H_k=1), \ldots, I(H_k=L) \right)$ be a vector of $L$ indicators that individual $k$ belongs to each of groups $1, \ldots, L$, and let $x_k = x_{1k}$ and $\sigma^2_k=1$.
From Sarndal et al.'s equations~(15.6.17) and~(15.6.18), we have
\[
\hat{y}_k = \hat{y}_{1k}
= \frac{ \sum_{i=1}^L I(H_i = H_k) \frac{ R^A_i }{ \pi^A_i } Y_i }
    { \sum_{i=1}^L I(H_i = H_k) \frac{ R^A_i }{ \pi^A_i } }
    = \hat{\bar{Y}}_{A,H_k}
\]
and from equations~(15.6.20)--(15.6.25), we have
\begin{eqnarray}
  \breve{E}_k & = & \frac{ y_k - \hat{\bar{Y}}_{A1,H_k} }{ \pi^{A1}_k }
  \label{eq:Sarndalnotation.start} \\
  B_s & = & (\hat{\bar{Y}}_{A1,1}, \ldots, \hat{\bar{Y}}_{A1,L})
  \\
  \breve{E}_{1k} & = & \frac{ y_k - \Ybar_{H_k} }{ \pi^{A1}_k }
  \\  
  B_1 & = & (\Ybar_1, \ldots, \Ybar_L)
  \\
  \breve{e}_{kr} & = & \frac{ y_k - \hat{\bar{Y}}_{A,H_k} }{ \pi^{A1}_k }
  \\
  \breve{e}_{1kr} & = & \frac{ y_k - \hat{\bar{Y}}_{A,H_k} }{ \pi^{A1}_k }
  \label{eq:Sarndalnotation.end} 
\end{eqnarray}
As Sarndal et al.\ point out, since $\hat{y}_k = \hat{y}_{1k}$, their equation~(15.6.16) for their estimator $\hat{t}_{\rm cr}$ of the population total reduces to their equation (15.6.19b), which is
\begin{eqnarray}
  \hat{t}_{\rm cr}
  & = &
  \sum_{k=1}^N \hat{y}_{1k}
  + \sum_{h=1}^L
  \frac{ n_h }{ m_h }
  \sum_{k=1}^N I(H_k=h) R^A_k \; \frac{ y_k - \hat{y}_{1k} }{ \pi^{A1}_k }
  \nonumber \\
  & = &
  \sum_{k=1}^N \hat{\bar{Y}}_{A,H_k}
  + \sum_{h=1}^L
  \frac{ n_h }{ m_h }
  \sum_{k=1}^N I(H_k=h) R^A_k \; \frac{ y_k - \hat{\bar{Y}}_{A,h} }{ \pi^{A1}_k }.
  \label{eq:tcr}
\end{eqnarray}
It follows from equations~(\ref{eq:piAi.calibration}), (\ref{eq:Yhatbar.calibration}) and~(\ref{eq:tcr}) that
\begin{eqnarray*}
  \hat{t}_{\rm cr}
  & = &
  \sum_{k=1}^N \hat{\bar{Y}}_{A,H_k}
  + \sum_{h=1}^L \frac{ n_h }{ m_h } \left\{
  \sum_{k=1}^N I(H_k=h) \frac{ R^A_k }{ \pi^{A1}_k } \; y_k 
  - \hat{\bar{Y}}_{A,h}
  \sum_{k=1}^N I(H_k=h) \frac{ R^A_k }{ \pi^{A1}_k }
 \right\}
  \\
  & = &
  \sum_{k=1}^N \hat{\bar{Y}}_{A,H_k}
  + \sum_{h=1}^L \frac{ n_h }{ m_h } \left\{
  \hat{\bar{Y}}_{A,h}
  \sum_{k=1}^N I(H_k=h) \frac{ R^A_k }{ \pi^{A1}_k } 
  - \hat{\bar{Y}}_{A,h}
  \sum_{k=1}^N I(H_k=h) \frac{ R^A_k }{ \pi^{A1}_k }
 \right\}
  \\
  & = &
  \sum_{k=1}^N \hat{\bar{Y}}_{A,H_k}
  \\
  & = &
  \sum_{h=1}^L N_k \hat{\bar{Y}}_{A,h}
  \\
  & = &
  \sum_{h=1}^L N_h
  \frac{ \sum_{i=1}^N I(H_i=h) \frac{ R^A_i }{ \pi^{A1}_i } Y_i }
       { \sum_{i=1}^N I(H_i=h) \frac{ R^A_i }{ \pi^{A1}_i } }
  \\
  & = &
  \sum_{h=1}^L N_h \sum_{i=1}^N I(H_i=h) \frac{ R^A_i }{ \pi^A_i } Y_i \frac{ 1 }{ N_h }
  \\
  & = &
  \sum_{i=1}^N \frac{ R^A_i }{ \pi^A_i } Y_i
  \\
  & = &
  N \times \hat{\mu}_{{\rm reg},Y}.
\end{eqnarray*}
This shows that Sarndal et al.'s estimator $\hat{t}_{\rm cr} / N$ of the population mean is the same as our estimator $\hat{\mu}_{{\rm reg},Y}$ of the population mean.

The approximate variance of $\hat{t}_{\rm cr}$, or equivalently $N \times \hat{\mu}_{{\rm reg},Y}$, is given by Sarndal et al.'s equation (15.6.26).
Using equations~(\ref{eq:Sarndalnotation.start})--(\ref{eq:Sarndalnotation.end}), it can be verified that equation (15.6.26) is the same as our equation~(\ref{eq:calibration.var}) multiplied by $N^2$.
Likewise, the estimator of the variance of $\hat{t}_{\rm cr}$ is given by Sarndal et al.'s expression (15.6.27).
Again, it can be verified that this expression is the same as our expression~(\ref{eq:calibration.varest}) multiplied by $N^2$.

\end{document}